\documentclass{article}

\usepackage[utf8]{inputenc}
\usepackage{geometry}
\usepackage{setspace}
\usepackage{titling}
\usepackage{booktabs}
\usepackage{bm}
\usepackage{amsfonts,amssymb,amsbsy,amsmath,amsthm}
\allowdisplaybreaks
\usepackage{commath}
\usepackage{physics}
\usepackage{multirow}
\usepackage{latexsym,dsfont}
\usepackage{gensymb}
\usepackage{relsize}
\usepackage{bbm}
\usepackage{graphicx}
\usepackage{subcaption}
\usepackage{array}
\usepackage{adjustbox}
\usepackage{float}
\usepackage{caption}
\usepackage{subcaption}
\usepackage{wrapfig}
\usepackage{lipsum}
\usepackage[bottom]{footmisc}
\usepackage{multicol}
\usepackage{rotating}
\usepackage{textcomp}
\usepackage[dvipsnames,table,xcdraw]{xcolor}
\usepackage{tcolorbox}
\usepackage{mathtools}
\mathtoolsset{showonlyrefs}
\usepackage{natbib}
\usepackage{hyperref}
\usepackage{authblk}
\usepackage{fancyvrb}

\title{Geometric modelling of spatial extremes}
\author[1,*]{L. Kakampakou}
\author[1]{J. L. Wadsworth}

\affil[1]{\small School of Mathematical Sciences, Lancaster University LA1 4YF, United Kingdom}
\affil[*]{Correspondence to:  \href{mailto:l.kakampakou1@lancaster.ac.uk}{l.kakampakou1@lancaster.ac.uk}}

\date{\today}

\begin{document}

\maketitle

\begin{abstract}
Recent developments in extreme value statistics have established the so-called geometric approach as a powerful modelling tool for multivariate extremes. We tailor these methods to the case of spatial modelling and examine their efficacy at inferring extremal dependence and performing extrapolation. The geometric approach is based around a limit set described by a gauge function, which is a key target for inference. We consider a variety of spatially-parameterised gauge functions and perform inference on them by building on the framework of \citet{Wadsworth2024}, where extreme radii are modelled via a truncated gamma distribution. We also consider spatial modelling of the angular distribution, for which we propose two candidate models. Estimation of extreme event probabilities is possible by combining draws from the radial and angular models respectively. We compare our method with two other established frameworks for spatial extreme value analysis and show that our approach generally allows for unbiased, albeit more uncertain, inference compared to the more classical models. We illustrate the methodology on a space weather dataset of daily geomagnetic field fluctuations.
\end{abstract}

\noindent%
\vspace{1cm}

\section{Introduction}
\label{sec:intro}

The study of geometric extremes is based on the idea that the shape of multivariate sample clouds can be informative about the extremal dependence structure of multivariate vectors. Its roots stem mainly from the late 1980s, when the first general studies on the behaviour of scaled multivariate sample clouds and their convergence to \textit{limit sets} were published \citep{Davis1988,Kinoshita1991}. These investigated appropriate sequences by which sample clouds with light-tailed margins must be scaled in order to converge to a limiting set, conditions for such convergence to occur, and properties of these limit sets. However, due to their probabilistic nature and, in part, because alternative methods based on \textit{multivariate regular variation} \citep{deHaan1970,deHaan1977,resnick1987} were already gaining popularity for inference on multivariate extremes, their utility in a statistical context was not immediately understood. More recently, as the limitations of methods based on multivariate regular variation have come to light, the same ideas were revisited and clearer connections have been established between the shape of the limit set and the underlying extremal dependence structure of the random vector in question. 
For instance, 
\citet{Balkema2010a} focus on so-called meta densities
and provide necessary and sufficient conditions for the convergence of the corresponding scaled sample cloud to its limit set. 
\citet{Balkema2010b} and \citet{Balkema2012} investigate the dependence properties of limit sets and establish criteria through which one can check for the presence of asymptotic independence and dependence, respectively.
\citet{Nolde2014} also explores extremal dependence properties of limit sets and establishes a link between the limit set and the so-called residual tail dependence coefficient of \citet{Ledford1996}. 
Finally, \citet{Nolde2020} extend the work of \citet{Nolde2014}, drawing links between the boundary of the limit set and several extremal dependence coefficients, thus connecting representations of multivariate extremes that were previously considered disjoint.  

The insights of this recent body of literature has led to the increased popularity of the geometric framework for multivariate extremes. Inference methods for multivariate geometric extremes are relatively recent, but represent a very active area of research. All methods use a radial-angular decomposition of a light-tailed random vector $\boldsymbol{X}$, whose margins are fixed and known. 
\citet{Simpson2024}, focus on the study of bivariate extremes and develop a semi-parametric estimator for the boundary of the limit set, by assuming that large radii can be appropriately described by a generalised Pareto distribution (GPD).  
\citet{Wadsworth2024}, develop a new asymptotically motivated framework for the modelling of the tails of general $d$-dimensional multivariate vectors. It is based on the assumption that large radial exceedances of a high threshold, under suitable conditions, can be described by a truncated gamma distribution.  
Inference is based on parameterised gauge functions which in turn allow one to obtain parametric estimates of the limit set shape. In addition to this, the proposed modelling framework provides a strategy for extrapolation in the tail.   
Similarly to \citet{Simpson2024}, \citet{Majumder2025} also focus on the bivariate case and develop a semi-parametric estimation technique for the limit set boundary, using B\'{e}zier splines for modelling and operating within a Bayesian framework for inference. 
\citet{Papastathopoulos2023} also embrace a Bayesian approach, using INLA methodology \citep{Rue2009} for dimension up to three. Unlike the aforementioned works, which focus on exponential margins, they use Laplace margins, which has advantages in the presence of negative dependence, and use a GPD to model large radial exceedances of a high threshold. 
\citet{MurphyBarltrop2024} and \citet{demonte2025} employ deep learning techniques to estimate the shape of the limit set, in up to $8$ and $10$ dimensions, respectively. Similarly to \citet{Papastathopoulos2023}, these authors focus on Laplace margins.

All these efforts have naturally concentrated on multivariate extremes where this approach shines, for its ability to capture and infer complex dependence features. Seeing its many advantages in the multivariate setting, it is natural to contemplate the potential utility of the framework for modelling spatial extremes. Specifically, we are interested in adopting ideas from the geometric framework for multivariate extremes to create a novel, self-sufficient framework for spatial extreme value modelling. The contribution of this paper is to introduce statistical methodology for spatial extreme value analysis, based on the geometric approach. We aim to explore and catalogue advantages and disadvantages of the methodology by comparing its performance to more standard techniques for spatial modelling. An abundance of approaches for spatial extreme value analysis has been developed over the years, but no universally superior method exists. 
For example, some models are suited only to one class of extremal dependence, while some are difficult to fit in high dimensions and many suffer both of these drawbacks.
For a detailed account of the most prevalent models for spatial extremes and their key advantages and shortcomings, see \citet{Huser2022}.

This paper is organised as follows. Section \ref{sec:background} provides background information on geometric extremes. Section \ref{sec:spatgeommod} introduces the spatial geometric modelling framework, outlining models for radii and angles derived from a finite-dimensional measurement of a spatial process. In Section \ref{sec:prediction} we detail the sampling of new extreme observations from our model and using those for predictive tasks, such as extreme probability set estimation. Section \ref{sec:simstudy} comprises simulation study results, while in Section \ref{sec:application} we apply our method to a dataset of geomagnetic field fluctuations. We conclude in Section \ref{sec:discussion}. 

\section{Multivariate geometric extremes}
\label{sec:background}

Let $\bm{X}$ be a random vector in $\mathbb{R}^d$ of which $n$ independent copies $\bm{X}_i, i=1,\ldots,n$, exist. A standard assumption within the geometric framework is that each $\bm{X}_i$ has identically distributed light-tailed margins; in practice, this is typically achieved via transformation. 
If we apply a suitable common scaling, $r_n$, to each of the margins, then the shape of the resulting $n$-point scaled sample cloud, $N_n = \{\bm{X}_1/r_{n}, \ldots, \bm{X}_n/r_{n}\}$, can be informative about the extremal dependence of $\bm{X}$ \citep{Balkema2010a,Nolde2014,Nolde2020}. The sequence $r_n$ depends on the choice of margins and a suitable choice is $r_n \sim F_j^{-1}(1-1/n)$, $n\to\infty$, with $F_j$ the marginal cumulative distribution function (cdf), $j=1,\ldots,d$. 
Under certain conditions (see \citet{Balkema2010a} for details), 
$N_n$ will converge onto a limit set $G$ characterised by its gauge function $g(\cdot)$ through the relationship $G=\{\bm{x}\in \mathbb{R}^d: g(\bm{x})\leq 1\}$. The gauge function $g(\cdot)$ is $1$-homogeneous, that is $g(t\bm{x})=tg(\bm{x})$ for all $t>0$, and the limit set $G$ is star-shaped \citep{Kinoshita1991}, which means that if $\bm{x}\in G$ then $t\bm{x}\in G$ for any $t\in[0,1]$. 

In the case where $\bm{X}$ has a Lebesgue density, $f_{\bm{X}}$, \cite{Balkema2010b} and \citet{Nolde2020} provide conditions for convergence of $N_n$ to $G$, based on this density. Specifically, if $\bm{X}$ has exponential margins and 
\begin{equation}
    g(\bm{x}) = \lim_{t\to\infty}-\log f_{\bm{X}}(t\bm{x})/t,
    \label{eq:conv_cond_NW}
\end{equation}
for a continuous gauge function $g$, then $N_n$ converges onto $G=\{\bm{x} \in \mathbb{R}_+^d: g(\bm{x})\leq1\}$.
The choice of standard exponential marginals is common in geometric extreme value literature (e.g.\ \citet{Wadsworth2024}, \citet{Simpson2024}, \citet{Campbell2024}), because it leads to simple modelling assumptions and is well suited to data exhibiting only positive association. Since this will typically be the case for spatial data, we also assume standard exponential margins. The coordinatewise supremum of $G$ is the vector $(1,\ldots,1)^{\text{T}}$, while the coordinatewise infimum is $(0,\ldots,0)^{\text{T}}$. 

Assumption \eqref{eq:conv_cond_NW} can equivalently be expressed as 
\begin{equation}
    f_{\bm{X}}(t\bm{x}) = \exp\{-tg(\bm{x})[1+o(1)]\}, \quad t\to\infty,
\end{equation}
for $g(\bm{x})>0$.
It is convenient to decompose $\bm{X}$ into radial and angular components; $R=\|\bm{X}\| \in \mathbb{R}_+$, where $\|\cdot\|$ is a norm --- typically taken to be the $\ell_1$ norm, i.e.\ $R=\Sigma_{i=1}^d X_i$, for simplicity --- and $\bm{W}=\bm{X}/R \in \mathcal{S}_+^{d-1}$, where $\mathcal{S}_+^{d-1}=\{\bm{x}\in\mathbb{R}_+^d: \|\bm{x}\|=1\}$ is the $d$-dimensional unit simplex, respectively.
Given the radial-angular decomposition we have 
\begin{equation}
    f_{R,\bm{W}}(r,\bm{w}) = r^{d-1}f_{\bm{X}}(r\bm{w}) = r^{d-1}\exp\{-rg(\bm{w})[1+o(1)]\},
\end{equation}
as $r\to\infty$. So, $f_{R\vert \bm{W}}(r\vert \bm{w}) \propto r^{d-1}\exp\{-rg(\bm{x})[1+o(1)]\}$. \citet{Wadsworth2024} show that in many cases a stronger assumption holds, namely $f_{R\vert \bm{W}}(r\vert \bm{w}) \propto r^{d-1}\exp\{-rg(\bm{w})\}[1+o(1)]$. Hence, they propose the following model for the conditional distribution of high radial threshold exceedances given an angle $\bm{W}=\bm{w}$,
\begin{equation}
\label{eq:WCmodel}
    R \mid \{\bm{W}=\bm{w}, R>r_{\tau}(\bm{w})\} \mathrel{\dot\sim} \text{truncGamma}(a, g(\bm{w})),
\end{equation}
where $r_{\tau}(\bm{w})$ is a high radial threshold. That is, large radial exceedances, given some angle $\bm{W}=\bm{w}$, are modelled via a left-truncated gamma distribution with shape parameter $a>0$, sometimes taken to be equal to the dimension, $d$, of $\bm{X}$, and rate parameter $g(\bm{w})>0$, the gauge function evaluated at angle $\bm{w}$.

\section{Spatial geometric modelling framework}
\label{sec:spatgeommod}

Consider a spatial process $\{X(\bm{s}): \bm{s}\in \mathcal{S}\subset\mathbb{R}^2 \}$ and let $\bm{X} = (X(\bm{s}_1),\ldots,X(\bm{s}_d))$ be the finite-dimensional version of  $X(\bm{s})$ comprised of the realisations of the process in the observed locations, $\bm{s}_1,\ldots,\bm{s}_d$.
In spatial extremes, interest lies in modelling the joint tail of $\bm{X}$. Inherently embedded in this problem is the task of characterising the dependence structure of the data at extreme levels. That is, we are interested in knowing whether extreme events can occur simultaneously in all $d$ locations --- a phenomenon referred to as asymptotic dependence (AD). The converse behaviour is termed asymptotic independence (AI). 
Environmental data often belong to the latter category, as extreme events become more and more localised with increasing level of extremity \citep{Huser2022}. What this means from a statistical standpoint is that at those most extreme levels, extremes do not tend to occur simultaneously over a large area. Examples of environmental data that display AD are much less common and generally correspond to geographically small spatial domains.

Our starting point is to consider the spatial structure of a limit set derived from $\bm{X}$, the finite-dimensional realisation of $X(\bm{s})$. Since many parametric multivariate distributions lead to known parametric forms for the gauge function $g$, we can expect this to be the case for parametric spatial processes as well.
In order to develop useful models for spatial data, we need to consider gauge functions that are at least able to accommodate AI or, more generally, that can interpolate between both extremal dependence regimes while remaining flexible and computationally efficient in relatively high dimensions. 

\subsection{Spatially parameterised gauge functions}
\label{sec:spat gauges}

\begin{figure}[h]
    \centering
\begin{subfigure}{0.24\textwidth}
\centering
\includegraphics[width=1\linewidth]{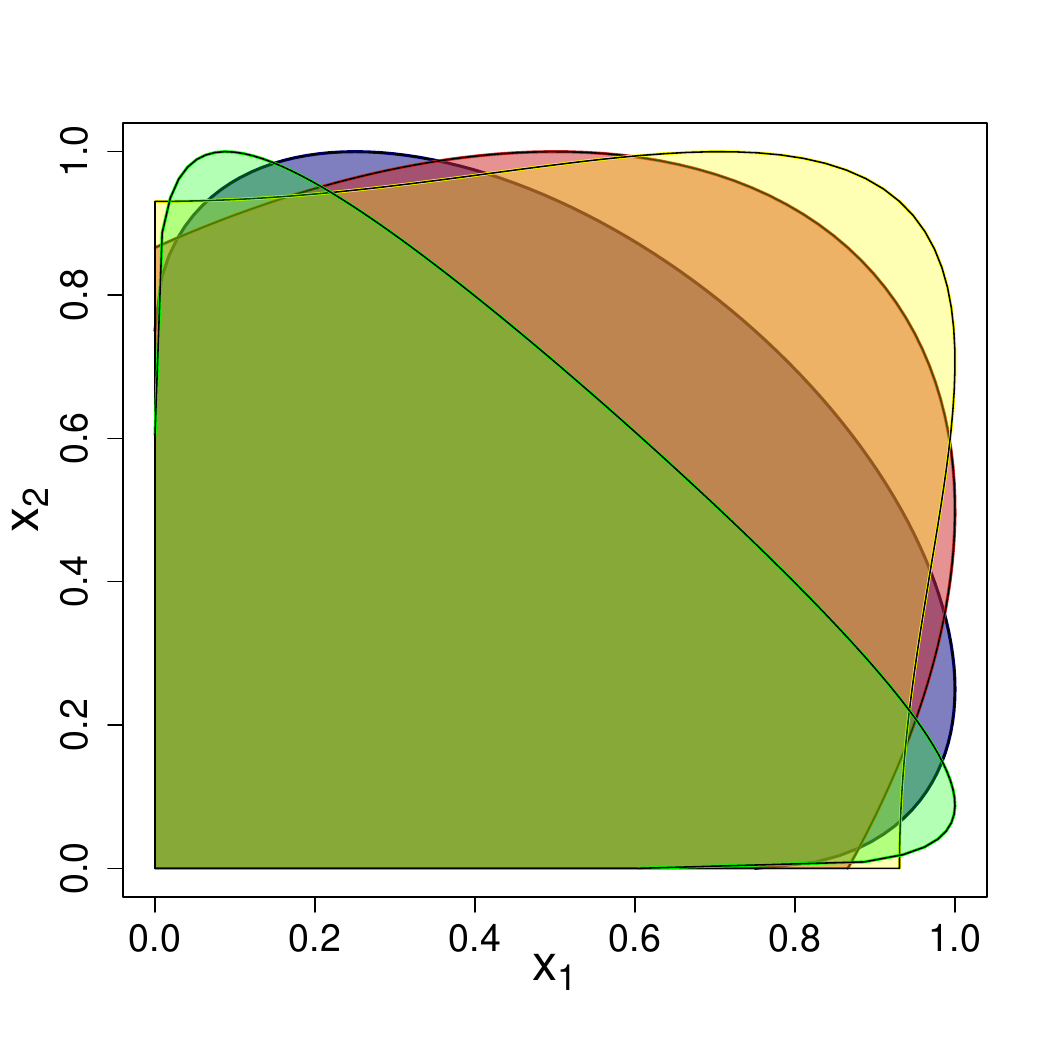}       
\caption{} 
\label{fig:example_lm_glg}
\end{subfigure}%
\hfill
\begin{subfigure}{0.24\textwidth}
\centering
\includegraphics[width=1\linewidth]{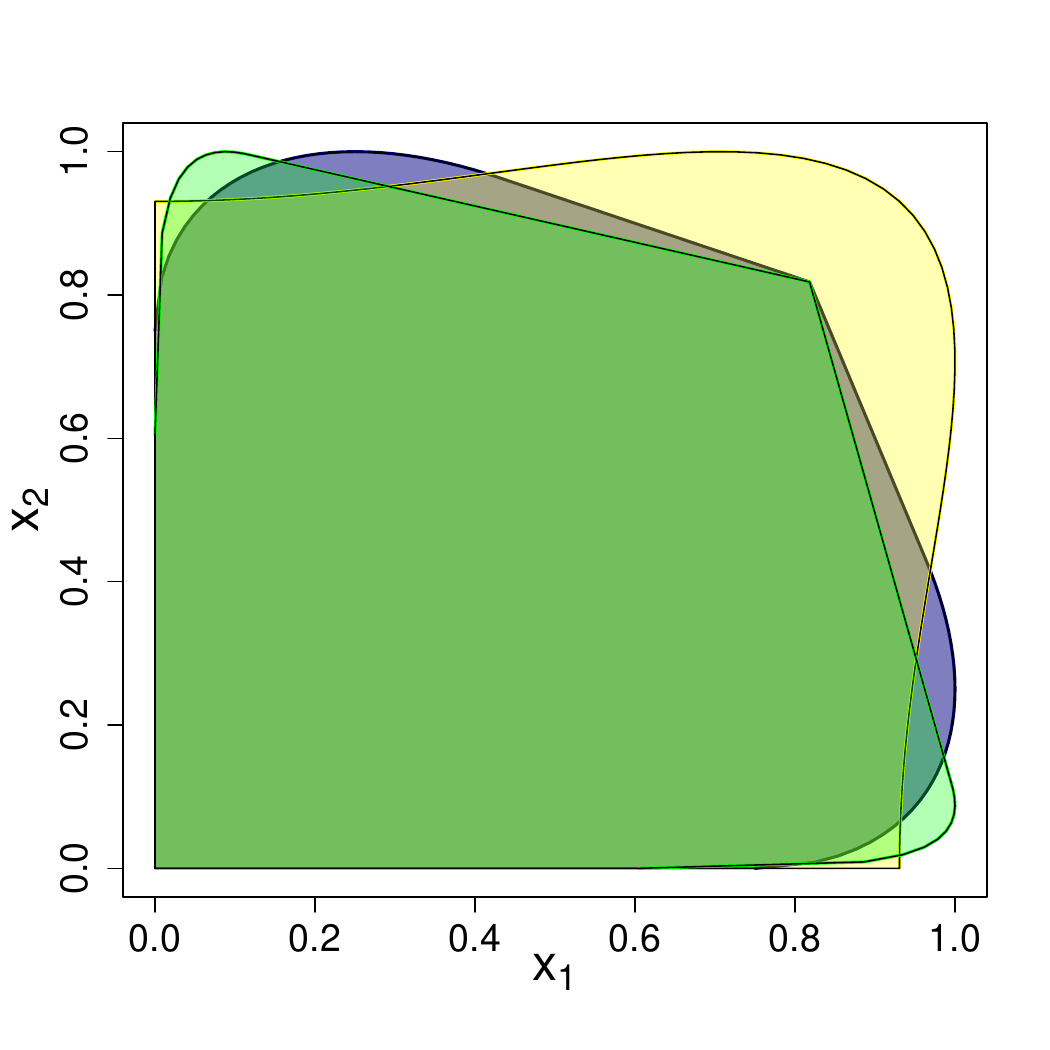}       
\caption{} 
\label{fig:example_lm_hwai}
\end{subfigure}
\hfill
\begin{subfigure}{0.24\textwidth}
\centering
\includegraphics[width=1\linewidth]{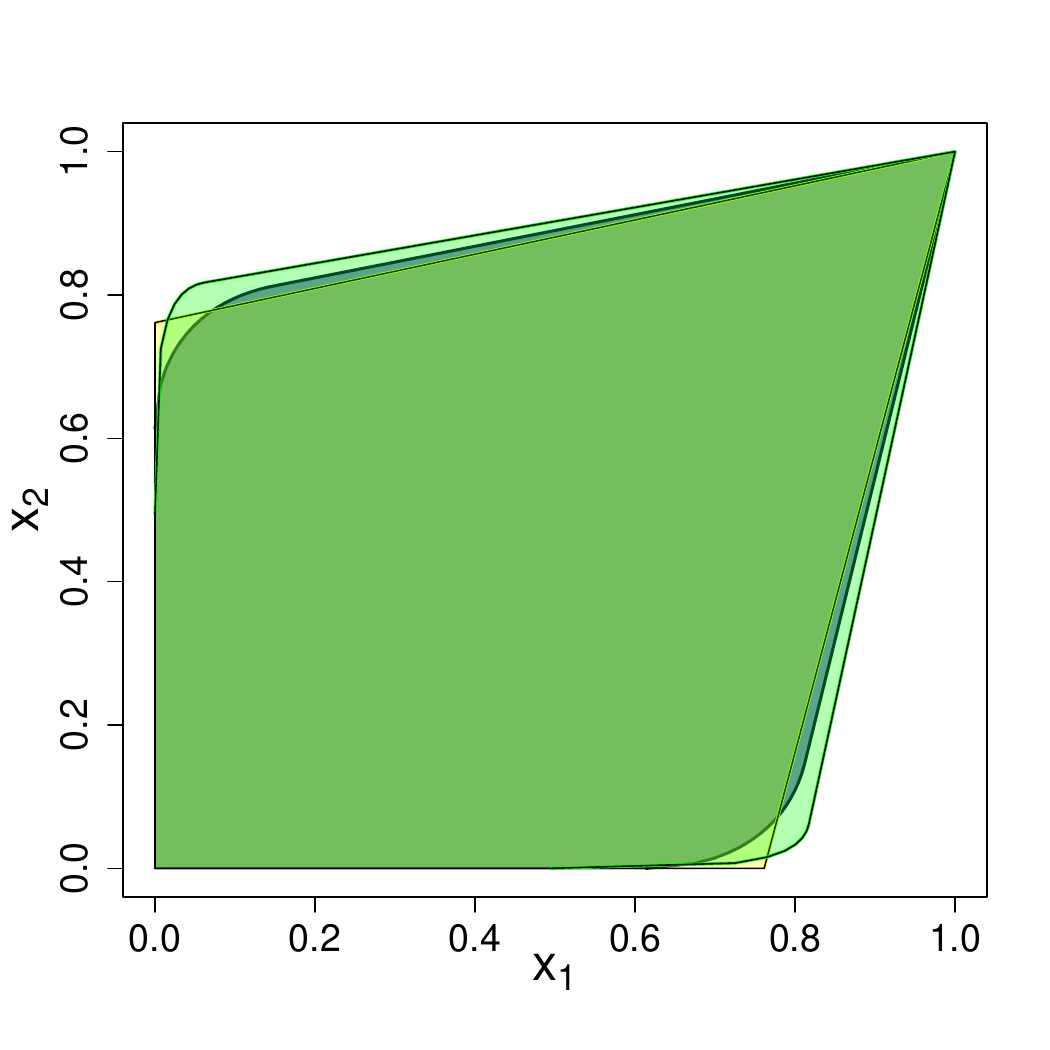}       
\caption{} 
\label{fig:example_lm_hwad}
\end{subfigure}       
\hfill
\begin{subfigure}{0.24\textwidth}
\centering
\includegraphics[width=1\linewidth]{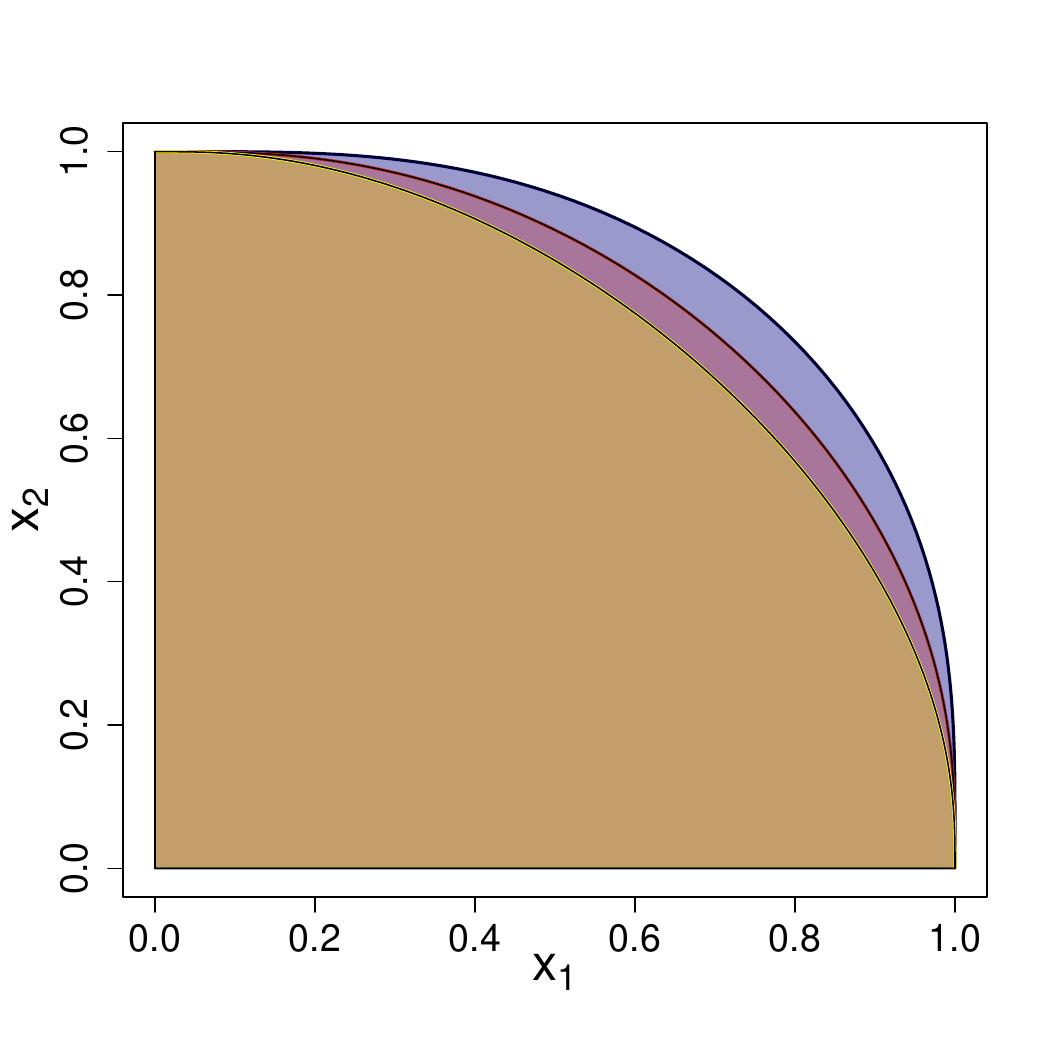}      
\caption{} 
\label{fig:example_lm_ibr}
\end{subfigure} 
\caption{Examples of bivariate limit sets linked to the gauge functions introduced in Section \ref{sec:spat gauges}.  Limit sets of (a) generalised Gaussian gauges with correlation parameter $\rho=0.5$, and $\nu=0.5,1,2,3.5$ in yellow, red, blue and green, respectively; (b) HW gauges with an underlying generalised Gaussian gauge, correlation parameter $\rho=0.5$, $\nu=0.5,2,3.5$ in yellow, blue and green, respectively, and mixing parameter $\delta=0.45$; (c) HW gauges with an underlying generalised Gaussian gauge, correlation parameter $\rho=0.5$, $\nu=0.5,2,3.5$ in yellow, blue and green, respectively, and $\delta=0.55$; (d) IBR gauge with semivariogram parameter $\gamma = 0.3,0.5,0.7$ in blue, red and yellow, respectively.}
\label{fig:unitlevelsets}
\end{figure}

Below we detail spatially parameterised gauge functions that arise as those describing the limit sets of $\bm{X}=(X(\bm{s}_1), \ldots, X(\bm{s}_d))$ for different processes $X(\bm{s})$ with exponential margins. Here we assume $X(\bm{s})$ to be stationary and isotropic, but note that these assumptions could be relaxed. Examples of bivariate unit level sets of these gauge functions can be found in Figure \ref{fig:unitlevelsets}.

\subsubsection{Spatial Gaussian gauge}
\label{sec:G_gauge}
Let $\{\bm{X}(\bm{s}): \bm{s} \in S \subset \mathbb{R}^2 \}$ be a Gaussian process, so $\bm{X} = (X(\bm{s}_1), \ldots, X(\bm{s}_d)) \sim \text{MVN}(\bm{0},\Sigma)$, where $\Sigma$ is a spatial correlation matrix. \citet{Nolde2020} show that when the entries of $\Sigma$ are non-negative, the corresponding gauge function takes the form 
\begin{equation}
\label{eq:g_gauge}
    g_\text{G}(\bm{x}) = (\bm{x}^{1/2})^T \Sigma^{-1} \bm{x}^{1/2},
\end{equation} 
where $\bm{x}^{1/2}$ is the componentwise square root of $\bm{x}$. In the spatial case, we let the entries of $\Sigma$ be determined by a suitable positive-definite correlation function $\rho(h)$. In this paper we will usually take $\rho$ to be a powered exponential correlation function, i.e.\ $\Sigma = \exp\{-(H/\lambda)^\kappa\}$, where $H$ is the distance matrix between locations $\{\bm{s}_1, \ldots, \bm{s}_d\}$, $\lambda>0$, $\kappa\in(0,2]$. We opt for this correlation function for its mathematical simplicity and practical flexibility, however other parametric forms, such as the Mat\'ern function, could be used instead, leading to similar inferential conclusions. We note that the Gaussian gauge is only suitable for the case of AI. However, it is very easy and fast to compute, even in large dimensions and has a flexible enough shape that renders it a useful model for many dependence structures.

\subsubsection{Spatial Laplace gauge}
\label{sec:L_gauge}
Let $\{X(\bm{s}): \bm{s} \in S \subset \mathbb{R}^2 \}$ be a Laplace random field \citep{OPITZ2016}, so $\bm{X} \sim \text{Laplace}(\Sigma)$, where $\Sigma$ is again a spatial correlation matrix. The corresponding gauge is 
\begin{equation}
\label{eq:l_gauge}
    g_{\text{L}}(\bm{x}) = (\bm{x}^T \Sigma^{-1} \bm{x})^{1/2}.
\end{equation}
A detailed derivation of \eqref{eq:l_gauge} can be found in \citet{Papastathopoulos2023} and is also provided in Section \ref{app:laplace} of the Supplementary Material for completeness. Once more, we typically take $\Sigma = \exp\{-(H/\lambda)^\kappa\}$, as outlined in Section \ref{sec:G_gauge}.

\subsubsection{Spatial generalised Gaussian gauge}
\label{sec:GLG_gauge}
Unlike the Gaussian and Laplace cases, the gauge function introduced here is not derived from an existing spatial process. 
The form given in equation \eqref{eq:glg_gauge} below is a generalisation of the Gaussian and Laplace spatial gauges. Its additional parameter, $\nu$, introduces increased modelling flexibility in a parsimonious manner.
\begin{equation}
\label{eq:glg_gauge}
    g_{\text{GG}}(\bm{x}) = \left((\bm{x}^{1/\nu})^T \Sigma^{-1} \bm{x}^{1/\nu}\right)^{\nu/2}.
\end{equation} 
Notice that the Gaussian gauge is recovered by setting $\nu=2$, while the Laplace by setting $\nu = 1$. As before, we take the correlation matrix $\Sigma = \exp\{-(H/\lambda)^\kappa\}$.

\subsubsection{Spatial Huser--Wadsworth gauge}
\label{sec:HW_gauge}
The \citet{Huser2019} (HW) model belongs to the broad class of random scale mixture models. It consists of two components; a heavy-tailed standard-Pareto-distributed random variable $R_{\text{P}}$ and a spatial process $\{W_{\text{P}}(\bm{s}): \bm{s}\in \mathcal{S}\in\mathbb{R}^2\}$ satisfying the dependence assumption of hidden regular variation \citep{Ledford1996,resnick2002}. The process $W_{\text{P}}(\bm{s})$ also has standard Pareto margins and is independent of the variable $R_{\text{P}}$. The two are combined through the parameter $\delta\in[0,1]$ as follows:
\begin{equation}
    Z(\bm{s}) = R_{\text{P}}^{\delta}W_{\text{P}}(\bm{s})^{1-\delta}.
    \label{eq:hw_orig}
\end{equation}
Transforming equation \eqref{eq:hw_orig} such that its components have unit exponential marginals and reparameterising slightly, by introducing the parameter $\zeta=\delta/(1-\delta)\in\mathbb{R}_+$ we let 
\begin{equation}
   \widetilde{Z}(\bm{s})= 
   \begin{cases}
    \zeta R_{\text{E}} +W_{\text{E}}(\boldsymbol{s}), \quad &\zeta \in [0,1] \\
    R_{\text{E}} + \frac{W_{\text{E}}(\boldsymbol{s})}{\zeta}, \quad &\zeta >1.
   \end{cases}
\end{equation}
\citet{Nolde2020} and \citet{Lee2025} show that the gauge function of $\widetilde{\bm{Z}}$, the finite-dimensional version of $\widetilde{Z}(\bm{s})$, is 
\begin{equation}
    g_{\text{HW}}(\bm{x})=
    \begin{cases}
    \underset{s\in\left[0,\frac{\min(\bm{x})}{\zeta}\right]}\min s+g_{\bm{W}_{\text{E}}}(\bm{x}-\zeta s), \quad & \zeta\in[0,1] \\
    \underset{s\in\left[0,\min(\bm{x})\right]}\min s+g_{\bm{W}_{\text{E}}}(\zeta\bm{x}-\zeta s), \quad & \zeta>1.
    
    \end{cases}
    \label{eq:hw_gauge}
\end{equation}
While $\widetilde{Z}(\bm{s})$ does not have exactly exponential margins, the gauge function $g_{\text{HW}}(\bm{x})$ is still suited to standard exponential margins. 
Note that useful models can arise by assuming any viable gauge for $g_{\bm{W}_{\text{E}}}(\cdot)$. In this work we use the Gaussian and generalised Gaussian gauges for their flexibility and computational efficiency; the minimisation in \eqref{eq:hw_gauge} is performed numerically. Where the former is used we will denote the resulting gauge by $g_{\text{HW}_\text{G}}$, whereas we will use $g_{\text{HW}_\text{GG}}$ when the latter is employed. Finally, we note that the HW model is capable of capturing both extremal dependence scenarios. When $\delta\leq0.5$ ($\zeta\in[0,1]$) we have AI, while $\delta>0.5$ ($\zeta>1$) signifies AD. 

\subsubsection{Spatial inverted Brown--Resnick gauge}
\label{sec:IBR_gauge}
Let $\{Z^{\text{MS}}(\boldsymbol{s}): \boldsymbol{s}\in \mathcal{S}\}$ be a max-stable process (MS) with standard Fr\'{e}chet margins, $\Pr(Z^{\text{MS}}(\boldsymbol{s})\leq z)=\exp(-1/z)$, $z>0$, and corresponding exponent function $V_{\text{MS}}(\cdot)$. \citet{Wadsworth2012} define the class of inverted max-stable processes (IMS), which is obtained by ``flipping'' the joint tails of the process $Z^{\text{MS}}(\boldsymbol{s})$; that is $Z^{\text{IMS}}(\bm{s}) = 1/Z^{\text{MS}}(\bm{s})$. The resulting process has exponential margins, joint survival function $\Pr(Z^{\text{IMS}}(\bm{s})>\bm{z})=\exp\{-V_{\text{MS}}(1/\bm{z})\}$, and exhibits AI. \citet{Nolde2020} show that the gauge function of an IMS process with exponential margins is simply $g_{\text{IMS}}(\bm{x}) = V_{\text{MS}}(1/\bm{x})$.

In this manner, we can choose any MS model of our liking and construct its corresponding spatial IMS gauge. A much popularised MS model in the extreme value literature is the Brown--Resnick (BR) model \citep{Brown_Resnick_1977,Kabluchko2009}. Its corresponding IMS model will then be the inverted Brown--Resnick (IBR) process with associated gauge function $g_{\text{IBR}}(\bm{x})=V_{\text{BR}}(1/\bm{x})$. 

\subsection{Spatial radial modelling}
\label{sec:radmod}
\subsubsection{Radial threshold}
\label{sec:thresh_sel}
\citet{Wadsworth2024} and \citet{Campbell2024} propose methods for obtaining a high radial threshold, $r_{\tau}(\bm{w})$, based on empirical quantiles of $R\vert \bm{W}=\bm{w}$ and a kernel density estimation technique, respectively. However, both techniques are limited to use in low/moderate dimensions as their implementation becomes increasingly laborious when higher dimensions are considered. To overcome this issue we leverage the observation that, if we wish $r_{\tau}(\bm{w})$ to approximately represent the $\tau$-quantile of $R|\bm{W}=\bm{w}$, then, for $\tau$ near $1$, the threshold function can be approximated by
\begin{equation}
\label{eq:thresh}
    r_{\tau}(\bm{w}) \approx C_\tau / g(\bm{w}),
\end{equation}
where $C_\tau$ is a constant such that the proportion of threshold exceedances $r>r_{\tau}(\bm{w})$ is $1-\tau$ \citep{Wadsworth2024}. 
Adopting a flexible spatial gauge function for $g(\bm{w})$, we can estimate its parameters via composite likelihood, as this requires us to specify pairwise thresholds initially. Specifically, we fit the bivariate truncated gamma model of \citet{Wadsworth2024} to (a subset of) the $d\choose 2$ possible pairs, using the authors' proposed quantile regression methods to obtain threshold exceedances and a spatial Gaussian gauge function $g_{\text{G}}$ for the gamma rate parameter. For a pair indexed by $(j,k)$ denote by $r^{jk} = x(\bm{s}_j)+x(\bm{s}_k)$ the pairwise radius, $\bm{w}^{jk}=(x(\bm{s}_j),x(\bm{s}_k))/r^{jk}$ the pairwise angle, $r_{\tau}^{jk}(\cdot)$ the pairwise threshold and $H^{jk}$ the $(j,k)$ element of the distance matrix $H$. Combining each pairwise likelihood contribution together we obtain a composite likelihood function,
\begin{equation}
\label{eq:copmlh}
    L_{\text{pw}}(\lambda_{\text{pw}},\kappa_{\text{pw}}) = \prod_{j<k}\prod_{i=1}^n\frac{f\left(r_i^{jk};a_{\text{pw}}, g_{\text{G}}(\bm{w}_i^{jk};\lambda_{\text{pw}},\kappa_{\text{pw}}, H^{jk})\right)}{\overline{\text{F}}\left(r_{\tau}^{jk}(\bm{w}_i^{jk});a_{\text{pw}}, g_{\text{G}}(\bm{w}_i^{jk};\lambda_{\text{pw}},\kappa_{\text{pw}}, H^{jk})\right)}, 
\end{equation}
where $f(\cdot)$ is the gamma density function, $\overline{\text{F}}$ the gamma survival function, and the gamma shape parameter is parameterised as a multiple of the dimension (in this case equal to $2$), $a_{\text{pw}}=2\alpha_{\text{pw}}$. The likelihood in equation \eqref{eq:copmlh} is minimised to obtain estimates for the $\lambda, \kappa$ parameters of the Gaussian gauge, denoted by $\lambda_{\text{pw}}, \kappa_{\text{pw}}$ respectively. Having obtained those, we can use the result of equation \eqref{eq:thresh}. We set $r_{\tau}(\bm{w})= C_{\tau}/g_{\text{G}}(\bm{w};\widehat{\lambda}_{\text{pw}},\widehat{\kappa}_{\text{pw}})$, where $C_{\tau}$ is obtained using a root finder such that the proportion of exceedances of $r_{\tau}(\bm{w})$ is as close to $1-\tau$ as possible. Although $g_{\text{G}}$ may not be the correct gauge function, it is not critical that $r_{\tau}(\bm{w})$ represents an equivalent $\tau$-quantile in each direction $\bm{w}$, hence we believe this simple approach to be adequate. However, we note that in high dimensions it is difficult to assess the threshold quality and its potential inferential implications with certainty. See Section \ref{sec:discussion} for further discussion of this matter.


\subsubsection{Model for radial exceedances}
Given $\bm{X}=(X(\bm{s}_1),\ldots,X(\bm{s}_d))$, we derive its radial and angular components, $R, \bm{W}$, and specify the exceedances of the radial threshold as detailed in Section \ref{sec:thresh_sel}. We can then fit the truncated gamma model of equation \eqref{eq:WCmodel} to the $r$ exceedances of $r_{\tau}(\boldsymbol{w})$, using likelihood \eqref{eq:fulllh}: 
\begin{equation}
\label{eq:fulllh}
    L_{d}(\bm{\theta}) = \prod_{i=1}^n\frac{f(r_i;a, g(\bm{w}_i;\bm{\theta}, H))}{\overline{\text{F}}(r_{\tau}(\bm{w}_i);a, g(\bm{w}_i;\bm{\theta}, H))}. 
\end{equation}
As in Section \ref{sec:thresh_sel}, we set the gamma shape parameter $a$ to be a multiple of the dimension $d$, $a=\alpha d$, where $\alpha>0$. In many cases, the true gamma shape is equal to the dimension $d$ \citep{Wadsworth2024} so, to reduce the number of model parameters, we also considered using a fixed shape parameter. However, we found an estimated shape to perform better so we adopt this strategy, parameterising as $a=\alpha d$, instead of simply $a=\alpha$, the former being more numerically stable than the latter. This modelling choice has the added benefit of enabling inference on observations with varying dimension $d_i$ per replicate $i$, $i=1,\ldots, n$, of the process, which is particularly useful when modelling data with missing values.

To perform inference on our radial model, we consider the parametric gauge forms presented in Section \ref{sec:spat gauges} as candidate models for the gamma rate parameter, and optimise the likelihood in equation \eqref{eq:fulllh}. The resulting fitted models are compared using Akaike's information criterion (AIC); the model with the smallest AIC is selected and its performance further assessed using diagnostic tools such as probability-probability (P-P) plots for the conditional distribution of $R\vert \bm{W}$.

\subsection{Spatial angular models}
\label{sec:angmod}
In our discussion so far, we have presented a model for the conditional distribution of extreme radii given some angles, which are only implicitly contributing to the inference procedure by being treated as covariates. However, regardless of dimension, the modelling of the radii ultimately boils down to a univariate problem, and so the focus for accurate modelling in high dimensions necessarily shifts to accurate modelling of the angular variable $\bm{W}$. Having a working model for both $f_{R\vert \bm{W}}(r\vert \bm{w})$ and $f_{\bm{W}}(\bm{w})$, the conditional radial and angular densities, respectively, is highly desirable because combining the two together provides a way to model the entire joint distribution of $(R,\bm{W})$, $f_{R,\bm{W}}(r,w) = f_{R\vert \bm{W}}(r\vert \bm{w})f_{\bm{W}}(\bm{w})$, which encodes the complete information about the spatial vector of interest, $\bm{X}$. In this section we discuss potential spatially-parameterised models for the angular density, $f_{\bm{W}}(\bm{w})$. 

\subsection{Process-based angular model}
\label{sec:am1}
Perhaps the simplest strategy for obtaining a model for $f_{\bm{W}}(\bm{w})$ is to start by assuming a model for some spatial process $\{\widetilde{X}(\bm{s}): \bm{s}\in \mathcal{S}\}$, represented in finite dimensions by the quantity $\widetilde{\bm{X}} = (\widetilde{X}(\bm{s}_1), \ldots, \widetilde{X}(\bm{s}_d))$. From $\widetilde{\bm{X}}$ we obtain its radial-angular decomposition, $(\widetilde{R}, \widetilde{\bm{W}})$, and calculate the spatially parameterised marginal density $f_{\widetilde{\bm{W}}}(\bm{w})$. The latter can then be used as a model for the quantity $f_{\bm{W}}(\bm{w})$ of interest. To make these ideas more concrete, let $Z(\bm{s})$ be a standard Gaussian process and $\widetilde{X}(\bm{s})= -\log{(1-\Phi(Z(\bm{s})))}$ be that process on standard exponential margins. This gives
\begin{equation}
\label{eq:am1joint}
    f_{\widetilde{R},\widetilde{\bm{W}}}(r,\bm{w}) = f_{\widetilde{\bm{X}}}(r\bm{w})r^{d-1}
    = \frac{r^{d-1}}{(2\pi)^{d/2}|\Sigma|^{1/2}}\exp{-\frac{1}{2}(\bm{z}(r,\bm{w}))^{\text{T}}\Sigma^{-1} (\bm{z}(r,\bm{w}))}\prod_{j=1}^d\frac{1-\Phi(z_j(r_j,w_j))}{\phi(z_j(r_j,w_j))},
\end{equation}
where $\phi, \Phi, \Phi^{-1}$ are the standard Gaussian density, probability and inverse probability functions respectively, $r^{d-1}$ is the Jacobian of the transformation $\widetilde{\bm{X}}\rightarrow(\widetilde{R},\widetilde{\bm{W}})$, $\prod_{j=1}^d\frac{1-\Phi(z_j(r_j,w_j))}{\phi(z_j(r_j,w_j))}$ is the Jacobian of the transformation from standard Gaussian to standard exponential margins and $\bm{z}(r,\bm{w})=\Phi^{-1}(1-\exp{-r\bm{w}})$. The covariance matrix $\Sigma$ in \eqref{eq:am1joint} corresponds to that of $\bm{Z}=(Z(\bm{s}_1),\ldots,Z(\bm{s}_d))$, and is parameterised spatially. As with the spatial gauge models in Section \ref{sec:spat gauges}, we typically use a powered exponential correlation function.
Finally, we obtain the marginal density for the angles $\widetilde{\bm{W}}$ from equation \eqref{eq:am1joint} through integration, that is
\begin{equation}
\label{eq:am1}
    f_{\widetilde{\bm{W}}}(\bm{w}) = \int_0^{\infty}f_{\widetilde{R},\widetilde{\bm{W}}}(r,\bm{w})\mathrm{d} r.
\end{equation}
We perform the integration step by numerical means and use standard optimisation techniques for inference on this model. We use the model in \eqref{eq:am1} as a candidate for the density $f_{\bm{W}}(\bm{w})$.

\subsubsection{Gauge-based angular model}
\label{sec:am2}
An alternative model for the angular distribution can be obtained by considering those derived from certain types of  \textit{homothetic} light-tailed densities \citep{Balkema2010b}. Homothetic densities are unimodal densities whose unit level sets are of the same shape.  Inspired by these results, \citet{Papastathopoulos2023} and \citet{Campbell2024} have used densities of the form 
\begin{equation}
\label{eq:am2}
   f_{\boldsymbol{W}}(\boldsymbol{w}) = g(\boldsymbol{w})^{-d}/\{d\text{vol}(G)\}, \quad \boldsymbol{w}\in \mathcal{S}^{d-1},
\end{equation}
where $g$ is the gauge function of a limit set $G$. In our case, the gauge function in \eqref{eq:am2} is taken to be a separate entity to the one used in the radial model, since using the same $g$ is only appropriate if the underlying distribution is close to homothetic. The challenge with using the above formulation is the fact that the volume of the limit set $G$, $\text{vol}(G)$, appearing in the normalising constant is generally unknown. For $d=2,3$, \citet{Papastathopoulos2023} estimate this quantity as part of their inference procedure. \citet{Campbell2024} propose a piecewise-linear formulation of the gauge function for which exact computation of the limit set volume is possible. 

To circumvent the problem of dealing with the normalising constant, we explore alternative inference avenues based on score matching and, specifically, the gradient score methodology of \citet{Hyvarinen2005}.
Let $p(\bm{y};\bm{\theta})$ be the non-normalised version of some density $f_{\bm{Y}}(\bm{y};\bm{\theta})$. Inference on $p(\bm{y};\bm{\theta})$ can be achieved by minimising the following objective function
\begin{equation}
\label{eq:hyv_objfun}
    J(\bm{\theta}) = \frac{1}{n}\sum_{i=1}^n\sum_{j=1}^d \left[\frac{\partial^2 \log p(\bm{y}_i;\bm{\theta})}{\partial y_{ij}^2} + \frac{1}{2}\left(\frac{\partial \log p(\bm{y}_i;\bm{\theta})}{\partial y_{ij}}\right)^2\right],
\end{equation}
which is proven to lead to (locally) consistent parameter estimation \citep{Hyvarinen2005}. The Hyv\"arinen score is restricted to situations where $\boldsymbol{Y}\in \mathbb{R}^d$. However, it is not unusual in applications to encounter random variables that only live in a subspace of $\mathbb{R}^d$, e.g.\ $\mathbb{R}_+^d$ or $\mathcal{S}^{d-1}$. \citet{Hyvarinen2007} adapted the methodology of \citet{Hyvarinen2005} for $\bm{Y}\in\mathbb{R}_+^d$, while \citet{deFondeville2018} adapted to the support of multivariate Pareto-type distributions. In the case of data living in the unit simplex, further adaptations of this methodology are necessary. To that end, \citet{scealy2024} perform an additive log-ratio transform \citep{Aitchison1982} to compositional data and then apply standard \citet{Hyvarinen2005} methodology for inference on the model of \citet{Scealy2023}. Very recently, \citet{Xu2025} obtain theoretical results for score matching under various data configuration scenarios, including continuous and discrete data. 
We also adopt an additive log-ratio transformation, $T:\mathcal{S}_+^{d-1}\rightarrow \mathbb{R}^{d-1}$, setting $\bm{V}=T(\bm{W})$, where the $k$th component of the mapping is
\begin{equation}
    V_k = T_k(\bm{W}) = \log\frac{W_k}{1-\sum_{j=1}^{d-1}W_j} = \log\frac{W_k}{W_d}, \quad k=1,\ldots, d.
\end{equation}
The inverse transformation, $\bm{W}=T^{-1}(\bm{V})$, is given by $W_k = T_k^{-1}(\bm{V}) = e^{V_k}/\sum_{j=1}^d e^{V_j}$, and the Jacobian determinant is $\prod_{j=1}^{d}T_j^{-1}(\bm{v})$.
Alternative transformations achieving the same purpose have also been studied in  \citet{Aitchison1982}. Note that $V_d\equiv0$, so we have $(V_1,\ldots, V_{d-1})\in\mathbb{R}^{d-1}$. We will sometimes write $v_d$ below for notational convenience, in the understanding that it is identically zero. Under this transformation we have
\begin{align}
    f_{\bm{V}}(\bm{v}) 
    =& f_{\bm{W}}(T^{-1}(\bm{v})) \prod_{j=1}^{d}T_j^{-1}(\bm{v}) \\
    =&f_{\bm{W}}\left(\frac{\exp(\bm{v})}{\sum_{j=1}^d\exp(v_j)}\right)\prod_{j=1}^{d}\frac{\exp(v_j)}{\sum_{l=1}^d\exp(v_l)} \\
    =& f_{\bm{W}}\left(\frac{\exp(\bm{v})}{\sum_{j=1}^d\exp(v_j)}\right) \exp\left\{\sum_{j=1}^{d-1}v_j\right\} \left(\sum_{j=1}^d\exp(v_j)\right)^{-d}.
\end{align} 
Recall that $f_{\bm{W}}(\bm{w}) \propto g(\bm{w})^{-d}$ and that $g(\bm{w})$ is $1$-homogeneous. This means that
\begin{equation}
    f_{\bm{W}}\left(\frac{\exp(\bm{v})}{\sum_{j=1}^d\exp(v_j)}\right) \propto   g\left(\frac{\exp(\bm{v})}{\sum_{j=1}^d \exp(v_j)}\right)^{-d} =     g(\exp(\bm{v}))^{-d} \left(\sum_{j=1}^d\exp(v_j)\right)^d.
\end{equation}
Putting everything together we have that the logarithm of the un-normalised version of $f_{\bm{V}}$, denoted $p$, is
\begin{equation}
\label{eq:f_v}
    \log p(\bm{v})=-d \log g(\exp(\bm{v})) + \sum_{j=1}^{d-1}v_j .
\end{equation}
Obtaining first and second derivatives of the expression in equation \eqref{eq:f_v} and substituting them in equation \eqref{eq:hyv_objfun}, allows us to leverage the standard \citet{Hyvarinen2005} method for inference on this angular model. Analytical expressions for the quantities $\frac{\partial \log p(\bm{v};\bm{\theta})}{\partial v_j}$ and $\frac{\partial^2 \log p(\bm{v};\bm{\theta})}{\partial v_j^2}$, with $g$ taken to be the generalised Gaussian gauge function presented in Section \ref{sec:GLG_gauge}, can be found in Section \ref{app:am2} of the Supplement. 

\section{Simulating from the model}
\label{sec:prediction}

A desirable feature in any statistical modelling setting is the ability to simulate new data. In the context of studying extremes, it is of particular interest to extrapolate further into the tails of a distribution. Extreme data, simulated above a high threshold, can be used to perform model checking tasks as well as to calculate extreme set probabilities. To simulate new extreme data $\bm{X} \vert R>r_{\tau}(\bm{W})$ or, equivalently, $\bm{X} \vert R'>1$ where $R'=R/r_{\tau}(\bm{W})$, Wadsworth and Campbell (2024) use the following scheme, which we also adopt here: 

\begin{enumerate}
    \item Simulate draws $\bm{w}^*$ from the distribution of $\bm{W} \vert R'>1$;
    \item Given the draws $\bm{w}^*$ from step $1$, simulate new radii $r^*$ from the distribution of $R\vert\{\bm{W}=\bm{w}^*, R'>1\}$ detailed in Section \ref{sec:radmod};
    \item Set $\bm{x}^*=r^* \bm{w}^*$. 
\end{enumerate}
In step $1$, we consider the empirical distribution of angles, alongside simulating from the fitted angular distributions \eqref{eq:am1} and \eqref{eq:am2}. 
Sampling new angles from \eqref{eq:am1} involves sampling new data $\widetilde{\boldsymbol{x}}$ from the distribution of $\widetilde{\boldsymbol{X}}$ and setting $\boldsymbol{w}^* = \widetilde{\boldsymbol{x}}/\|\widetilde{\boldsymbol{x}}\|$.
To sample new angles from \eqref{eq:am2}, we firstly simulate from the density $f_{\widetilde{\bm{X}}}(\bm{x})=\exp{-g(\bm{x})}/[d!\text{vol}(G)]$, for which $\widetilde{\bm{W}}=\widetilde{\bm{X}}/\|\widetilde{\bm{X}}\|$ has density \eqref{eq:am2}. Simulating from $f_{\widetilde{\bm{X}}}$ is done via Markov Chain Monte Carlo. In low dimensions ($d\leq5$) an independence sampler works well for this task, but in higher dimensions we found a random-walk Metropolis scheme to work much better. Draws are thinned to reduce autocorrelation. 
Finally we note that one can extrapolate further into the joint tail by altering the conditioning event to $\{R'>k\}$, $k>1$. This case is covered in detail in \citet{Wadsworth2024} and so we refer the interested reader there.

Now that we have a way to simulate new data we want to use draws from $\bm{X} \vert R'>1$ to calculate extreme set probabilities. In other words, we want to calculate $\Pr(\bm{X}\in B)$, where $B$ is an extreme set of interest. We note that set $B$ can be of any desirable shape and size and, importantly, can lie anywhere in $\mathbb{R}^d_+$. Let $\mathbb{B}_k := \{\bm{x}:\|\bm{x}\|>kr_{\tau}(\bm{x}/\|\bm{x}\|)\}$ and $\mathbb{B}_k^{c} := \{\bm{x}:\|\bm{x}\|\leq kr_{\tau}(\bm{x}/\|\bm{x}\|)\}$, the complement of $\mathbb{B}_k$. Provided $B$ lies entirely in $\mathbb{B}_1$, it is sufficient to solely rely on simulated $\bm{X}\vert R'>1$ values to calculate this probability since
\begin{equation}
\label{eq:probB_AT}    \Pr(\bm{X}\in B) = \Pr(\bm{X}\in B\cap \mathbb{B}_1) = \Pr(\bm{X}\in B \vert \bm{X}\in \mathbb{B}_1) \Pr(\bm{X}\in \mathbb{B}_1).
\end{equation}
We calculate $\Pr(\bm{X}\in B\cap \mathbb{B}_1)$ empirically from the simulated data and use $\Pr(\bm{X}\in \mathbb{B}_1) = 1-\tau$, where $\tau$ is the quantile level we used to define the threshold function $r_0(\bm{w})$ (see Section \ref{sec:thresh_sel} for details).
However, when $B\not\subset\mathbb{B}_1$, we cannot rely on simulated data alone to compute $\Pr(\bm{X}\in B)$. In this case we adopt a mixed estimation approach whereby we use the original data $\bm{X}$ to estimate the part of the probability where $B$ lies below the threshold and use the simulated extreme data to estimate the part of the probability where $B$ is above the threshold, as shown below. A similar scheme has been used in \citet{Papastathopoulos2023} and \citet{demonte2025}.
\begin{align}
\label{eq:probB_BT}    \Pr(\bm{X}\in B) 
=& \Pr(\bm{X}\in B\cap \mathbb{B}_1^c)+\Pr(\bm{X}\in B\cap \mathbb{B}_1) \\ 
=& \Pr(\bm{X}\in B\cap\mathbb{B}_1^c)+\Pr(\bm{X}\in B | \bm{X}\in \mathbb{B}_1) \Pr(\bm{X}\in \mathbb{B}_1).
\end{align}

It is often of interest in spatial extreme value analyses to examine estimates of pairwise coefficients of tail dependence, $\chi(\bm{s}_l,\bm{s}_m)$. This coefficient encodes information about the extremal dependence properties of the process $X(\bm{s})$ and is given by $\chi(\bm{s}_l,\bm{s}_m) = \lim_{u \to 1} \chi_u(\bm{s}_l,\bm{s}_m)$, with
\begin{equation}
\label{spat chi}
   \chi_u(\bm{s}_l,\bm{s}_m) = \Pr \left(X(\bm{s}_l)>-\log(1-u), X(\bm{s}_m)>-\log(1-u)\right)/(1-u),\quad u\in[0,1].
\end{equation}
In this scenario, the extreme set of interest is $B^u=(-\log(1-u),\infty)^2 \times (0,\infty)^{d-2}$, with a slight abuse of notation assuming $l=1$, $m=2$. We make extensive use of both equations \eqref{eq:probB_AT} and \eqref{eq:probB_BT} in Section \ref{sec:simstudy} to calculate pairwise $\chi_u$ coefficients in a variety of simulation settings.

To assess whether the condition $B^u\subset\mathbb{B}_k$ is satisfied for this task, we need to consider the shape of the sets $B^u$ and $\mathbb{B}_k$. 
The $x_l$-$x_m$ face of set $B^u$ is symmetric about the line $x_l = x_m$, as it is defined by both $x_l$ and $x_m$ coordinates being above the same threshold level,  $-\log(1-u)$. As a result, the vertex of set $B^u$ --- with coordinates $(-\log(1-u),-\log(1-u), 0, \ldots, 0)$ --- lies on the diagonal $x_l=x_m$ line and constitutes the point in $B^u$ closest to the origin $\bm{0}$ at distance $x(\bm{s}_l)+x(\bm{s}_m) +0+ \ldots+0=r^{lm}=-2\log(1-u)$. 
As for the set $\mathbb{B}_k$, this is bounded below by the threshold surface --- a rescaled version of the spatial Gaussian gauge; see Section \ref{sec:thresh_sel} for details. We know that lower dimensional projections of a gauge function can generally be obtained via minimisation of the $d$-dimensional gauge with respect to the coordinates to be marginalised \citep{Nolde2020}. In particular, the projection of the Gaussian gauge on the $x_l$-$x_m$ face of interest, is a $2$-dimensional Gaussian gauge with the same parameters and distance equal to the $(l,m)$ entry of the original distance matrix $H$. Therefore, the projection of the threshold surface on the $x_l$-$x_m$ face is a multiple of this bivariate Gaussian gauge, which is convex and symmetric about the $x_l=x_m$ line. Hence, the first point of contact of the set $B^u$ with set $\mathbb{B}^c_k$ is the point $kC_{\tau}(1/2,1/2)/g_{\text{G}}((1/2,1/2); \widehat{\lambda}_{\text{pw}}, \widehat{\kappa}_{\text{pw}}, H^{lm})$ on the $x_l=x_m$ diagonal. The distance of this point from the origin is $kC_{\tau}/g_{\text{G}}((1/2,1/2); \widehat{\lambda}_{\text{pw}}, \widehat{\kappa}_{\text{pw}}, H^{lm})$; that is, the $k$ multiple of the bivariate threshold function $r_{\tau}^{lm}(\bm{w}^{lm})$ evaluated at $\bm{w}^{lm}=(x(\bm{s}_l),x(\bm{s}_m))/r^{lm}=(1/2,1/2)$ --- see Figure \ref{fig:chicheck} of the Supplement for a $3$-dimensional illustration of these concepts. Therefore, if $-2\log(1-u) > kC_{\tau}/g_{\text{G}}((1/2,1/2); \widehat{\lambda}_{\text{pw}}, \widehat{\kappa}_{\text{pw}}, H^{lm})$, then $B^u$ lies in $\mathbb{B}_k$ in its entirety. We refer to assessment of this inequality as the ``$\chi_u$-check".

\section{Simulation study}
\label{sec:simstudy}

This section comprises simulation study results in order to assess the performance of our model. We examine various data configurations --- described in detail in each of the following subsections --- with standard exponential margins, on which we implement the modelling procedure described in Sections \ref{sec:radmod} and \ref{sec:angmod}. Although we initially considered using $g_{\text{IBR}}(\bm{x})$ for inference on the radial model, it is highly computationally intensive to fit --- likewise all MS and IMS processes \citep{Huser2022} --- and was ultimately excluded. Moreover, in unreported investigations, fitting $g_{\text{IBR}}$ to IBR data did not seem to result in noticeable improvements, compared to other simpler gauge functions such as $g_{\text{G}}$ or $g_{\text{GG}}$, to justify the non-negligible computational cost of its implementation.

Our ultimate goal is to simulate new data $\bm{X}\vert R'>k$ according to Section \ref{sec:prediction}, from which to calculate pairwise $\chi_u$ estimates. We assess the performance of our model relative to the true $\chi_u$ quantities and in comparison to $\chi_u$ estimates from two more standard models, namely the censored Gaussian (cG) tail model \citep{Bortot2002} and the Huser--Wadsworth (HW) model \citep{Huser2019}, both of which are separately fitted to the same simulated datasets. These represent models for the extremal dependence structure of $X(\bm{s})$, fitted to the tail of $\bm{X}=(X(\bm{s}_1), \ldots,X(\bm{s}_d))$ via censored likelihood. Details on both these models, and their fitting, can be found in \citet{Huser2019}; we use the \texttt{SpatialADAI} \texttt{R} package that accompanies that paper for estimation of these models.

In all cases through Sections \ref{sec:gausdata} - \ref{sec:hw6data} we set the number of independent replicates to be $n=5000$, the threshold $\tau=0.7$ and consider irregularly spaced data of dimensions $d=5,10,20$. We note that similarly low quantile levels $\tau$ have been adopted by other authors, including \citet{Simpson2024}, \citet{MurphyBarltrop2024} and \citet{Campbell2024}, without inducing any noticeable biases, and diagnostic plots presented in Section \ref{app:ss_diagnostics} of the Supplement show satisfactory agreement of the radial exceedances with the truncated gamma model, even with this lower threshold. We revisit the discussion of $\tau$ in Section~\ref{sec:am2bad_main}.
Two parameter settings, resulting in different dependence strengths, are explored. All simulations are repeated $200$ times. For each simulation iteration, a total of $n^*=50000$ new realisations of $\bm{X}\vert R'>k$ are simulated for the task of $\chi_u$ estimation, using a potentially different $k$ value for each $u$. Specifically, starting from $k=1$ and moving up by increments of $0.1$, we select the largest value of $k$ such that the $\chi_u$-check described in Section \ref{sec:prediction} is satisfied for all $200$ simulated datasets $\bm{X}$. If the $\chi_u$-check is not satisfied for any $k$ then we use $k=1$ and the mixed estimation procedure in equation \eqref{eq:probB_BT}. When fitting the cG and HW models a different threshold, $\tau'$, is used. This corresponds to the marginal threshold that gives proportion $1-\tau$ of being extreme in at least one coordinate; $\tau'$ is calculated using a root finder.

\subsection{Gaussian data}
\label{sec:gausdata}

Let $\{X(\boldsymbol{s}):\boldsymbol{s}\in \mathcal{S}\subset[0,10]^2\}$ be a Gaussian process with powered exponential correlation function $\rho(h)=e^{-(h/\lambda)^\kappa}$. Table \ref{tab:mvn_modsel} below provides information on the gauge type contributing to the preferred radial fit across all simulated datasets, when the dependence parameters are $\bm{\theta}_1 = (\lambda_1,\kappa_1)=(10,1)$ and $\bm{\theta}_2 = (\lambda_2,\kappa_2)=(6,0.5)$, respectively. It is worth noting that although the Gaussian gauge model is the true model in this simulation scenario, the more flexible generalised Gaussian gauge is often favoured, especially as the dimension increases.
\begin{table}[h]
\centering
\begin{tabular}{cc|ccccc|ccccc}
\toprule
\multicolumn{2}{l|}{\textbf{Parameter}} & \multicolumn{5}{c|}{$\bm{\theta}_1$}    & \multicolumn{5}{c}{$\bm{\theta}_2$}                 \\ \hline 
\multicolumn{2}{c|}{\textbf{Model}} & GG & G & L & $\text{HW}_{\text{G}}$ & $\text{HW}_{\text{GG}}$ & GG & G & L & $\text{HW}_{\text{G}}$ & $\text{HW}_{\text{GG}}$ \\ \hline 
\multicolumn{1}{c|}{}           & 5   & 51.5 & 43.0 & 0 & 1.5 & 4.0  &  39.0 & 57.5 & 0 & 1.0 & 2.5   \\
\multicolumn{1}{c|}{$\bm{d}$} & 10  & 88   & 12   & 0 & 0   & 0    &  56.5 & 43.5 & 0 & 0   & 0  \\
\multicolumn{1}{c|}{}           & 20  & 100  & 0    & 0 & 0   & 0    &  88.5 & 11.0 & 0 & 0   & 0.5 \\  \bottomrule                        
\end{tabular}
\caption{Percentage of times the generalised Gaussian, Gaussian, Laplace, HW-Gaussian and HW-generalised Gaussian gauges were selected via AIC in the radial fit across all $200$ simulations when $X(\bm{s})$ is a Gaussian process with correlation parameters $\bm{\theta}_1=(10,1)$ and $\bm{\theta}_2=(6,0.5)$.}
\label{tab:mvn_modsel}
\end{table}

\begin{figure}[p!]
\centering
\begin{subfigure}[t]{1\textwidth}
\centering
\includegraphics[width=0.3\textwidth]{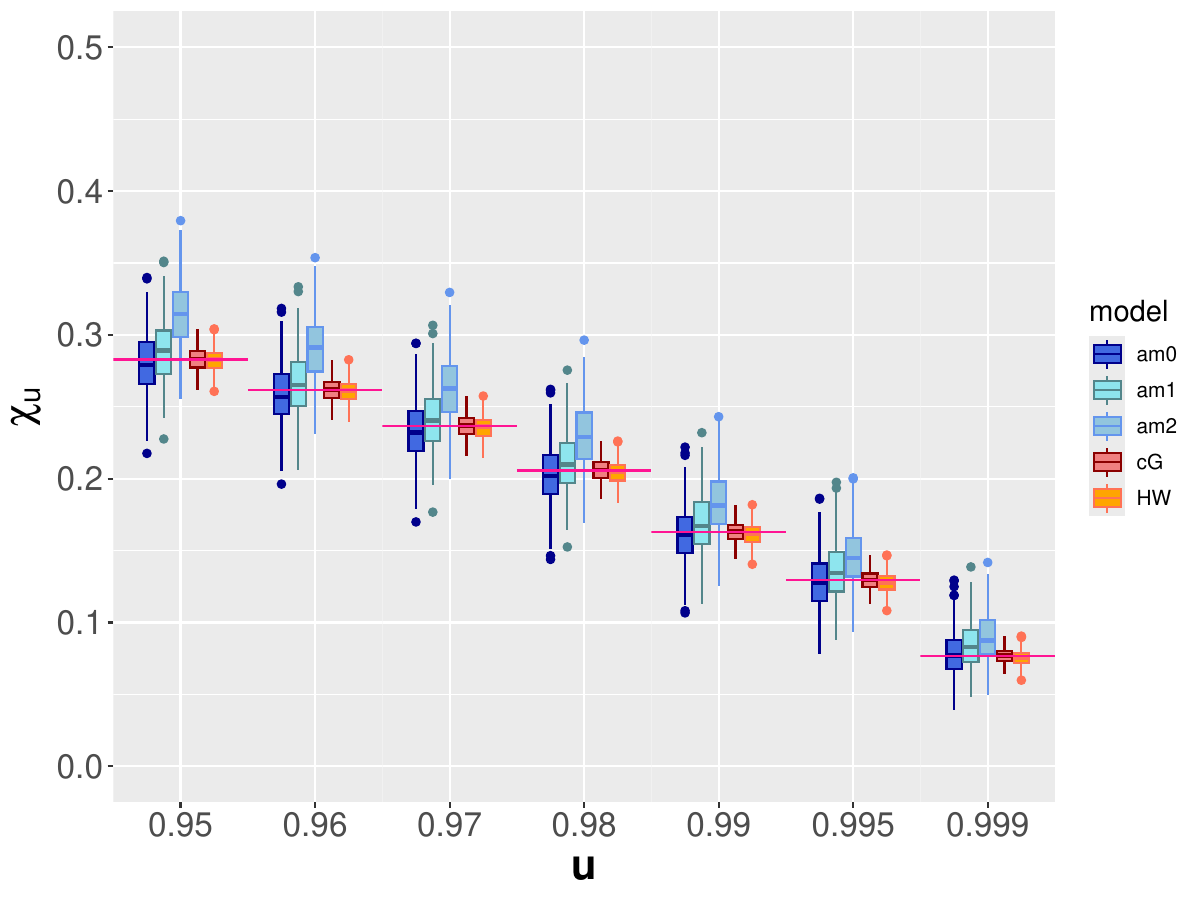}
  \hfill
  \includegraphics[width=0.3\textwidth]{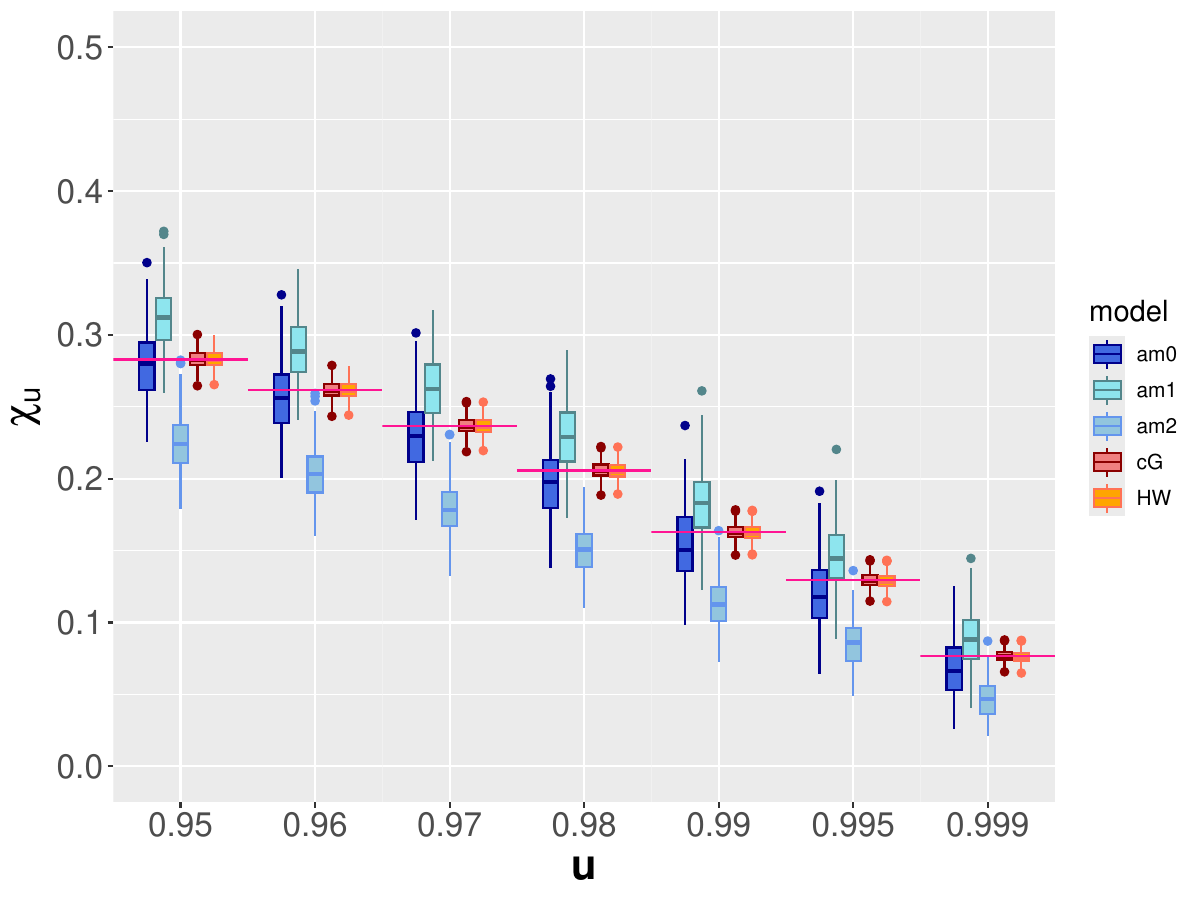}
  \hfill
  \includegraphics[width=0.3\textwidth]{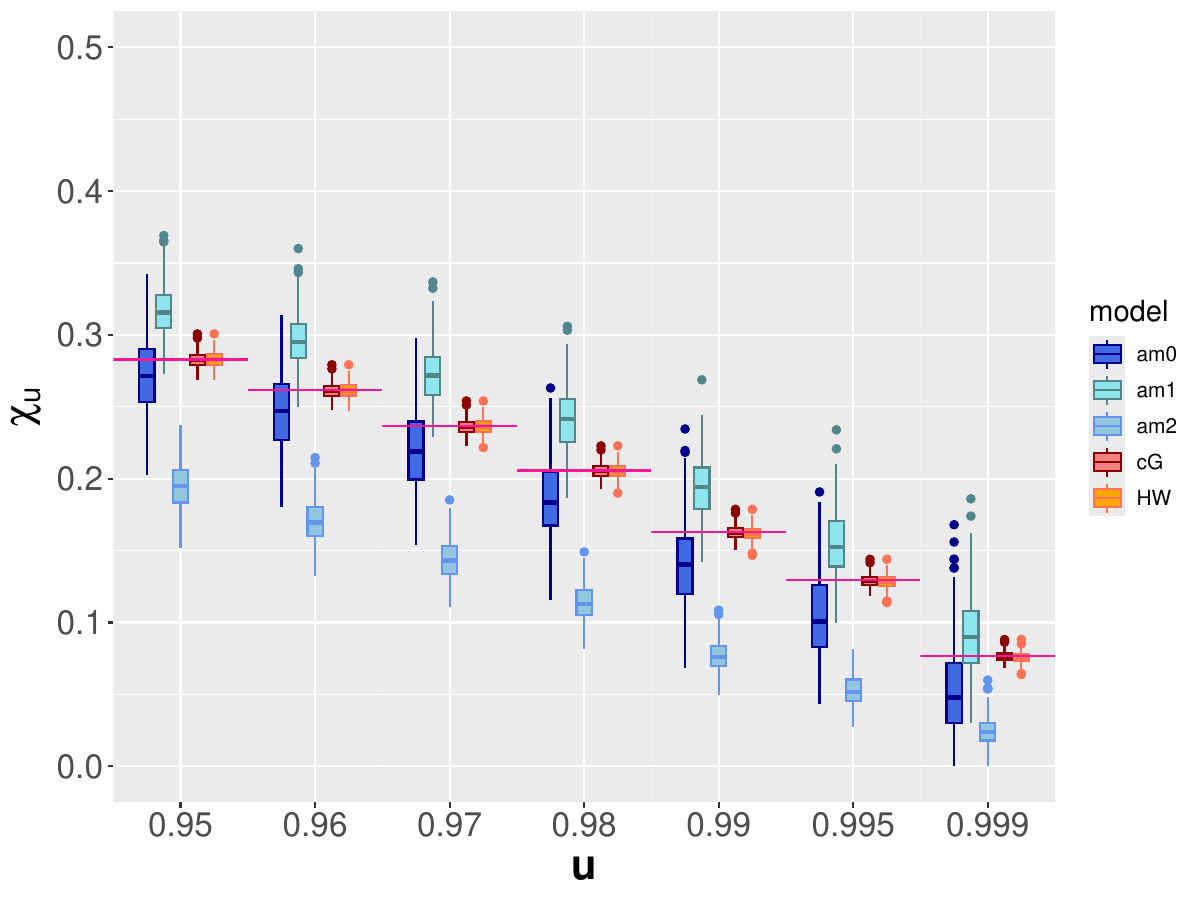} 
\caption{$X(\bm{s})$ simulated from a Gaussian process} \label{fig:box1_mvn}
\end{subfigure}

\begin{subfigure}[t]{1\textwidth}
\centering
\includegraphics[width=0.3\textwidth]{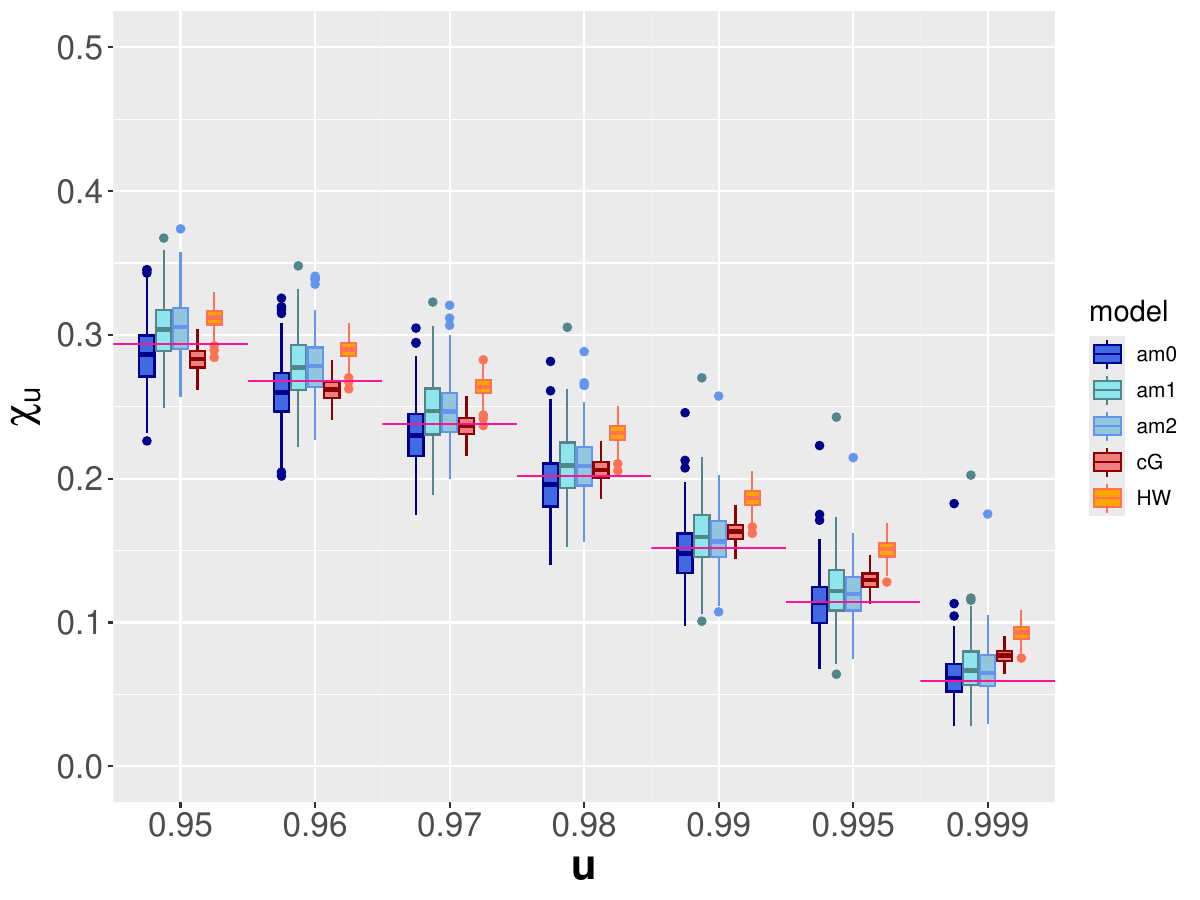}
  \hfill
  \includegraphics[width=0.3\textwidth]{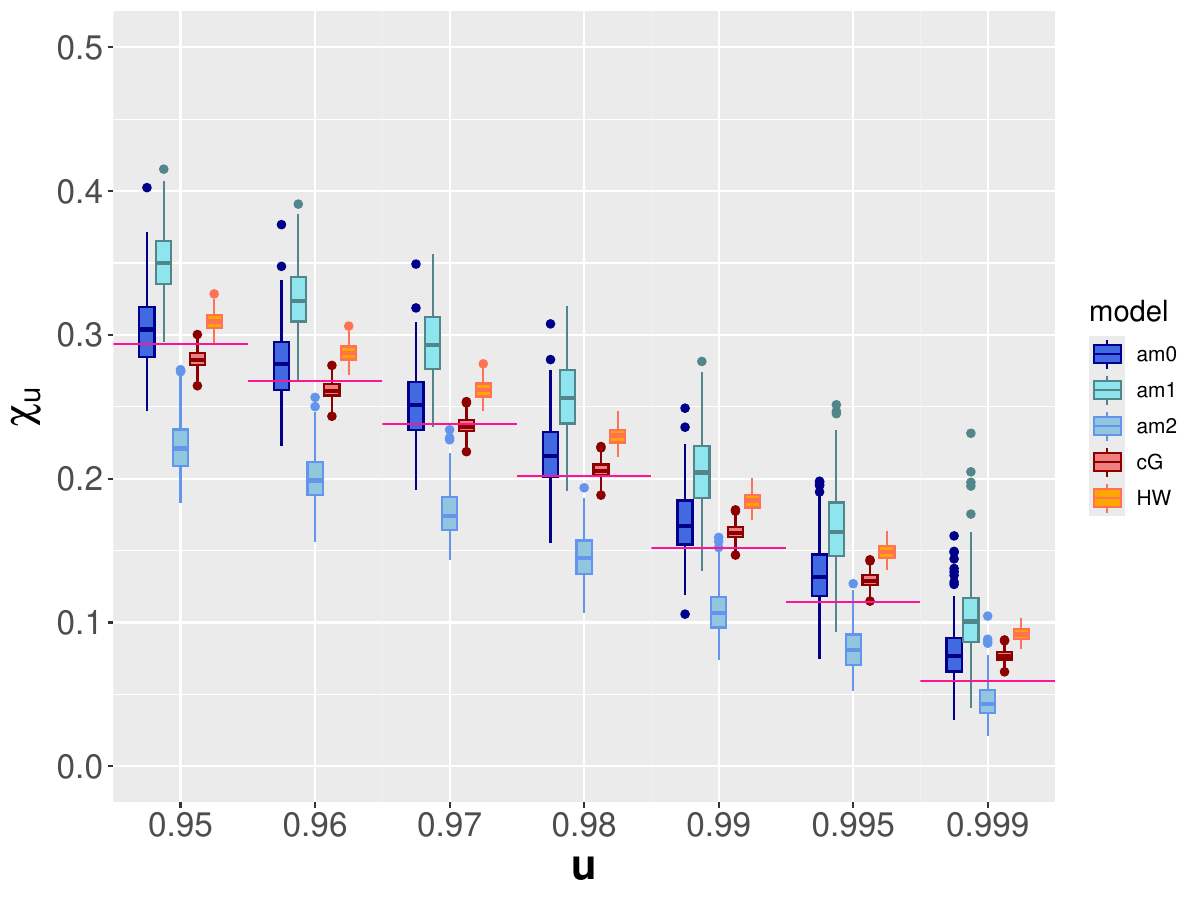}
  \hfill
  \includegraphics[width=0.3\textwidth]{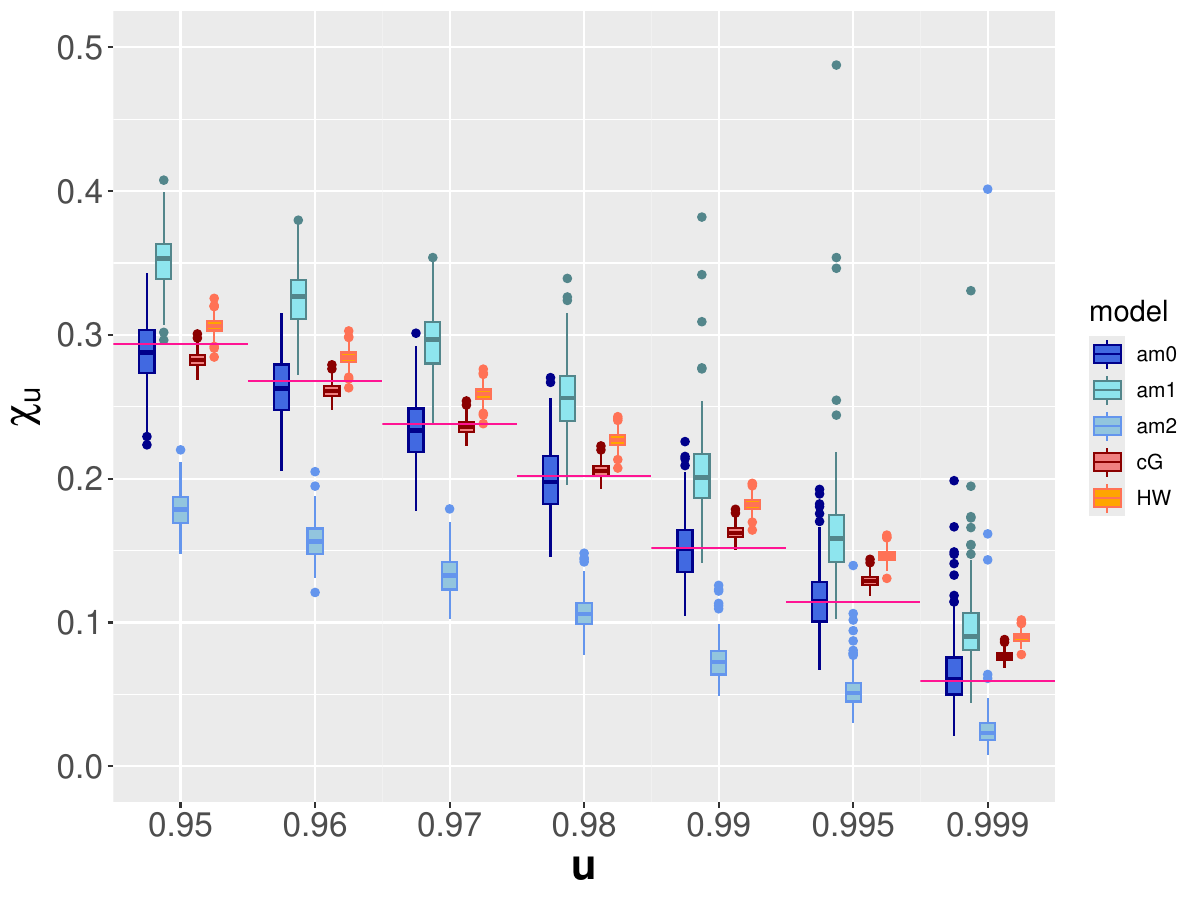} 
\caption{$X(\bm{s})$ simulated from an IBR process} \label{fig:box1_ibr}
\end{subfigure}

\begin{subfigure}[t]{1\textwidth}
\centering
\includegraphics[width=0.3\textwidth]{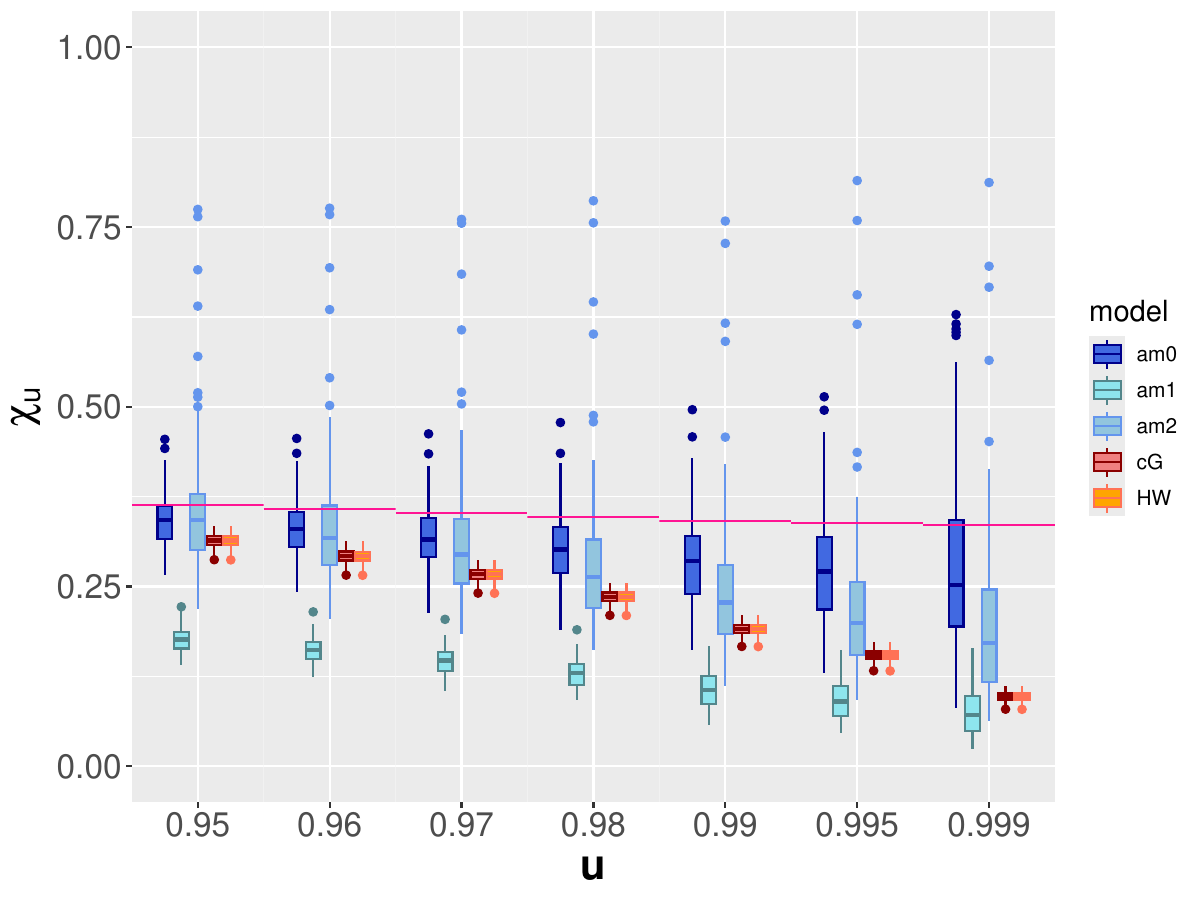}
  \hfill
  \includegraphics[width=0.3\textwidth]{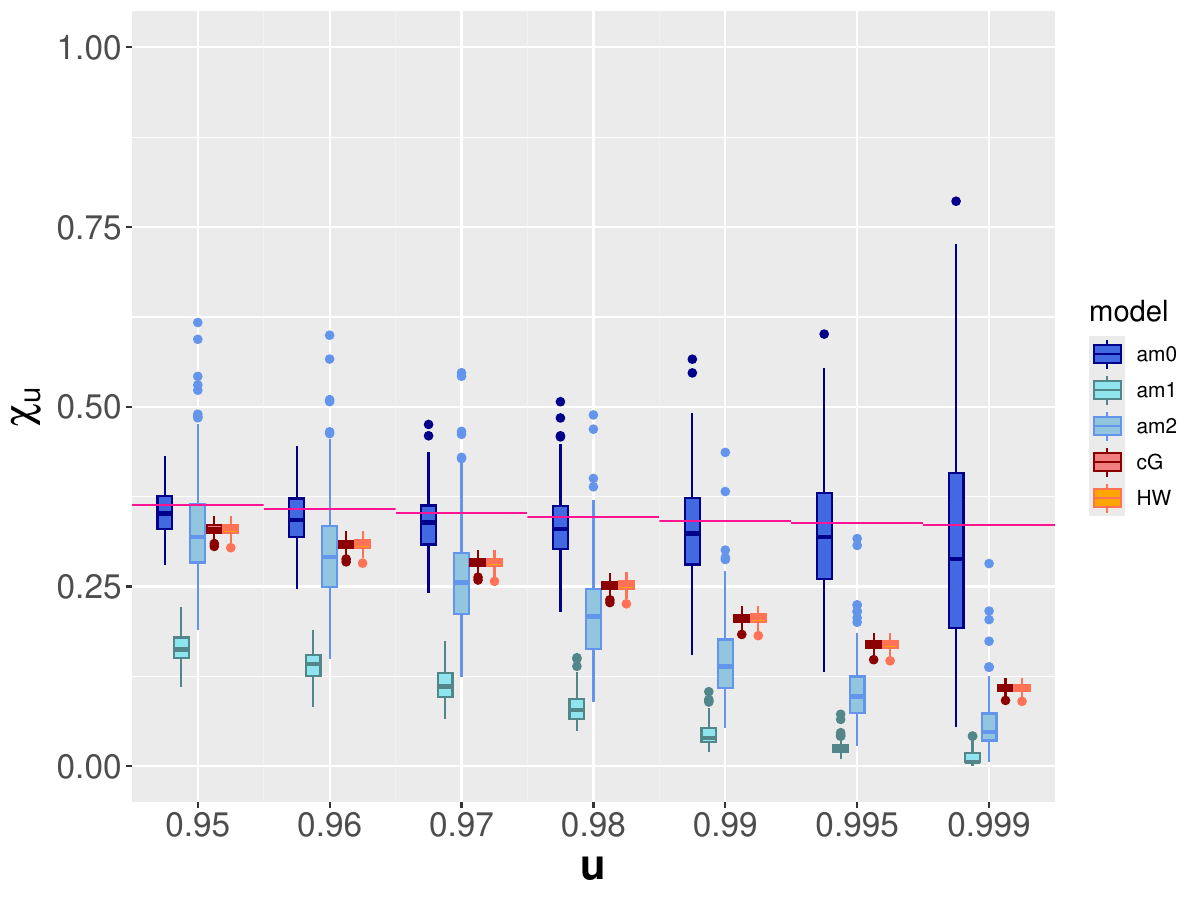}
  \hfill
  \includegraphics[width=0.3\textwidth]{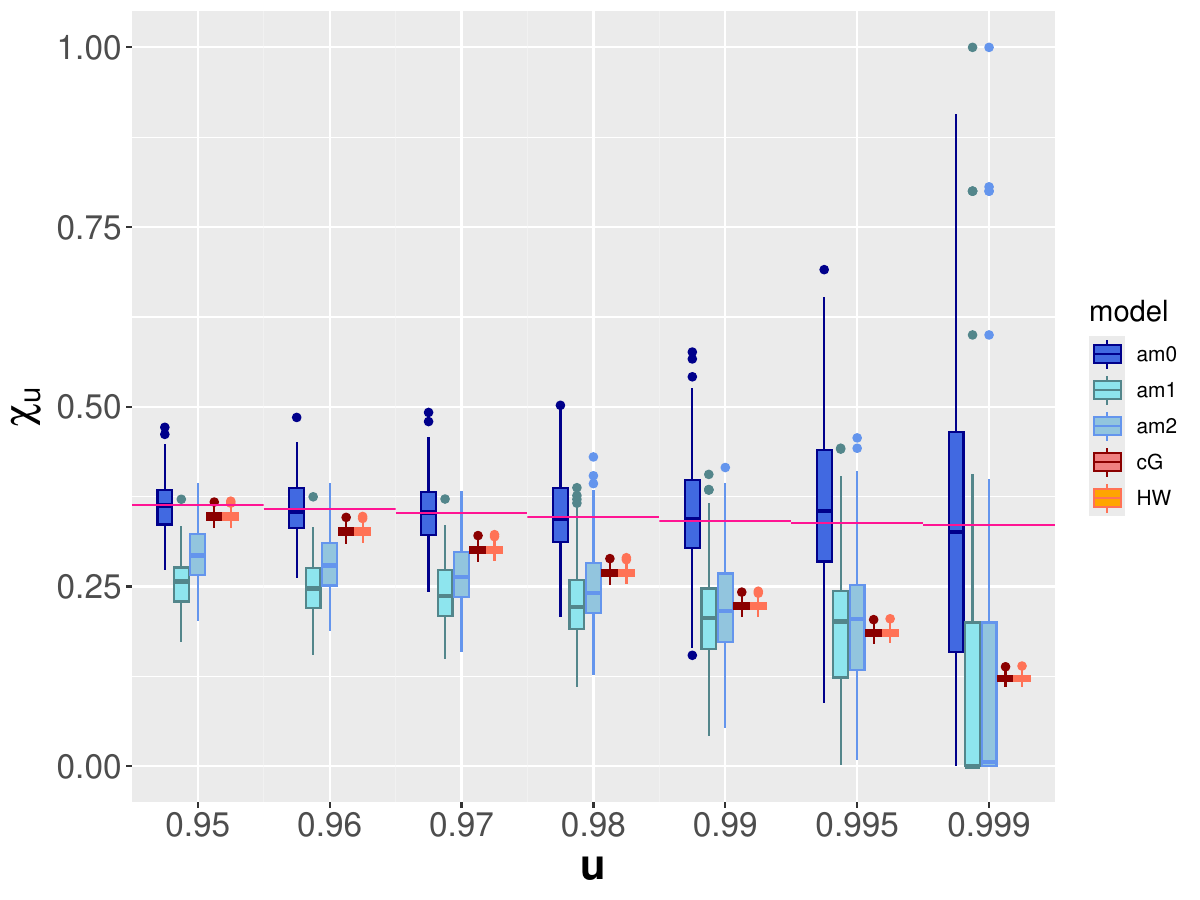}
\caption{$X(\bm{s})$ simulated from a BR process} \label{fig:box1_br}
\end{subfigure}

\begin{subfigure}[t]{1\textwidth}
\centering
\includegraphics[width=0.3\textwidth]{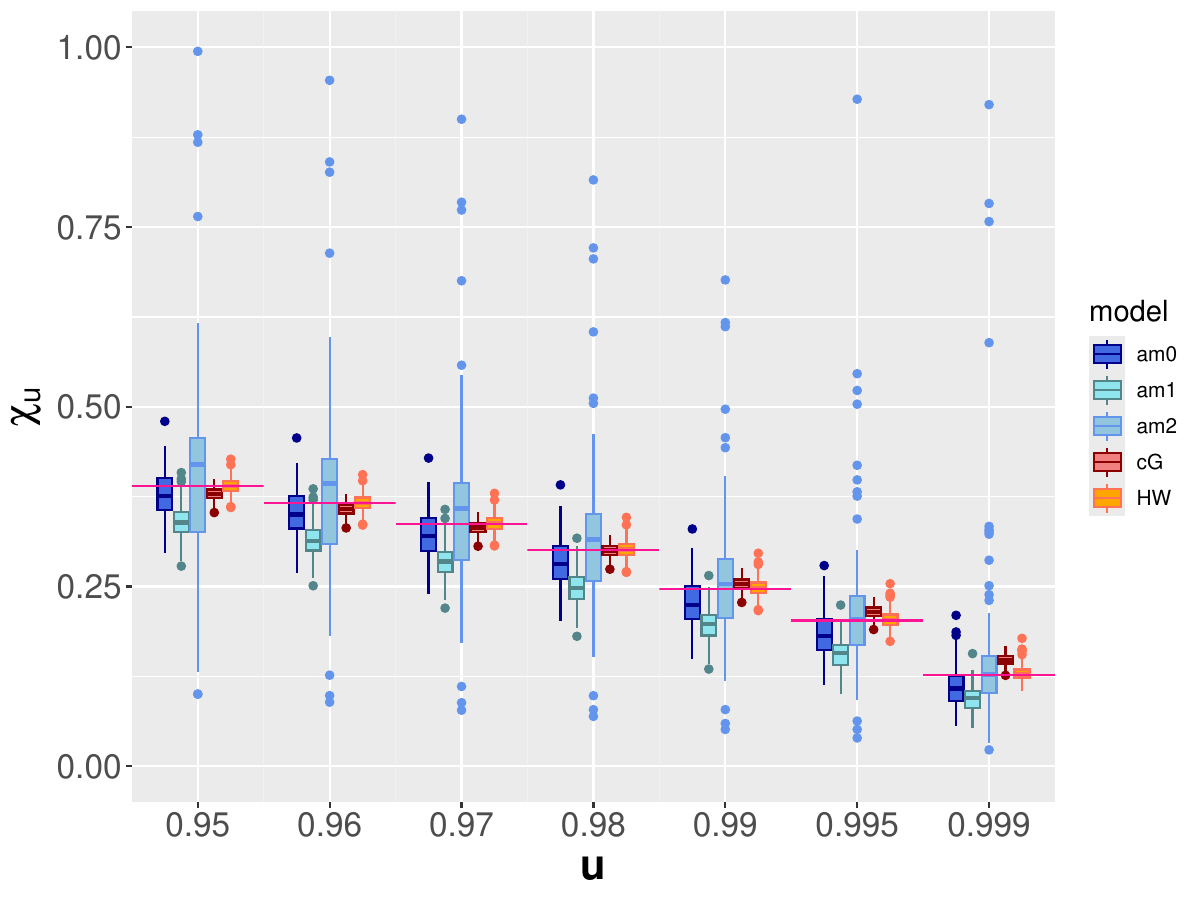}
  \hfill
  \includegraphics[width=0.3\textwidth]{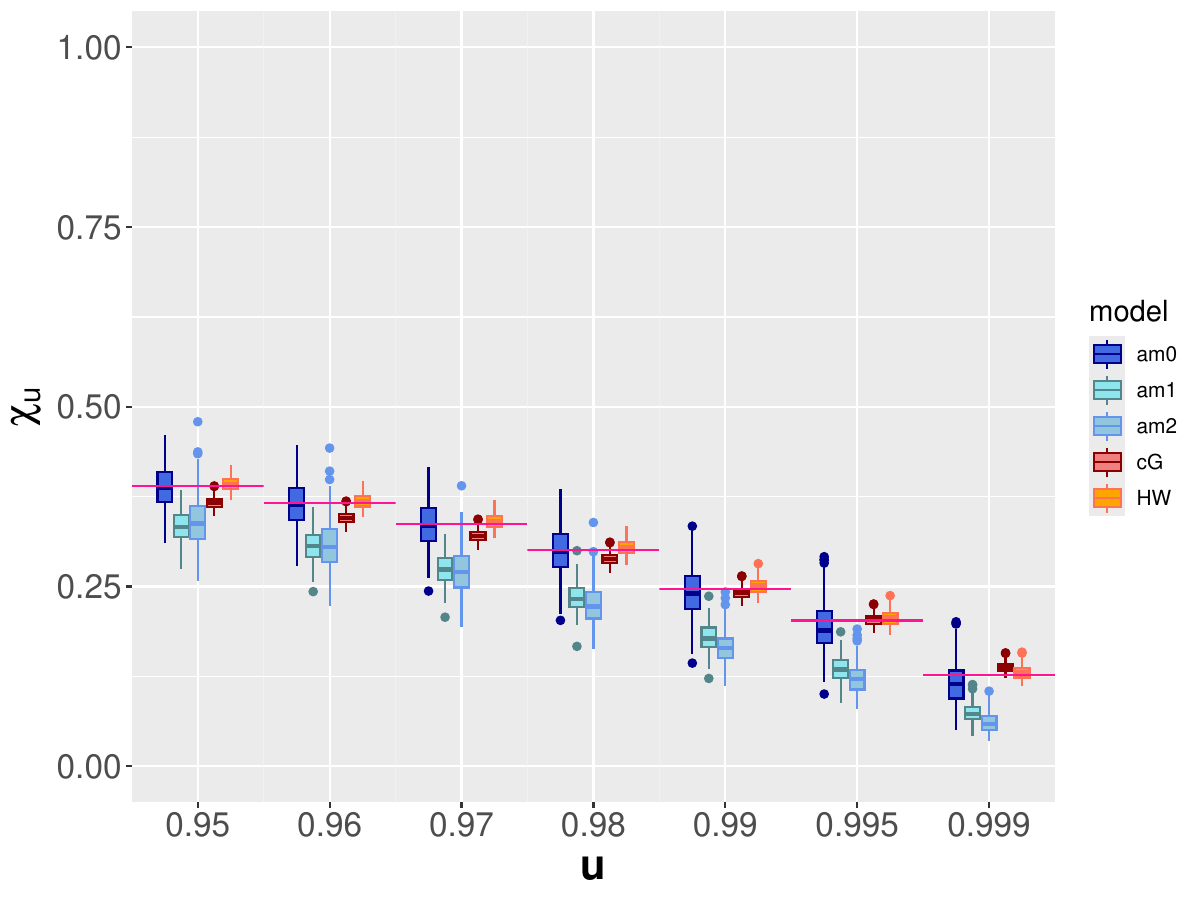}
  \hfill
  \includegraphics[width=0.3\textwidth]{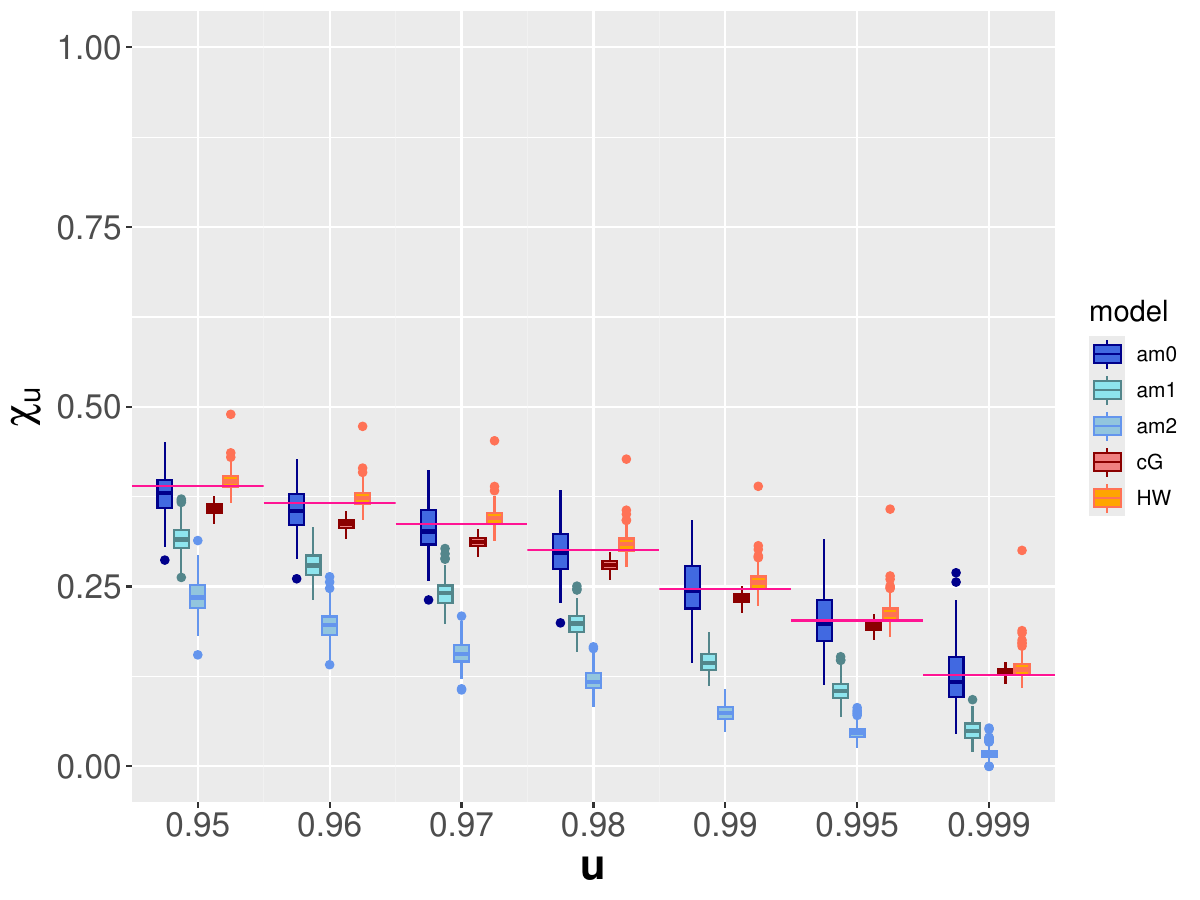}
\caption{$X(\bm{s})$ simulated from a HW process with $\delta=0.4$} \label{fig:box1_hw4}
\end{subfigure}

\begin{subfigure}[t]{1\textwidth}
\centering
\includegraphics[width=0.3\textwidth]{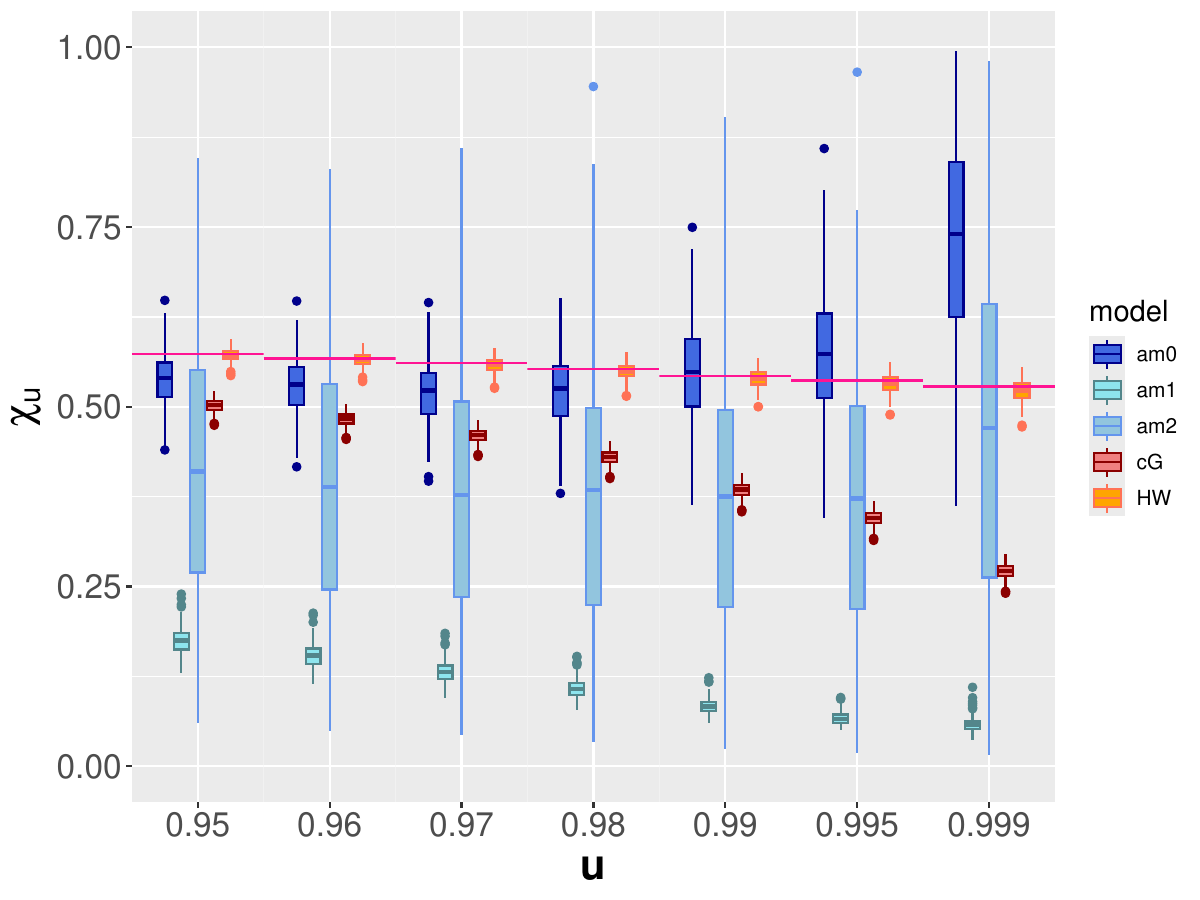}
  \hfill
  \includegraphics[width=0.3\textwidth]{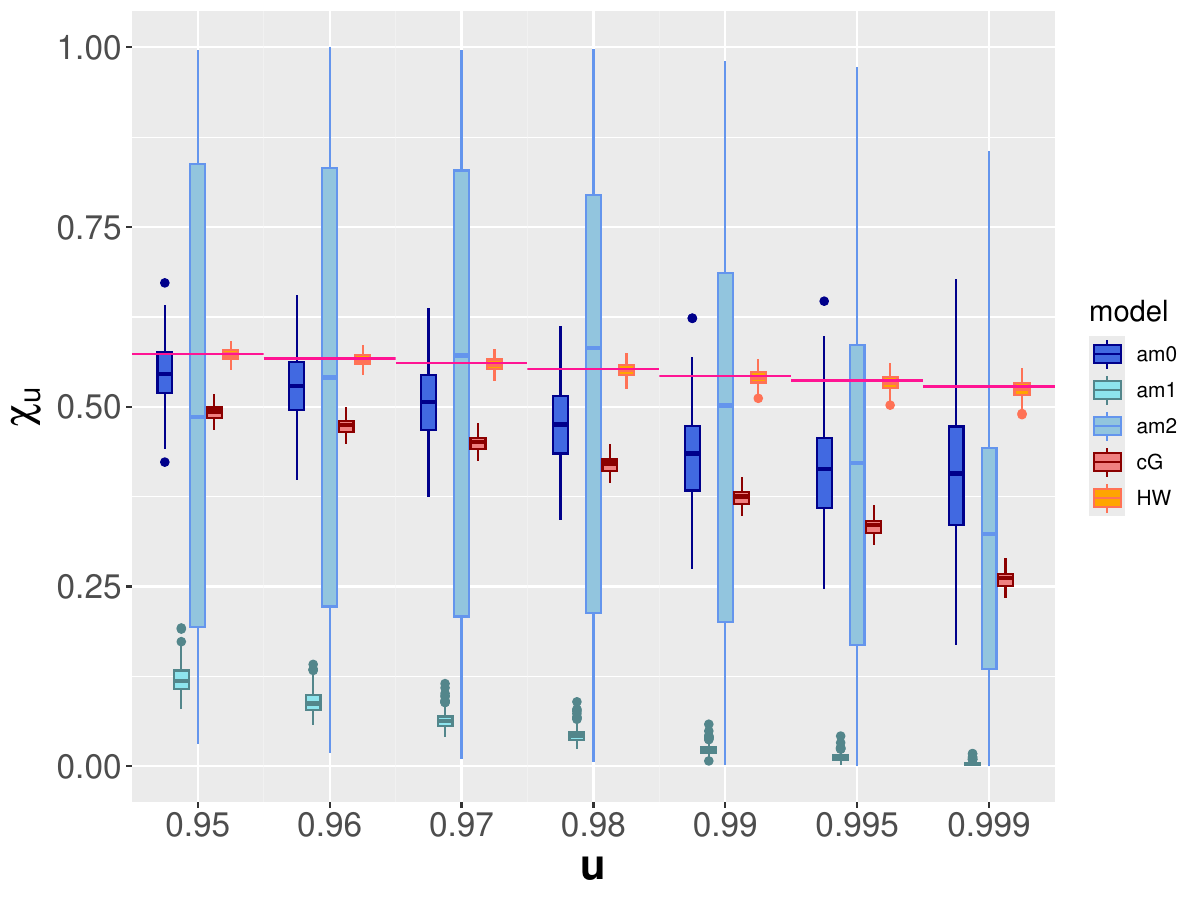}
  \hfill
  \includegraphics[width=0.3\textwidth]{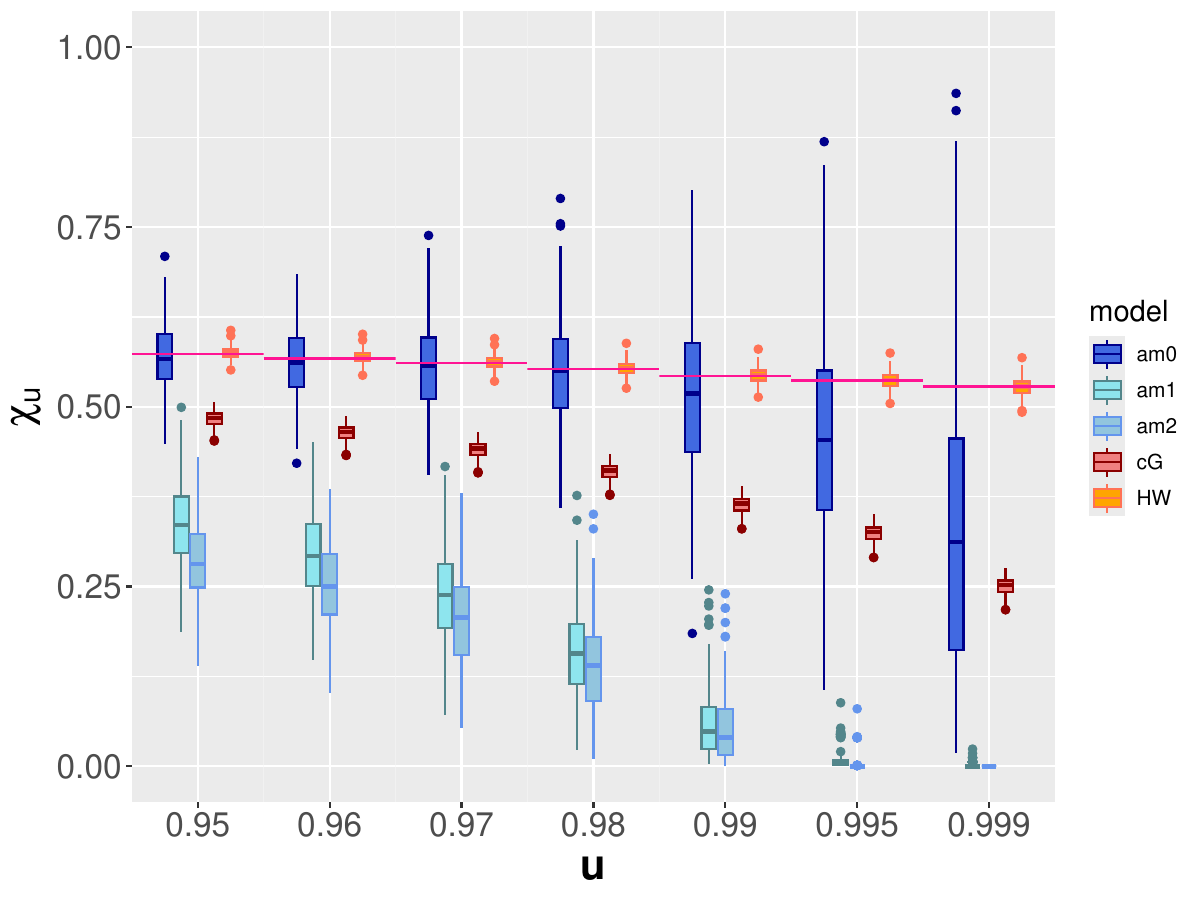}
\caption{$X(\bm{s})$ simulated from a HW process with $\delta=0.6$} \label{fig:box1_hw6}
\end{subfigure}

 \caption{Boxplots of pairwise $\chi_u$ estimates for the pair $\left(X(\bm{s}_2), X(\bm{s}_3)\right)$ at distance $5.78$ apart, calculated for a range of $u$ values and using all $200$ simulated datasets. $X(\bm{s})$ is simulated as specified in (\subref{fig:box1_mvn}) - (\subref{fig:box1_hw6}) with parameter $\bm{\theta}_1$. Boxplots on the left correspond to $d=5$, $d=10$ in the middle and $d=20$ on the right. Estimates in dark blue (labelled am0), turquoise blue (am1) and light blue (am2) are obtained via the empirical angular distribution, the angular distribution in \eqref{eq:am1} and the angular distribution in \eqref{eq:am2}, respectively. Estimates in red come from cG fits, while estimates in orange from HW fits. The pink horizontal line corresponds to the simulated truth.}
 \label{fig:box1}
\end{figure}

Pairwise $\chi_u$ estimates can be found in Figures \ref{fig:box1}(\subref{fig:box1_mvn}) and \ref{fig:box2}(\subref{fig:box2_mvn}) when parameters $\bm{\theta}_1$ and $\bm{\theta}_2$ are used, respectively. These depict boxplots of pairwise $\chi_u$ estimates for a particular location pair, calculated across all simulated datasets and for a variety of $u$ values. Additional results on pairwise $\chi_u$ estimates for all possible location pairs versus distance, are provided in Section \ref{app:ss_extrachis} of the Supplementary Material.

It is evident from these figures that our model does a reasonable job at capturing the extremal dependence of the simulated data. Both estimates stemming from the empirical angular distribution and those from the angular model in \eqref{eq:am1} are unbiased when $d=5$, with an increase in bias in both with increasing dimension; this is more pronounced for $\bm{\theta}_1$ than $\bm{\theta}_2$. Estimates from the angular model in \eqref{eq:am2} exhibit low bias when $d=5$, but become increasingly biased with dimension. The cG and HW models perform very well here, as the cG represents the truth, and the the Gaussian process is also a special case of the HW model. Compared to the cG and HW estimates, the spatial geometric estimates are more variable.

\subsection{Inverted Brown--Resnick data}
\label{sec:ibrdata}

Let $\{X(\bm{s}):\bm{s}\in \mathcal{S}\subset[0,10]^2\}$ be an inverted Brown--Resnick random field, constructed as defined in Section \ref{sec:IBR_gauge}. The Brown--Resnick process from which $X(\bm{s})$ is constructed is parameterised by a semivariogram $\gamma(h) = (h/\lambda)^{\kappa}$, with range parameter $\lambda$ and smooth parameter $\kappa$. Data from this process is simulated in \texttt{R} via the package \texttt{SpatialExtremes} \citep{SpatialExtremes}. We consider parameter settings $\bm{\theta}_1=(\lambda_1,\kappa_1)=(10,1)$ and $\bm{\theta}_2=(\lambda_2,\kappa_2)=(6,0.5)$, as in the Gaussian case. Table \ref{tab:ibr_modsel} below provides information on the gauge type contributing to the preferred radial fit across all simulated datasets. The Gaussian gauge seems to be favoured in lower dimensions, while the Laplace gauge is preferred as $d$ increases.
\begin{table}[h]
\centering
\begin{tabular}{cc|ccccc|ccccc}
\toprule
\multicolumn{2}{l|}{\textbf{Parameter}} & \multicolumn{5}{c|}{$\bm{\theta}_1$}    & \multicolumn{5}{c}{$\bm{\theta}_2$}                 \\ \hline 
\multicolumn{2}{c|}{\textbf{Model}} & GG & G & L & $\text{HW}_{\text{G}}$ & $\text{HW}_{\text{GG}}$ & GG & G & L & $\text{HW}_{\text{G}}$ & $\text{HW}_{\text{GG}}$ \\ \hline 
\multicolumn{1}{c|}{}           & 5   & 9.5 & 48.0 & 37.5 &  1.5 & 3.5  &  11.0 & 54.0 & 34.5 & 0.5 & 0    \\
\multicolumn{1}{c|}{$\bm{d}$} & 10  & 16.0 & 13.5 & 70.0 & 0.5  & 0    &  21.0 & 39.5 & 39.5 & 0 & 0  \\
\multicolumn{1}{c|}{}           & 20  & 29.5 & 0.5 & 69.5 & 0 & 0.5    &  42 & 3 & 55 & 0 & 0 \\  \bottomrule                        
\end{tabular}
\caption{Percentage of times the generalised Gaussian, Gaussian, Laplace, HW Gaussian and HW generalised Gaussian gauges were selected via AIC in the radial fit across all $200$ simulations when $X(\bm{s})$ is an IBR process with range and smooth parameters $\bm{\theta}_1=(10,1)$ and $\bm{\theta}_2=(6,0.5)$.}
\label{tab:ibr_modsel}
\end{table}

Boxplots of pairwise $\chi_u$ estimates --- as described in Section \ref{sec:gausdata} --- can be found in Figures \ref{fig:box1}(\subref{fig:box1_ibr}) and \ref{fig:box2}(\subref{fig:box2_ibr}) for $\bm{\theta}_1$ and $\bm{\theta}_2$, respectively, while Figures \ref{fig:cloud195}(\subref{fig:cloud195_ibr}), \ref{fig:cloud198}(\subref{fig:cloud198_ibr}), \ref{fig:cloud295}(\subref{fig:cloud295_ibr}) and \ref{fig:cloud298}(\subref{fig:cloud298_ibr}) of the Supplement provide additional information on the behaviour of pairwise $\chi_u$ estimates with distance.

This is a more interesting case compared to the Gaussian presented in Section \ref{sec:gausdata} because now the cG model is misspecified. Our model again seems to be able to capture the extremal dependence of the data well, giving $\chi_u$ estimates with low bias when the empirical angular distribution is used to simulate new data. This is in spite of the fact that all gauge functions are misspecified. Conversely, both cG and HW estimates seem to suffer from some degree of bias, although they remain less variable compared to our model. The Gaussian-process angular model and the gauge-based angular model appear relatively unbiased for $d=5$, but both become more biased as $d$ increases.

\subsection{Brown--Resnick data}
\label{sec:brdata}
Let $\{X(\bm{s}):\bm{s}\in \mathcal{S}\subset[0,10]^2\}$ be a Brown--Resnick random field with range parameter $\lambda$, smooth parameter $\kappa$, and transformed to standard exponential margins. The parameter settings are $\bm{\theta}_1=(\lambda_1, \kappa_1) = (3,1)$ and $\bm{\theta}_2=(\lambda_2, \kappa_2) = (1,0.5)$, in order to achieve similar strength of extremal dependence compared to the Gaussian and IBR cases. It is worth noting here that the geometric framework is ill-suited for BR data as the true gauge function corresponding to BR processes is degenerate, concentrating entirely along the diagonal \citep{Nolde2020}. Nevertheless, we include it as a special case of extreme model misspecification. 

Table \ref{tab:br_modsel} below provides information on the gauge type contributing to the preferred radial fit across all simulated datasets, when the simulated range and smooth parameters are $\bm{\theta}_1$ and $\bm{\theta}_2$, respectively. The choice of gauge function is split between the generalised Gaussian and the HW gauge with an underlying generalised Gaussian when $\bm{\theta}_1$ is used, but the generalised Gaussian seems preferable when $\bm{\theta}_2$ is used --- an interesting observation, given the HW model can accommodate AD while the generalised Gaussian one cannot.
\begin{table}[h]
\centering
\begin{tabular}{cc|ccccc|ccccc}
\toprule
\multicolumn{2}{l|}{\textbf{Parameter}} & \multicolumn{5}{c|}{$\bm{\theta}_1$}    & \multicolumn{5}{c}{$\bm{\theta}_2$}                 \\ \hline 
\multicolumn{2}{c|}{\textbf{Model}} & GG & G & L & $\text{HW}_{\text{G}}$ & $\text{HW}_{\text{GG}}$ & GG & G & L & $\text{HW}_{\text{G}}$ & $\text{HW}_{\text{GG}}$ \\ \hline 
\multicolumn{1}{c|}{}           & 5   & 49.5 & 0 & 0 & 0 & 50.5  &  88 & 0 & 1 & 2 & 9    \\
\multicolumn{1}{c|}{$\bm{d}$} & 10  & 58.5 & 0 & 0 & 0 & 41.5  &  100 & 0 & 0 & 0 & 0  \\
\multicolumn{1}{c|}{}           & 20  & 35.5 & 0 & 0 & 0 & 64.5  &  89 & 1 & 0 & 0 & 10 \\  \bottomrule                        
\end{tabular}
\caption{Percentage of times the generalised Gaussian, Gaussian, Laplace, HW Gaussian and HW generalised Gaussian gauges were selected via AIC in the radial fit across all $200$ simulations when $X(\bm{s})$ is a BR process with range and smooth parameters $\bm{\theta}_1=(3,1)$ and $\bm{\theta}_2=(1,0.5)$.}
\label{tab:br_modsel}
\end{table}

\begin{figure}[p!]
\centering
\begin{subfigure}[t]{1\textwidth}
\centering
\includegraphics[width=0.3\textwidth]{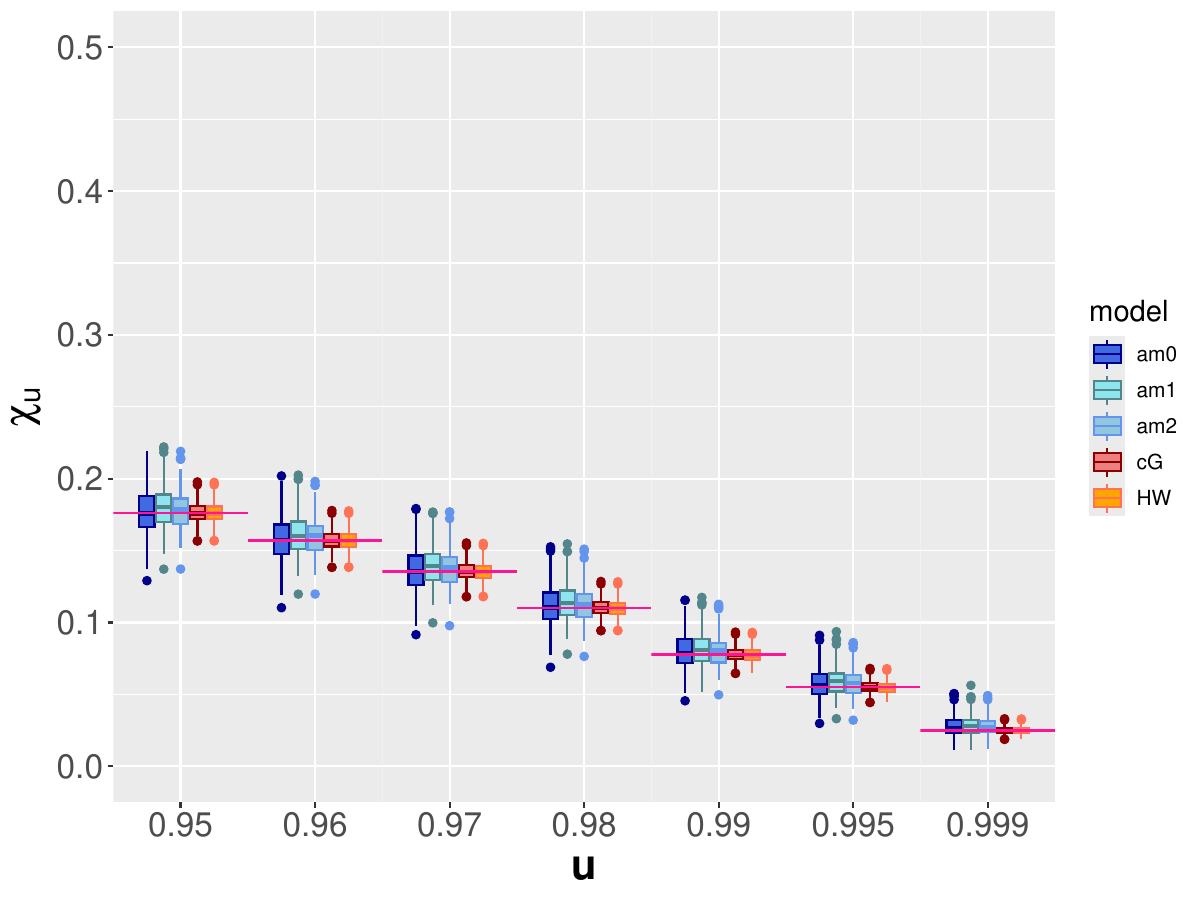}
  \hfill
  \includegraphics[width=0.3\textwidth]{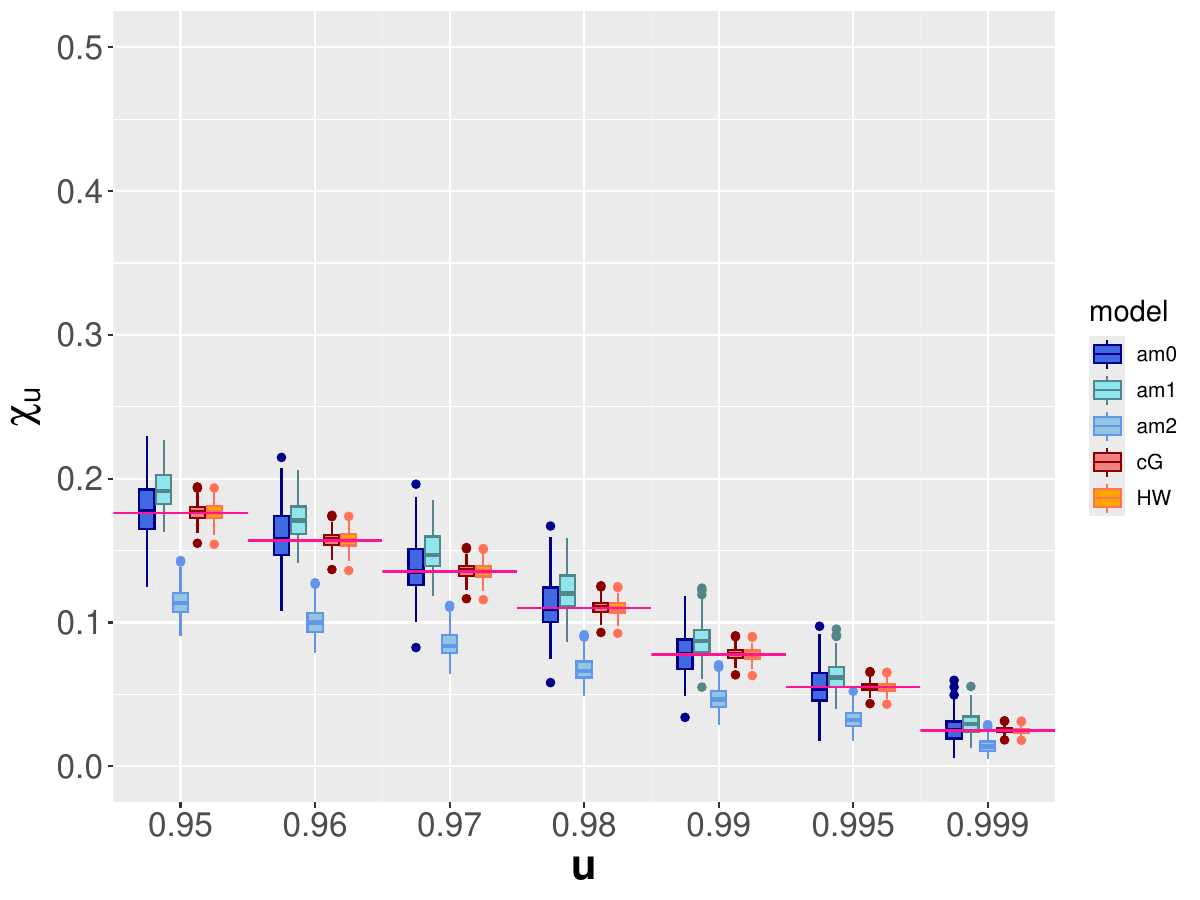}
  \hfill
  \includegraphics[width=0.3\textwidth]{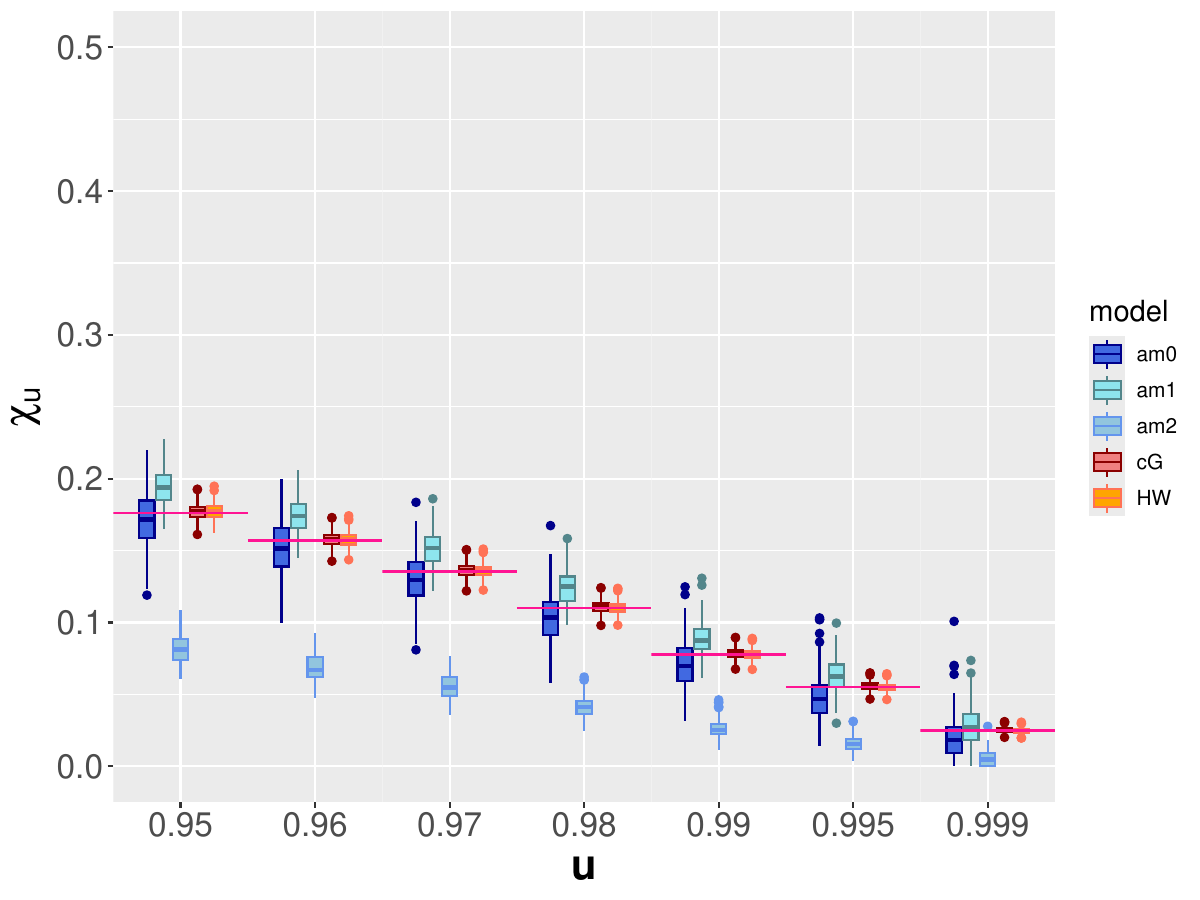} 
\caption{$X(\bm{s})$ simulated from a Gaussian process} \label{fig:box2_mvn}
\end{subfigure}

\begin{subfigure}[t]{1\textwidth}
\centering
\includegraphics[width=0.3\textwidth]{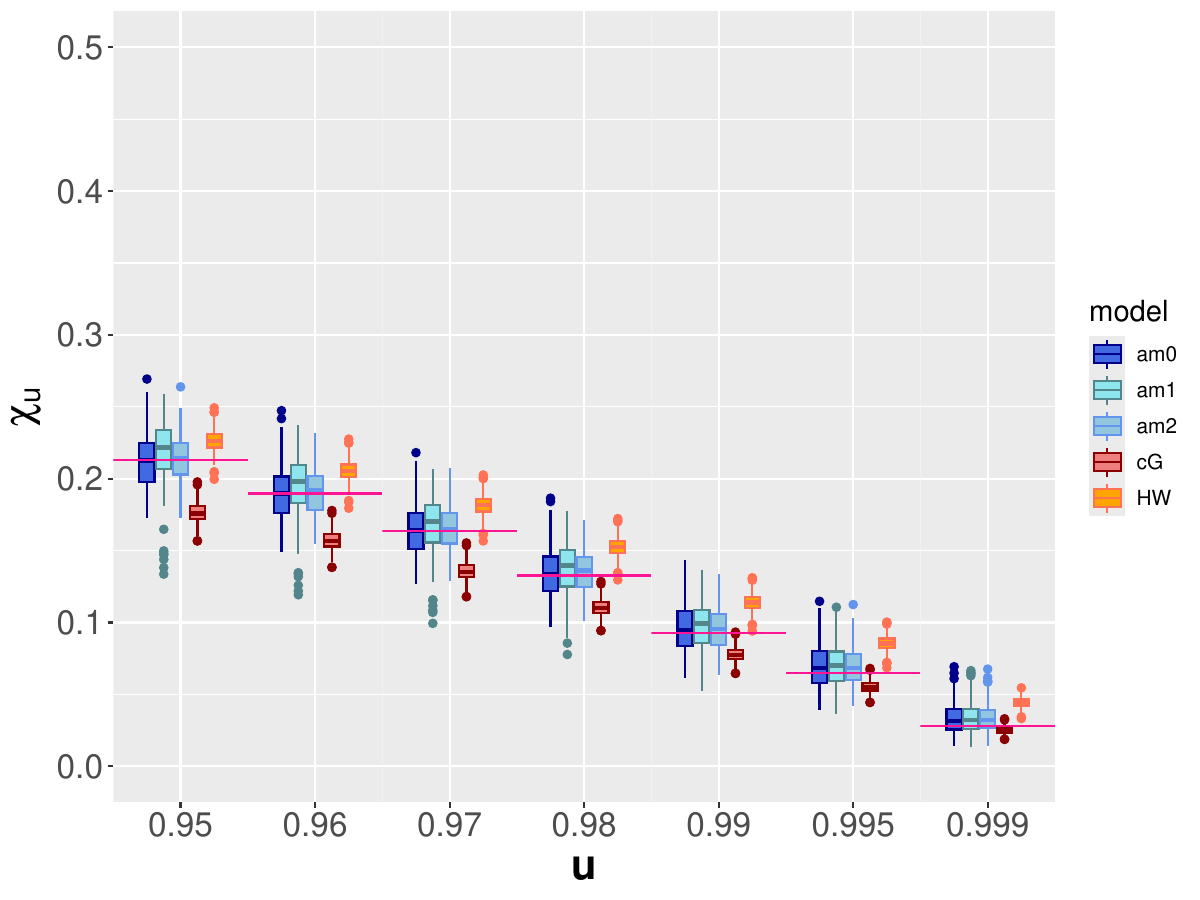}
  \hfill
  \includegraphics[width=0.3\textwidth]{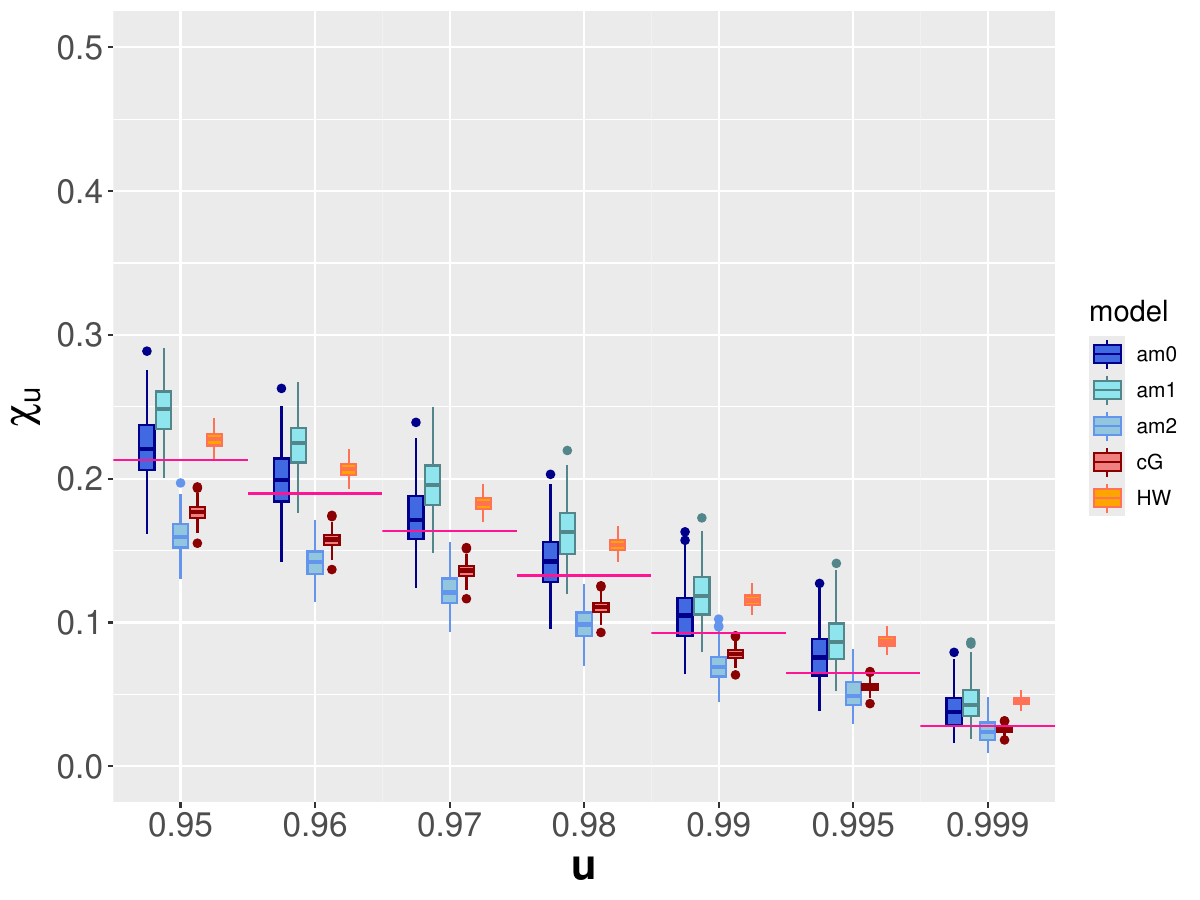}
  \hfill
  \includegraphics[width=0.3\textwidth]{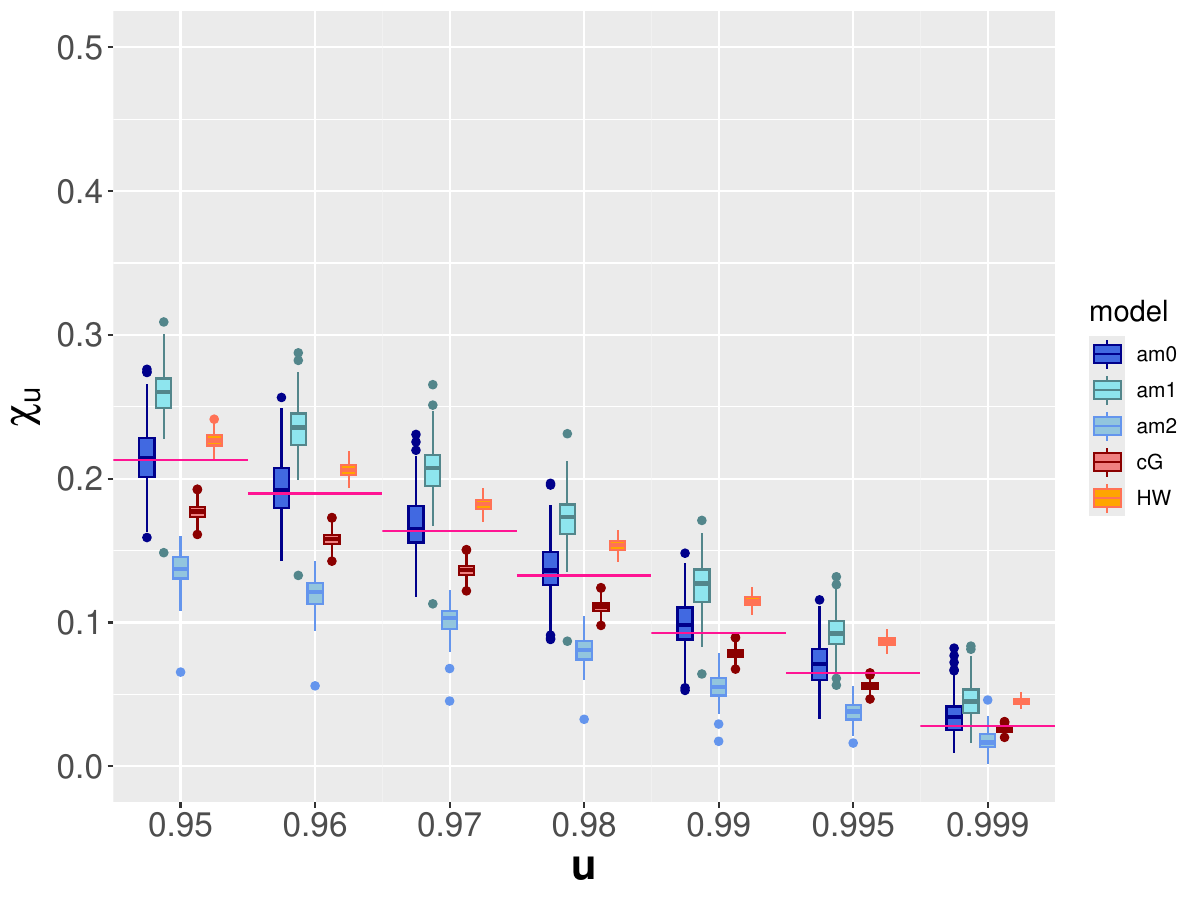} 
\caption{$X(\bm{s})$ simulated from an IBR process} \label{fig:box2_ibr}
\end{subfigure}

\begin{subfigure}[t]{1\textwidth}
\centering
\includegraphics[width=0.3\textwidth]{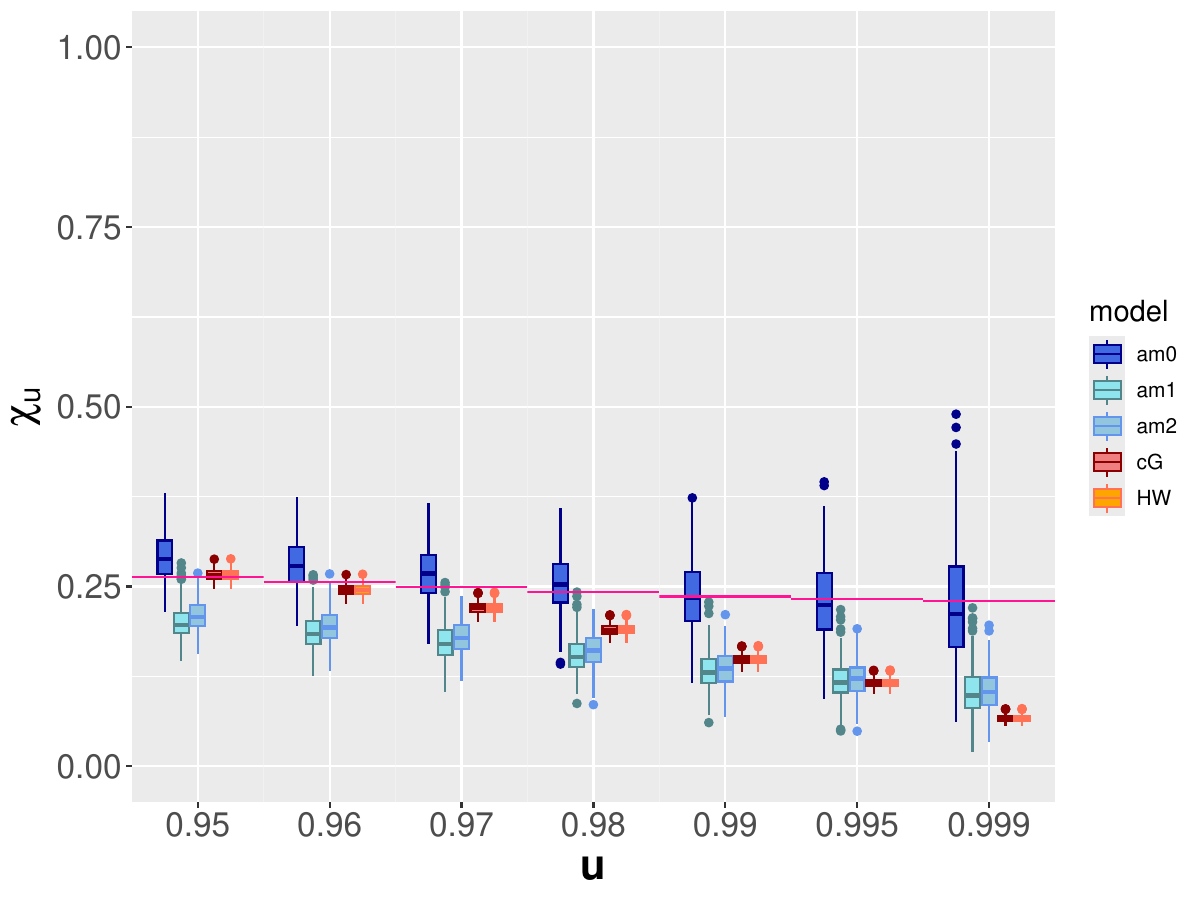}
  \hfill
  \includegraphics[width=0.3\textwidth]{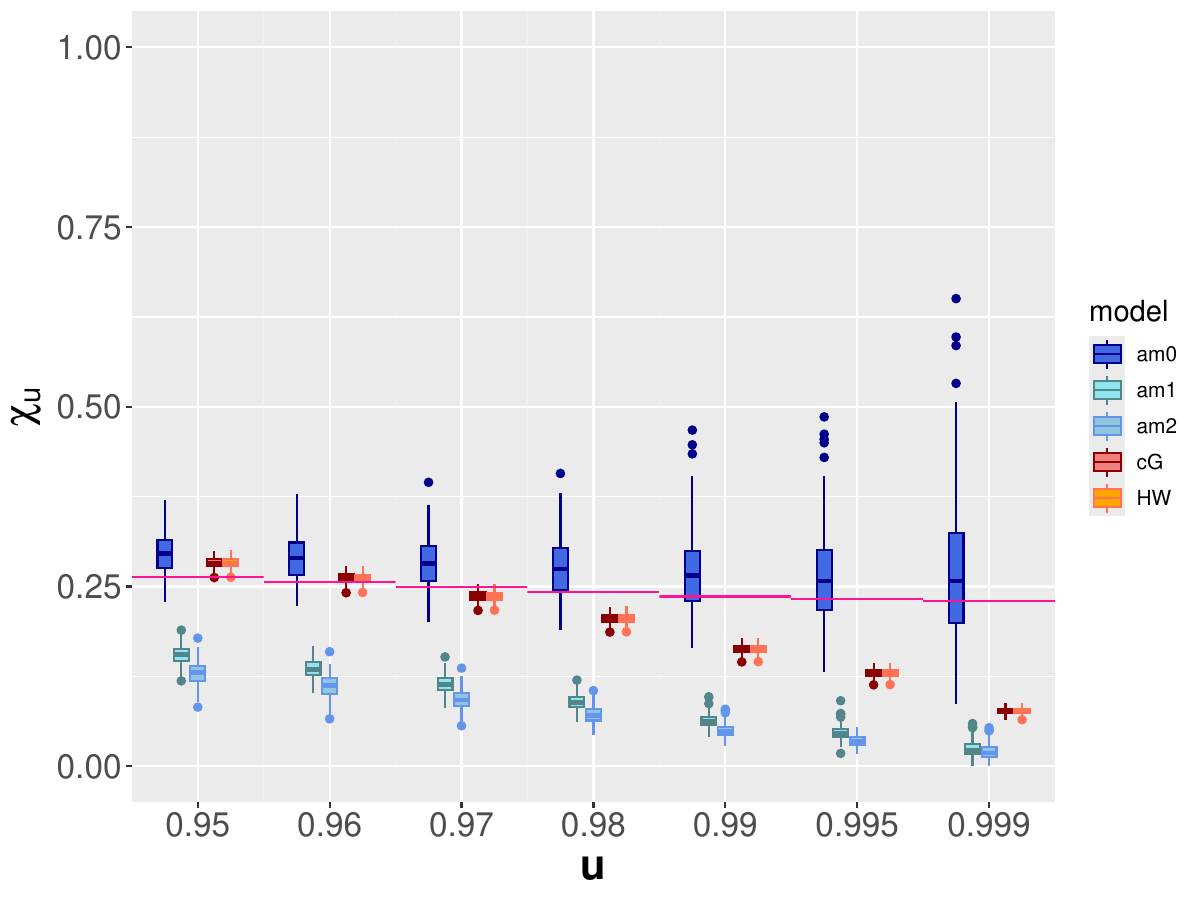}
  \hfill
  \includegraphics[width=0.3\textwidth]{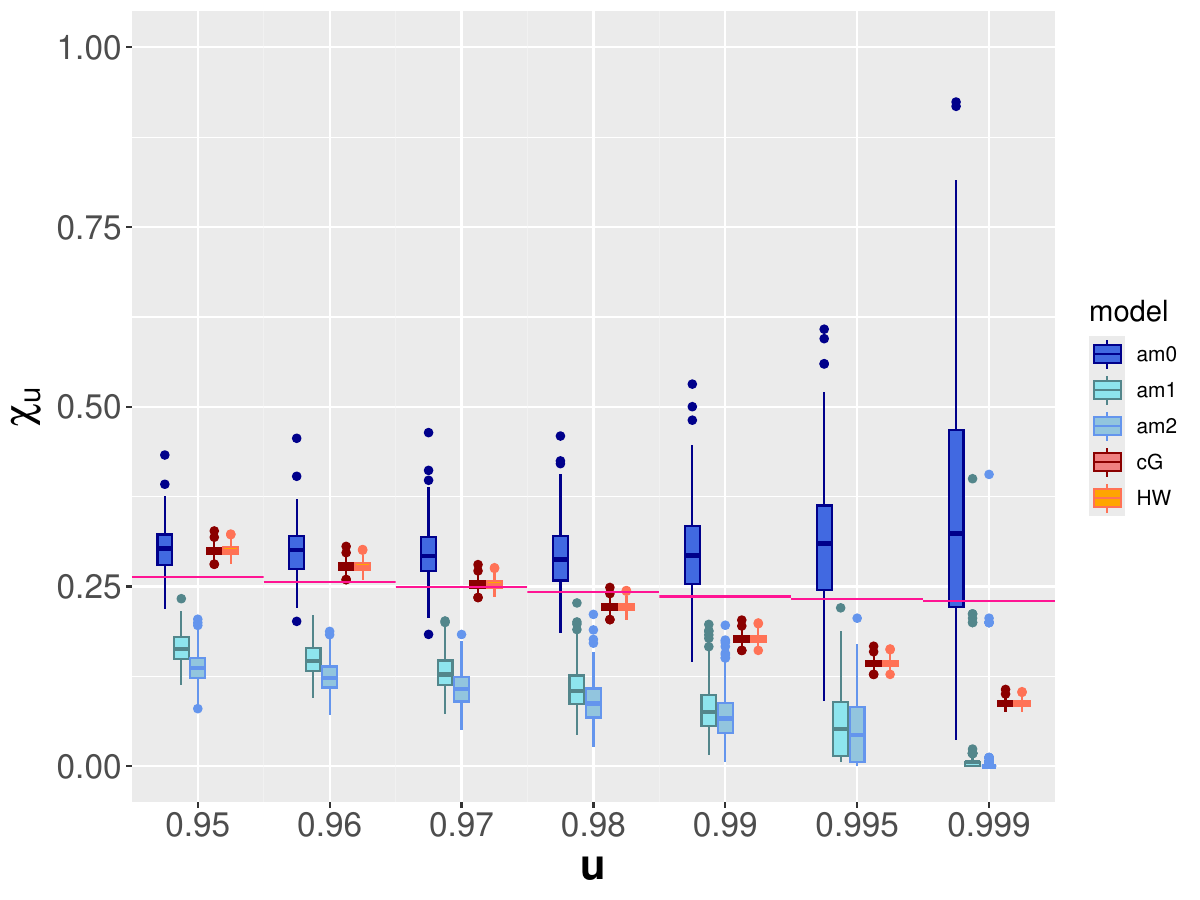}
\caption{$X(\bm{s})$ simulated from a BR process} \label{fig:box2_br}
\end{subfigure}

\begin{subfigure}[t]{1\textwidth}
\centering
\includegraphics[width=0.3\textwidth]{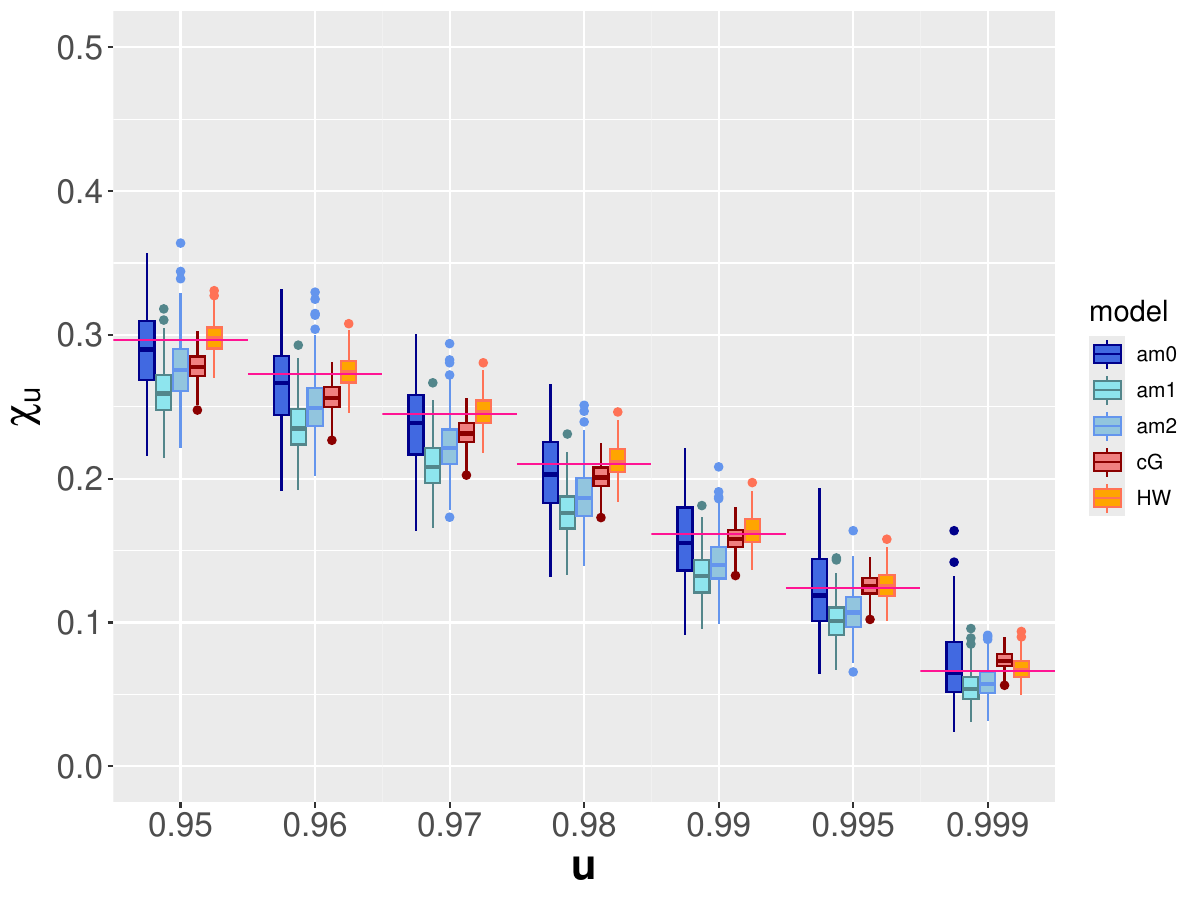}
  \hfill
  \includegraphics[width=0.3\textwidth]{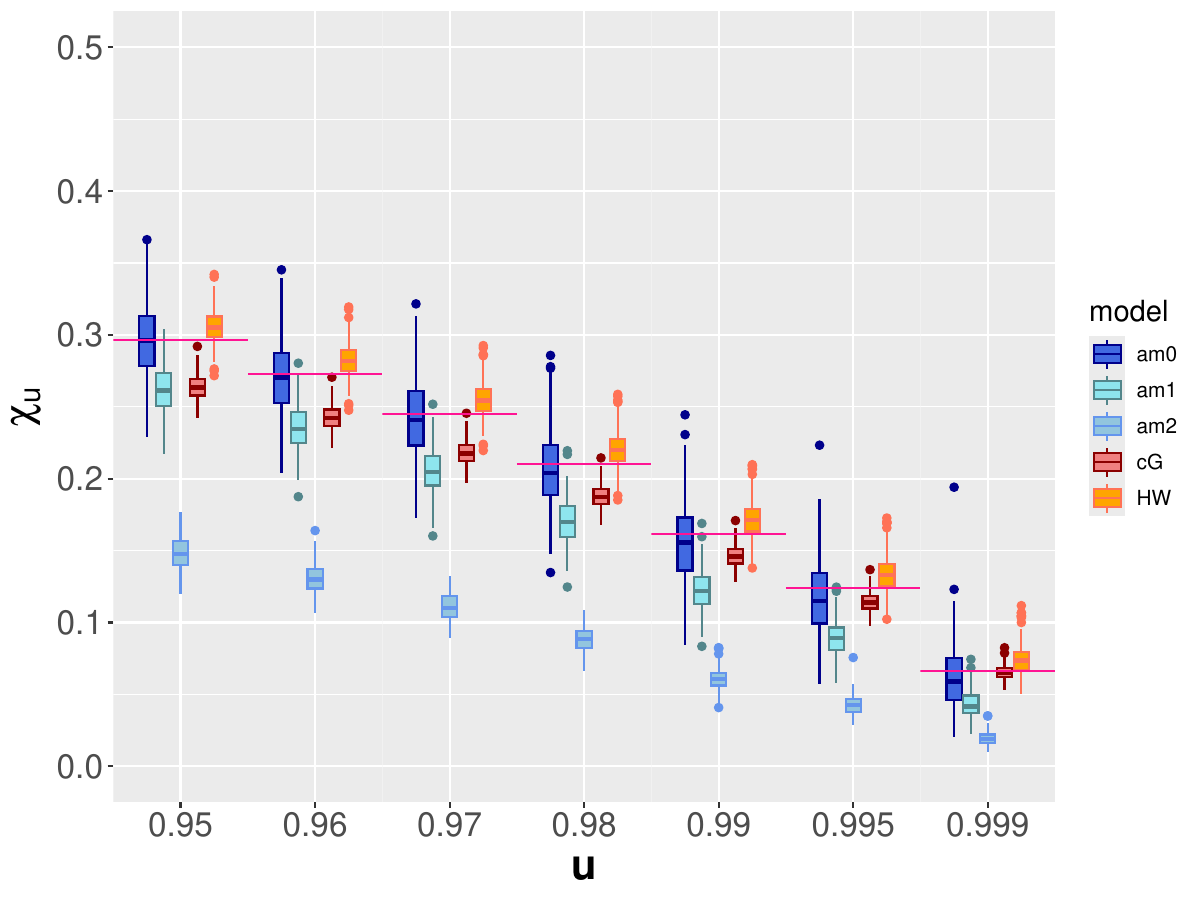}
  \hfill
  \includegraphics[width=0.3\textwidth]{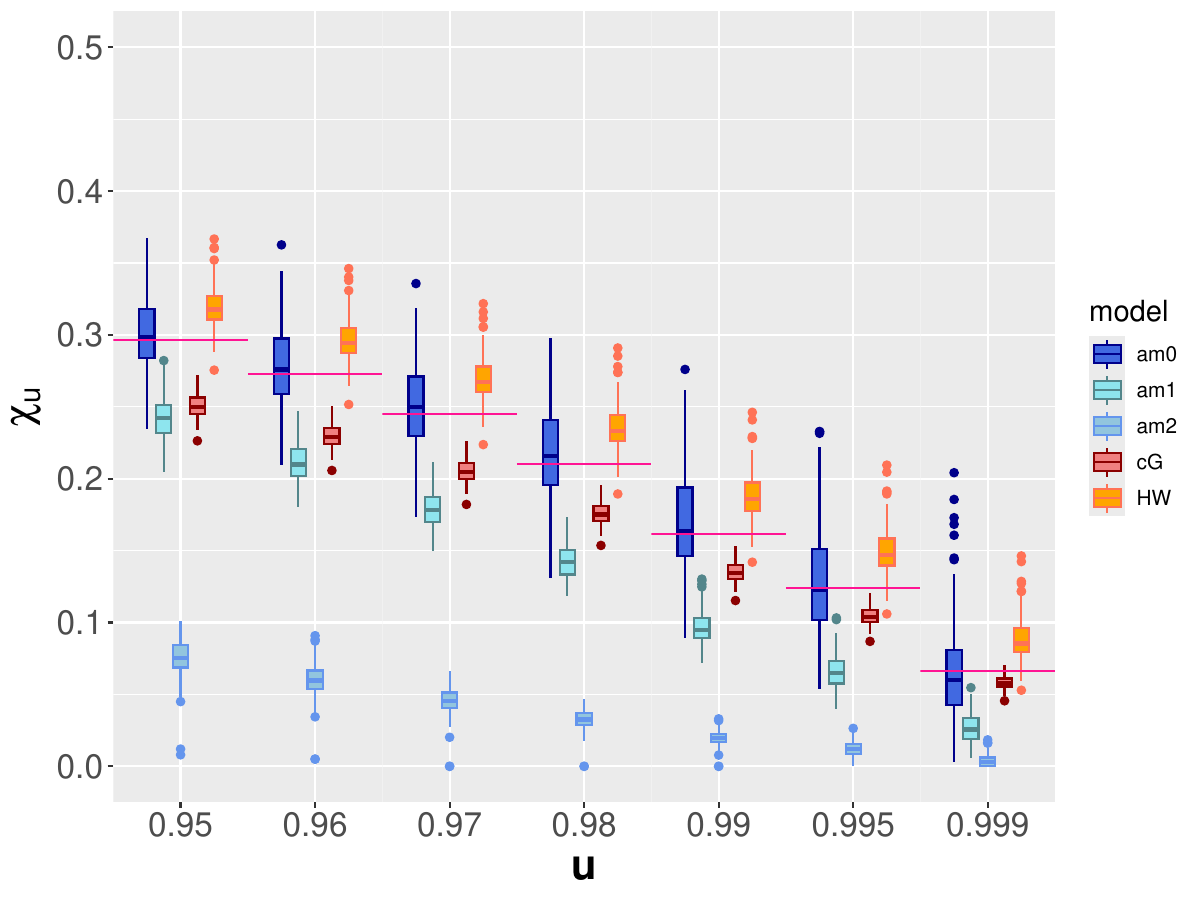}
\caption{$X(\bm{s})$ simulated from a HW process with $\delta=0.4$} \label{fig:box2_hw4}
\end{subfigure}

\begin{subfigure}[t]{1\textwidth}
\centering
\includegraphics[width=0.3\textwidth]{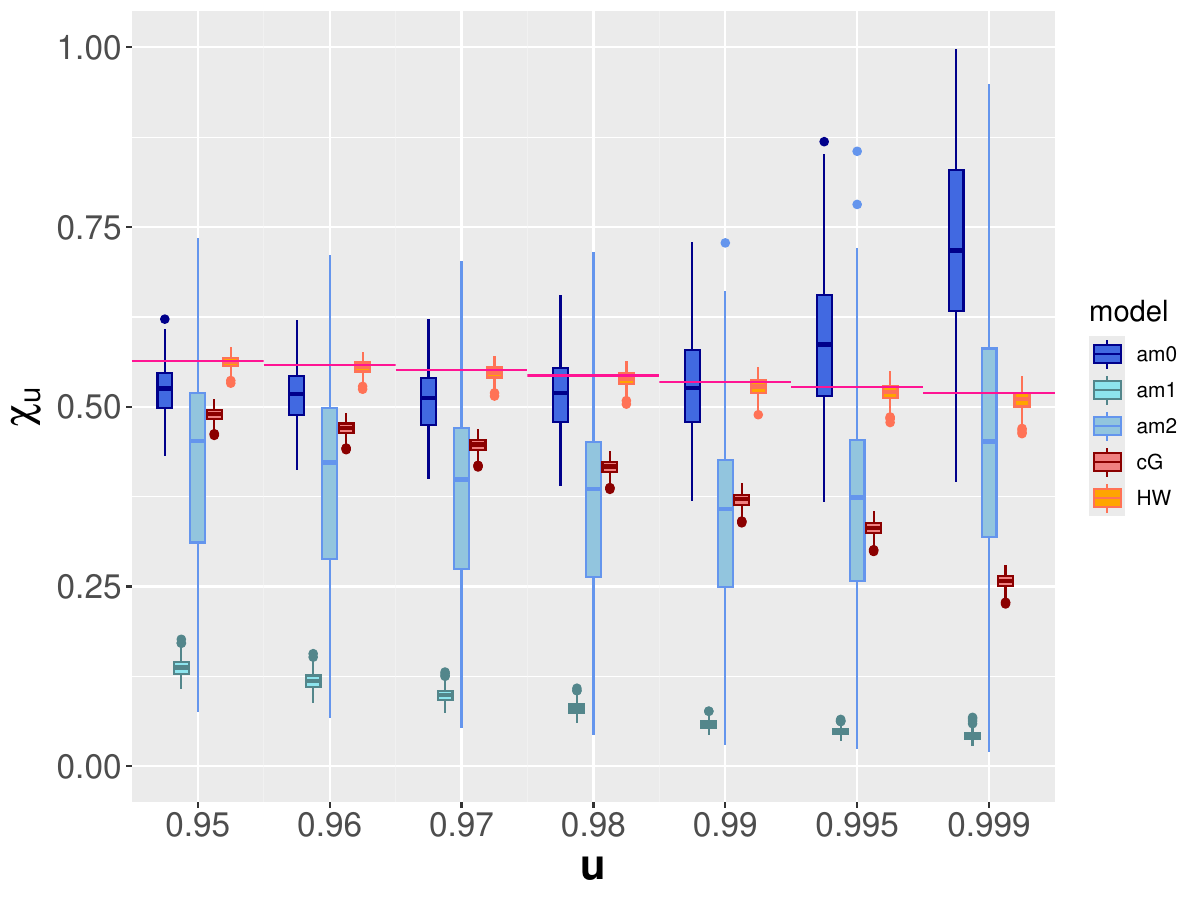}
  \hfill
  \includegraphics[width=0.3\textwidth]{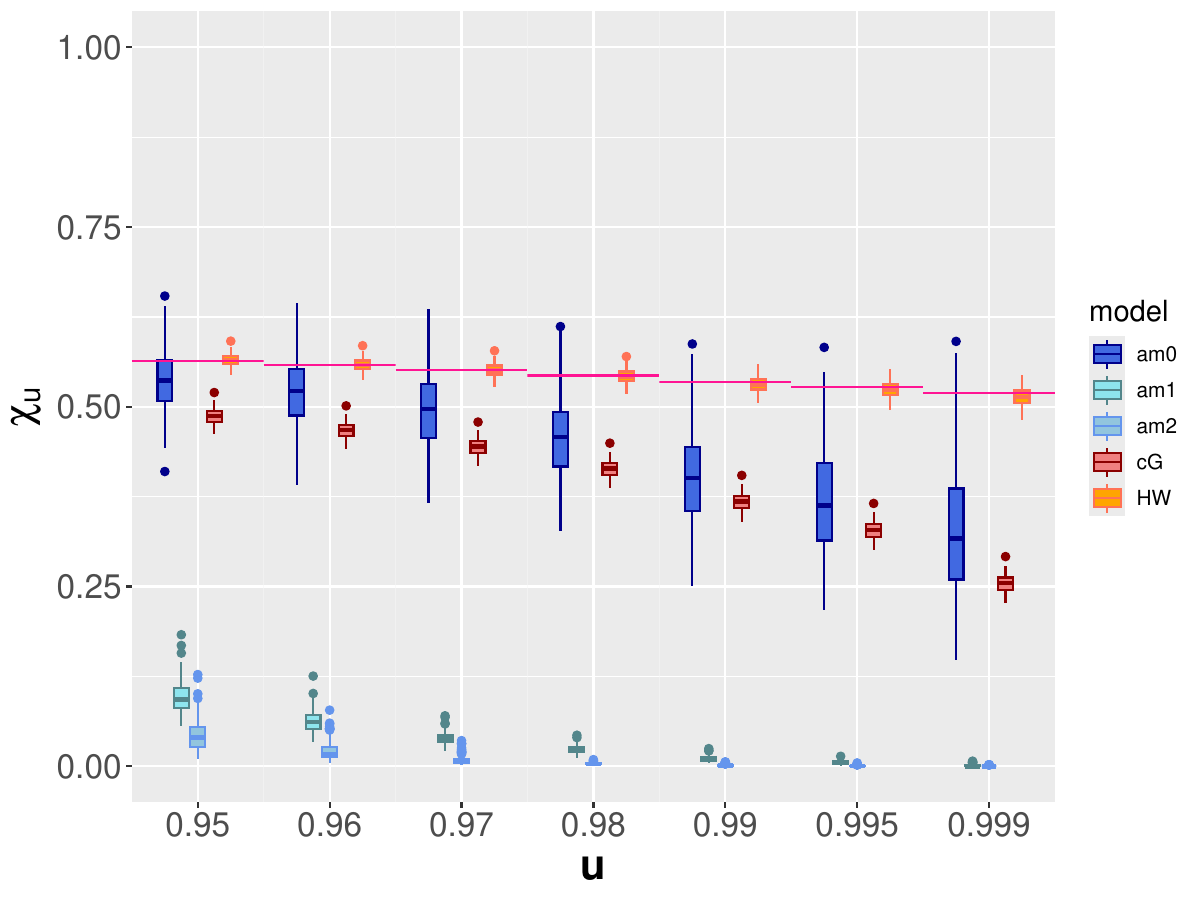}
  \hfill
  \includegraphics[width=0.3\textwidth]{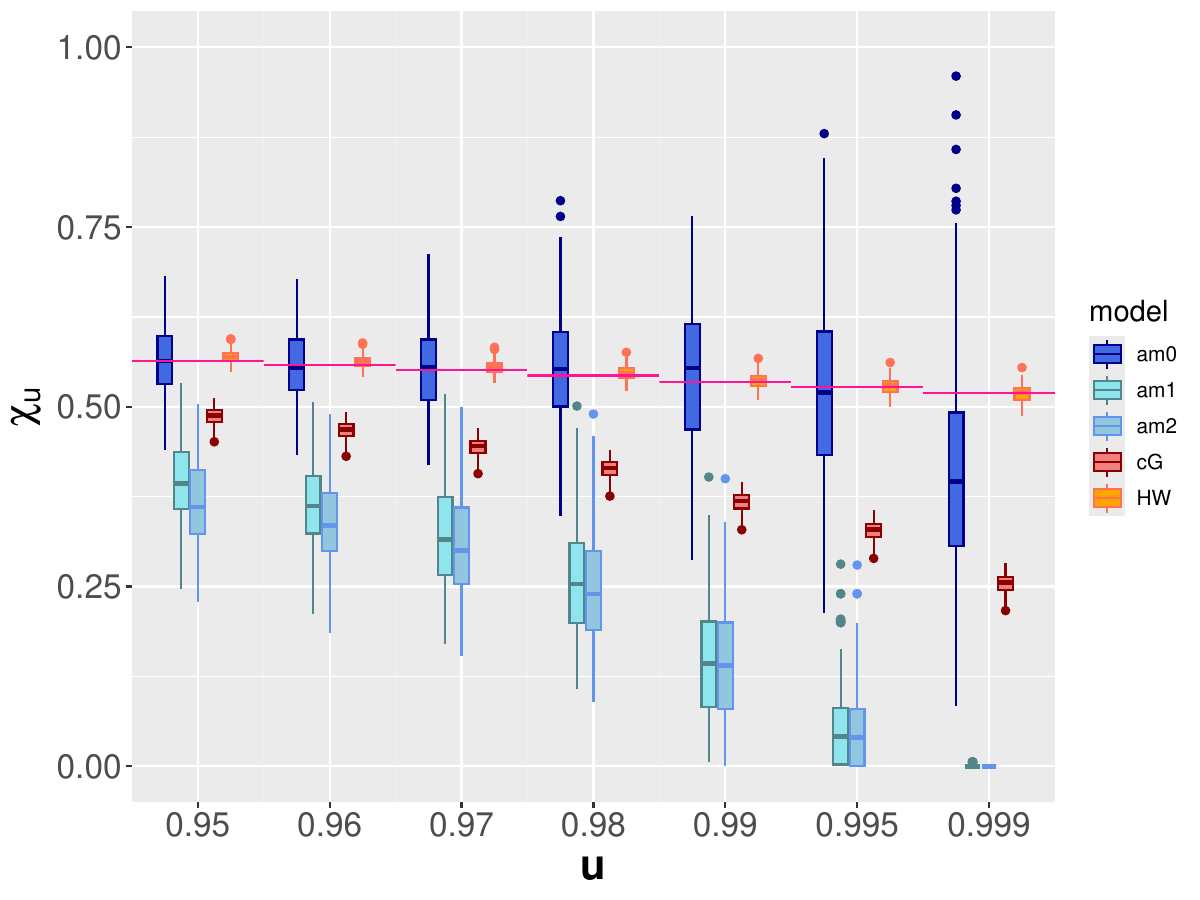}
\caption{$X(\bm{s})$ simulated from a HW process with $\delta=0.6$} \label{fig:box2_hw6}
\end{subfigure}

 \caption{Boxplots of pairwise $\chi_u$ estimates for the pair $\left(X(\bm{s}_2), X(\bm{s}_3)\right)$ at distance $5.78$ apart, calculated for a range of $u$ values and using all $200$ simulated datasets. $X(\bm{s})$ is simulated as specified in (\subref{fig:box2_mvn}) - (\subref{fig:box2_hw6}) with parameter $\bm{\theta}_2$. Boxplots on the left correspond to $d=5$, $d=10$ in the middle and $d=20$ on the right. Estimates in dark blue (labelled am0), turquoise blue (am1) and light blue (am2) are obtained via the empirical angular distribution, the angular distribution in \eqref{eq:am1} and the angular distribution in \eqref{eq:am2}, respectively. Estimates in red come from cG fits, while estimates in orange from HW fits. The pink horizontal line corresponds to the simulated truth.}
 \label{fig:box2}
\end{figure}

As before, boxplots of pairwise $\chi_u$ estimates can be found in Figures \ref{fig:box1}(\subref{fig:box1_br}) and \ref{fig:box2}(\subref{fig:box2_br}) for $\bm{\theta}_1$ and $\bm{\theta}_2$, respectively. Figures \ref{fig:cloud195}(\subref{fig:cloud195_br}), \ref{fig:cloud198}(\subref{fig:cloud198_br}), \ref{fig:cloud295}(\subref{fig:cloud295_br}) and \ref{fig:cloud298}(\subref{fig:cloud298_br}) in Section \ref{app:ss_extrachis} of the Supplementary Material provide information on pairwise $\chi_u$ estimates as a function of distance.

Interestingly, the geometric model based on the empirical angular distribution does a decent job at capturing the extremal dependence of the simulated datasets. This is not the case however for the Gaussian-process angular model, which consistently underestimates the dependence. The gauge-based angular model also underestimates the extremal dependence, the bias once again increasing with dimension. Variability of the geometric estimates remains high compared to the cG and HW models. 
The cG and HW models are consistently biased, failing to capture the dependence strength of the simulated data. This is somewhat unexpected of the HW model which can accommodate AD as well as AI. To the best of our knowledge, the performance of the HW model has not been extensively tested on BR data, but it appears to perform poorly. 

\subsection{Huser--Wadsworth data exhibiting AI}
\label{sec:hw4data}
We simulate datasets from the process of \citet{Huser2019}, briefly introduced in Section \ref{sec:HW_gauge} and transformed to have standard exponential margins, using the \texttt{SpatialADAI} package in \texttt{R}. We use a Gaussian dependence structure for the process $W_{\text{E}}(\bm{s})$ and the same dimension and parameter configurations as in Section \ref{sec:gausdata}, namely $\bm{\theta}_1=(\lambda_1,\kappa_1)=(10,1)$, $\bm{\theta}_2=(\lambda_2,\kappa_2)=(6,0.5)$. The mixing parameter $\delta$ is set to $0.4$, thus inducing AI. Table \ref{tab:hw4_modsel} below provides information on the gauge type contributing to the preferred radial fit across all simulated datasets, when the simulated correlation parameters are $\bm{\theta}_1$ and $\bm{\theta}_2$, respectively. Interestingly, the simpler, misspecified models based on the Gaussian and generalised Gaussian gauges seem to be preferred over the more complex, yet correctly specified, HW gauges. 

Figures \ref{fig:box1}(\subref{fig:box1_hw4}) and \ref{fig:box2}(\subref{fig:box2_hw4}) show boxplots of pairwise $\chi_u$ estimates for $\bm{\theta}_1$ and $\bm{\theta}_2$, respectively. Additionally, Figures \ref{fig:cloud195}(\subref{fig:cloud195_hw4}), \ref{fig:cloud198}(\subref{fig:cloud198_hw4}), \ref{fig:cloud295}(\subref{fig:cloud295_hw4}) and \ref{fig:cloud298}(\subref{fig:cloud298_hw4}) in Section \ref{app:ss_extrachis} of the Supplement provide information on pairwise $\chi_u$ estimates as a function of distance. As expected, the estimates stemming from the HW model are performing best, both in terms of variance and bias, although a slight deterioration is observed when $d=20$. We have investigated this anomaly, finding that $\lambda$ is generally slightly underestimated, and $\delta$ overestimated, but it is not clear why this is the case. Estimates from the spatial geometric model with resampled angles are unbiased in all dimensions, but are considerably more variable. Once again, the process-based angular model seems to underestimate the extremal dependence, a feature that becomes more prominent with increasing dimension. The gauge-based angular model showcases a similar, but slightly worse, behaviour compared to the process-based angular model, except perhaps in the case $d=5$. Finally, cG estimates also exhibit a degree of bias, which becomes worse the higher the dimension.
\begin{table}[h]
\centering
\begin{tabular}{cc|ccccc|ccccc}
\toprule
\multicolumn{2}{l|}{\textbf{Parameter}} & \multicolumn{5}{c|}{$\bm{\theta}_1$}    & \multicolumn{5}{c}{$\bm{\theta}_2$}                 \\ \hline 
\multicolumn{2}{c|}{\textbf{Model}} & GG & G & L & $\text{HW}_{\text{G}}$ & $\text{HW}_{\text{GG}}$ & GG & G & L & $\text{HW}_{\text{G}}$ & $\text{HW}_{\text{GG}}$ \\ \hline 
\multicolumn{1}{c|}{}           & 5   & 17.5 & 78.0 & 0 & 4.0 & 0.5  &  7.0 & 54.0 & 5.0 & 26.5 & 7.5    \\
\multicolumn{1}{c|}{$\bm{d}$} & 10  & 25   & 75   & 0 & 0   & 0    &  35.0 & 57.0 & 1.5 & 2.0 & 4.5  \\
\multicolumn{1}{c|}{}           & 20  & 68.5 & 31.5 & 0 & 0   & 0    &  35.5 & 60.5 & 0.5 & 1.0 & 2.5 \\  \bottomrule                        
\end{tabular}
\caption{Percentage of times the generalised Gaussian, Gaussian, Laplace, HW Gaussian and HW generalised Gaussian gauges were selected via AIC in the radial fit across all $200$ simulations when $X(\bm{s})$ is a HW process with correlation parameters $\bm{\theta}_1=(10,1)$ and $\bm{\theta}_2=(6,0.5)$ and mixing parameter $\delta=0.4$.}
\label{tab:hw4_modsel}
\end{table}

\subsection{Huser--Wadsworth data exhibiting AD}
\label{sec:hw6data}
We use the same underlying model as in Section \ref{sec:hw4data}, but set the mixing parameter $\delta$ to $0.6$, introducing AD. We take the other parameters to be $\bm{\theta}_1=(\lambda_1,\kappa_1)=(3,1)$, $\bm{\theta}_2=(\lambda_2,\kappa_2)=(1,0.5)$. Table \ref{tab:hw6_modsel} below provides information on the gauge type contributing to the preferred radial fit across all simulated datasets, when the dependence parameters are $\bm{\theta}_1$ and $\bm{\theta}_2$, respectively. Contrary to the misspecified BR case, here there is a clear preference for the (correctly specified) HW gauges. 
\begin{table}[h]
\centering
\begin{tabular}{cc|ccccc|ccccc}
\toprule
\multicolumn{2}{l|}{\textbf{Parameter}} & \multicolumn{5}{c|}{$\bm{\theta}_1$}    & \multicolumn{5}{c}{$\bm{\theta}_2$}                 \\ \hline 
\multicolumn{2}{c|}{\textbf{Model}} & GG & G & L & $\text{HW}_{\text{G}}$ & $\text{HW}_{\text{GG}}$ & GG & G & L & $\text{HW}_{\text{G}}$ & $\text{HW}_{\text{GG}}$ \\ \hline 
\multicolumn{1}{c|}{}           & 5   & 0   & 0    & 0   & 46   & 54    &  0 & 0 & 0 & 27 & 73    \\
\multicolumn{1}{c|}{$\bm{d}$} & 10  & 0   & 0    & 0   & 73.5 & 26.5  &  0 & 0 & 0 & 49.5 & 50.5  \\
\multicolumn{1}{c|}{}           & 20  & 2.5 & 13.5 & 1.0 & 59.5 & 23.5  &  0.5 & 0 & 0.5 & 7.5 & 91.5 \\  \bottomrule                        
\end{tabular}
\caption{Percentage of times the generalised Gaussian, Gaussian, Laplace, HW Gaussian and HW generalised Gaussian gauges were selected via AIC in the radial fit across all $200$ simulations when $X(\bm{s})$ is a HW process with correlation parameters $\bm{\theta}_1=(3,1)$ and $\bm{\theta}_2=(1,0.5)$ and mixing parameter $\delta=0.6$.}
\label{tab:hw6_modsel}
\end{table}

Figures \ref{fig:box1}(\subref{fig:box1_hw6}) and \ref{fig:box2}(\subref{fig:box2_hw6}) show boxplots of pairwise $\chi_u$ estimates for $\bm{\theta}_1$ and $\bm{\theta}_2$, respectively, while, Figures \ref{fig:cloud195}(\subref{fig:cloud195_hw6}), \ref{fig:cloud198}(\subref{fig:cloud198_hw6}), \ref{fig:cloud295}(\subref{fig:cloud295_hw6}) and \ref{fig:cloud298}(\subref{fig:cloud298_hw6}) in the Supplementary Material provide information on pairwise $\chi_u$ estimates versus distance. In this case only the HW model provides an accurate description of the extremal dependence of the data. The geometric model with resampled angles, although highly variable, is the second best performing model, especially for lower $u$ values. Finally, the cG model as well as the process- and gauge-based angular models all fare poorly at this estimation task. The gauge-based angular model in particular has a similarly poor performance to its process-based counterpart at larger dimensions, yet seems to struggle significantly in the case $d=5$, where not only bias, but also increased variability are present. This is because the estimates from this model are themselves highly variable, particularly for the parameter $\lambda$. We further discuss any additional issues relating to the performance of this model in Section \ref{sec:am2bad_main} that follows. 

\subsection{Performance of proposed angular models}
\label{sec:am2bad_main}

This section provides discussion on the performance of the process- and gauge-based angular models introduced in Section \ref{sec:angmod}, with a particular focus on the latter model.

As has already been discussed in Sections \ref{sec:gausdata}-\ref{sec:hw6data}, both process-based and gauge-based angular models often do a reasonable job at capturing the extremal dependence of data with low dimensionality, but their performance quickly deteriorates as dimension increases. In the case where AD is present, there is the additional disadvantage of the models failing to sufficiently capture the amount of mass that the angular distribution displays close to the centre of the simplex, thus underestimating the extremal dependence regardless of dimension. 
What this effectively means is that the proportion of angles approximately equal to $1/d$ in all of their components --- that is, that lie in the centre of the simplex --- is generally smaller for the angles sampled from either the gauge- or process-based angular models compared to the underlying sample of $\bm{W}\vert R'>1$. Table \ref{tab:angmod_spike} illustrates this phenomenon for the BR case with $\bm{\theta}_2$, as described in Section \ref{sec:brdata}, and similar conclusions hold for the other AD scenarios considered. In Table \ref{tab:angmod_spike}, we calculate the proportion of angles with all of their components approximately equal to $1/d$ from the underlying sample, as well as samples from the fitted process-based and gauge-based models. The process- and gauge-based models tend to yield less mass in this central region of the simplex. 
\begin{table}[ht]
\centering
\begin{tabular}{cccc}
\toprule
 $\bm{d}$ & \textbf{Sample} & \textbf{Process-based} & \textbf{Gauge-based} \\ 
\midrule
  5  & 0.0100 & 0.0032 & 0.0049 \\ 
  10 & 0.0153 & 0.0021 & 0.0018 \\ 
  20 & 0.121 & 0.0485 & 0.0764 \\ 
\bottomrule
\end{tabular}
\caption{Proportion of angles $\bm{w}$ from the sample of exceedance angles $\bm{W}\vert R'>1$, and process- and gauge-based angular models, such that $w_j \in ((1/d)-0.05,(1/d)+0.05)$, for all $j = 1,\ldots,d$. Figures are given in up to $3$ s.f.\ when available. All estimates correspond to the BR case with $\bm{\theta}_2$.}
\label{tab:angmod_spike}
\end{table}

In the case of the process-based model this is likely due to the underlying asymptotic independence of $\widetilde{\bm{W}}$. A potential remedy for this defect is to use an AD process for $\widetilde{\bm{X}}$ in place of the currently used AI Gaussian process. Such an approach however would likely be computationally prohibitive via likelihood methods. Inference for this model is already burdened by the computational cost of numerical integration procedures and the additional hurdle of fitting AD process models --- generally computationally intensive even for moderate dimensions \citep{Huser2022} --- is likely to make this model too impractical if at all possible to use. In the case of the gauge-based model the problem likely arises from the fact that the generalised Gaussian gauge is used in the derivation of the gradient score objective function. This gauge, although highly flexible, is still only suited to AI. Considering an alternative gauge model suitable for AD might alleviate this issue.

Another possible explanation for the underperformance of both angular models in higher dimensions might be linked to the threshold quality. Our threshold choice in equation~\eqref{eq:thresh} is linked to an asymptotic result between the high radial quantiles and the gauge function, but it is possible that the threshold does not give an even set of exceedances over the domain of the angular distribution in high dimensions, which affects the angular model fit. Figure \ref{fig:app_AMthresh} of the Supplement compares some of the results of Section \ref{sec:simstudy} for $d=20$ with results obtained using the same data and parameter configurations but with threshold levels $\tau=0.8,0.9$ instead. Specifically, two AI and one AD cases are investigated. The AD case shows an apparent significant improvement in $\chi_u$ estimation, but this can be explained by a substantially greater empirical (and hence unbiased) contribution to the estimates at all $u$ levels for $\tau=0.9$. Smaller improvements are seen for AI distributions. This could have a similar explanation to the AD case, although the empirical contributions to the estimates are substantially lower in the AI case. Another possibility is that our threshold choice~\eqref{eq:thresh}, being based on an asymptotic result, gets a more even set of angle exceedances everywhere in the simplex for higher values of $\tau$, and this improves the model fits. We note that the radial model may also be improved at higher $\tau$ levels, and this feeds into $\chi_u$ estimation, although the fact that the estimates are usually unbiased using the empirical angular model with $\tau=0.7$ suggests that the issue lies with the angles.
 
\begin{figure}[h!]
    \centering
\begin{subfigure}{\textwidth}
\centering
\includegraphics[width=0.3\linewidth]{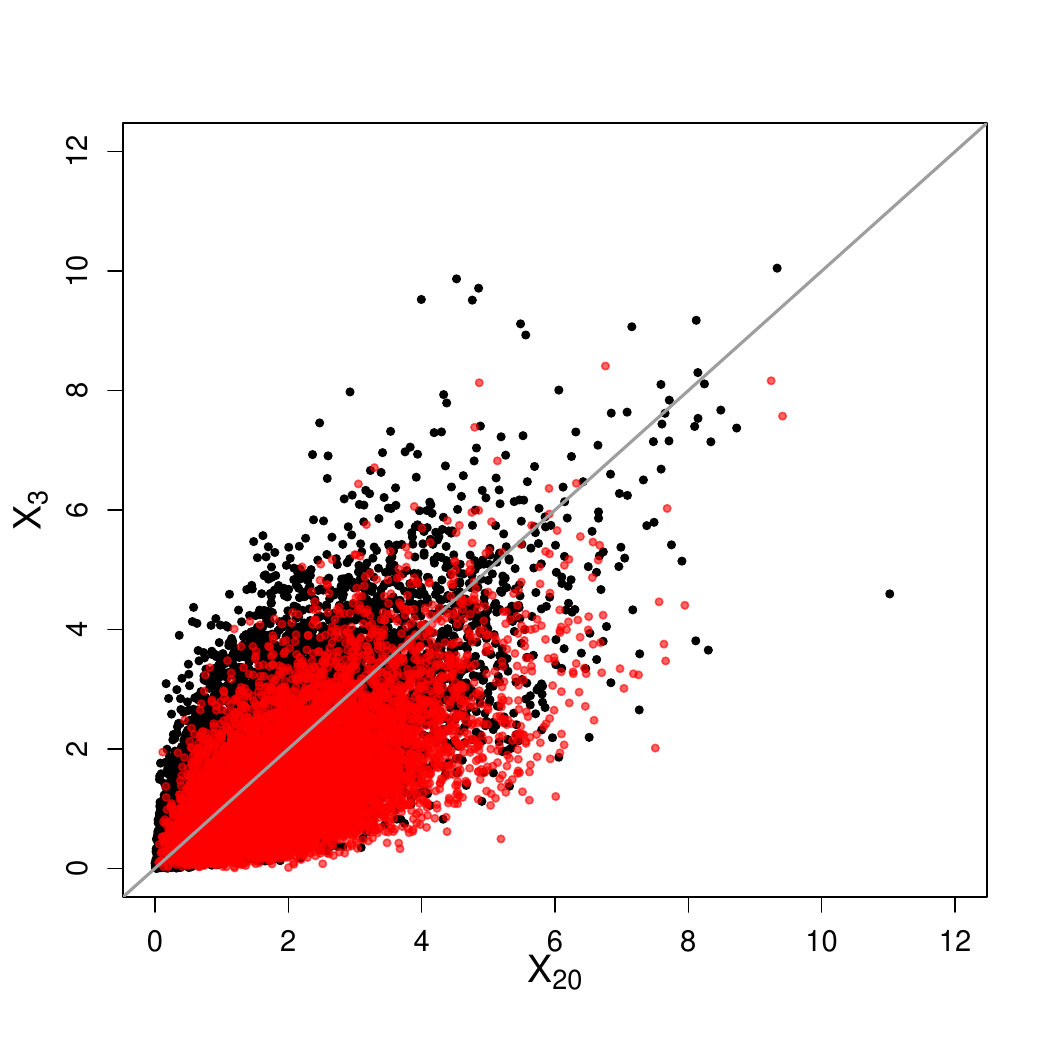}
\hfill        \includegraphics[width=0.3\linewidth]{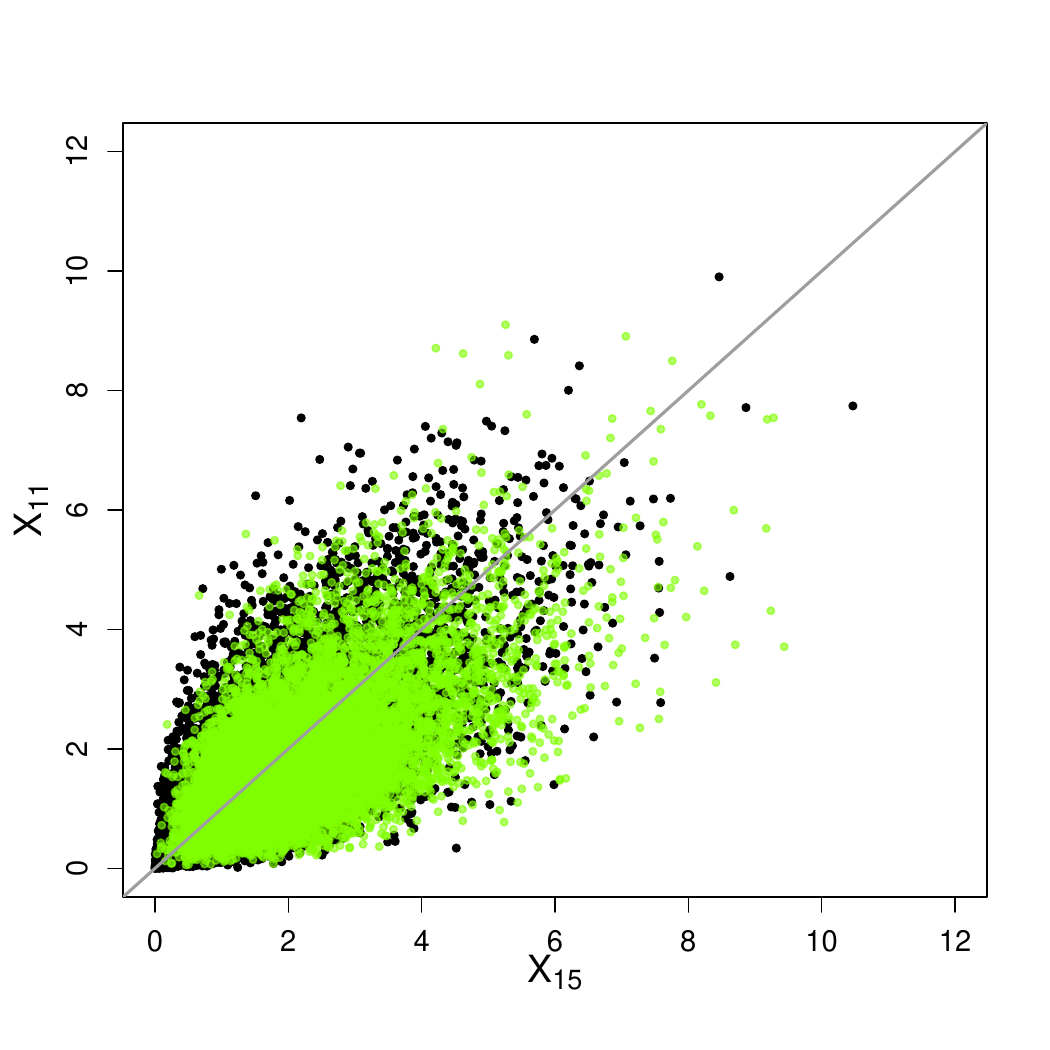}
\hfill        \includegraphics[width=0.3\linewidth]{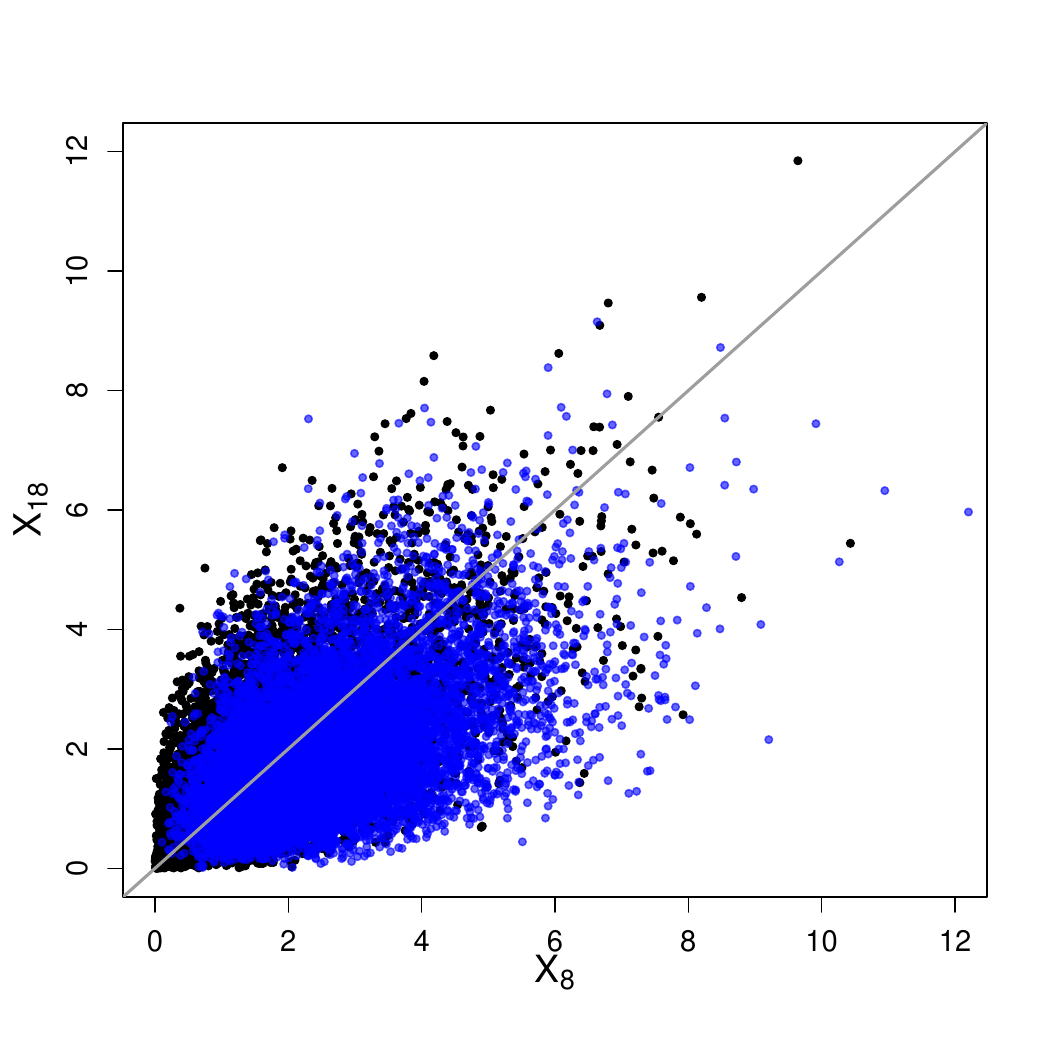}
\caption{} 
\label{fig:am2_scatter}
\end{subfigure}
\begin{subfigure}{\textwidth}
\centering
    \includegraphics[width=0.5\linewidth]{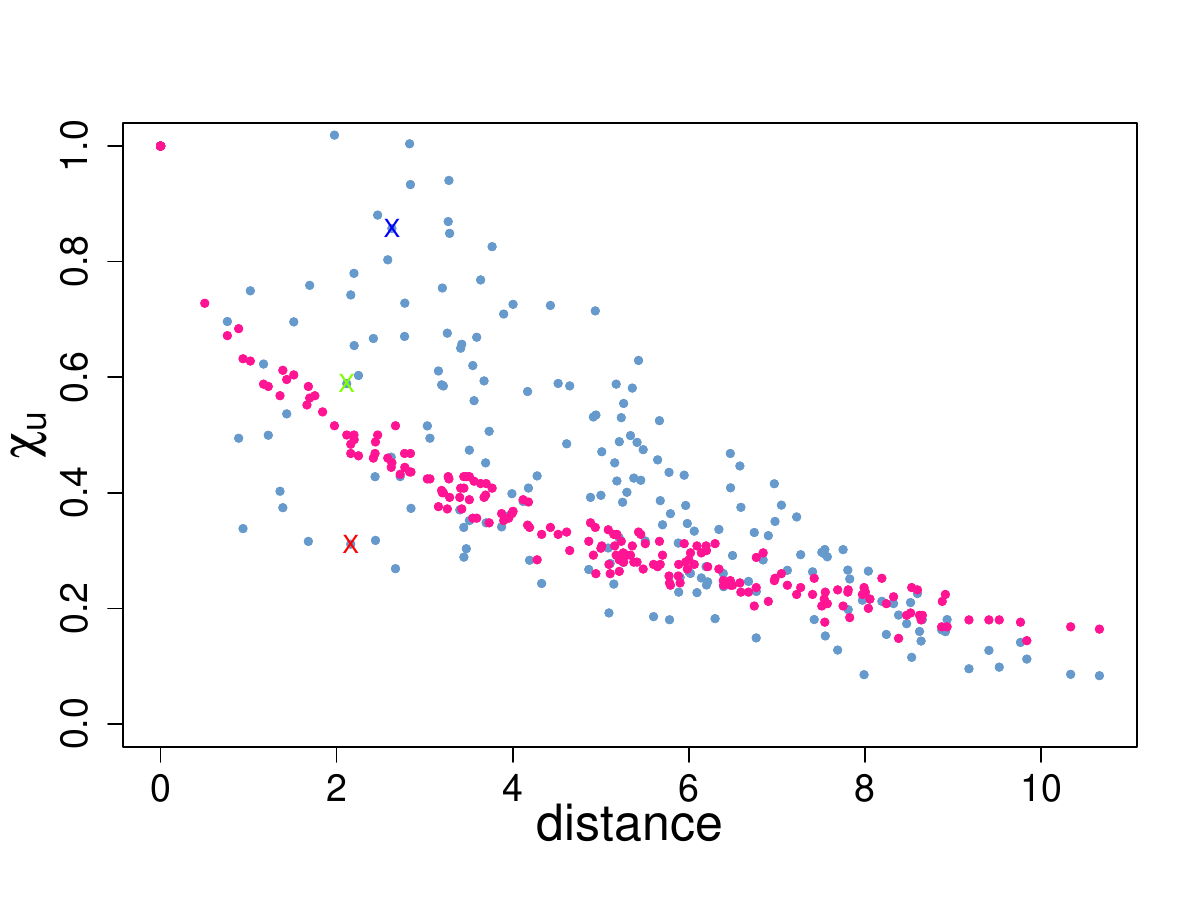}
    \caption{} 
    \label{fig:am2_chicloud}
\end{subfigure}
    \caption{(a) Pairwise scatter plots of $\bm{X}|R'>1$ data simulated using the process-based (black) and gauge-based (red, green, blue) angular models in \eqref{eq:am1} and \eqref{eq:am2}, respectively. The dashed line corresponds to the diagonal line, $x_j=x_k$, where $j=20,15,8$ and $k=3, 11, 18$. (b) Pairwise $\chi_u(\bm{s}_j,\bm{s}_k)$ estimates computed for all $d\choose2$ pairs and plotted over distance, for $d=20$. Estimates in light blue stem from the gauge-based angular model, while empirical $\chi_u$ are shown in pink. Differently coloured points correspond to estimates obtained from the $(X_j,X_k)$ pairs presented in Figure \ref{fig:am2perfplots}(\subref{fig:am2_scatter}) and are coloured accordingly.}
    \label{fig:am2perfplots}
\end{figure}
We now focus on particular issues with the gauge-based angular model. For the discussion that follows, we will showcase all results using the Gaussian example only, focusing specifically on the case with parameters $\bm{\theta}_1$ and dimension $d=20$, such as it is described in Section \ref{sec:gausdata}. However, the same observations apply to all other data scenarios considered throughout this simulation study. Another, more subtle, issue with the gauge-based angular model is that the density $f_{\widetilde{\bm{X}}} (\bm{x})\propto \exp(-g(\bm{x}))$ from which we simulate $\widetilde{\bm{X}}$ and thus obtain $\widetilde{\bm{W}}=\widetilde{\bm{X}}/\|\widetilde{\bm{X}}\|$ for this model is not necessarily symmetric in bivariate margins, even when the gauge function $g(\bm{x})$ itself is. This asymmetry is intrinsic to the model, because the density class $f_{\widetilde{\bm{X}}} (\bm{x})$ is generally not closed under taking margins of different dimensions, and can affect the quality of estimation tasks such as (lower-dimensional) $\chi_u$ calculation. Figure \ref{fig:am2perfplots}(\subref{fig:am2_scatter}) illustrates this issue. In particular, $10000$ simulated $\bm{X}\vert R'>1$ data using angular models \eqref{eq:am1} and \eqref{eq:am2} are considered, from which three pairs at similar distances apart are selected and plotted against each other. The points in black stem from the process-based model, while the coloured points come from the gauge-based model. The latter display an obvious asymmetry.

We note this asymmetry to be less obvious, if at all noticeable, in lower dimensions. The issue became evident in higher dimensions when pairwise $\chi_u$ estimates for all simulated pairs were considered. Figure \ref{fig:am2perfplots}(\subref{fig:am2_chicloud}) presents all $d\choose 2$ such estimates plotted against distance. The ones corresponding to the three nearly equidistant pairs shown in Figure \ref{fig:am2perfplots}(\subref{fig:am2_scatter}) are coloured accordingly to highlight their obvious disagreement. In fact, pairwise $\chi_u$ estimates stemming from this model are highly variable, with pairs at similar distances often implying very different extremal dependence strengths. For this reason, we decided to exclude the gauge-based model from our subsequent analyses, namely the results presented in Section \ref{sec:application} as well as those accompanying the present simulation study provided in Section \ref{app:ss_extrachis} of the Supplement. In conclusion, this asymmetry introduces additional challenges in the estimation tasks we have considered, especially in higher dimensions. However, we believe that it remains a fast, reliable and easy to use model for lower dimensions.

\section{Application to space weather data}
\label{sec:application}

In this section we illustrate how our method can be used in practice. Given the mixed performance of the proposed methods, especially those concerning the angular distribution, our prime objective here is to showcase how one might use the spatial geometric extreme value framework to perform statistical analyses, rather than to argue for its use currently over other models for spatial extremes.

The term space weather events is used to characterise the physical phenomena that occur as a result of the sun's explosive activity and include, for instance, solar flares, coronal mass ejections and others. It is possible for severe space weather events to reach the Earth and interact with its magnetosphere, causing fluctuations in the geomagnetic field and, thus, geomagnetically induced (electrical) currents (GIC). Such currents can disrupt or damage critical land infrastructure, such as electrical power grids \citep{gaunt2016}, transport \citep{boteler2021modeling}, communication \citep{nevanlinna2001breakdown} and satellite \citep{pirjola2005} systems as well as pose public health threats \citep{palmer2006solar}. Understanding the behaviour of extreme weather events and quantifying the risks associated with their occurrence is therefore important to help prevent and mitigate their effect. 

\citet{Rogers2021} analyse a space weather dataset of magnetic field measurements obtained from $125$ magnetometer stations around the globe. The data form part of the SuperMAG project \citep{Gjerloev2012} and comprise daily maximum absolute $1$-minute fluctuations in the horizontal component of the geomagnetic field, $\left|\text{d}B_h/\text{d}t\right|$, which is used as a proxy for GIC; see Section \ref{app:app_geodatainfo}  of the Supplement for more information on the geographic characteristics of the data. We choose to focus our analysis on a subset of the available locations, namely those originating from magnetometer stations above $70\degree$ latitude, where we expect large geomagnetic field fluctuations to occur \citep{Rogers2021}. 
This leaves us with $d=16$ spatial locations. We have a total of $17532$ daily measurements, recorded between $1969$ and $2017$, inclusive. However, there is a lot of missingness present. Aiming to retain as much of the available information as possible, we only remove observations having at most $1$ non-missing value; this leaves us with $15034$ data points, approximately $86\%$ of the original data. The daily observations may therefore differ in dimension. 

We fit our model in equation \eqref{eq:fulllh} to the above mentioned data --- using the spatial gauge functions of Section \ref{sec:spat gauges} --- but modify it appropriately to accommodate the varying dimensionality. In particular, equation \eqref{eq:fulllh} becomes
\begin{equation}
    L_{d}(\bm{\theta}) = \prod_{i=1}^n\frac{f\left(r_i;\alpha d_i, g(\bm{w}_i;\bm{\theta}, H_i)\right)}{\overline{\text{F}}\left(r_{\tau}(\bm{w}_i);\alpha d_i, g(\bm{w}_i;\bm{\theta}, H_i)\right)}. 
\end{equation}
The dimension $d_i$ takes values from $2$ to $16$, the radii, $r_i$, and angles, $\bm{w}_i$, correspond to observation $\bm{x}_i$ of dimension $d_i$ and the distance matrix $H_i$ now comprises of distances between the non-missing locations making up $\bm{x}_i$. The radial threshold is calculated using $r_{\tau}(\bm{w}_i) = C_{\tau,d_i} / g(\bm{w}_i)$; that is, we obtain a separate $C_{\tau,d}$ for each unique dimension $d \in \{2,\ldots, 16\}$, using only radii and angles corresponding to that dimension, where $g$ varies with dimension as appropriate. This is done such that the proportion of threshold exceedances is $1-\tau$, both overall and for each separate dimension, ensuring that observations of all dimension are represented in the likelihood function.
Table \ref{tab:mles} below, presents parameter estimates and Akaike/Bayesian Information Criterion (AIC/BIC) values for the fitted models, using quantile level $\tau=0.8$. This level was selected using diagnostics to assess the quality of the fit. 
\begin{table}[h]
\centering
\begin{tabular}{cccccccc}
\toprule
\textbf{Model}                        & $\bm{\alpha}$ & $\bm{\lambda}$ & $\bm{\kappa}$ & $\bm{\nu}$ & $\bm{\delta}$ & \textbf{AIC} & \textbf{BIC}  \\ \midrule
G & 0.345 & 29.7 & 0.201 & $-$ & $-$ & 14416 & 14434\\
L & 0.253 & 4.81 & 0.443 & $-$ & $-$ & 14406 & $\bm{14424}$\\
\textbf{GG} & $\bm{0.314}$ & $\bm{30.5}$ & $\bm{0.219}$ & $\bm{1.47}$ & $\bm{-}$ & $\bm{14403}$ & 14428\\
HW$_\text{G}$ & 0.347 & 40.1 & 0.185 & $-$ & 0.165 & 14419 & 14443\\
HW$_{\text{GG}}$ & 0.316 & 40.4 & 0.200 & 1.48 & 0.327 & 14407 & 14437\\ \bottomrule
\end{tabular}
\caption{Parameter estimates (reported in 3 s.f.) and corresponding AIC/BIC score (5 s.f.) for all five geometric spatial extremes models fitted.}
\label{tab:mles}
\end{table}

\begin{figure}[h]
    \centering
        \includegraphics[width=0.45\linewidth]{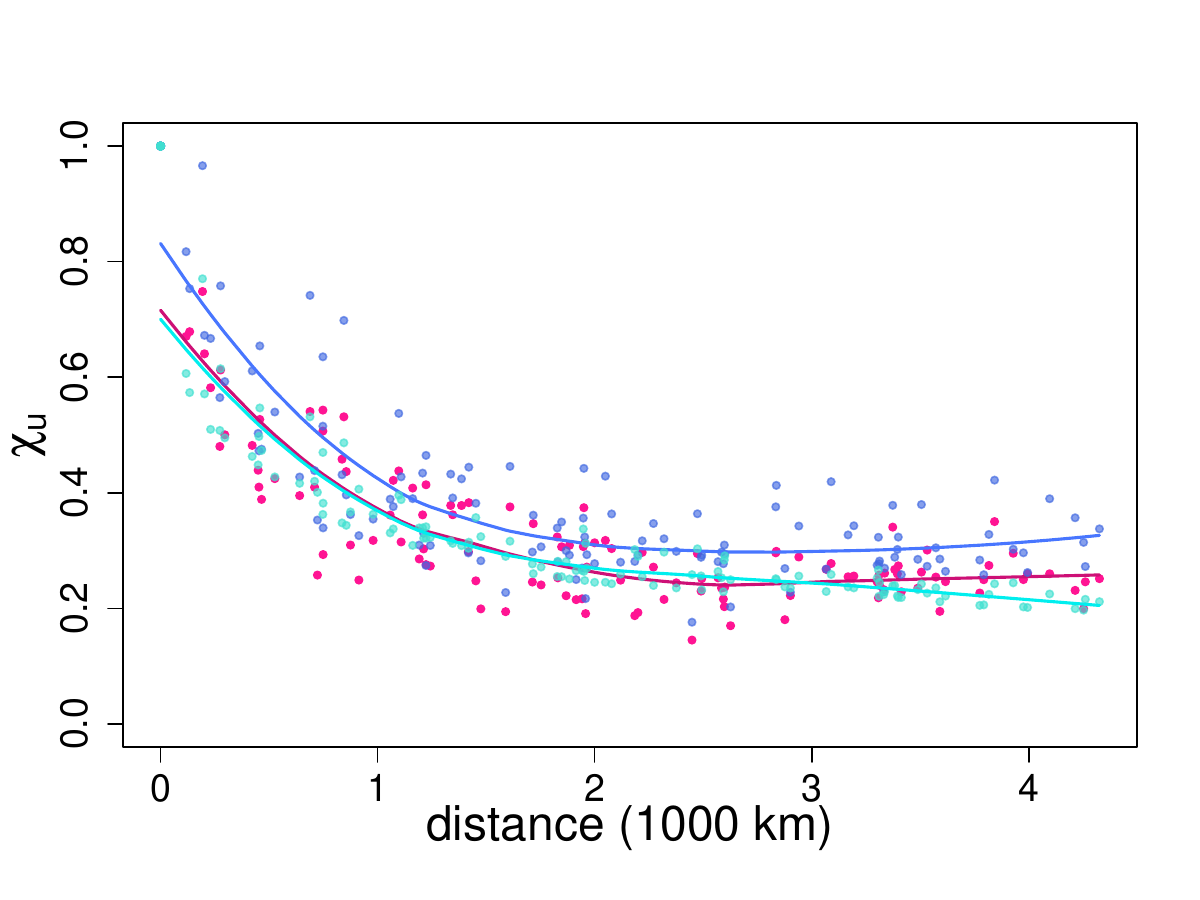}            
        \quad        \includegraphics[width=0.45\linewidth]{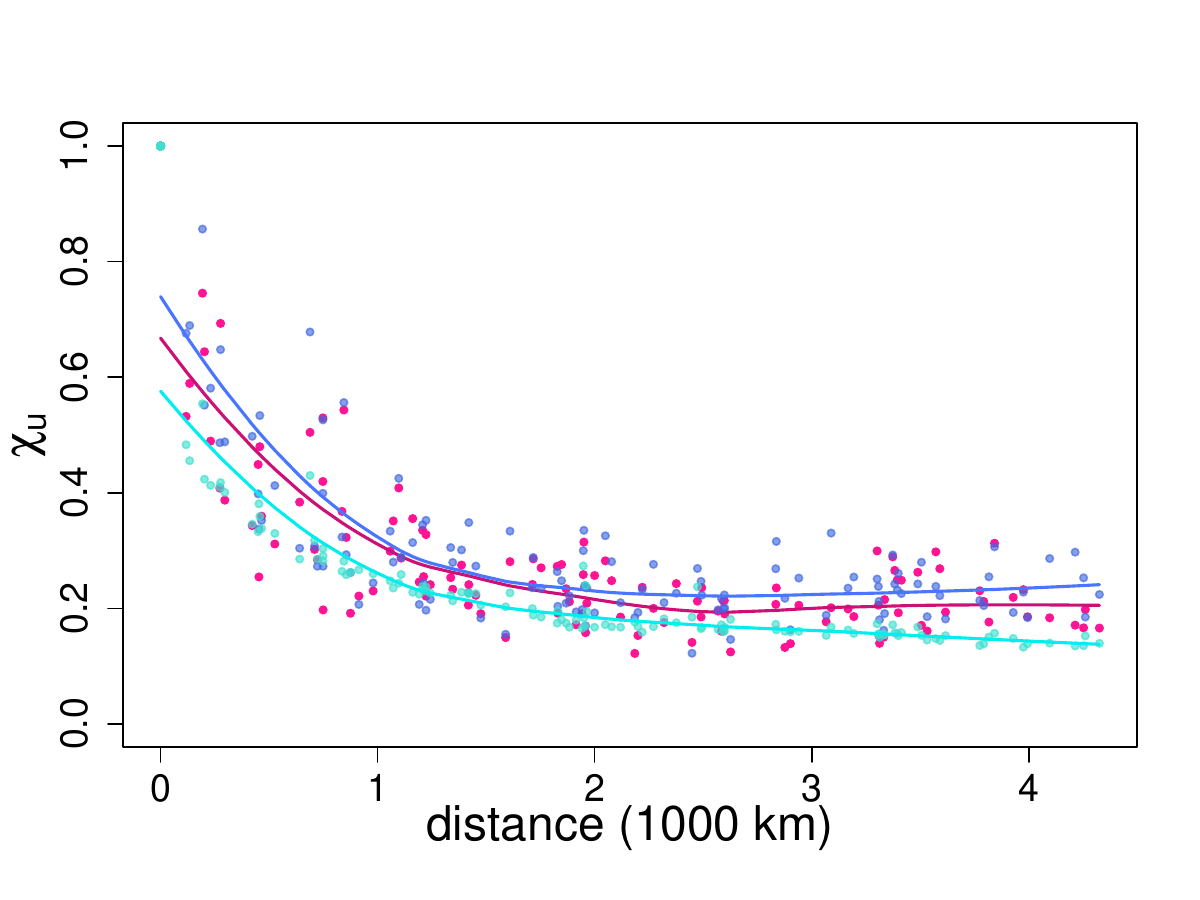}  
\caption{Pairwise $\chi(u)$ estimates over (geodesic) distance for $u=0.95$ (left) and $u=0.98$ (right). Pink points correspond to empirical estimates, dark blue to estimates with resampled angles and turquoise points to estimates from the angular model in \eqref{eq:am1}. Loess curves are fitted to each of these point clouds.}
\label{fig:chidistappl}
\end{figure}

The model based on the generalised Gaussian gauge has the smallest AIC, while that based on the Laplace gauge had the lowest BIC. The model diagnostics were similar for both cases, thus we only present results based on the generalised Gaussian gauge here. Model diagnostics for this case can be found in Section \ref{app:app_diagnost} of the Supplement, which illustrate that the fit is satisfactory, though not perfect. We proceed by fitting the angular model in \eqref{eq:am1} and use the obtained estimates to simulate $50000$ extreme observations $\bm{X}\vert R'>1$. Finally, we obtain pairwise $\chi_u$ estimates for all $\binom{16}{2}$ location pairs; we plot those over distance in Figure \ref{fig:chidistappl}. Overall, our model seems to be doing a reasonable job at describing the extremal dependence of the data, both when using the empirical and the process-based angular models. The same conclusion for the empirical angular model is supported by additional plots of $\chi_u(\bm{s}_j,\bm{s}_t)$ versus $u$ provided in Figure \ref{fig:chiuappl} of the Supplement. 

\section{Discussion}
\label{sec:discussion} 
In this paper we have presented a novel modelling framework for spatial extremes, by extending the geometric approach to cater for the spatial setting. In doing so, we have considered modelling of both the radial and angular components of the spatial vector of interest. 
The proposed method is flexible in its ability to capture the extremal dependence characteristics of the data under various, primarily misspecified, scenarios. It is suitable for both AI and AD data, although it seems to perform better in the case of the AI models studied. Moreover, it is generally fast to implement, with the main computational slowdown occurring when the HW gauges are used, due to a minimisation step involved in their implementation, or when fitting the angular model in \eqref{eq:am1}, due to the numerical integration procedure it requires. See Table \ref{tab:comptime} in the Supplementary Material for an overview of computational times for each model. 

The main drawbacks of the framework relate to the modelling of the angular distribution. In particular, specifying sensible models for $f_{\bm{W}}$ has proven to be challenging. The models we propose in this paper, detailed in Section \ref{sec:angmod}, both have serious shortcomings. The model in \eqref{eq:am1} seems to induce a bit of bias with increasing dimensionality. When AD is present especially, estimates from this model are incapable of capturing the proportion of mass of the angular data lying in the centre of the simplex. The same is true for the gauge-based model in \eqref{eq:am2} which performs poorly in higher dimensions and generally leads to both bias and variance increases in the estimation tasks we have considered; see Section \ref{sec:am2bad_main} for additional insights on the reasons behind the poor performance of this model. Another potential reason for reduced performance in higher dimensions is that the angles available for modelling depend on the threshold $r_{\tau}(\bm{w})$, and this may be further from our ideal of a $\tau$ quantile of the distribution of $R|\bm{W}=\bm{w}$, across all $\bm{w}\in \mathcal{S}^{d-1}$, as $d$ grows. Therefore the estimated threshold could potentially introduce irregularities in the distribution of $\bm{W}$ that are not well-captured by the models proposed. The empirical angular distribution currently provides the best representation of the angular component, however its intrinsic limitation that one can only generate data in certain parts of the angular space, instead of its entirety, is potentially problematic. It is therefore not the ideal solution to solely rely on it for angular modelling purposes and an alternative, parametric model for the angles is desired.

Limitations aside, it is important to stress not only the difficulty but also the necessity of developing good models for the angular distribution. This is of paramount importance, if we are to use the framework to effectively model spatial processes in high dimensions. As it currently stands, the radial model seems capable of scaling up to high dimensions, however this is evidently not yet the case for the angular models proposed. In brief, we believe that the geometric spatial extremes framework is promising for use in spatial extremes, but angular modelling needs to be improved before it can be routinely implemented in higher dimensions. This challenge presents a very interesting avenue for future work.

One possibility is to pursue further theoretical links between the angular density $f_{\bm{W}}$ and the gauge function $g$. The work of \citet{Papastathopoulos2023} and \citet{demonte2025} explores this prospect. As mentioned in Section \ref{sec:am2}, if $f_{\bm{X}}$ is homothetic, then there is a direct link between $f_{\bm{W}}$ and $g$. 
\citet{demonte2025} exploit not only this link, but also relaxations of it, to build semi-parametric models for $f_{\bm{W}}$ and use machine learning techniques, namely, the method of normalising flows, to perform inference on them. 
\citet{wessel2025} also explore machine learning techniques as a generative tool for angular values and, additionally, propose a series of diagnostic tools specifically tailored to angular modelling to assess the quality of generated angles. Such tools could be employed to test the effectiveness of any new angular model. Both papers adopt a multivariate setup and are able to accommodate moderate dimensions, up to $10$. Extending these ideas to spatial considerations might present a viable pathway towards improved spatial angular modelling. 

Finally, our setting is currently restricted to a pre-specified number of spatial locations and simulation at new locations is not yet possible. To do this, the threshold $r_{\tau}(\bm{w})$ needs adapting so that the constant $C_{\tau}$ varies with the number of locations considered. Because the radius $R$ is increasing when adding observations at new locations, so is the value of $C_{\tau}$. When fitting to data of different dimension, we dealt with this by estimating $C_{\tau}$ separately for each $d$, but a new modelling step would be required to simulate at new arbitrary locations.

\newpage
\subsection*{Acknowledgments}
This paper is based on work completed while Lydia Kakampakou was part of EPSRC funded project EP/W524438/1. Jennifer L. Wadsworth gratefully acknowledges EPSRC grant number EP/X010449/1. We thank Neil Rogers, Jim Wild and Emma Eastoe for providing the geomagnetometer data, and Neil Rogers for processing the data as described in \citet{Rogers2021}. We also gratefully acknowledge the SuperMAG collaborators: INTERMAGNET, Alan Thomson; CARISMA, PI Ian Mann; CANMOS, Geomagnetism Unit of the Geological Survey of Canada; AARI, PI Oleg Troshichev; The MACCS program, PI M. Engebretson; 210 Chain, PI K. Yumoto; IMAGE, PI Liisa Juusola; Finnish Meteorological Institute, PI Liisa Juusola; UiT the Arctic University of Norway, Troms{\o} Geophysical Observatory, PI Magnar G. Johnsen; DTU Space, Thom Edwards and PI Anna Willer; ICESTAR; RAPIDMAG; SuperMAG, PI Jesper W. Gjerloev.

\subsection*{Data Availability}
The raw ground magnetometer data were provided by SuperMAG and are available to download from \url{https://supermag.jhuapl.edu/}.

\subsection*{Code Availability}
Code to implement the methods in the article is available as Supplementary Material or from the corresponding author upon request.

\subsection*{Conflict of interest}
The authors declare no potential conflict of interests.

\bibliography{citations}

\newpage
\appendix
\section*{\centering Supplementary material for ``Geometric modelling of spatial extremes"}
\label{sec:appendix}

\renewcommand{\thesubsection}{S\arabic{subsection}}

\subsection{Laplace gauge function derivation}
\label{app:laplace}
\setcounter{equation}{0}
\renewcommand\theequation{S1.\arabic{equation}}
\setcounter{figure}{0}        
\renewcommand\thefigure{S1.\arabic{figure}} 
\setcounter{table}{0}
\renewcommand\thetable{S1.\arabic{table}}

Let $\{X(\bm{s}): \bm{s} \in S \subset \mathbb{R}^2 \}$ be a Laplace random field as defined in \citet{OPITZ2016}, and $\bm{X}=\left(X(\bm{s}_1), \ldots, X(\bm{s}_d)\right)$ its $d$-dimensional representation. Then $\bm{X}$ has density function $f_{\bm{X}}(\bm{x})$ given by
\begin{equation}
    f_{\bm{X}}(\bm{x}) = \frac{2^{1/2}}{(2\pi)^{d/2}|\Sigma|^{1/2}}\left(\frac{\bm{x}^{\text{T}}\Sigma^{-1}\bm{x}}{4}\right)^{\beta/2}\mathcal{K}_{\beta}\left(\left(\bm{x}^{\text{T}}\Sigma^{-1}\bm{x}\right)^{1/2}\right),
\end{equation}
where $\Sigma$ is the spatial correlation matrix, $\mathcal{K}_{\beta}$ is the modified Bessel function of the third kind and $\beta=1-d/2$ \citep{OPITZ2016}.

We can obtain the gauge function associated with density $f_{\bm{X}}(\bm{x})$ by using equation \eqref{eq:conv_cond_NW} of the main text, that is $g_{\text{L}}(\bm{x}) = \lim_{t\rightarrow\infty} -\log f_{\bm{X}}(t\bm{x})/t$. Note that the univariate tails of the Laplace distribution are by default exponential and so no marginal transformation is required in order to use equation \eqref{eq:conv_cond_NW}. We have: 
\begin{equation}
    -\frac{\log f_{\bm{X}}(t\bm{x})}{t} = -\frac{\log 2}{2t} + \frac{d}{2t}\log (2\pi) + \frac{\log|\Sigma|}{2t} - \frac{\log t}{t}\beta - \frac{\beta}{2t} \log\left(\frac{t^2\bm{x}^{\text{T}}\Sigma^{-1}\bm{x}}{4}\right) - \frac{1}{t}\log\left(\mathcal{K}_{\beta}\left(t \left(\bm{x}^{\text{T}}\Sigma^{-1}\bm{x}\right)^{1/2}\right)\right).
\end{equation}
In the limit, as $t\rightarrow\infty$, the first five summands become zero. Therefore, 
\begin{equation}
\label{eq:lapl_g_limit}
    \lim_{t\rightarrow\infty} -\frac{\log f_{\bm{X}}(t\bm{x})}{t} = \lim_{t\rightarrow\infty} -\frac{1}{t}\log\left(\mathcal{K}_{\beta}\left(t \left(\bm{x}^{\text{T}}\Sigma^{-1}\bm{x}\right)^{1/2}\right)\right)
\end{equation}
The asymptotic behaviour of the Bessel function of the third kind is $\mathcal{K}_{\beta}(z) \sim \left(\pi/(2z)\right)^{1/2}\exp(-z)$, $z\rightarrow \infty$ \citep{OPITZ2016}; hence $\mathcal{K}_{\beta}\left(t \left(\bm{x}^{\text{T}}\Sigma^{-1}\bm{x}\right)^{1/2}\right) \sim \left(\pi/\left(2t\left(\bm{x}^{\text{T}}\Sigma^{-1}\bm{x}\right)^{1/2}\right)\right)^{1/2}\exp\left\{-t\left(\bm{x}^{\text{T}}\Sigma^{-1}\bm{x}\right)^{1/2}\right\}$. Substituting this in \eqref{eq:lapl_g_limit} we get
\begin{align}
    \lim_{t\rightarrow\infty} -\frac{\log f_{\bm{X}}(t\bm{x})}{t} 
    &= \lim_{t\rightarrow\infty} -\frac{1}{t}\log\left( \left(\pi/\left(2t\left(\bm{x}^{\text{T}}\Sigma^{-1}\bm{x}\right)^{1/2}\right)\right)^{1/2}\exp\left\{-t\left(\bm{x}^{\text{T}}\Sigma^{-1}\bm{x}\right)^{1/2}\right\} \right) \\
    &= \lim_{t\rightarrow\infty} -\frac{\log\pi}{2t} + \frac{\log 2}{2t} + \frac{\log t}{2t} + \frac{\log\left(\bm{x}^{\text{T}}\Sigma^{-1}\bm{x}\right)}{4t} + \frac{
    t\left(\bm{x}^{\text{T}}\Sigma^{-1}\bm{x}\right)^{1/2}}{t} \\
    &= \left(\bm{x}^{\text{T}}\Sigma^{-1}\bm{x}\right)^{1/2}. 
\end{align}

\subsection{Derivations for score matching in Section \ref{sec:am2}}
\label{app:am2}
\setcounter{equation}{0}
\renewcommand\theequation{S2.\arabic{equation}}
\setcounter{figure}{0}        
\renewcommand\thefigure{S2.\arabic{figure}} 
\setcounter{table}{0}
\renewcommand\thetable{S2.\arabic{table}}

Recall that $g_{\text{GG}}(\bm{x}) = \left((\bm{x}^{1/\nu})^{\text{T}}\Sigma^{-1}\bm{x}^{1/\nu}\right)^{\nu/2}$ and that $f_{\bm{V}}(\bm{v}) \propto p(\bm{v})$ with $\log p(\bm{v}) = -d\log g_{\text{GG}}(e^{\bm{v}}) + \sum_{j=1}^{d-1}v_j=-d(\nu/2)\log \left( (e^{\bm{v}/\nu})^{\text{T}}\Sigma^{-1}e^{\bm{v}/\nu}\right) + \sum_{j=1}^{d-1}v_j$. Then the first order derivative of the un-normalised log-density of $\bm{V}$ with respect to $v_j$ is
\begin{align}
    \frac{\partial \log p(\bm{v})}{\partial v_j} 
    &= - \frac{d\nu}{2} \frac{\frac{\partial}{\partial v_j} \left( (e^{\bm{v}/\nu})^{\text{T}}\Sigma^{-1}e^{\bm{v}/\nu}\right)}{ (e^{\bm{v}/\nu})^{\text{T}}\Sigma^{-1}e^{\bm{v}/\nu}} + 1 \\
    &= -\frac{d\nu}{2} \frac{\frac{\partial}{\partial v_j} \left( \sum_{j=1}^d\sum_{k=1}^d e^{v_j/\nu}e^{v_k/\nu}(\sigma_{jk}^{-1})\right)}{(e^{\bm{v}/\nu})^{\text{T}}\Sigma^{-1}e^{\bm{v}/\nu}} +1 \\
    &= -\frac{d\nu}{2}\frac{2}{\nu} \frac{e^{v_j/\nu}\sum_{k=1}^d(\sigma_{jk}^{-1})e^{v_k/\nu}}{(e^{\bm{v}/\nu})^{\text{T}}\Sigma^{-1}e^{\bm{v}/\nu}} + 1\\
    &= - d \frac{e^{v_j/\nu}\sum_{k=1}^d(\sigma_{jk}^{-1})e^{v_k/\nu}}{(e^{\bm{v}/\nu})^{\text{T}}\Sigma^{-1}e^{\bm{v}/\nu}} + 1,
\end{align}
where $\frac{\partial}{\partial v_j} \left( (e^{\bm{v}/\nu})^{\text{T}}\Sigma^{-1}e^{\bm{v}/\nu}\right) = \frac{\partial}{\partial v_j} \left( \sum_{j=1}^d\sum_{k=1}^d e^{v_j/\nu}e^{v_k/\nu}(\sigma_{jk}^{-1})\right) = \frac{2}{\nu} e^{v_j/\nu}\sum_{k=1}^d(\sigma_{jk}^{-1})e^{v_k/\nu}$ and $(\sigma_{jk}^{-1})$ is the $(j,k)$ element of $\Sigma^{-1}$. 

The second derivative of $\log p(\bm{v})$ with respect to $v_j$ is 
\begin{align}
  \frac{\partial^2 \log p(\bm{v})}{\partial v_j^2}&=  \frac{-d\frac{\partial}{\partial v_j}\left(e^{v_j/\nu}\sum_{k=1}^d(\sigma_{jk}^{-1})e^{v_k/\nu}\right)(e^{\bm{v}/\nu})^{\text{T}}\Sigma^{-1}e^{\bm{v}/\nu} +de^{v_j/\nu}\sum_{k=1}^d(\sigma_{jk}^{-1})e^{v_k/\nu}\frac{\partial}{\partial v_j}\left((e^{\bm{v}/\nu})^{\text{T}}\Sigma^{-1}e^{\bm{v}/\nu}\right)}{\left((e^{\bm{v}/\nu})^{\text{T}}\Sigma^{-1}e^{\bm{v}/\nu}\right)^2}\\
  &= \frac{-d\frac{\partial}{\partial v_j}\left(e^{v_j/\nu}\sum_{k=1}^d(\sigma_{jk}^{-1})e^{v_k/\nu}\right)}{(e^{\bm{v}/\nu})^{\text{T}}\Sigma^{-1}e^{\bm{v}/\nu}} + \frac{\frac{d\nu}{2}\frac{2}{\nu}e^{v_j/\nu}\sum_{k=1}^d(\sigma_{jk}^{-1})e^{v_k/\nu}\frac{\partial}{\partial v_j}\left((e^{\bm{v}/\nu})^{\text{T}}\Sigma^{-1}e^{\bm{v}/\nu}\right)}{\left((e^{\bm{v}/\nu})^{\text{T}}\Sigma^{-1}e^{\bm{v}/\nu}\right)^2} \\
  &= \frac{-\frac{d}{\nu}e^{v_j/\nu}\left(\sum_{k=1}^d(\sigma_{jk}^{-1})e^{v_k/\nu}\right)-d e^{v_j/\nu} \left((\sigma_{jj}^{-1})\frac{e^{v_j/\nu}}{\nu}\right)}{(e^{\bm{v}/\nu})^{\text{T}}\Sigma^{-1}e^{\bm{v}/\nu}} + \frac{\frac{d\nu}{2}\left(\frac{\partial}{ \partial v_j}\left((e^{\bm{v}/\nu})^{\text{T}}\Sigma^{-1}e^{\bm{v}/\nu}\right)\right)^2}{\left((e^{\bm{v}/\nu})^{\text{T}}\Sigma^{-1}e^{\bm{v}/\nu}\right)^2} \\
  &= -\frac{d}{\nu}\frac{e^{v_j/\nu}\left[(\sigma_{jj}^{-1})e^{v_j/\nu}+\sum_{k=1}^d(\sigma_{jk}^{-1})e^{v_k/\nu}\right]}{(e^{\bm{v}/\nu})^{\text{T}}\Sigma^{-1}e^{\bm{v}/\nu}} + \frac{d\nu}{2}\left[\frac{\frac{\partial}{ \partial v_j}\left((e^{\bm{v}/\nu})^{\text{T}}\Sigma^{-1}e^{\bm{v}/\nu}\right)}{(e^{\bm{v}/\nu})^{\text{T}}\Sigma^{-1}e^{\bm{v}/\nu}}\right]^2.
\end{align}

\subsection{Additional plots for Section \ref{sec:prediction}}
\label{app:chicheck}
\setcounter{equation}{0}
\renewcommand\theequation{S3.\arabic{equation}}
\setcounter{figure}{0}       
\renewcommand\thefigure{S3.\arabic{figure}} 
\setcounter{table}{0}
\renewcommand\thetable{S3.\arabic{table}}

\begin{figure}[h]
    \centering
        \includegraphics[width=0.3\linewidth]{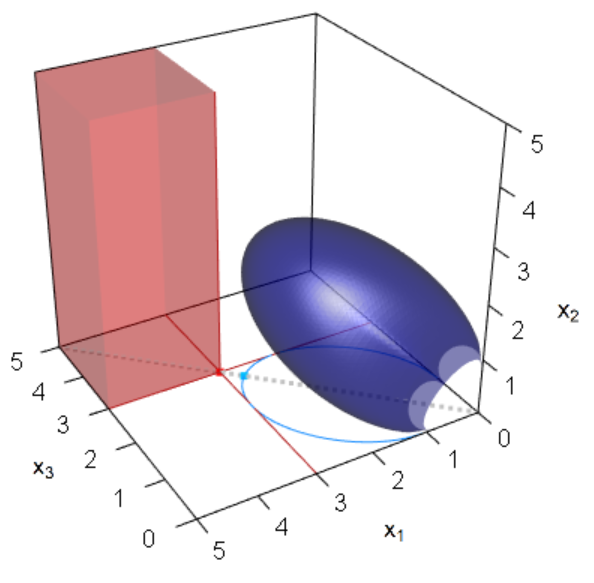} 
        \quad        \includegraphics[width=0.3\linewidth]{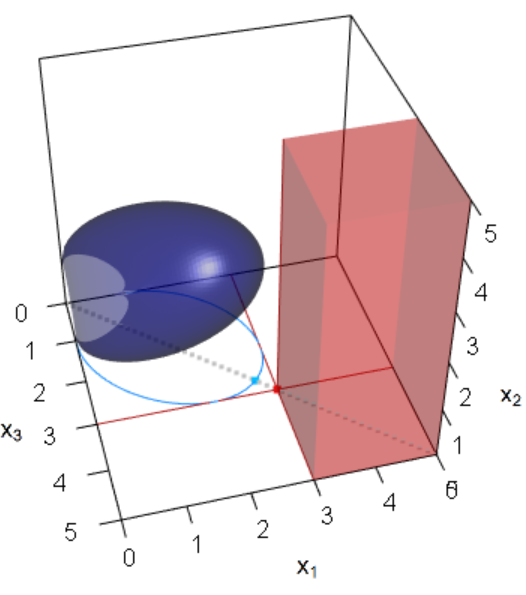}
        \quad        \includegraphics[width=0.3\linewidth]{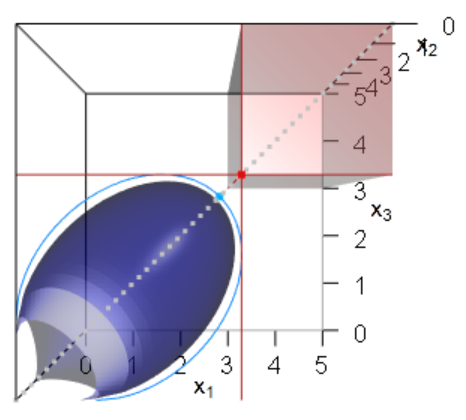}
\caption{Threshold surface in $3$ dimensions (in dark blue) and its $2$-dimensional projection on the $x_1$-$x_3$ face (light blue curve). Red lines correspond to the $X_1$ and $X_3$ marginal $u$-quantiles, which define the $\chi_u$ region of interest, shaded in red. The diagonal of the $x_1$-$x_3$ face is depicted with the grey dotted line. The red and light blue points correspond to the vertex of set $B^u$ and the point on the $2$-d projection of the threshold function which lies on the diagonal, respectively. The three panels show the same plot from different viewpoints.}
\label{fig:chicheck}
\end{figure}

Figure \ref{fig:chicheck} below provides a $3$-dimensional illustration of the threshold surface and the pairwise $\chi_u$ region of interest defined by set $B^u=(-\log(1-u),\infty)\times(0,\infty)\times(-\log(1-u),\infty)$. Because of the shape of the two objects, to assess whether $\chi_u$ estimation can solely be based on simulated $\bm{X}|R'>k$ points or not, it is sufficient to check if the distance of the vertex of set $B^u$ (in red) from the origin is greater that that of the point sitting on the $2$-dimensional projection of the threshold surface and the diagonal line $x_1=x_3$ (in light blue).

\subsection{Additional plots for Section \ref{sec:simstudy}}
\label{app:simstudy}
\subsubsection{Goodness of fit diagnostics}
\label{app:ss_diagnostics}

Figure \ref{fig:ss_pp_plots} shows complementary results for the simulation study presented in Section \ref{sec:simstudy} of the main paper. Specifically, rescaled Probability-Probability (P-P) plots of the preferred radial model are provided for a single iteration of the simulation study and for all data, dimension, and parameter configurations. The rescaling of the classic P-P diagnostic is such that the differences in empirical and model probabilities are plotted on the y-axis, similarly to \citet{Gabda2012}. This is to improve the readability of the plot.
All plots showcase decent goodness of fit, across all data and dimension configurations. 

\subsubsection{Additional pairwise $\chi_u$ plots}
\label{app:ss_extrachis}

Figures \ref{fig:cloud195}(\subref{fig:cloud195_mvn})-(\subref{fig:cloud195_hw6}), \ref{fig:cloud198}(\subref{fig:cloud198_mvn})-(\subref{fig:cloud198_hw6}), \ref{fig:cloud295}(\subref{fig:cloud295_mvn})-(\subref{fig:cloud295_hw6}) and \ref{fig:cloud298}(\subref{fig:cloud298_mvn})-(\subref{fig:cloud298_hw6}) depict $\chi_u(\bm{s}_j, \bm{s}_k)$ estimates calculated for all $j,k \in \{1,\ldots,d\}$ and plotted versus distance, for parameter configurations $\bm{\theta}_1$ and $u=0.95$, $\bm{\theta}_1$ and $u=0.98$, $\bm{\theta}_2$ and $u=0.95$, and $\bm{\theta}_2$ and $u=0.98$, respectively. Specifically, $\chi_u(\bm{s}_j, \bm{s}_k)$ estimates have been computed for all $200$ simulation iterations and their $0.025$ and $0.975$ quantiles have been calculated. Subsequently, envelopes have been created by connecting and colouring the area between the point $0.025$ and $0.975$ quantile estimates, in order to summarise the behaviour of $\chi_u$ over distance across all possible location pairs and all simulation iterations. Figure \ref{fig:app_AMthresh} accompanies Section \ref{sec:am2bad_main} of the main text. It provides additional pairwise $\chi_u$ plots for a range of $u$ values, as in Figures \ref{fig:box1} and \ref{fig:box2} of the main text, but calculated using higher threshold levels, namely $\tau = 0.8, 0.9$.

\subsection{Additional plots for Section \ref{sec:application}}
\label{app:app_extras}
\subsubsection{Spatial characteristics of magnetometer stations}
\label{app:app_geodatainfo}
Table \ref{tab:apl_loc_info} and Figure \ref{fig:geomap} present additional information on the spatial characteristics of the space weather dataset analysed in Section \ref{sec:application} of the main paper. In particular, Table \ref{tab:apl_loc_info} details the coordinates of the magnetometer stations analysed, while Figure \ref{fig:geomap} visualises these as well as the remaining stations in the form of a map.

\subsubsection{Additional $\chi_u$ plots}
\label{app:app_chius}

Figure \ref{fig:chiuappl} below is given as a complement to Figure \ref{fig:chidistappl} of the main text. It shows pairwise $\chi_u(\bm{s}_j,\bm{s}_k)$ plots for four different pairs at various distances apart. In most cases, the fit using the empirical angular model appears satisfactory, although there is some overestimation for one pair at lower values of $u$. The performance of the process-based angular model appears less robust, its quality varying depending on the pair considered.

\subsubsection{Goodness-of-fit diagnostics}
\label{app:app_diagnost}

Figure \ref{fig:app_rescaledPP} depicts a rescaled (as in Section \ref{app:ss_diagnostics}) P-P diagnostic plot assessing the quality of the spatial geometric model fit to the space weather data discussed in Section \ref{sec:application} of the main text. It suggest an imperfect, yet satisfactory goodness of fit. The tolerance limits in this plot are based on an assumption of independent data, therefore true tolerance limits are likely to be wider.

\subsection{Computational time}
\label{app:comptime}

Table \ref{tab:comptime} contains a guide to the computational times one might expect for each model and dimension scenario. The composite likelihood model generally takes longer to fit compared to the full likelihood one, the computational time increasing with increasing dimensionality, since likelihood contributions from more pairs need to be evaluated. Of the full radial models, the ones involving Gaussian-type gauge functions are quick to fit, while the HW ones are more computationally involved due to minimisation operations. As for the angular models, the process-based one is generally more computationally demanding because of the numerical integration step it entails, in contrast to the gauge-based model which is very fast to implement.

\setcounter{equation}{0}
\renewcommand\theequation{S4.1.\arabic{equation}}
\setcounter{figure}{0}       
\renewcommand\thefigure{S4.1.\arabic{figure}} 
\setcounter{table}{0}
\renewcommand\thetable{S4.1.\arabic{table}}

\begin{figure}[ht!]
\centering
\setlength{\tabcolsep}{5pt} 

\begin{tabular}{>{\centering\arraybackslash}m{0.5cm} m{0.45\textwidth} m{0.45\textwidth}}
   & \makebox[\linewidth][c]{\textbf{$\bm{\theta}_1$}} & \makebox[\linewidth][c]{\textbf{$\bm{\theta}_2$}} \\[2mm]
   
   \adjustbox{valign=c, rotate=90}{\textbf{$d=5$}} &
   \begin{subfigure}{0.45\textwidth}
       \includegraphics[width=\linewidth]{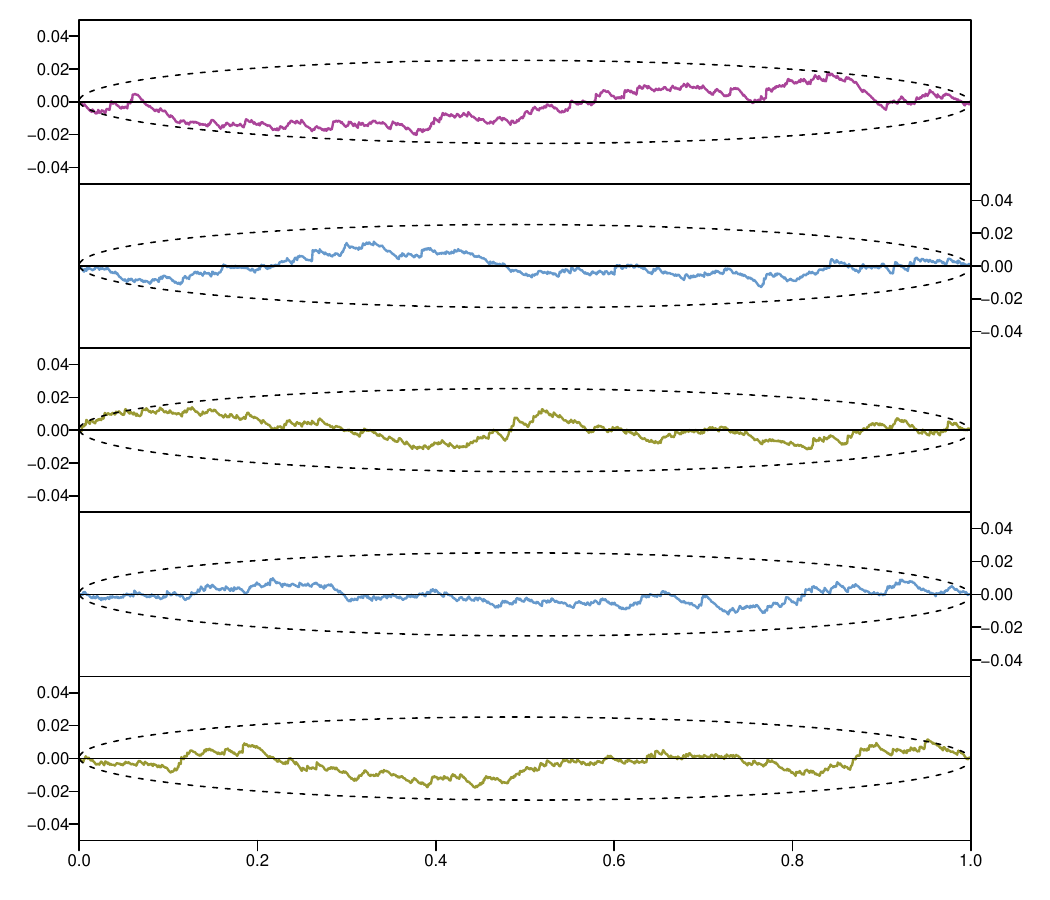}
   \end{subfigure} &
   \begin{subfigure}{0.45\textwidth}
       \includegraphics[width=\linewidth]{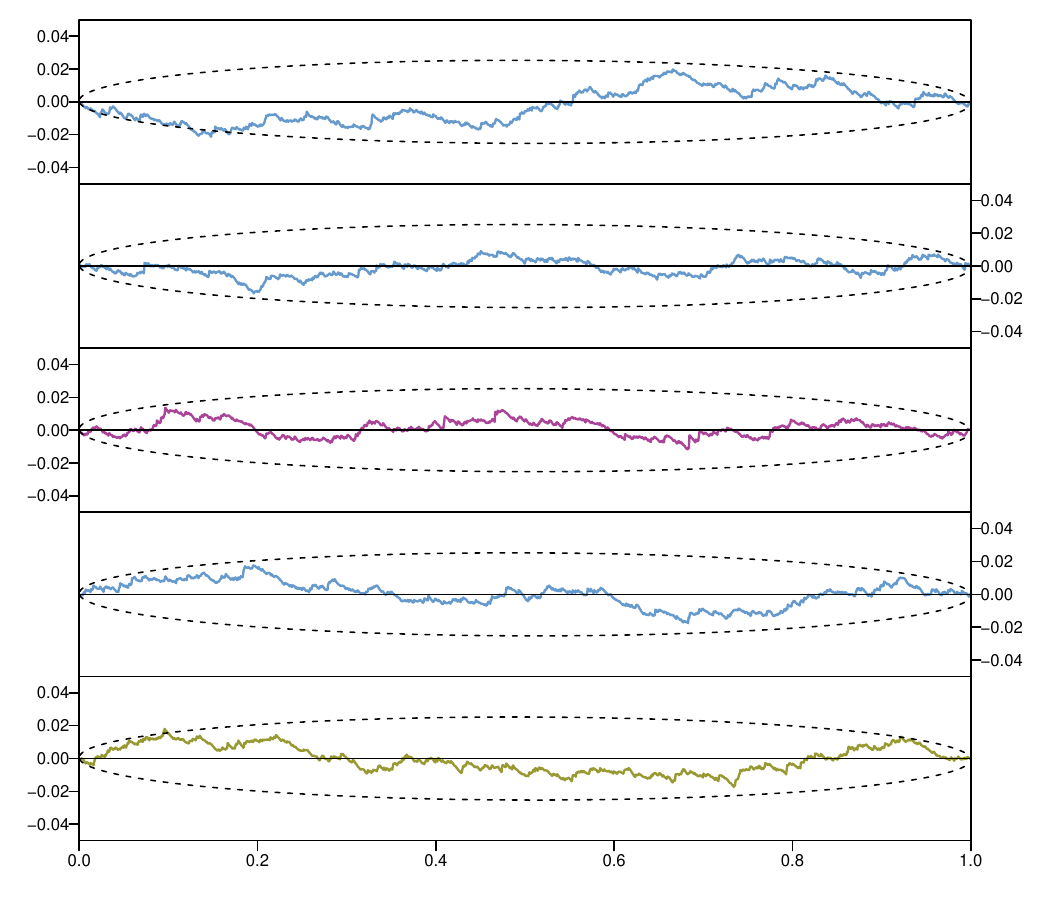}
   \end{subfigure} \\[2mm]

   \adjustbox{valign=c, rotate=90}{\textbf{$d=10$}} &
   \begin{subfigure}{0.45\textwidth}
       \includegraphics[width=\linewidth]{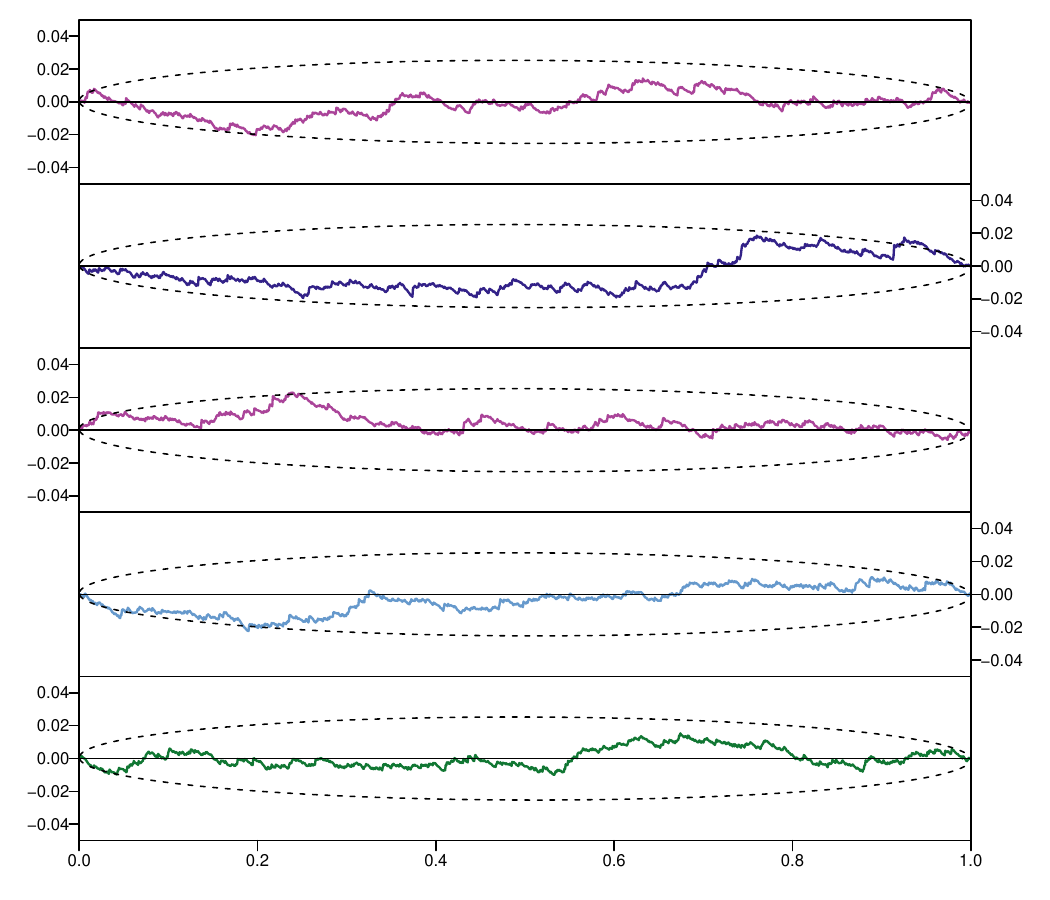}
   \end{subfigure} &
   \begin{subfigure}{0.45\textwidth}
       \includegraphics[width=\linewidth]{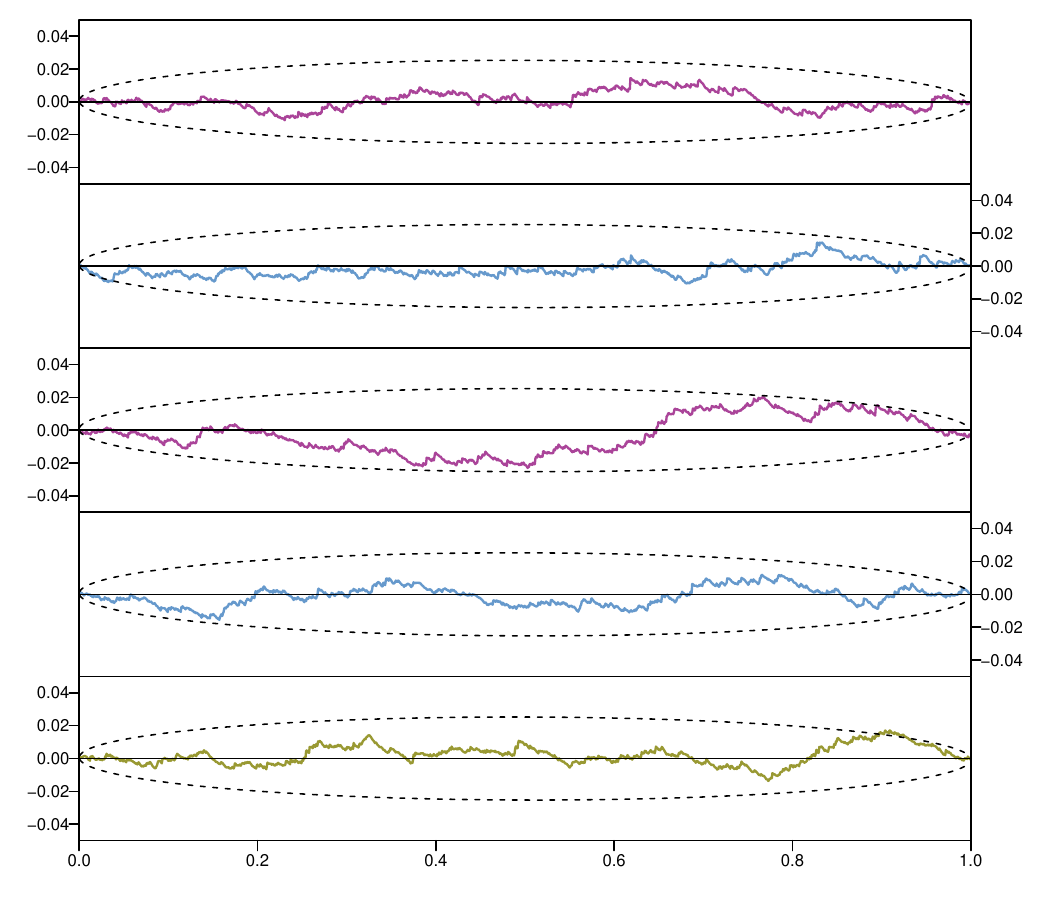}
   \end{subfigure} \\[2mm]

   \adjustbox{valign=c, rotate=90}{\textbf{$d=20$}} &
   \begin{subfigure}{0.45\textwidth}
       \includegraphics[width=\linewidth]{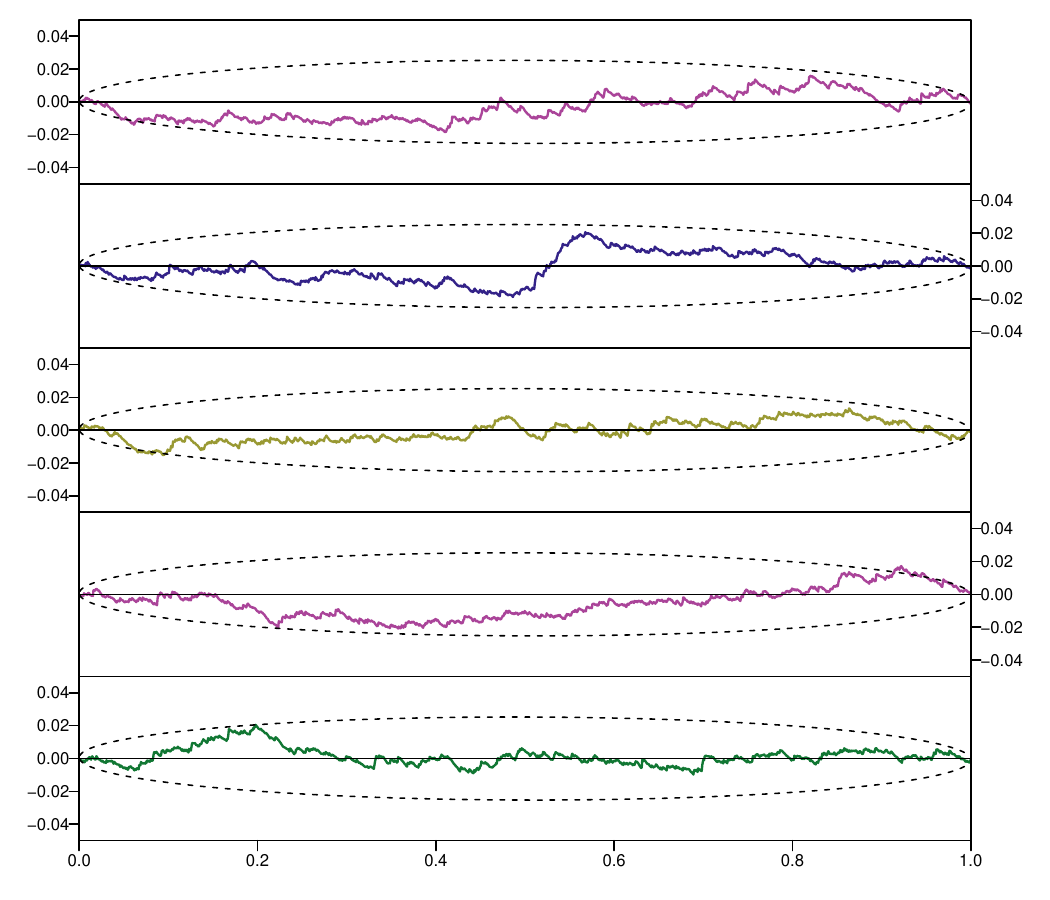}
   \end{subfigure} &
   \begin{subfigure}{0.45\textwidth}
       \includegraphics[width=\linewidth]{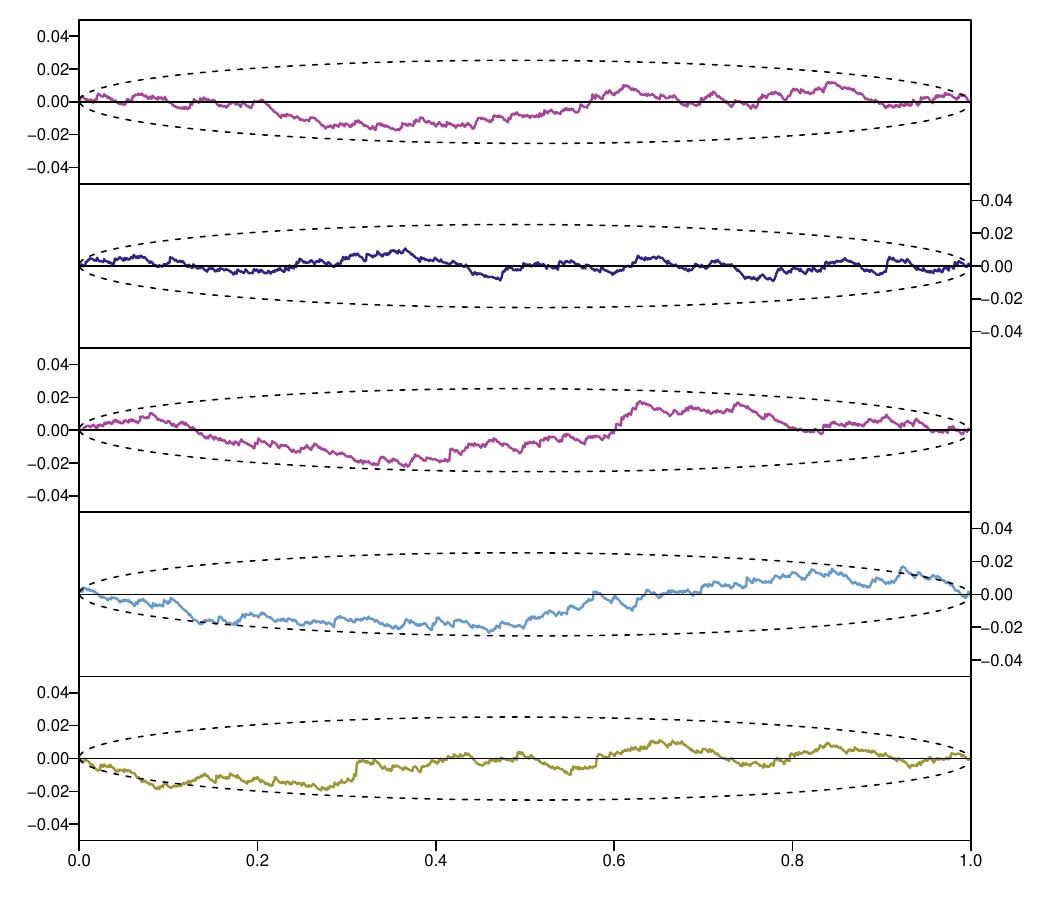}
   \end{subfigure} \\
\end{tabular}
\caption{Rescaled Probability-Probability plots for a single iteration of the simulation study discussed in Section \ref{sec:simstudy} of the main paper. In each panel, rescaled P-P plots are stacked in the following order, from top to bottom: $X(\bm{s})$ simulated from a Gaussian process, an IBR process, a BR process, a HW process with $\delta=0.4$, and a HW process with $\delta=0.6$. All plots depict the performance of the preferred radial model with associated gauge type encoded in the given colour. Specifically, lines in light blue correspond to the Gaussian, dark blue to the Laplace, purple to the generalised Gaussian, dark green to the HW with an underlying Gaussian and light green to the HW with an underlying generalised Gaussian gauge.} 
 \label{fig:ss_pp_plots} 
\end{figure}

\newpage

\setcounter{equation}{0}
\renewcommand\theequation{S4.2.\arabic{equation}}
\setcounter{figure}{0}       
\renewcommand\thefigure{S4.2.\arabic{figure}} 
\setcounter{table}{0}
\renewcommand\thetable{S4.2.\arabic{table}}
\begin{figure}[H]
\centering
\begin{subfigure}[t]{1\textwidth}
\centering
  \includegraphics[width=0.3\textwidth]{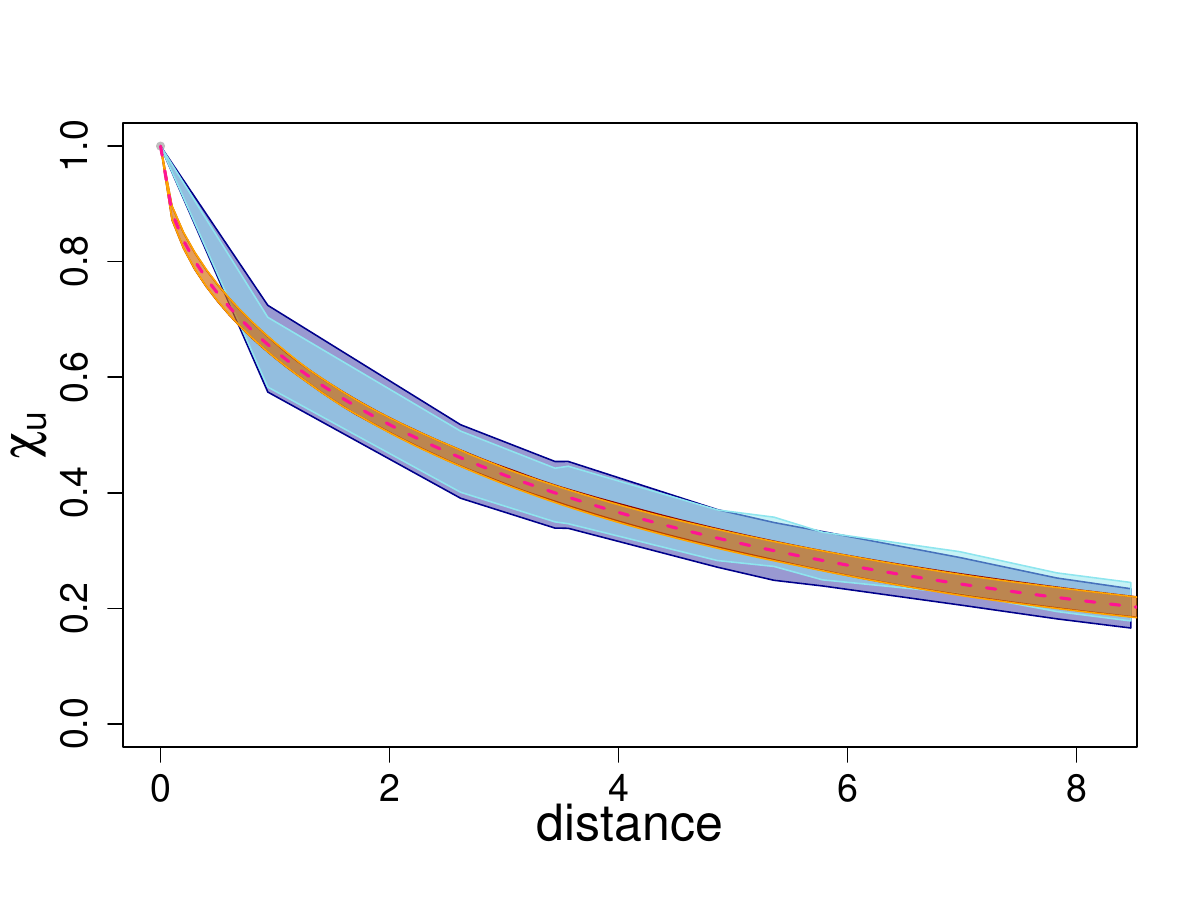}
  \hfill
  \includegraphics[width=0.3\textwidth]{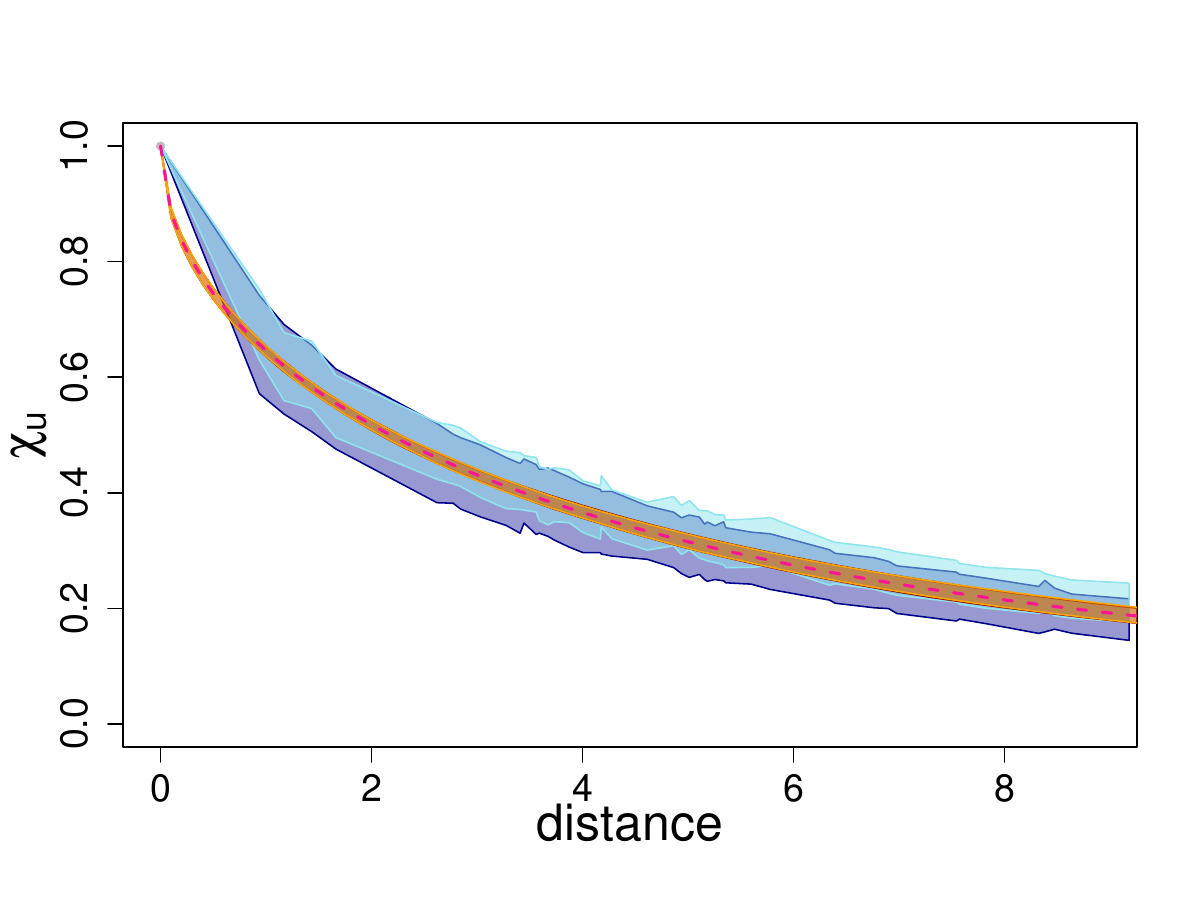}
  \hfill
  \includegraphics[width=0.3\textwidth]{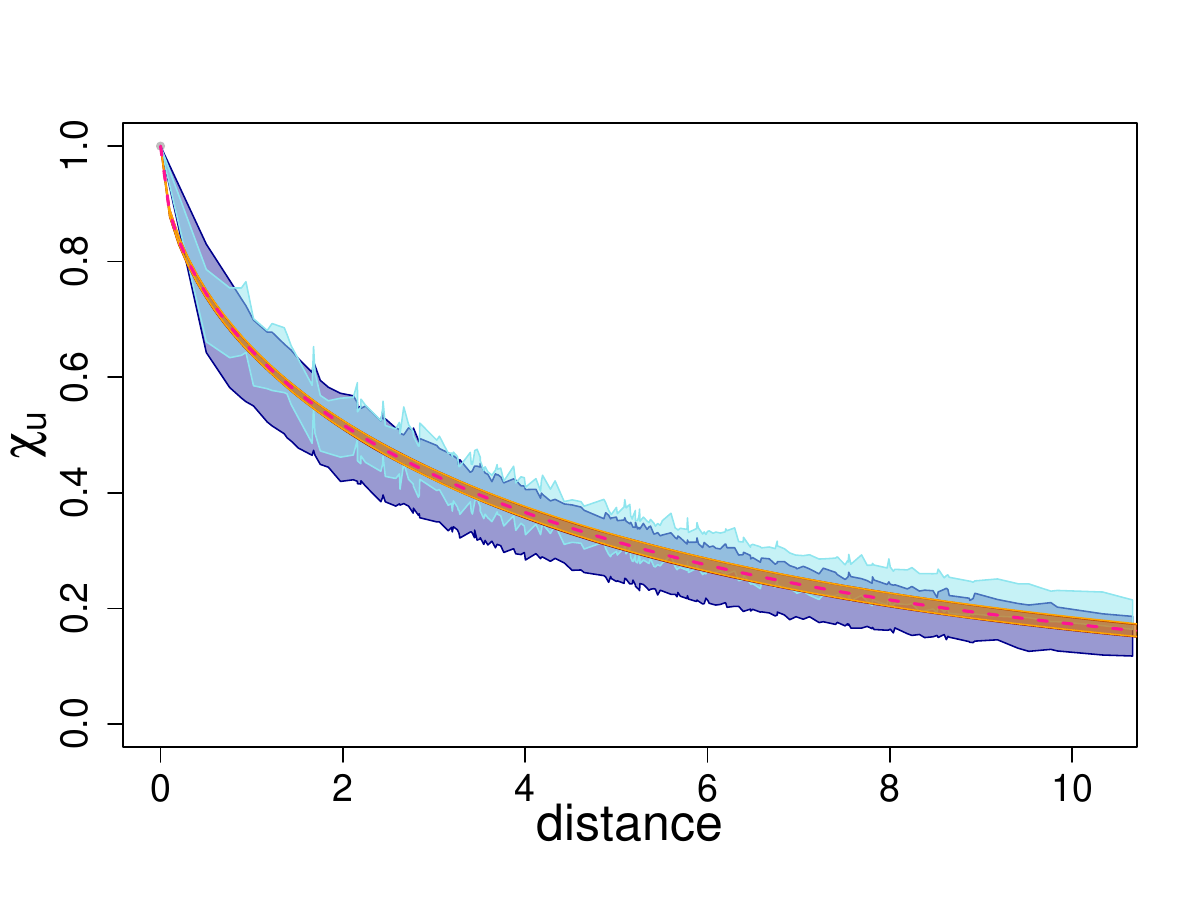}
\caption{$X(\bm{s})$ simulated from a Gaussian process} \label{fig:cloud195_mvn}
\end{subfigure}

\begin{subfigure}[t]{1\textwidth}
\centering
  \includegraphics[width=0.3\textwidth]{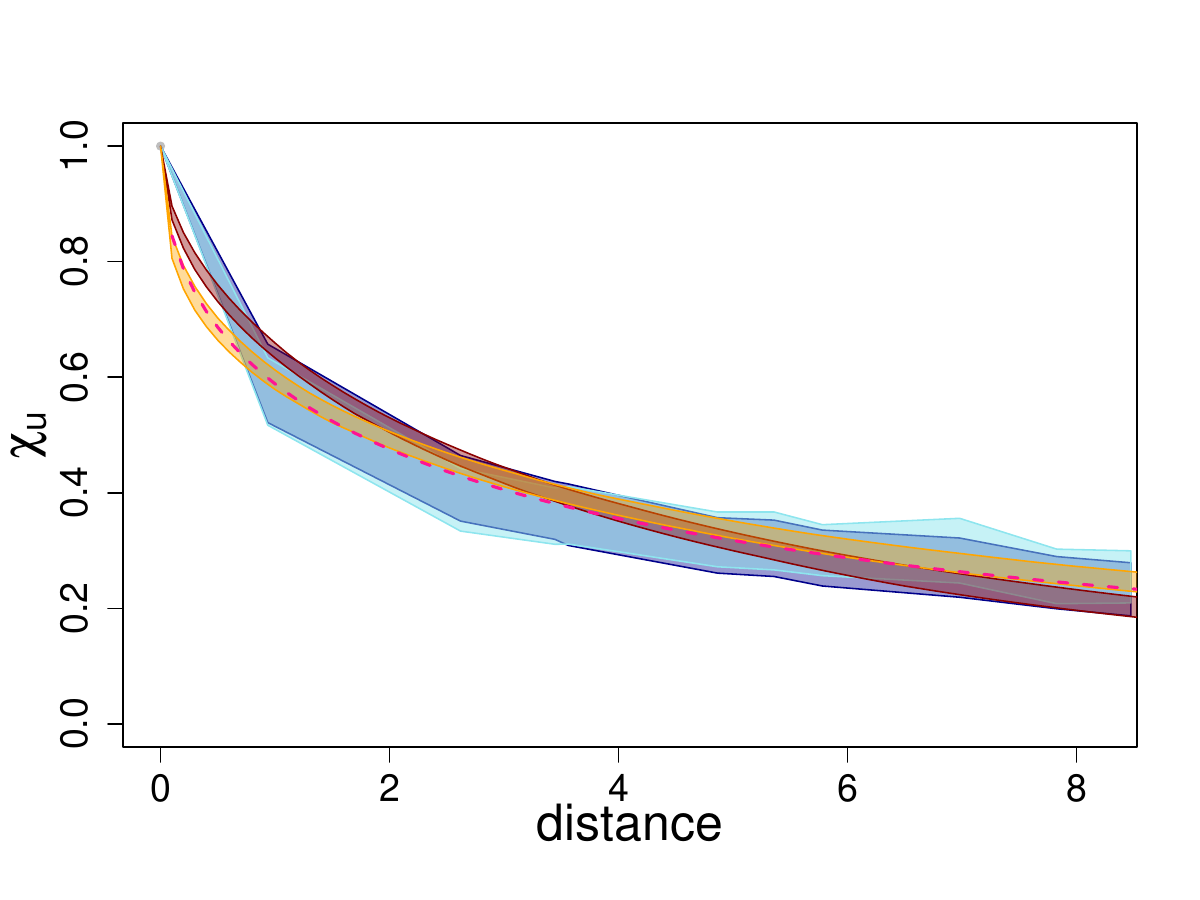}
  \hfill
  \includegraphics[width=0.3\textwidth]{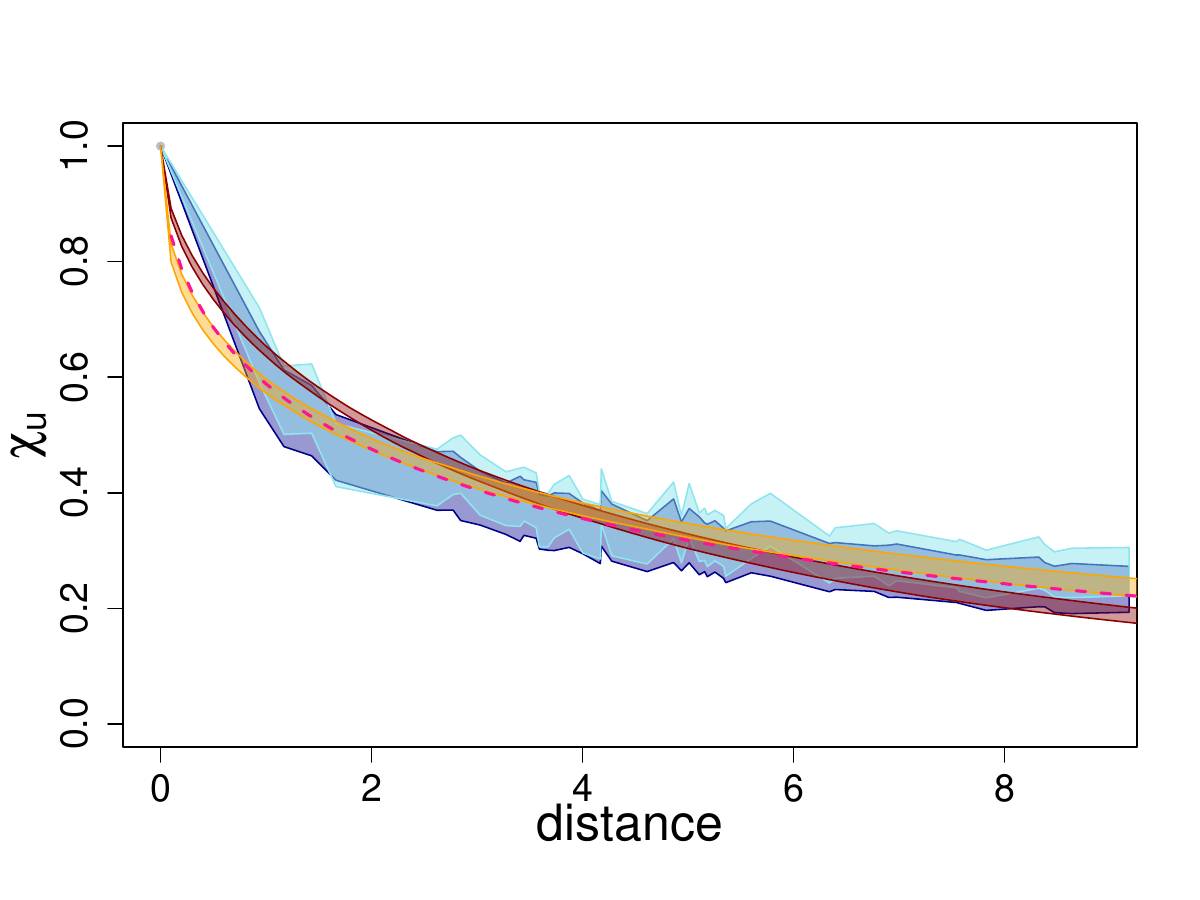}
  \hfill
  \includegraphics[width=0.3\textwidth]{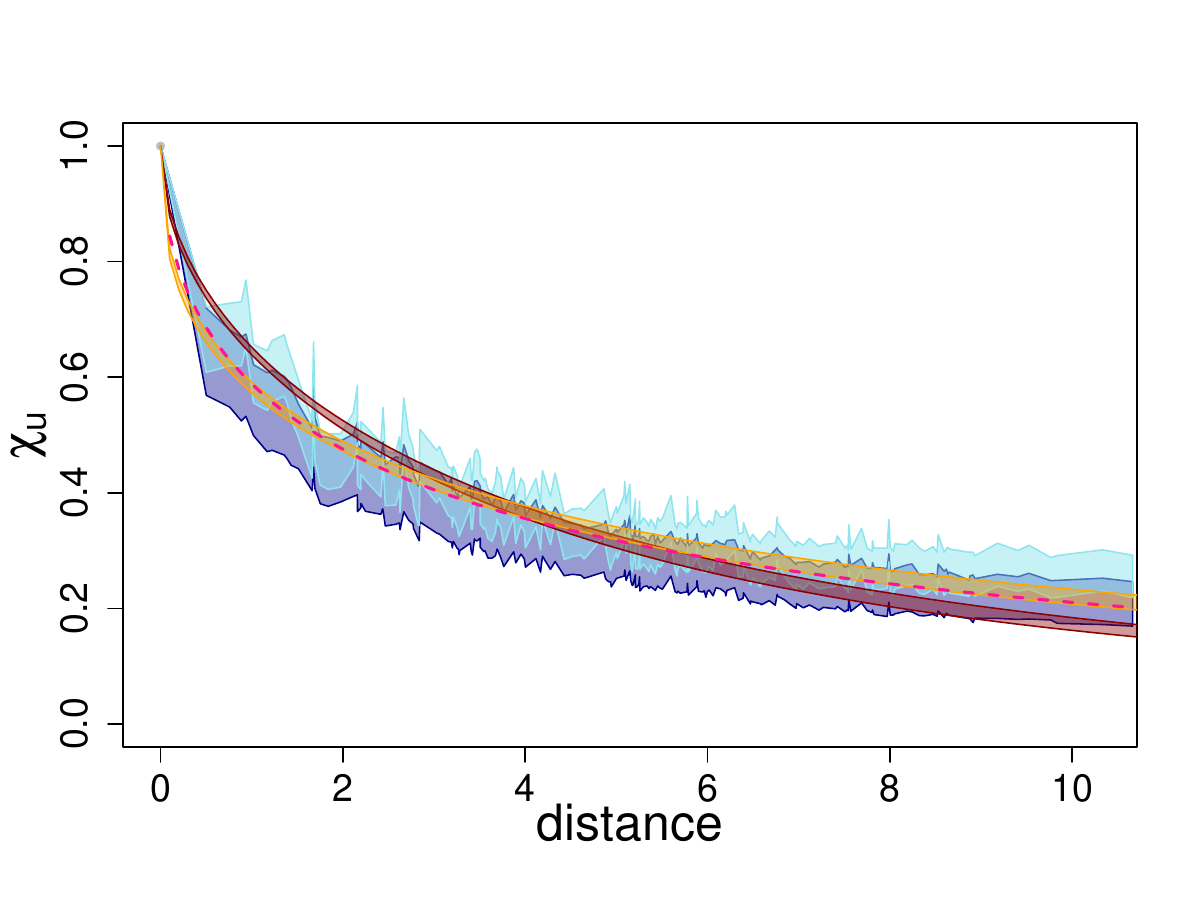} 
\caption{$X(\bm{s})$ simulated from an IBR process} \label{fig:cloud195_ibr}
\end{subfigure}

\begin{subfigure}[t]{1\textwidth}
\centering
  \includegraphics[width=0.3\textwidth]{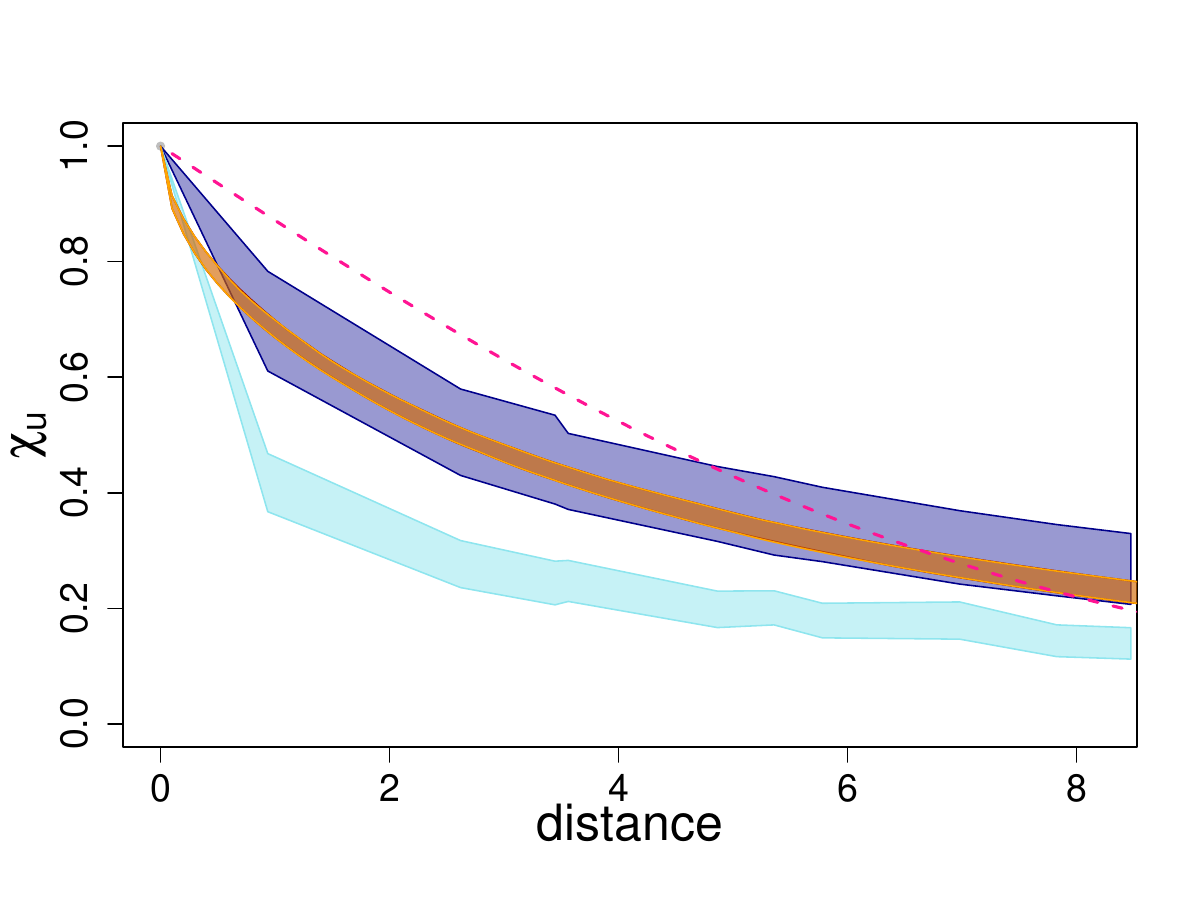}
  \hfill
  \includegraphics[width=0.3\textwidth]{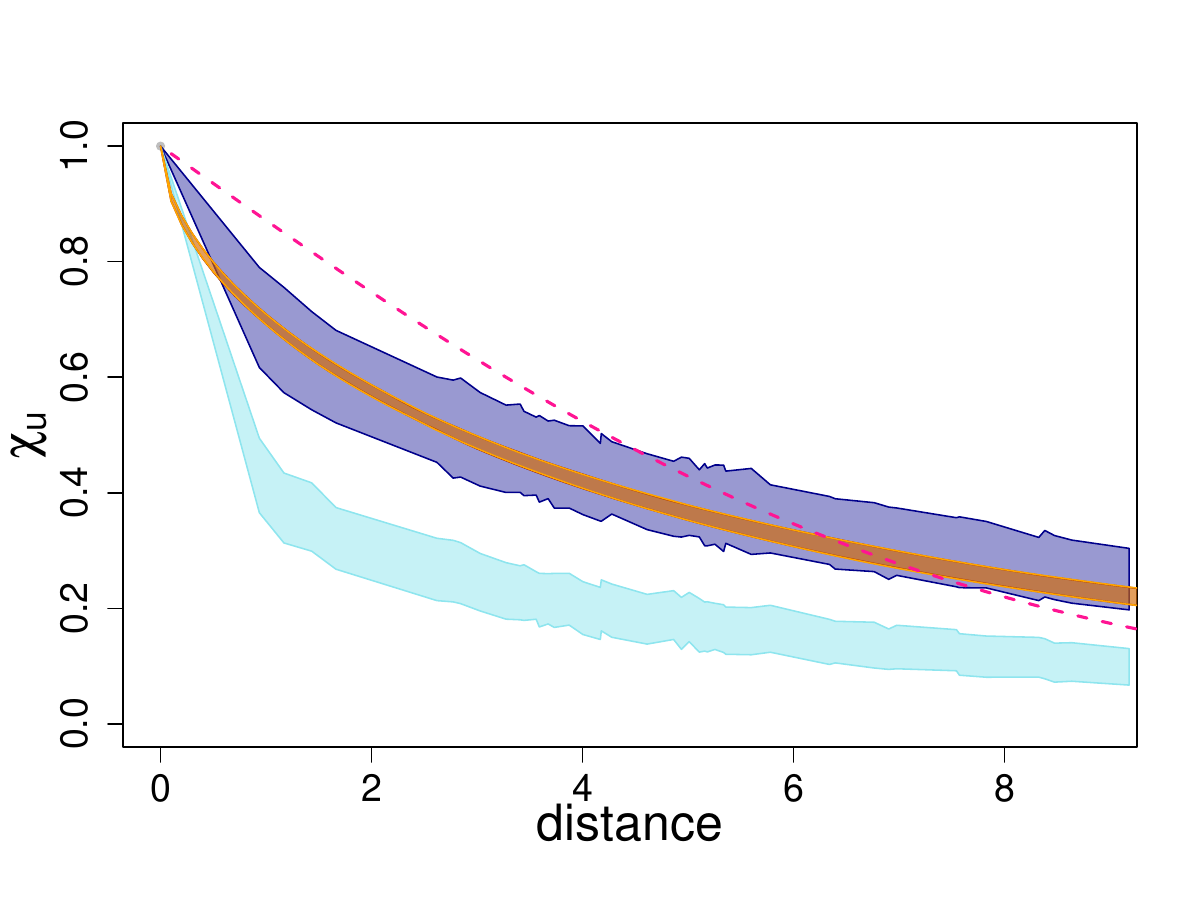}
  \hfill
  \includegraphics[width=0.3\textwidth]{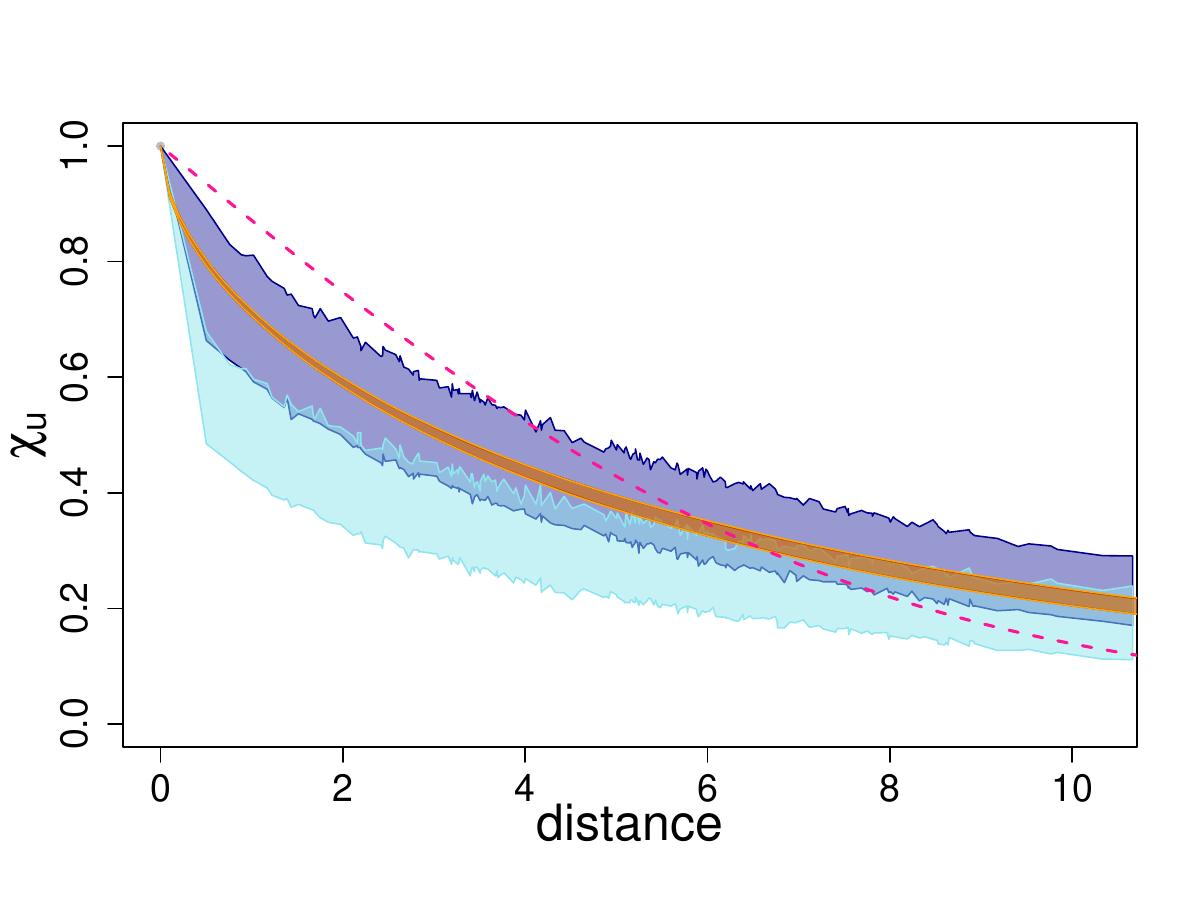}
\caption{$X(\bm{s})$ simulated from a BR process} \label{fig:cloud195_br}
\end{subfigure}

\begin{subfigure}[t]{1\textwidth}
\centering
  \includegraphics[width=0.3\textwidth]{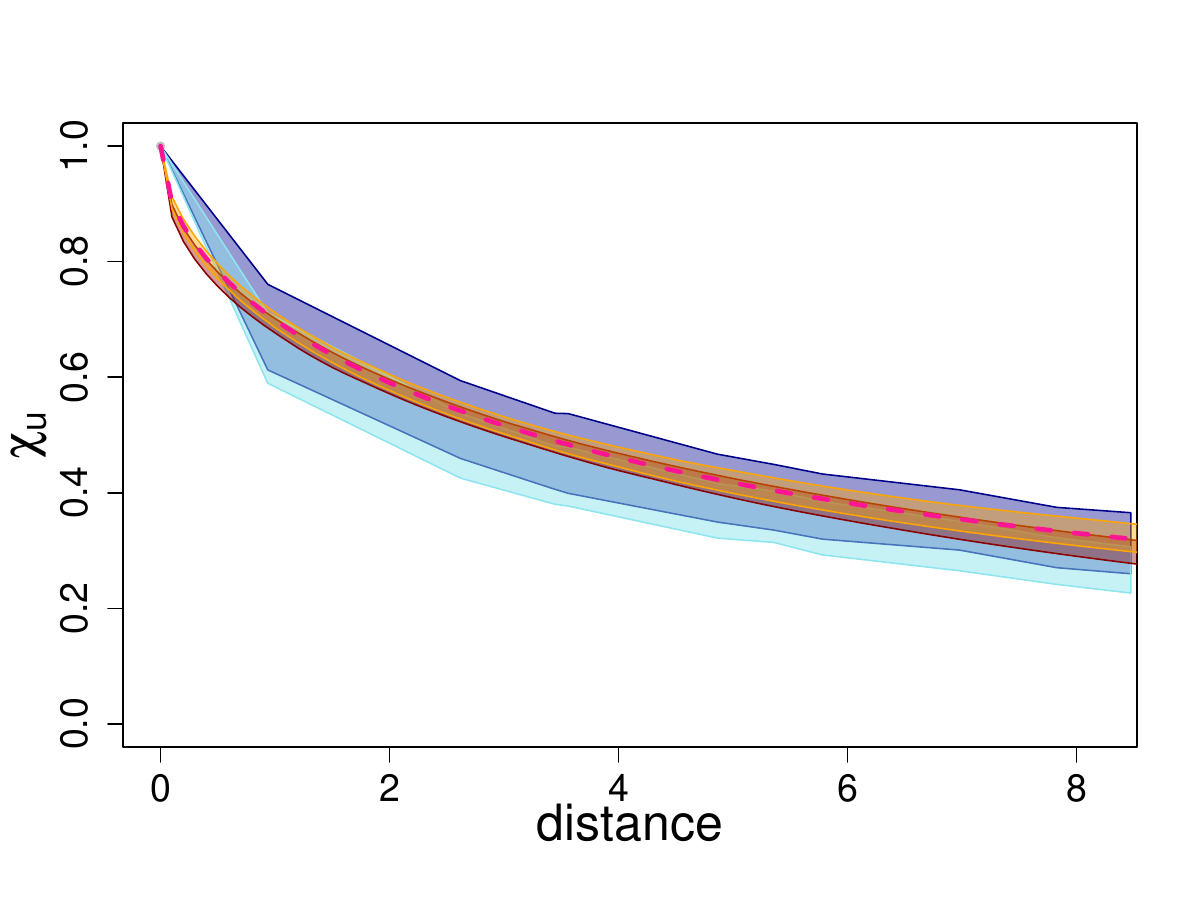}
  \hfill
  \includegraphics[width=0.3\textwidth]{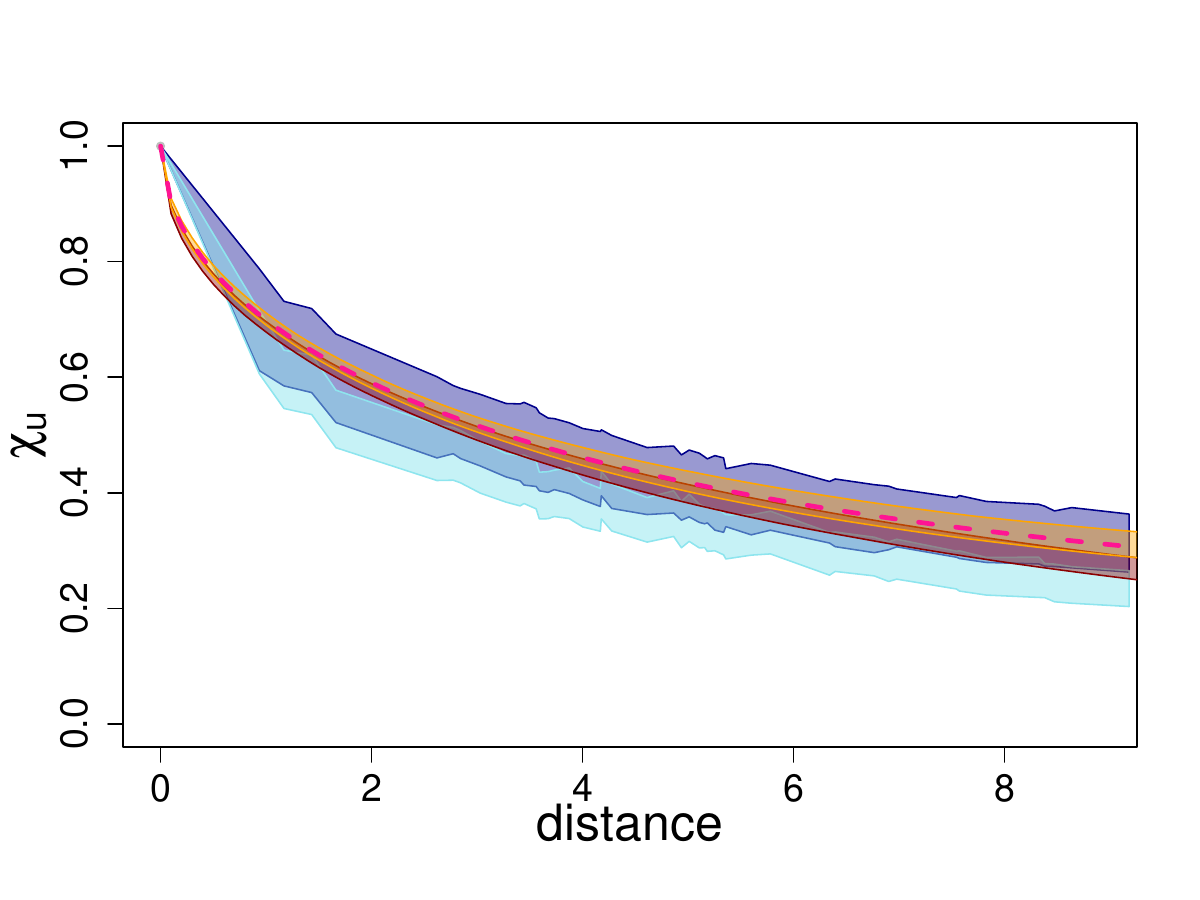}
  \hfill
  \includegraphics[width=0.3\textwidth]{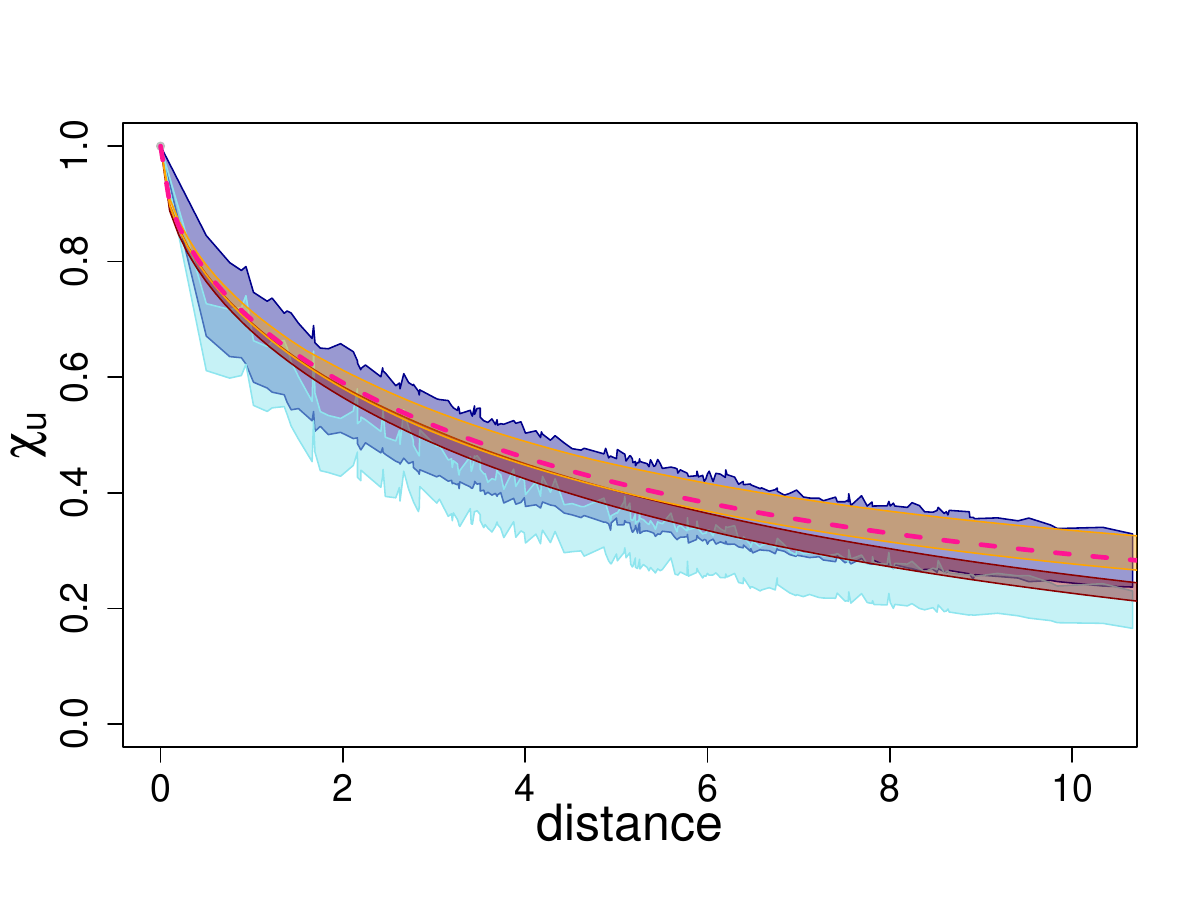}
\caption{$X(\bm{s})$ simulated from a HW process with $\delta=0.4$} \label{fig:cloud195_hw4}
\end{subfigure}

\begin{subfigure}[t]{1\textwidth}
\centering
  \includegraphics[width=0.3\textwidth]{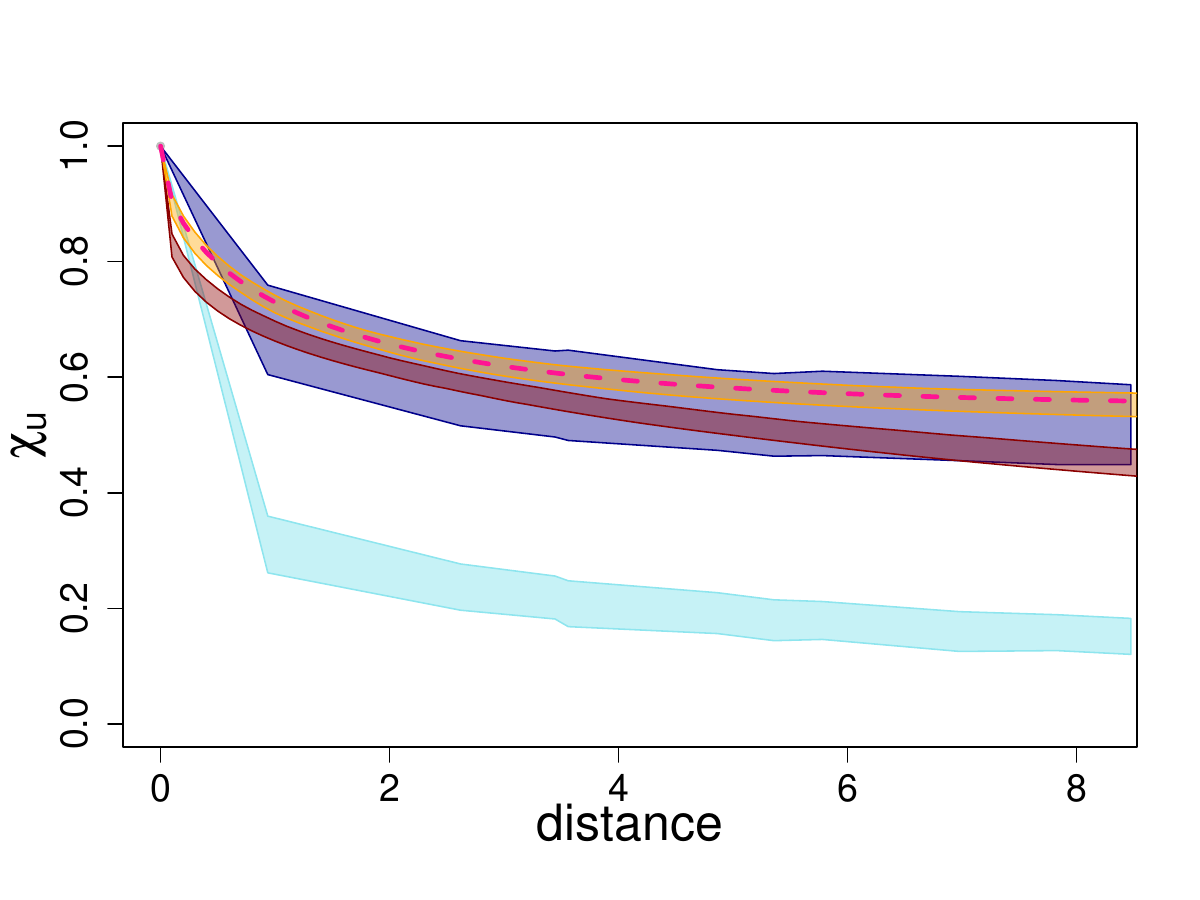}
  \hfill
  \includegraphics[width=0.3\textwidth]{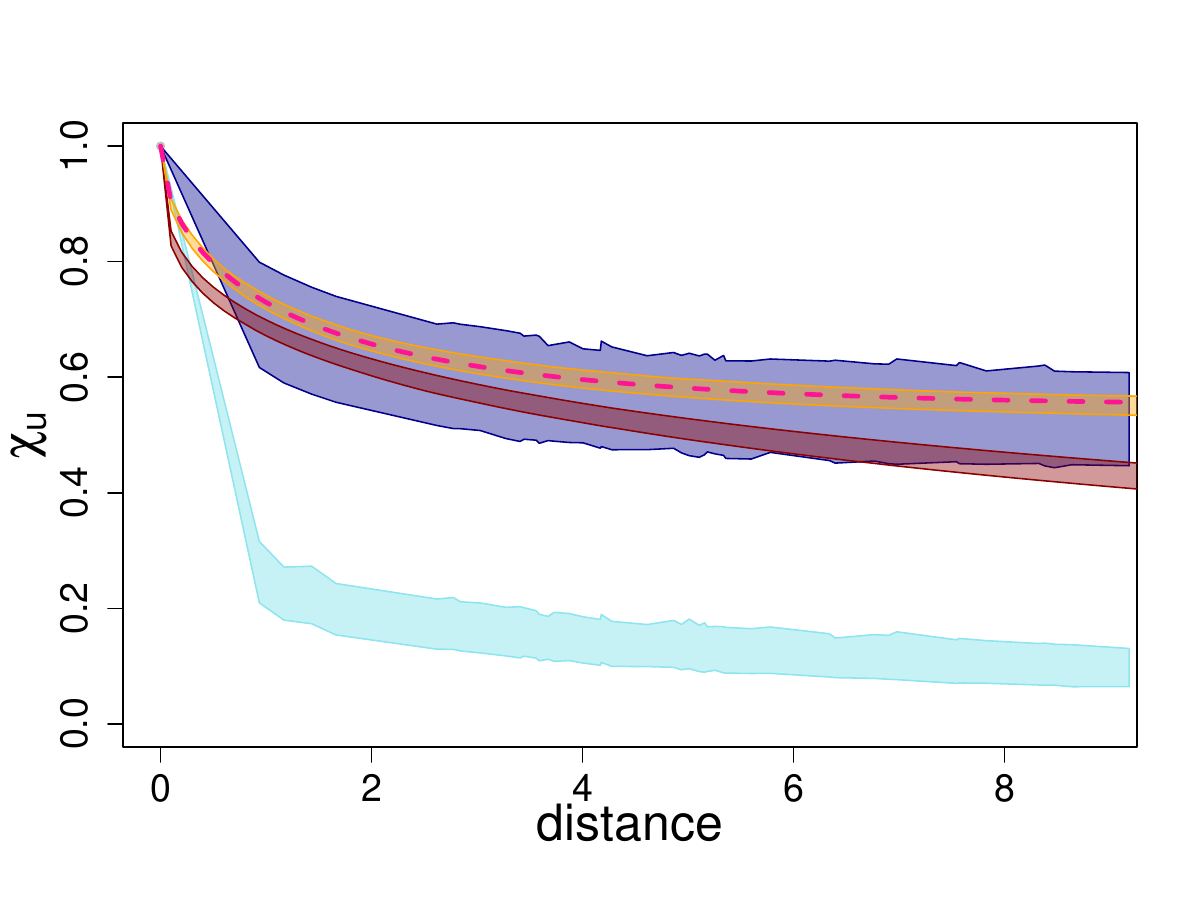}
  \hfill
  \includegraphics[width=0.3\textwidth]{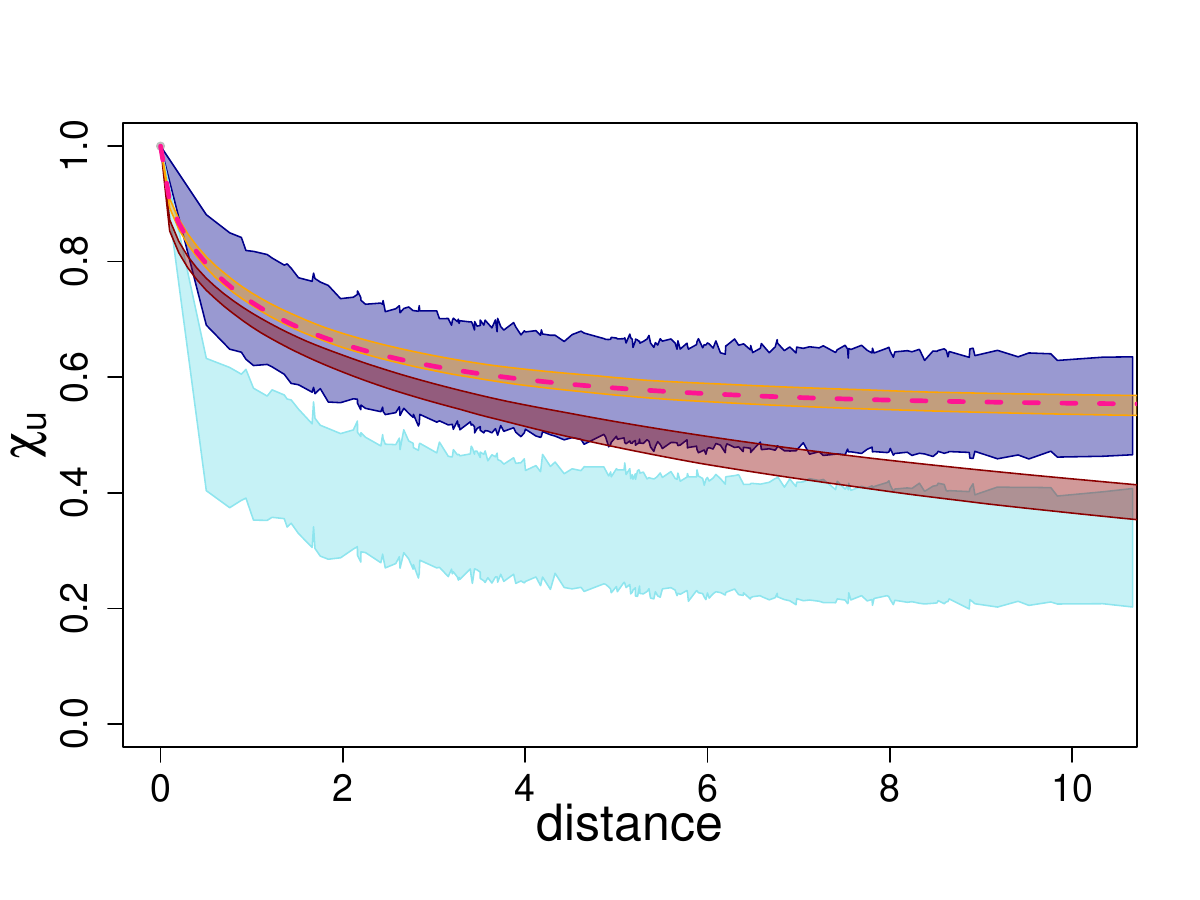}
\caption{$X(\bm{s})$ simulated from a HW process with $\delta=0.6$} \label{fig:cloud195_hw6}
\end{subfigure}

 \caption{Envelope plots of all $\binom{d}{2}$ pairwise $\chi_u$ estimates plotted over distance, calculated for $u=0.95$ using all $200$ simulated datasets. $X(\bm{s})$ is simulated as specified in panels (\subref{fig:cloud195_mvn}) - (\subref{fig:cloud195_hw6}) with parameter $\bm{\theta}_1$. Plots on the left correspond to $d=5$, $d=10$ in the middle and $d=20$ on the right. Envelopes in dark and light blue are obtained via the empirical angular distribution and the angular distribution in \eqref{eq:am1}, respectively. Envelopes in red come from cG fits, while envelopes in orange from HW fits. The dashed pink curve corresponds to the simulated truth.}
 \label{fig:cloud195} 
\end{figure}

\newpage

\begin{figure}[H]
\centering
\begin{subfigure}[t]{1\textwidth}
\centering
  \includegraphics[width=0.3\textwidth]{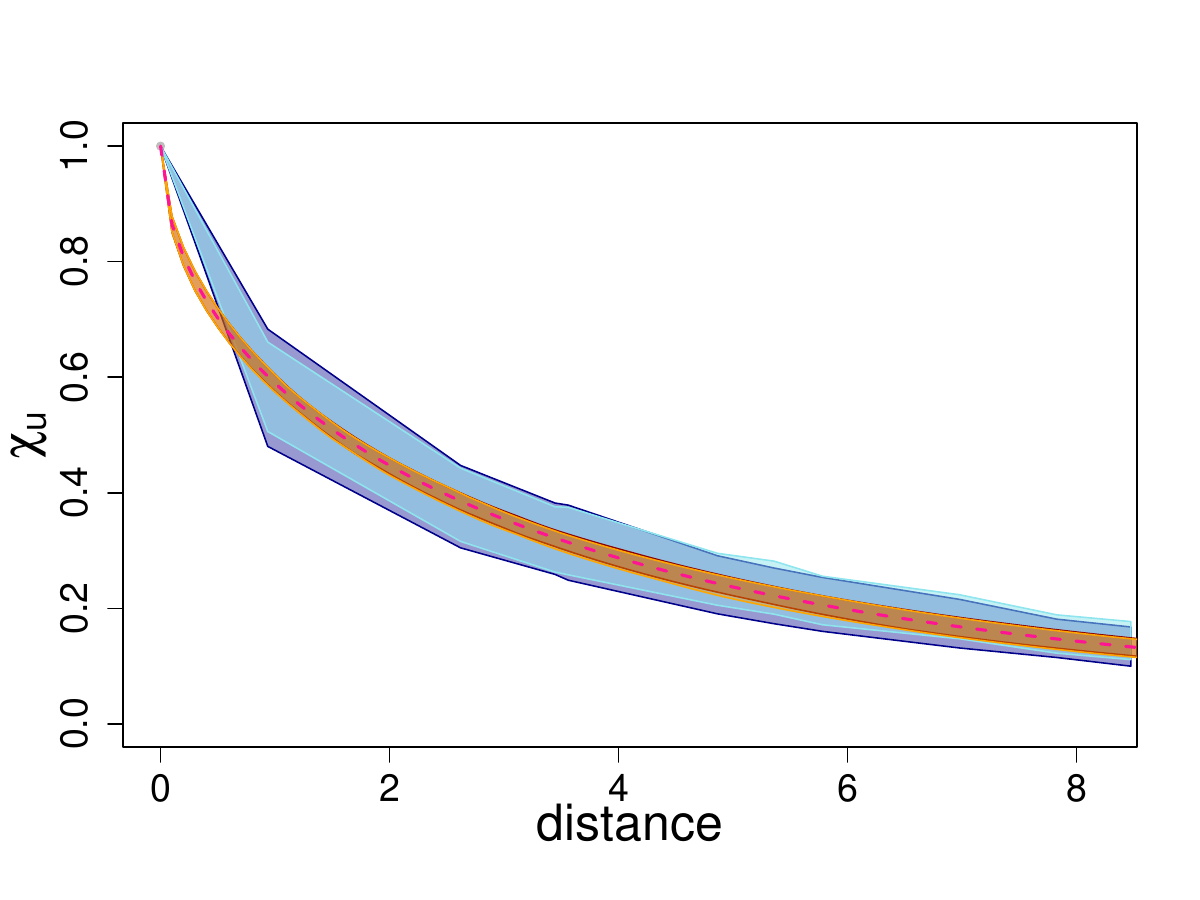}
  \hfill
  \includegraphics[width=0.3\textwidth]{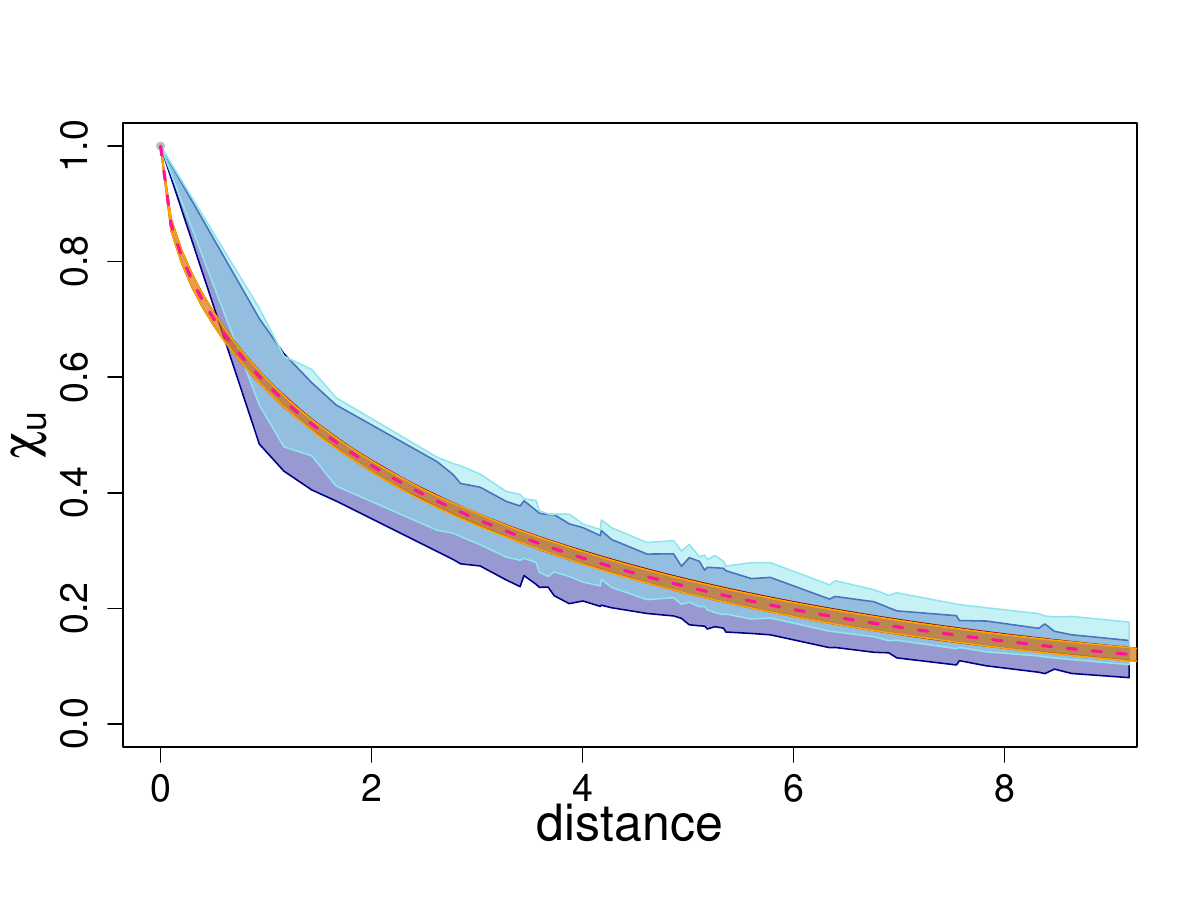}
  \hfill
  \includegraphics[width=0.3\textwidth]{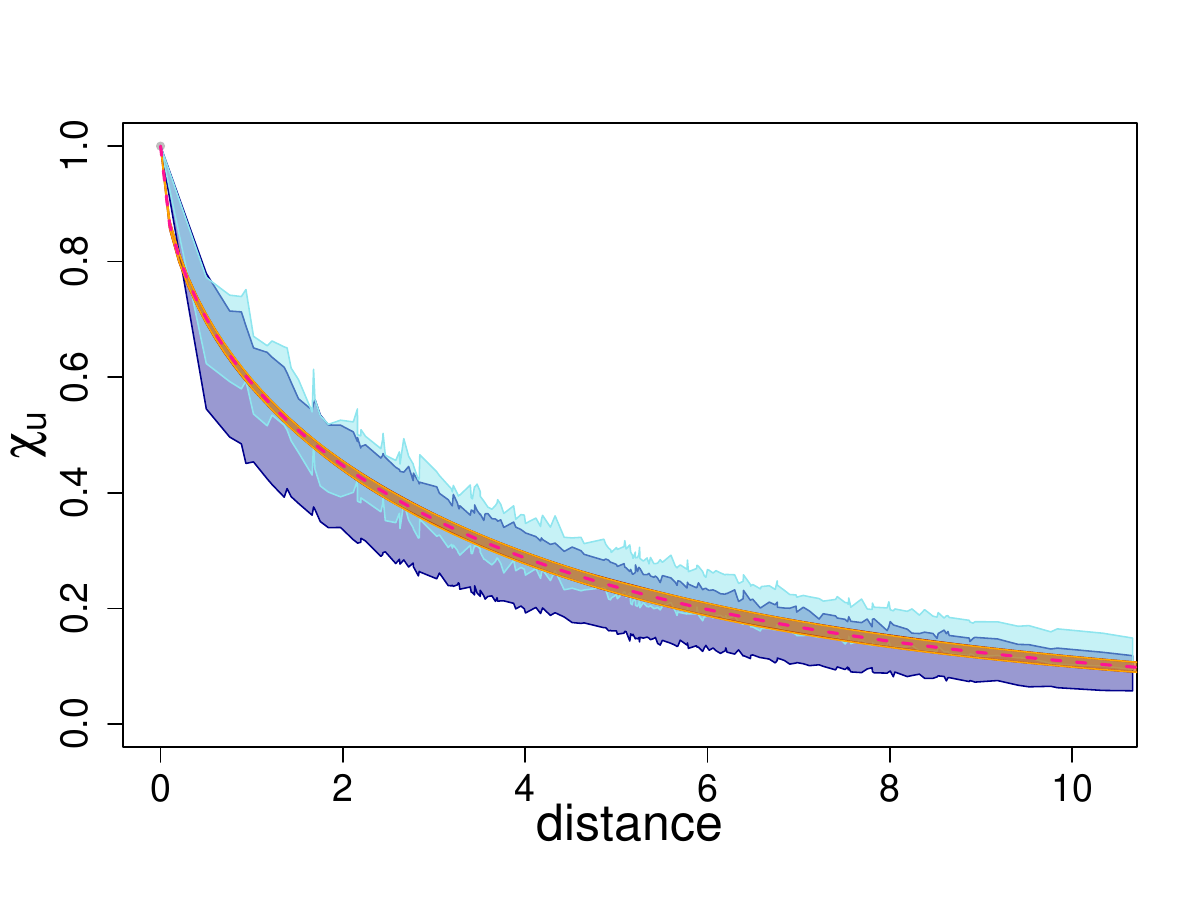}
\caption{$X(\bm{s})$ simulated from a Gaussian process} \label{fig:cloud198_mvn}
\end{subfigure}

\begin{subfigure}[t]{1\textwidth}
\centering
  \includegraphics[width=0.3\textwidth]{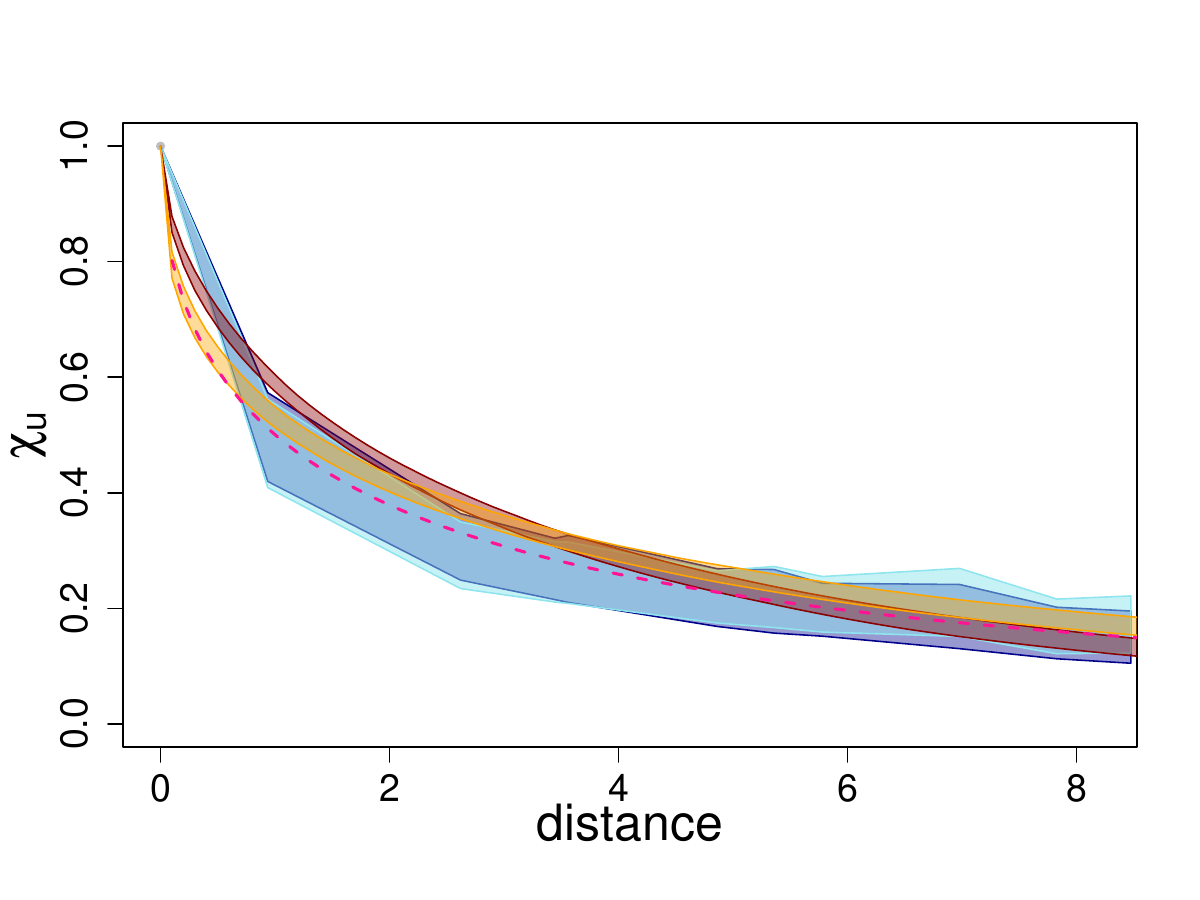}
  \hfill
  \includegraphics[width=0.3\textwidth]{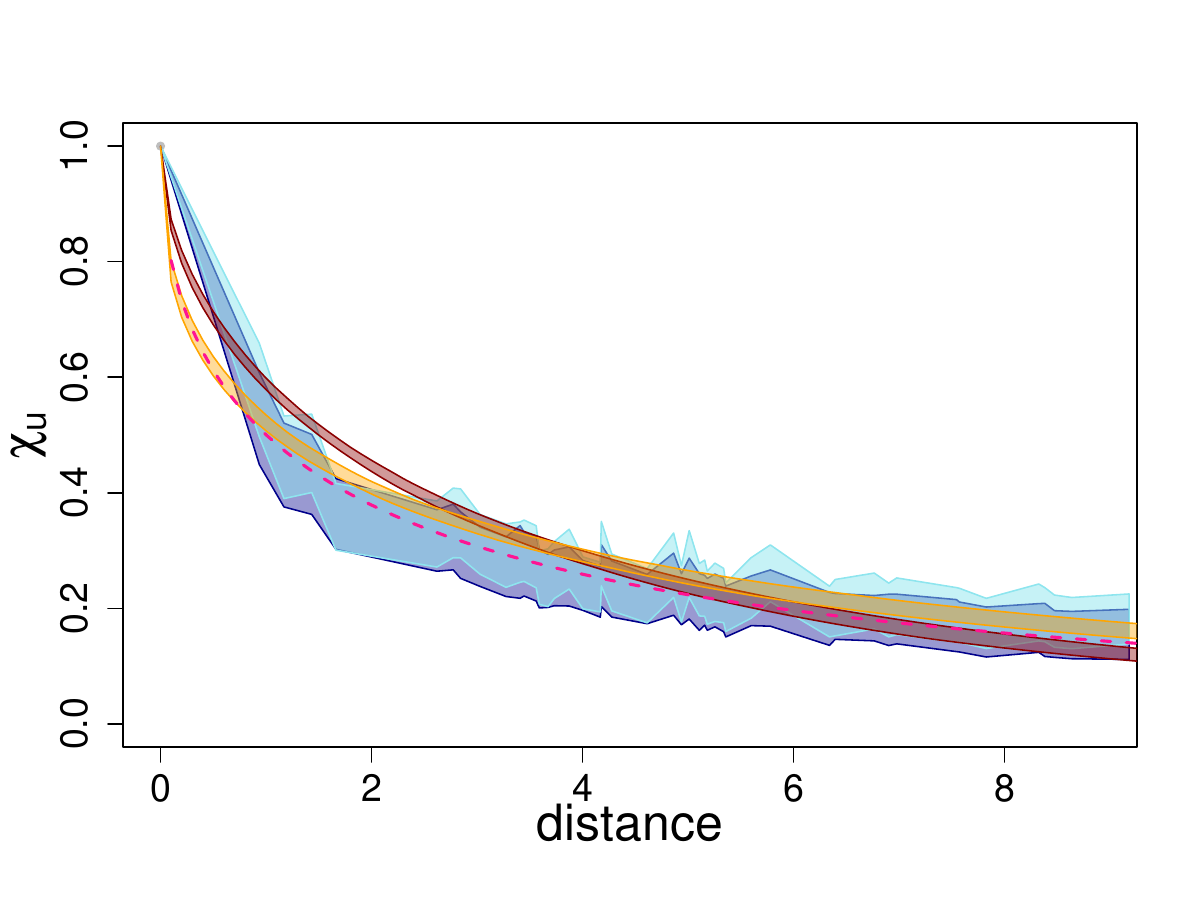}
  \hfill
  \includegraphics[width=0.3\textwidth]{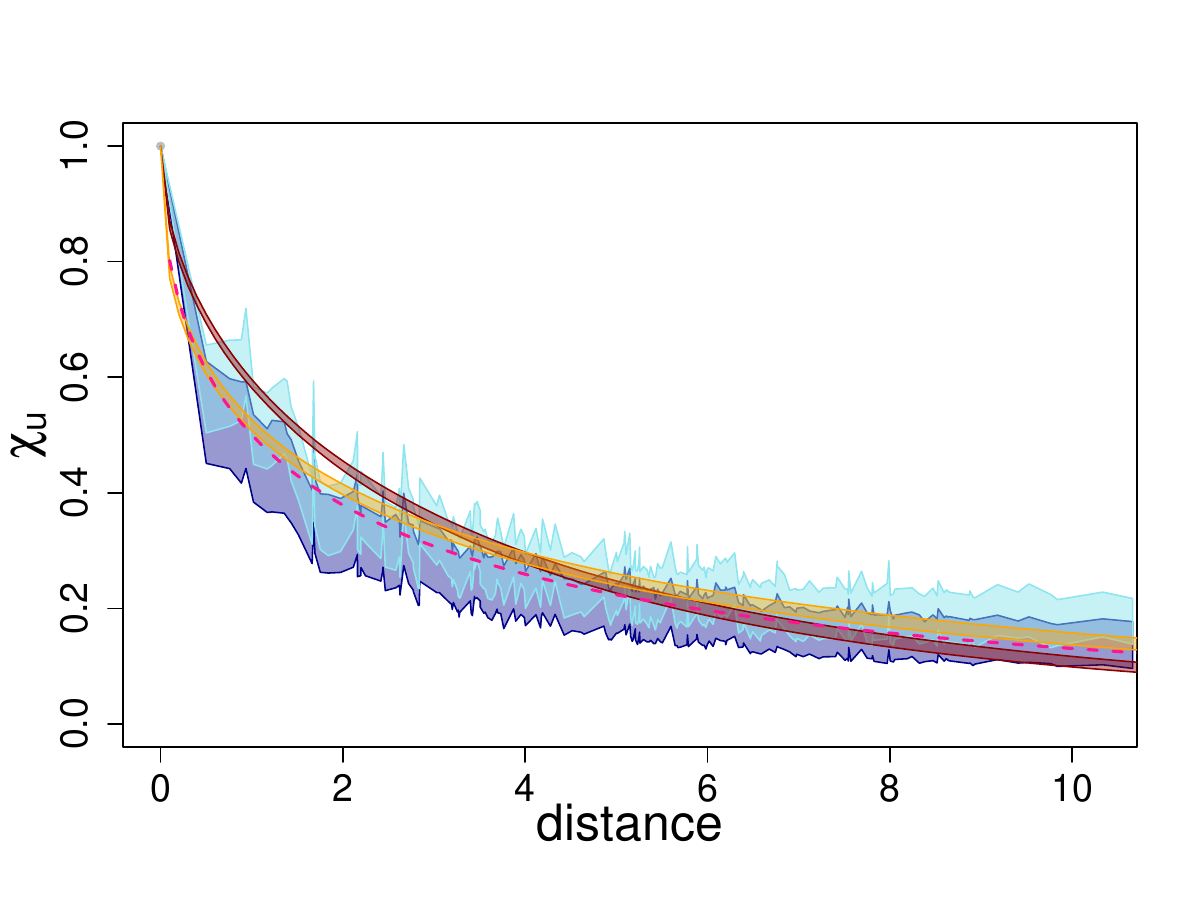} 
\caption{$X(\bm{s})$ simulated from an IBR process} \label{fig:cloud198_ibr}
\end{subfigure}

\begin{subfigure}[t]{1\textwidth}
\centering
  \includegraphics[width=0.3\textwidth]{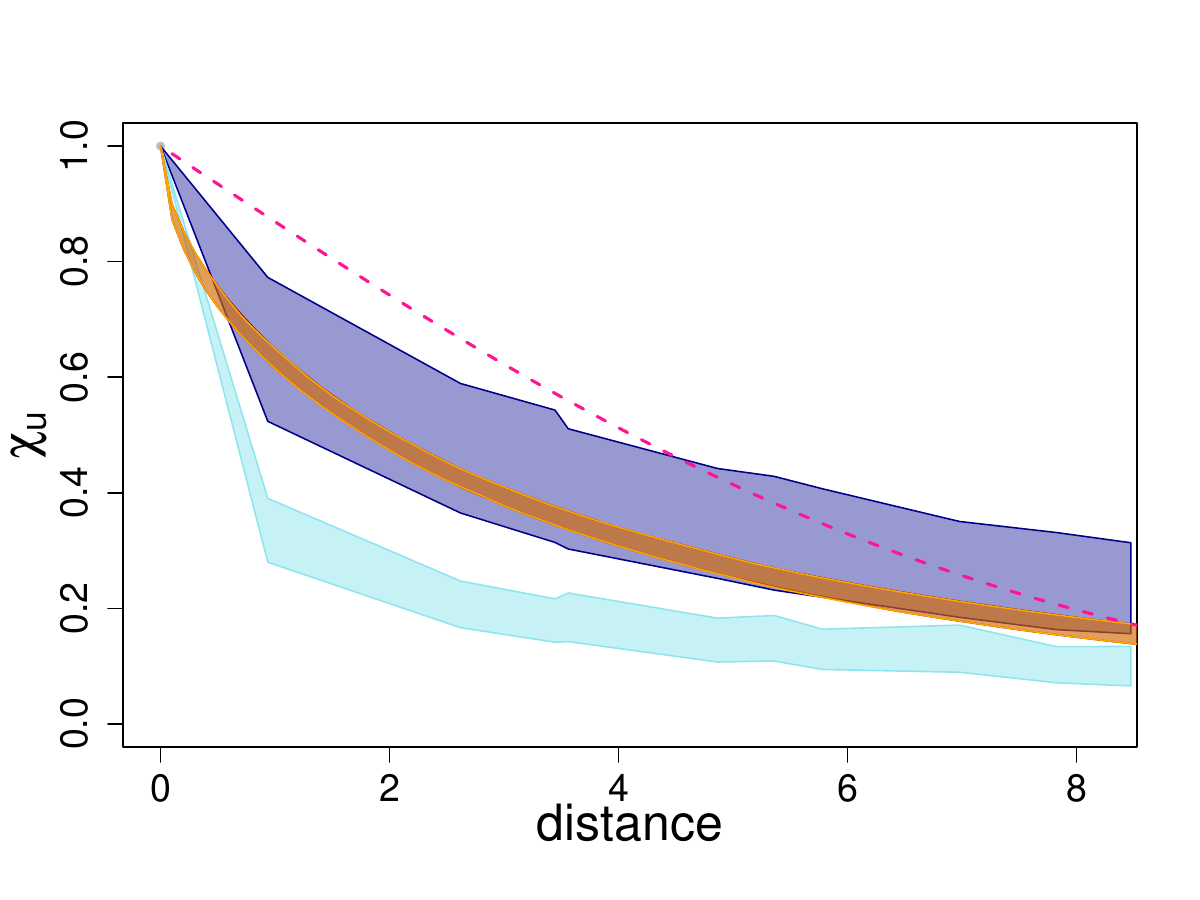}
  \hfill
  \includegraphics[width=0.3\textwidth]{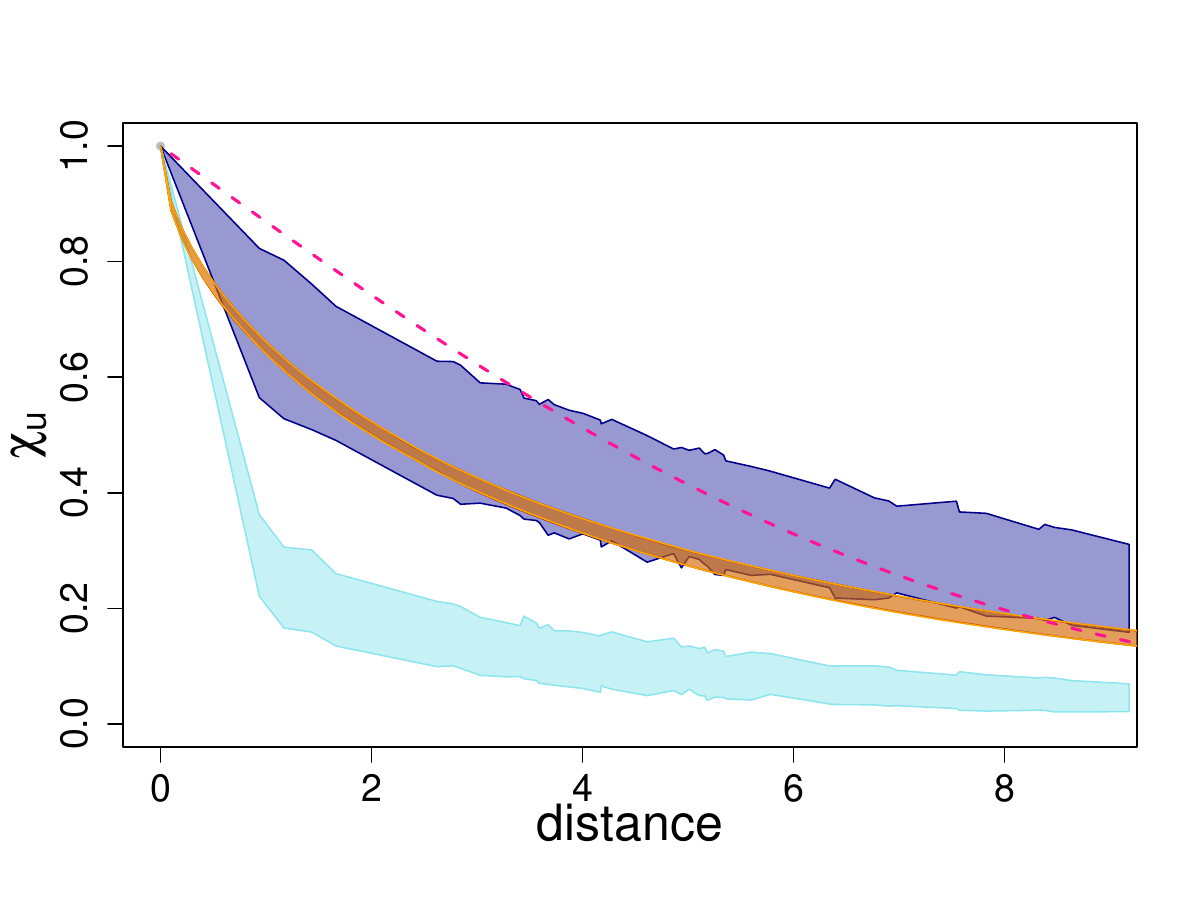}
  \hfill
  \includegraphics[width=0.3\textwidth]{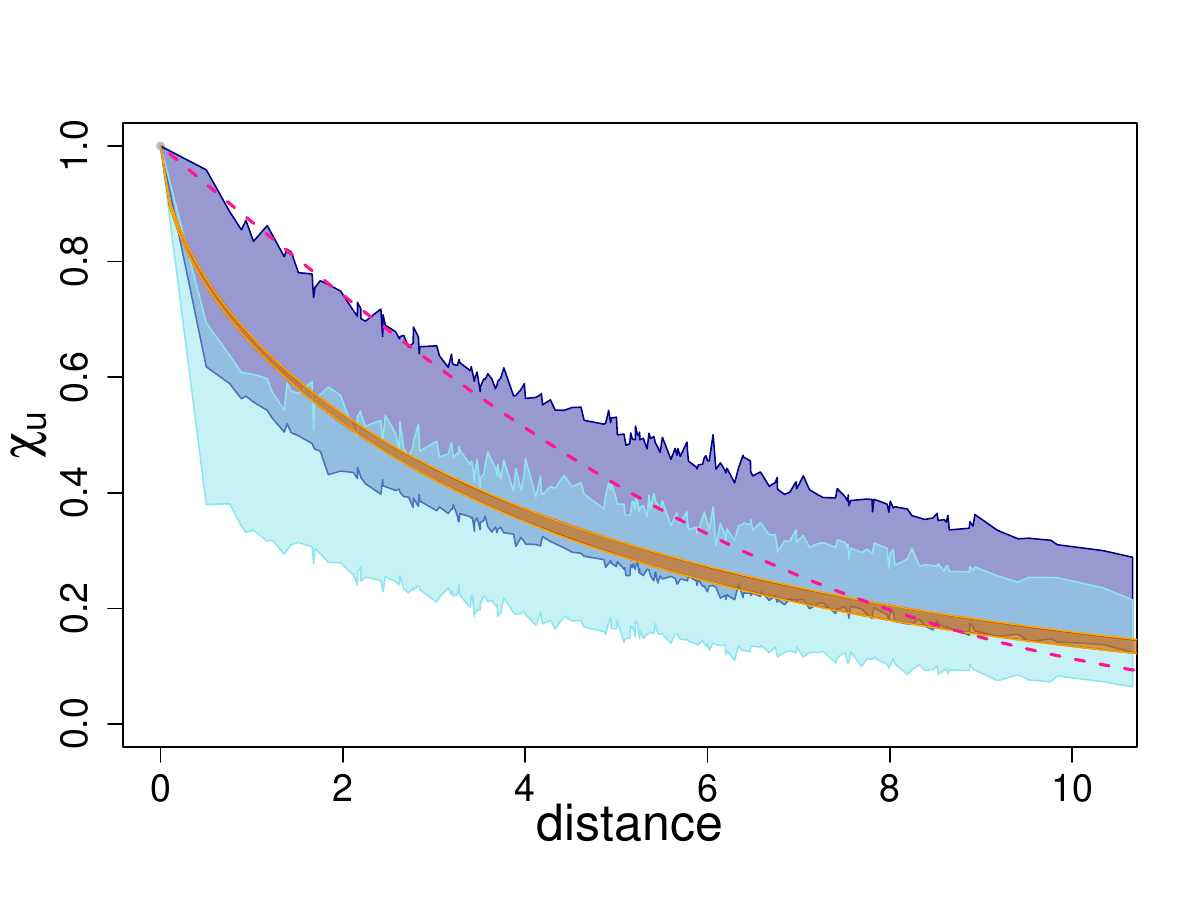}
\caption{$X(\bm{s})$ simulated from a BR process} \label{fig:cloud198_br}
\end{subfigure}

\begin{subfigure}[t]{1\textwidth}
\centering
  \includegraphics[width=0.3\textwidth]{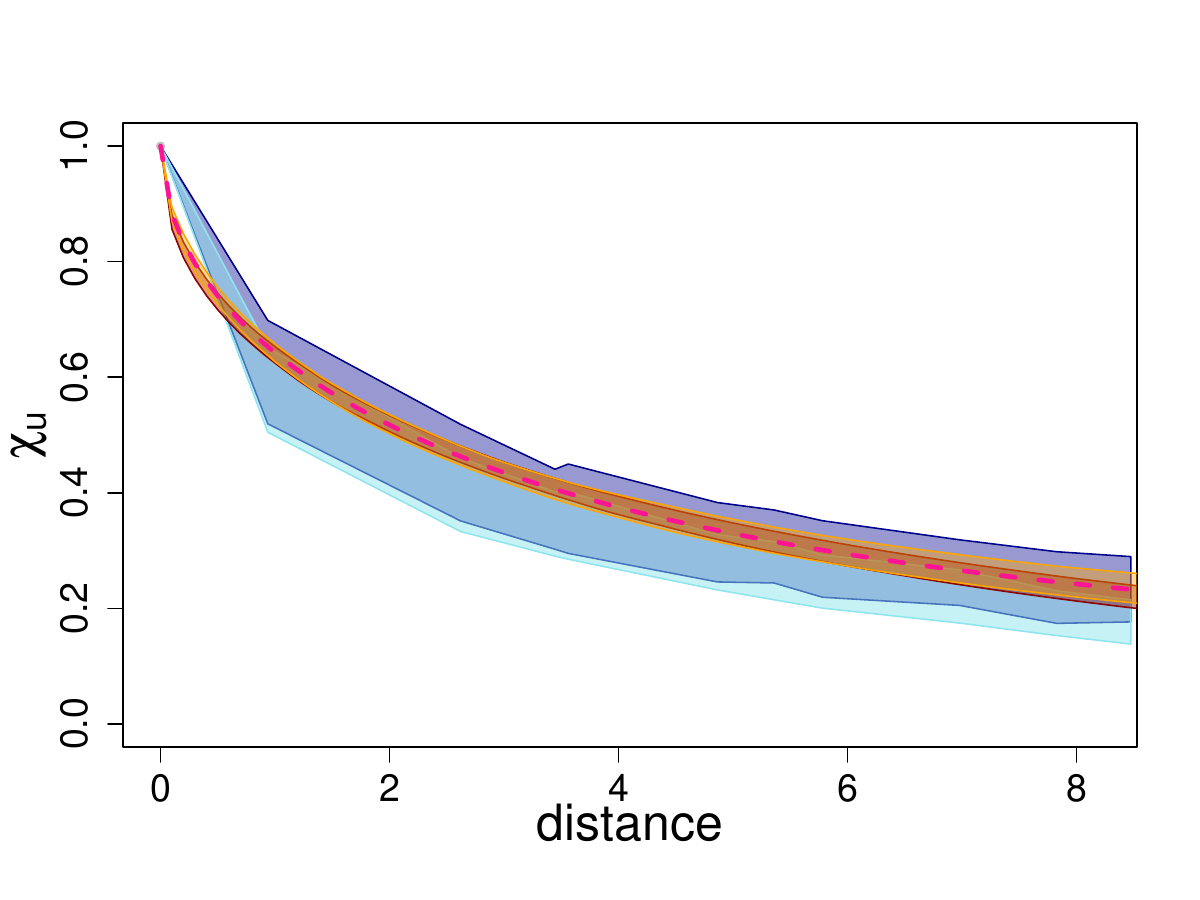}
  \hfill
  \includegraphics[width=0.3\textwidth]{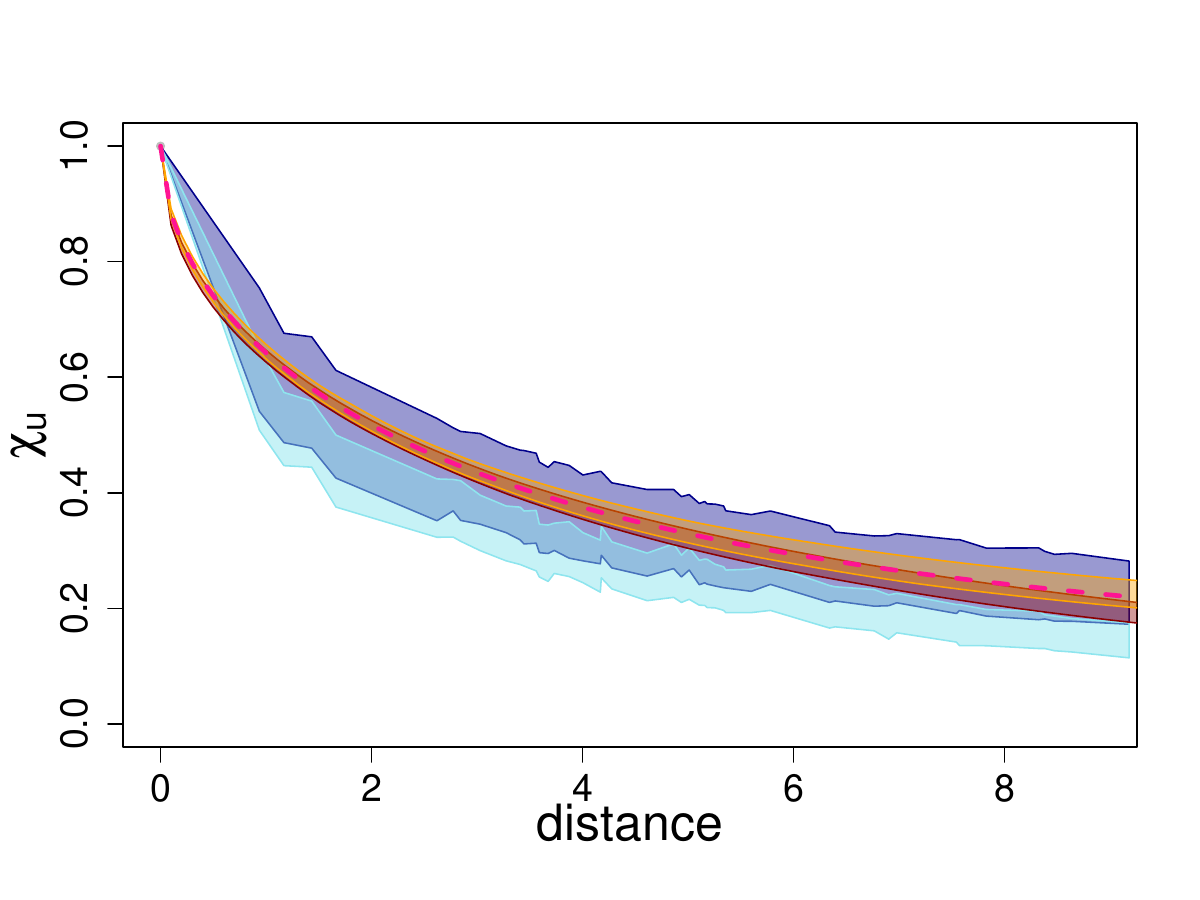}
  \hfill
  \includegraphics[width=0.3\textwidth]{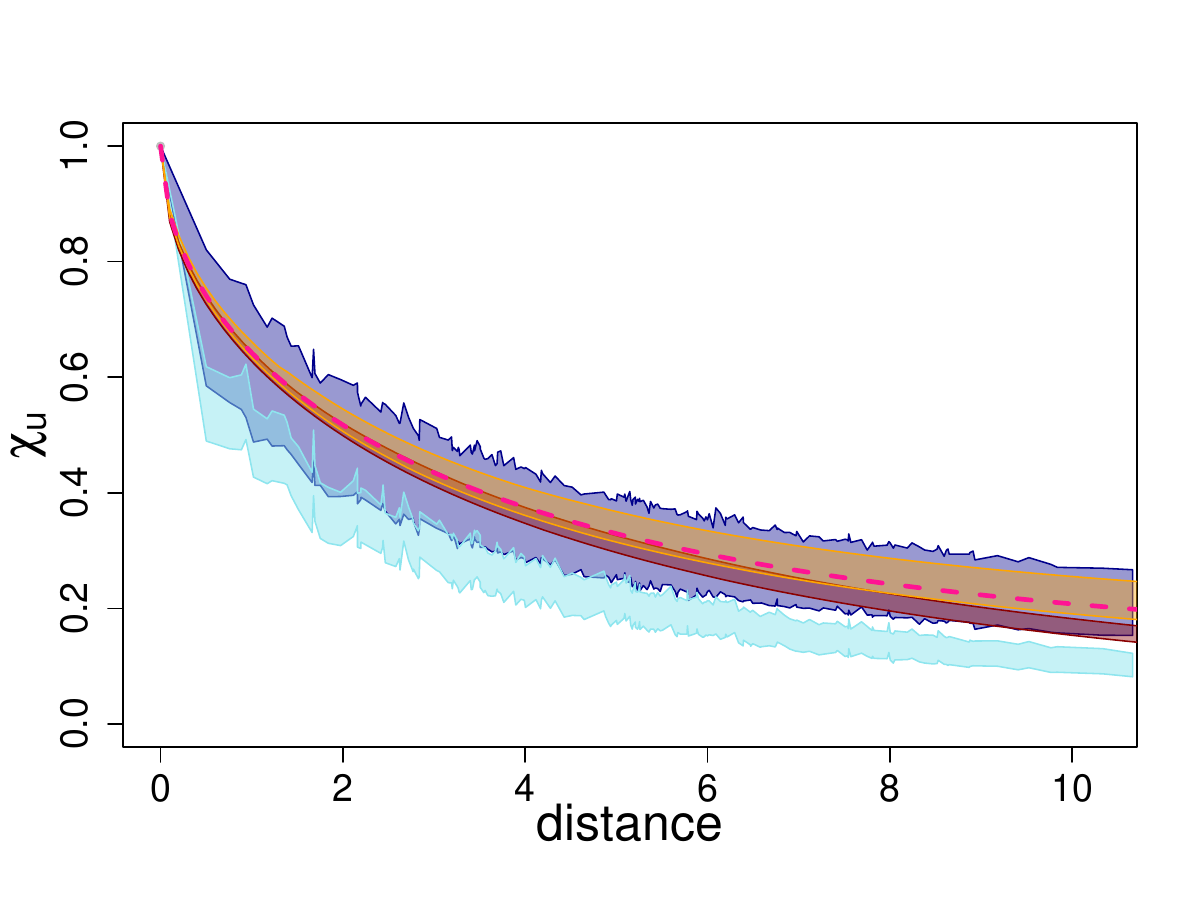}
\caption{$X(\bm{s})$ simulated from a HW process with $\delta=0.4$} \label{fig:cloud198_hw4}
\end{subfigure}

\begin{subfigure}[t]{1\textwidth}
\centering
  \includegraphics[width=0.3\textwidth]{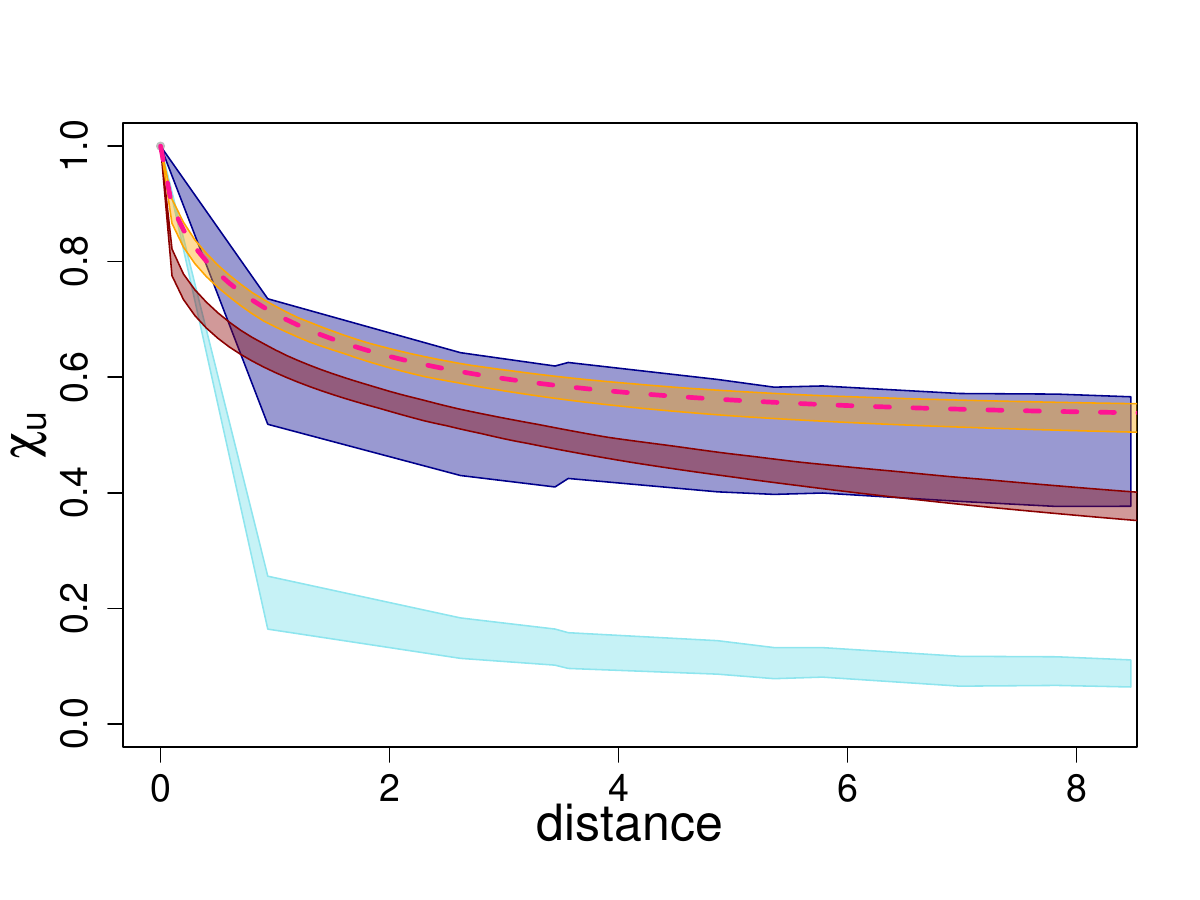}
  \hfill
  \includegraphics[width=0.3\textwidth]{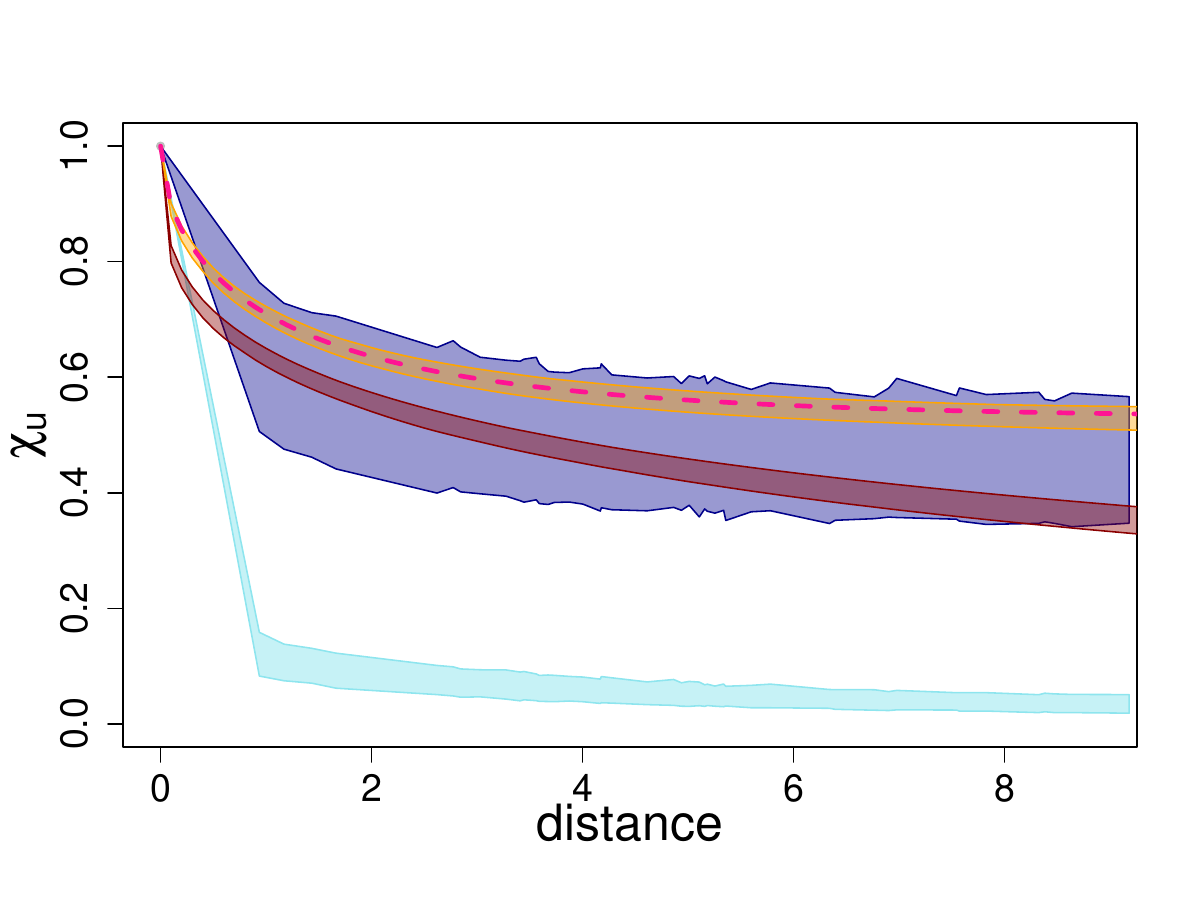}
  \hfill
  \includegraphics[width=0.3\textwidth]{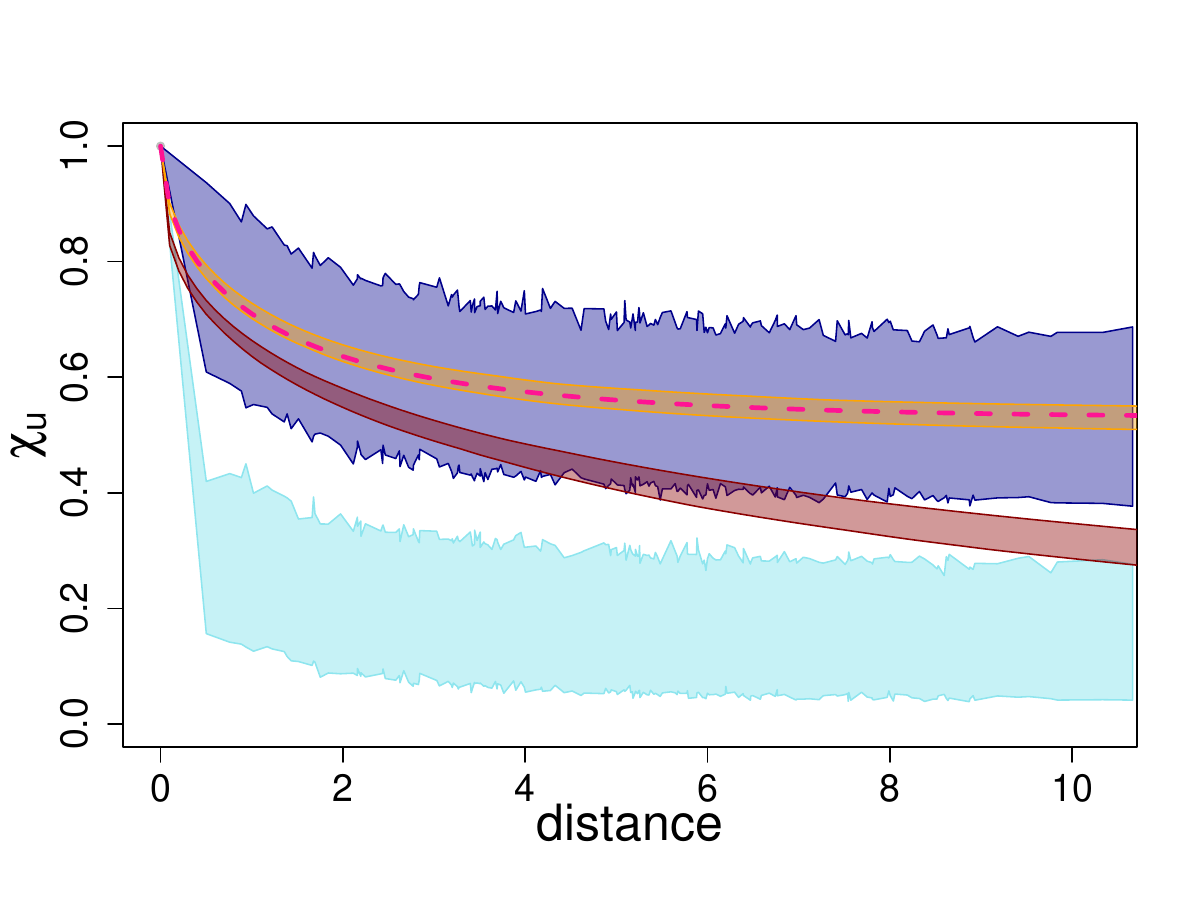}
\caption{$X(\bm{s})$ simulated from a HW process with $\delta=0.6$} \label{fig:cloud198_hw6}
\end{subfigure}

 \caption{Envelope plots of all $\binom{d}{2}$ pairwise $\chi_u$ estimates plotted over distance, calculated for $u=0.98$ using all $200$ simulated datasets. $X(\bm{s})$ is simulated as specified in panels (\subref{fig:cloud198_mvn}) - (\subref{fig:cloud198_hw6}) with parameter $\bm{\theta}_1$. Plots on the left correspond to $d=5$, $d=10$ in the middle and $d=20$ on the right. Envelopes in dark and light blue are obtained via the empirical angular distribution and the angular distribution in \eqref{eq:am1}, respectively. Envelopes in red come from cG fits, while envelopes in orange from HW fits. The dashed pink curve corresponds to the simulated truth.}
 \label{fig:cloud198} 
\end{figure}

\newpage

\begin{figure}[H]
\centering
\begin{subfigure}[t]{1\textwidth}
\centering
  \includegraphics[width=0.3\textwidth]{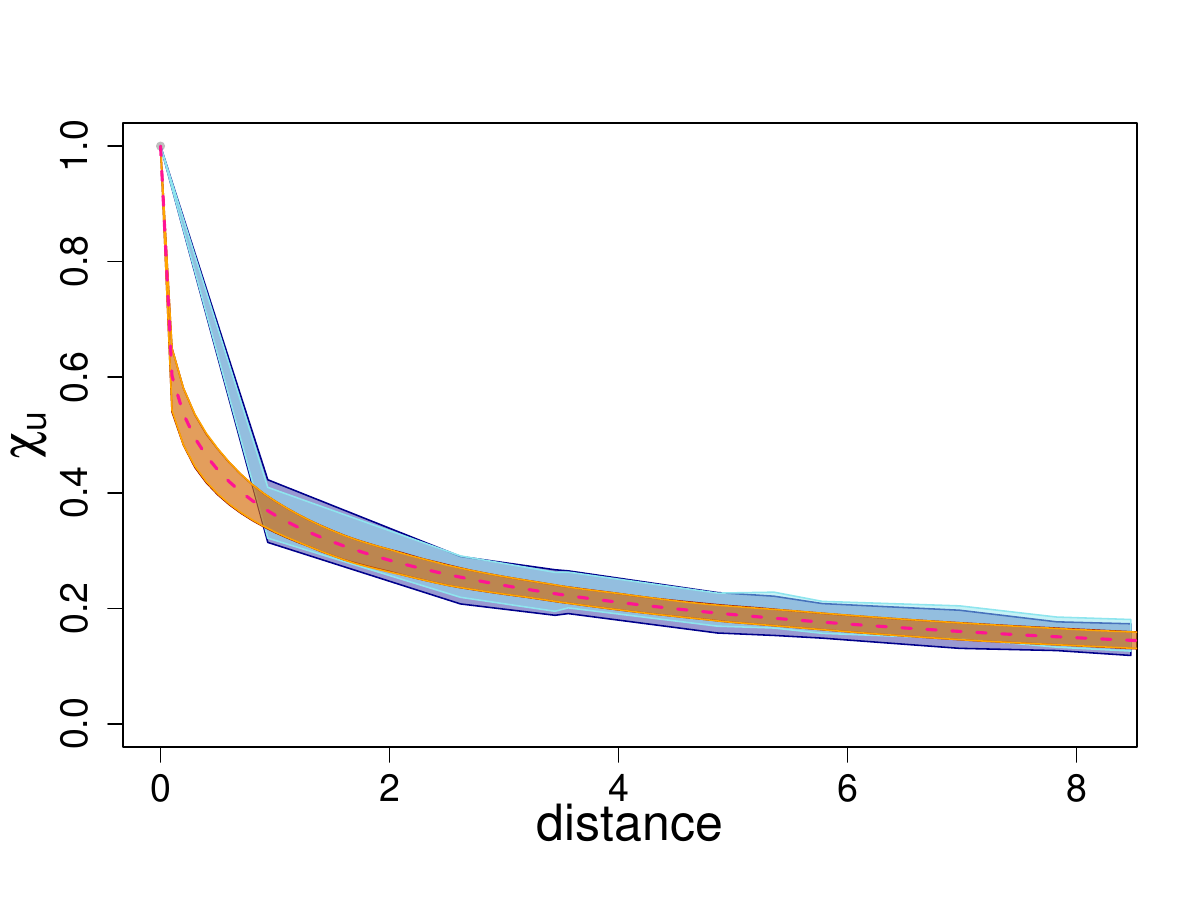}
  \hfill
  \includegraphics[width=0.3\textwidth]{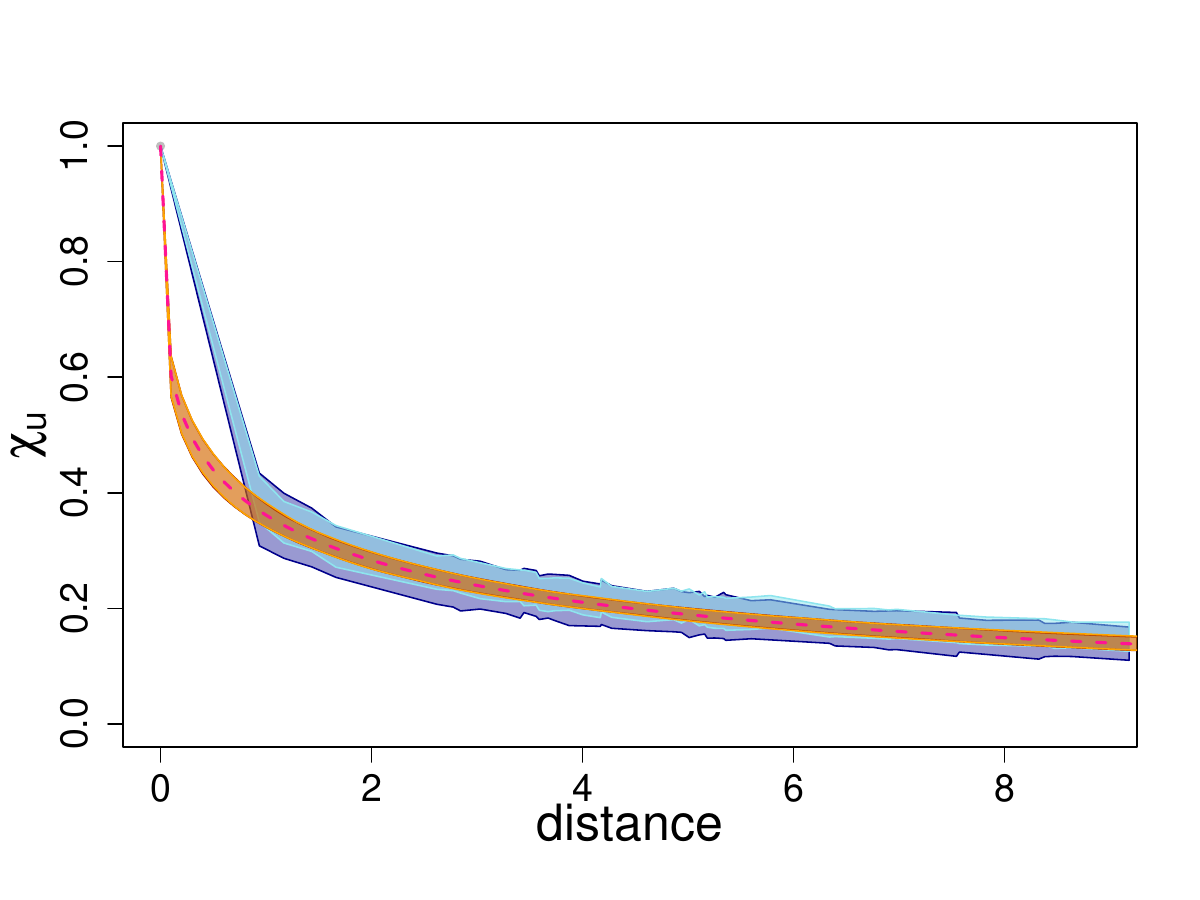}
  \hfill
  \includegraphics[width=0.3\textwidth]{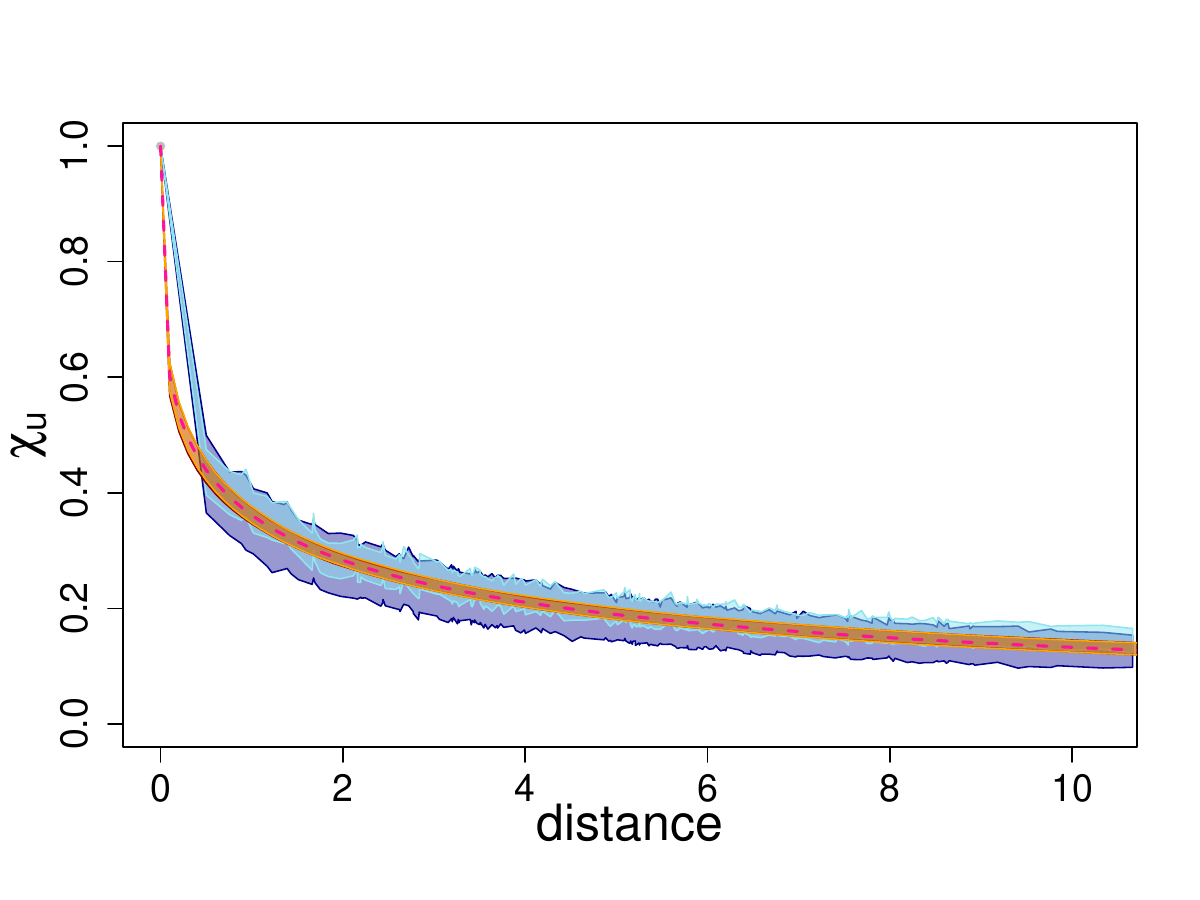}
\caption{$X(\bm{s})$ simulated from a Gaussian process} \label{fig:cloud295_mvn}
\end{subfigure}

\begin{subfigure}[t]{1\textwidth}
\centering
  \includegraphics[width=0.3\textwidth]{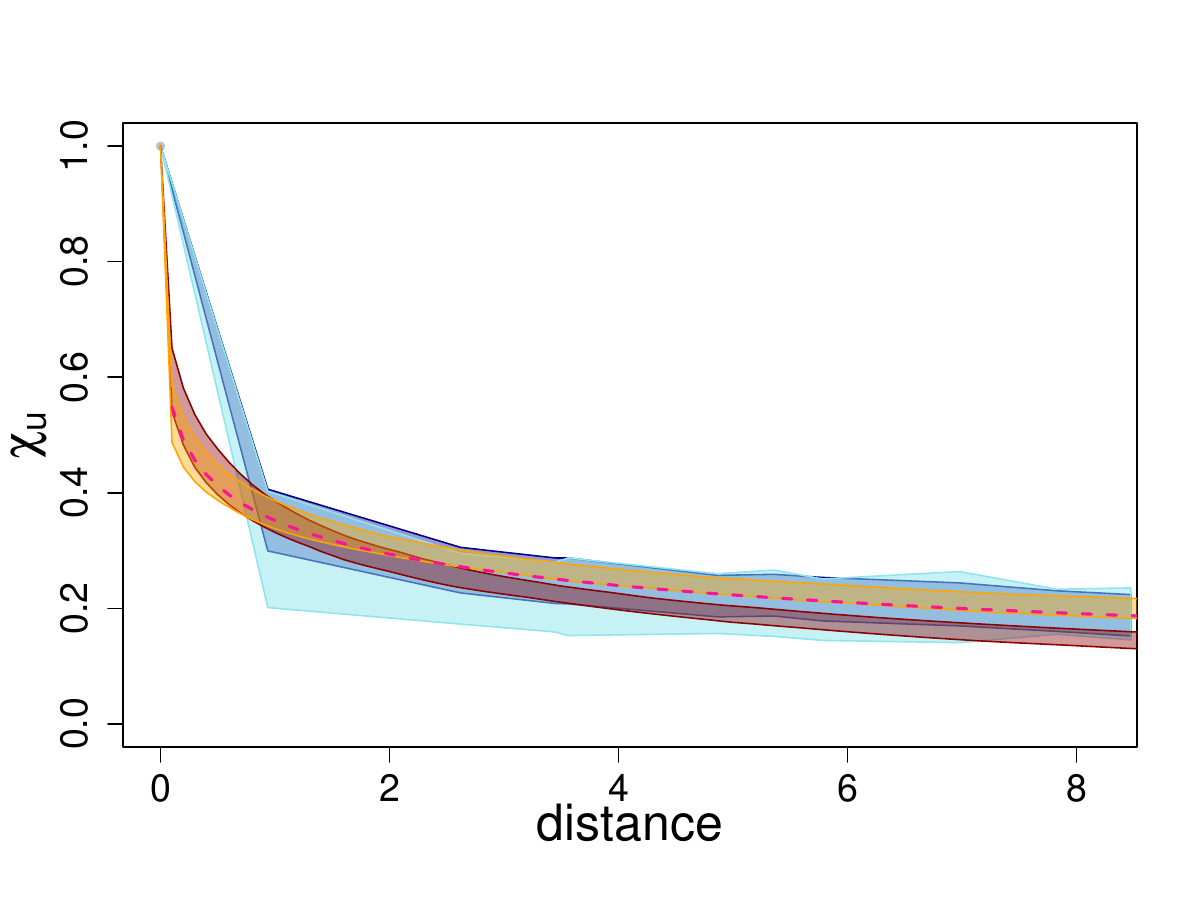}
  \hfill
  \includegraphics[width=0.3\textwidth]{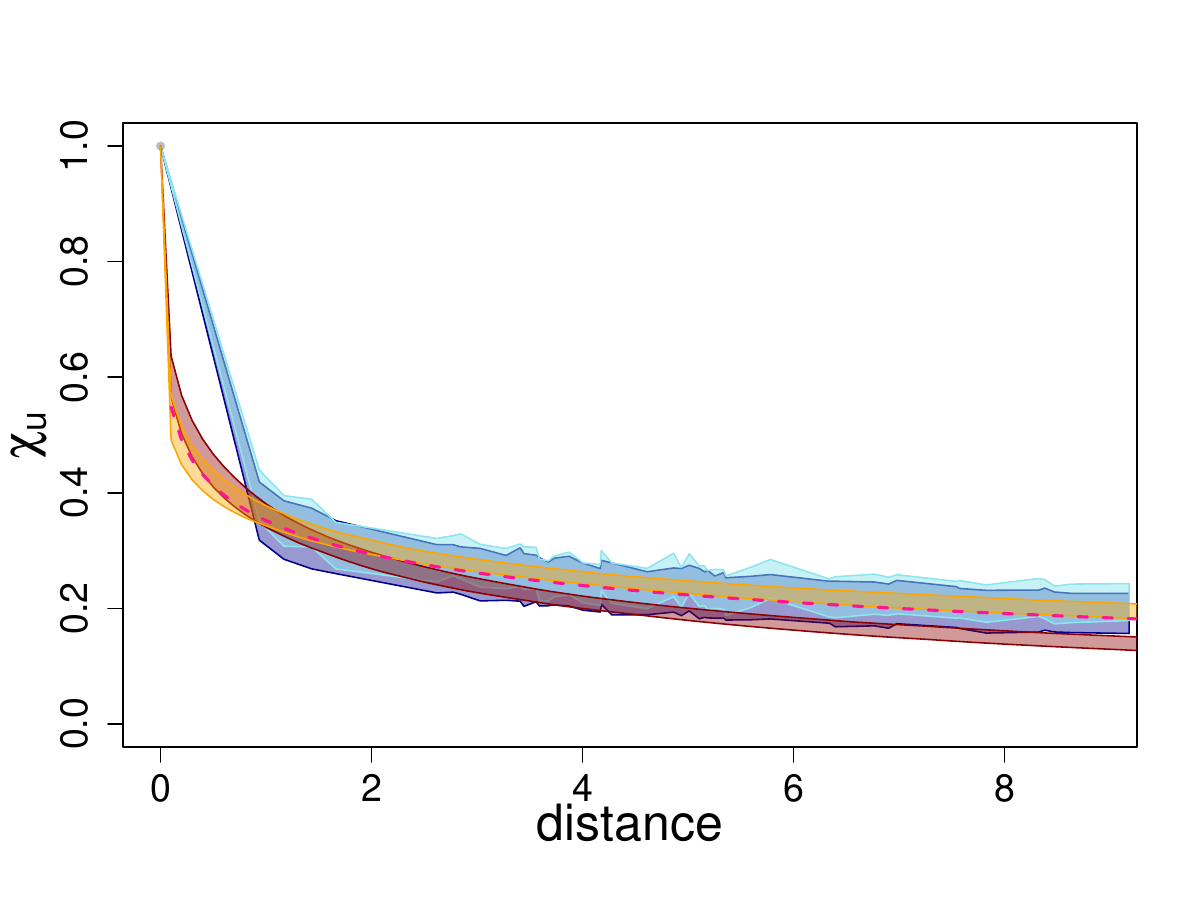}
  \hfill
  \includegraphics[width=0.3\textwidth]{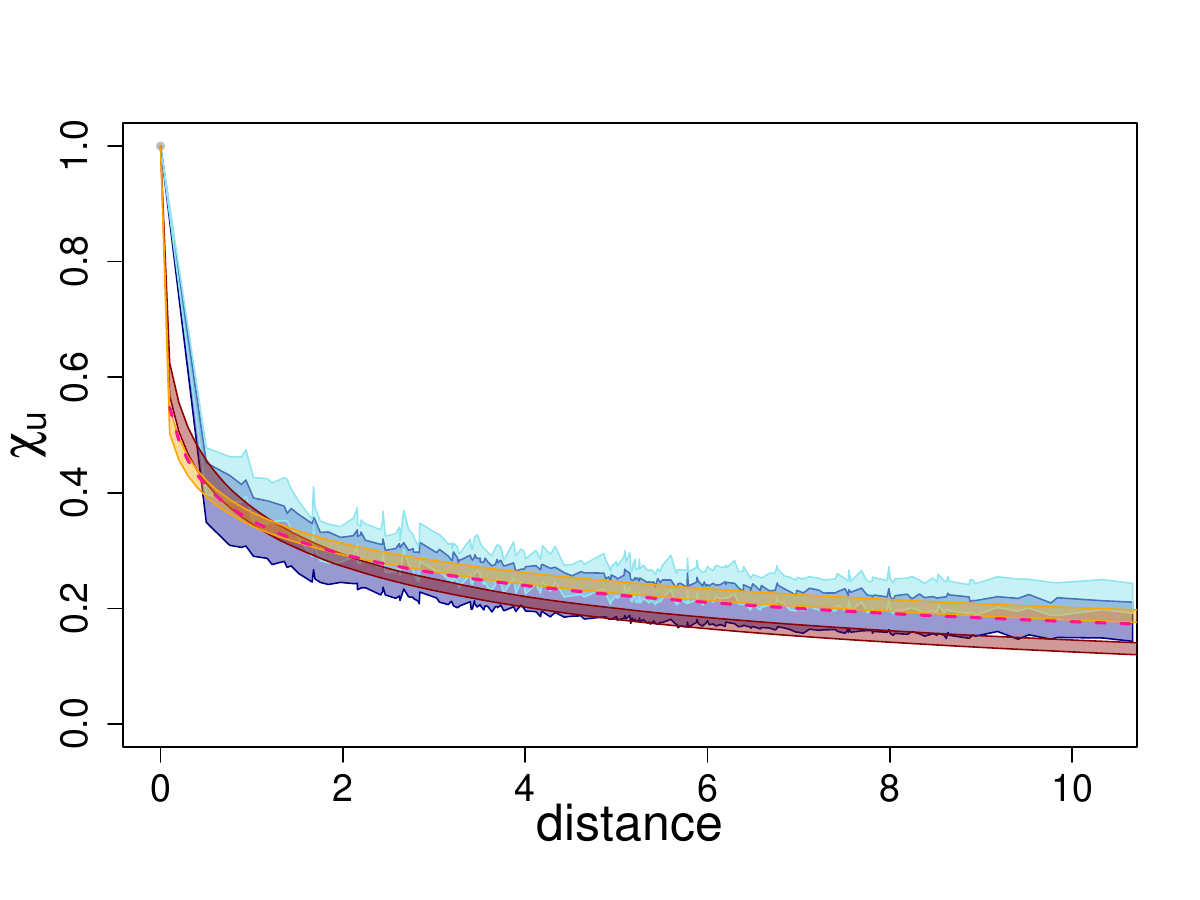} 
\caption{$X(\bm{s})$ simulated from an IBR process} \label{fig:cloud295_ibr}
\end{subfigure}

\begin{subfigure}[t]{1\textwidth}
\centering
  \includegraphics[width=0.3\textwidth]{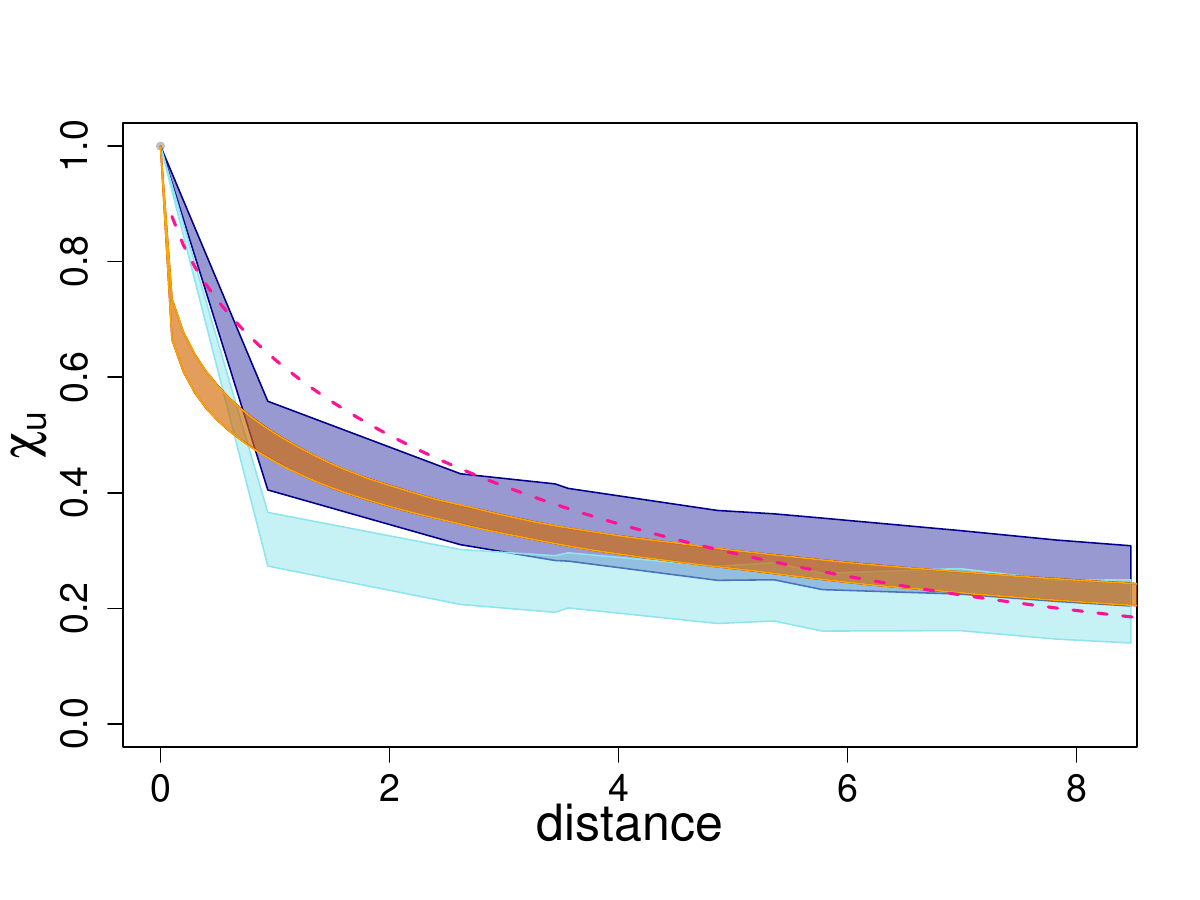}
  \hfill
  \includegraphics[width=0.3\textwidth]{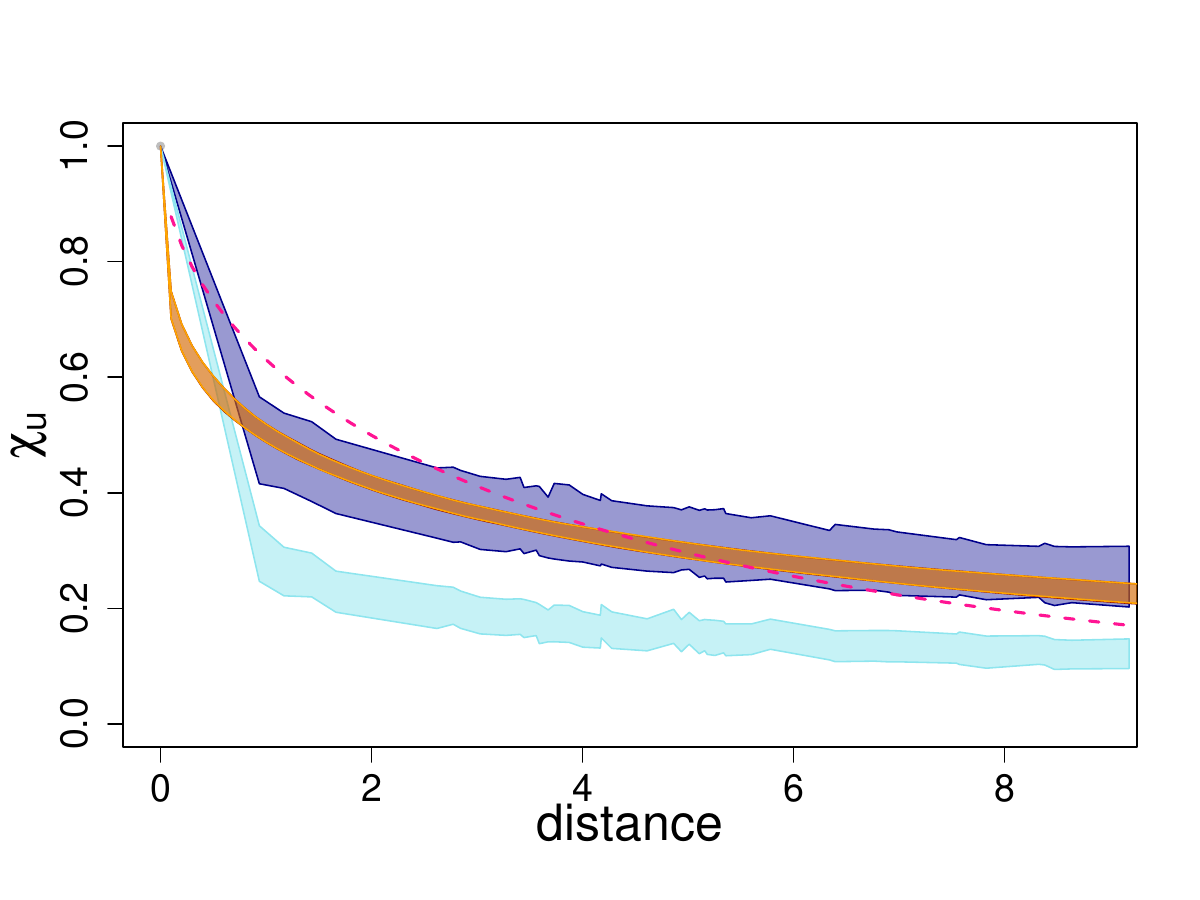}
  \hfill
  \includegraphics[width=0.3\textwidth]{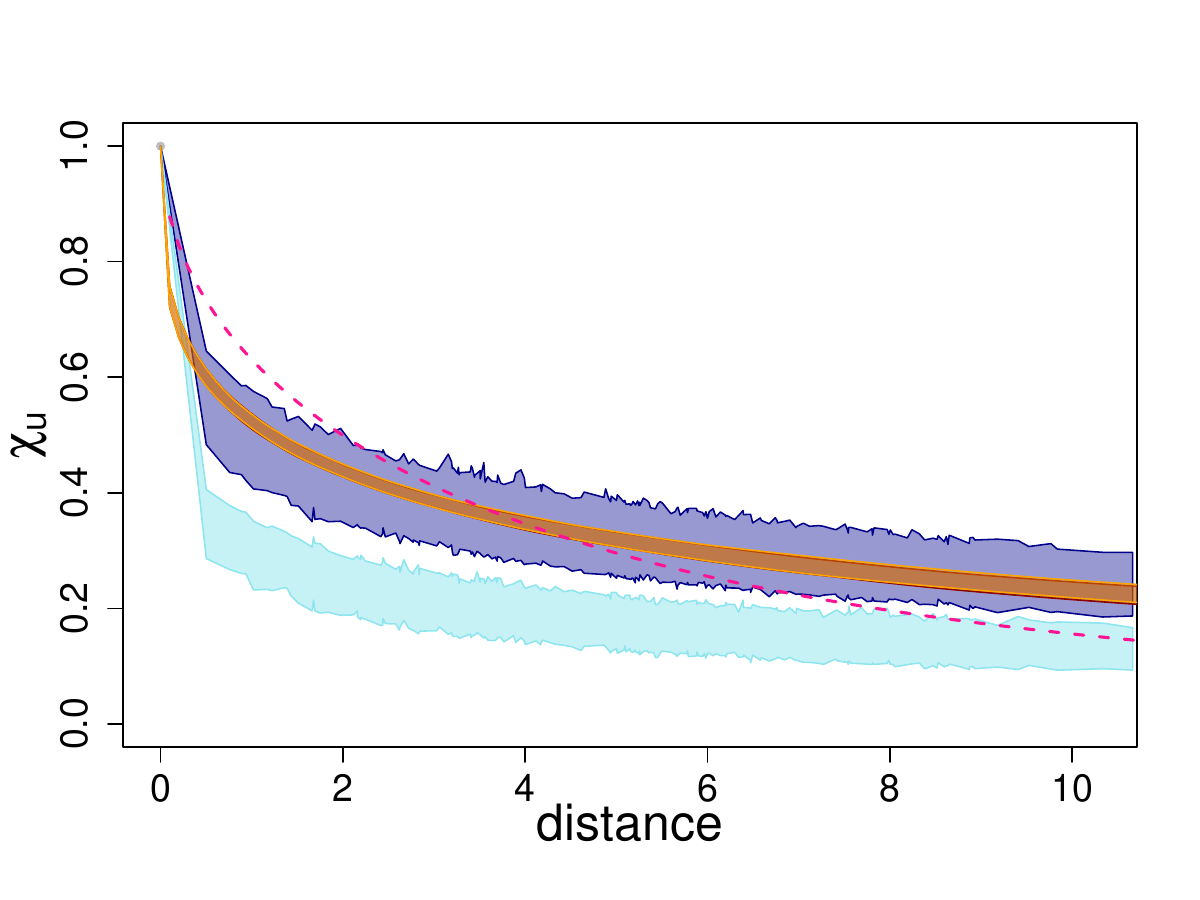}
\caption{$X(\bm{s})$ simulated from a BR process} \label{fig:cloud295_br}
\end{subfigure}

\begin{subfigure}[t]{1\textwidth}
\centering
  \includegraphics[width=0.3\textwidth]{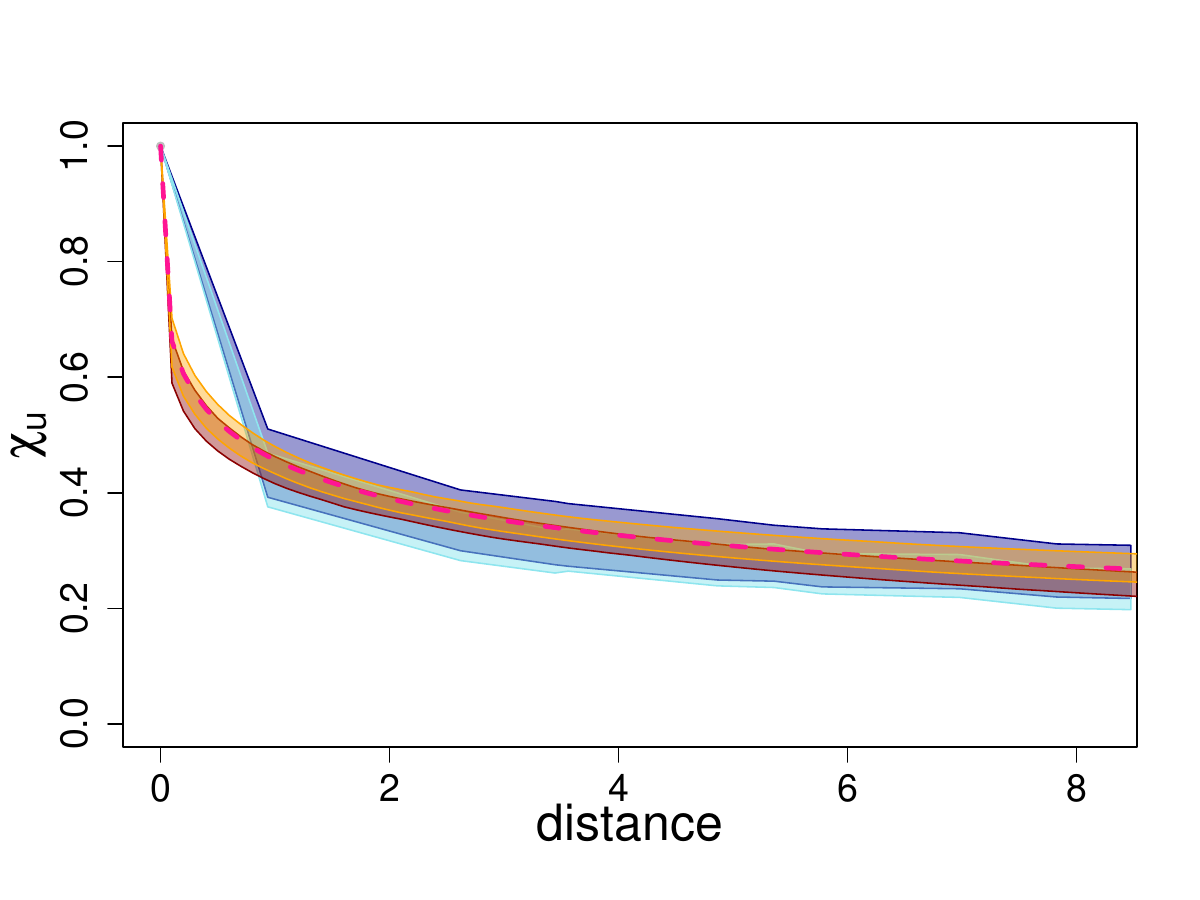}
  \hfill
  \includegraphics[width=0.3\textwidth]{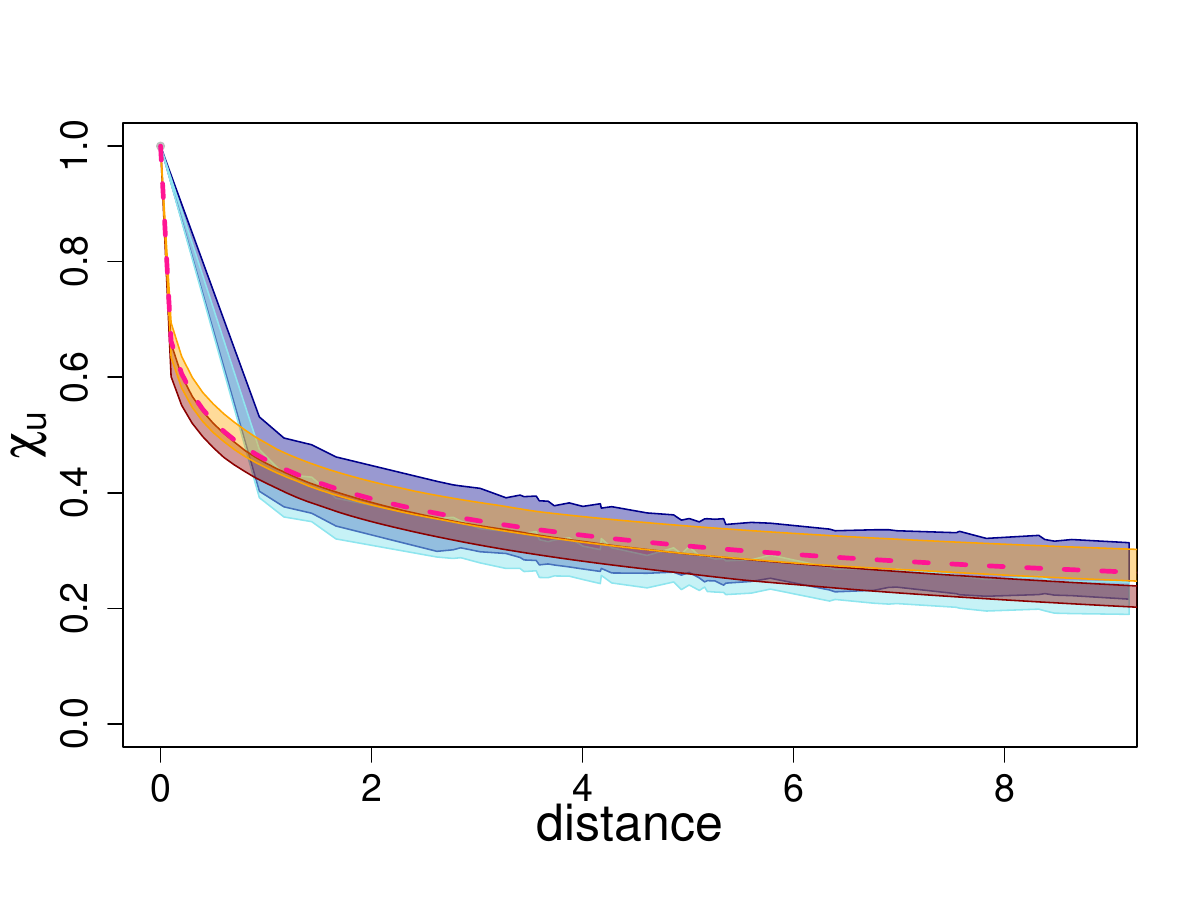}
  \hfill
  \includegraphics[width=0.3\textwidth]{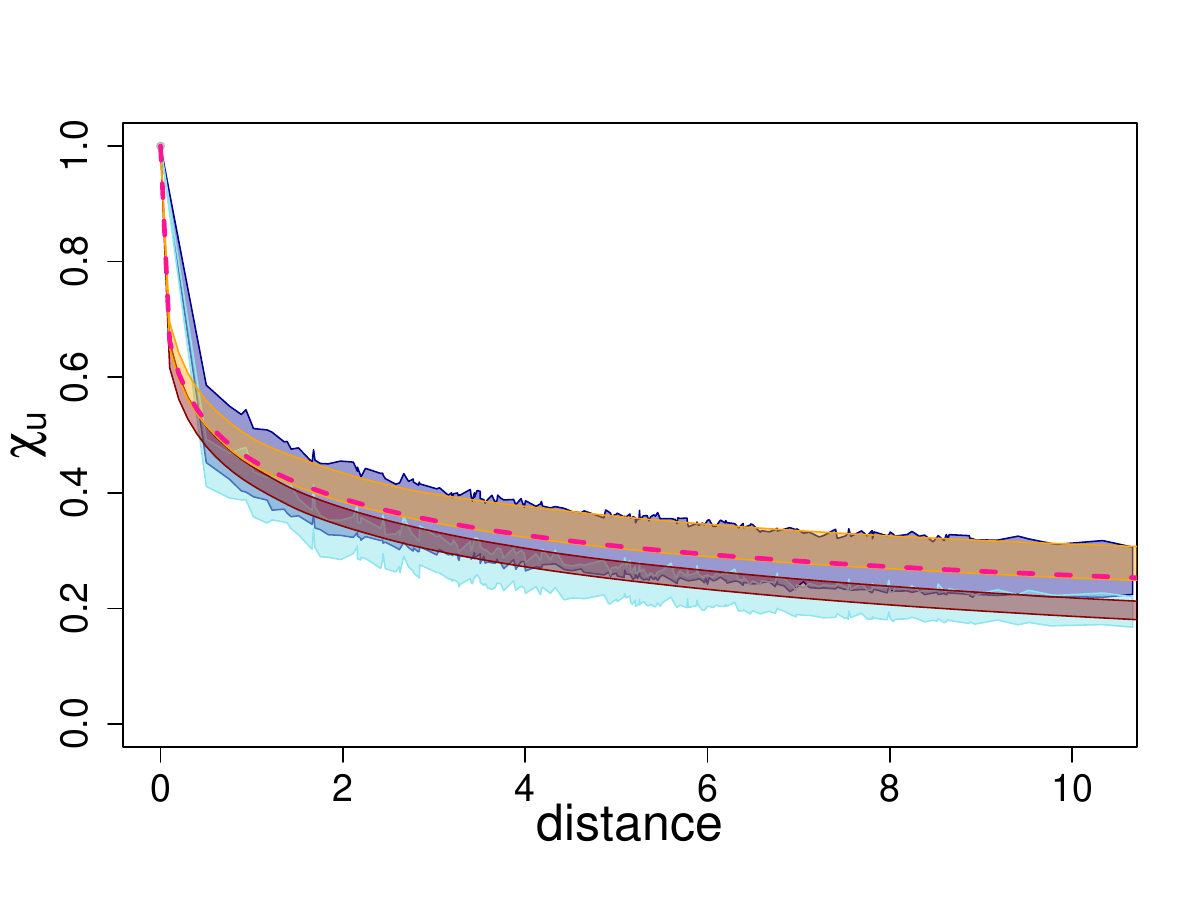}
\caption{$X(\bm{s})$ simulated from a HW process with $\delta=0.4$} \label{fig:cloud295_hw4}
\end{subfigure}

\begin{subfigure}[t]{1\textwidth}
\centering
  \includegraphics[width=0.3\textwidth]{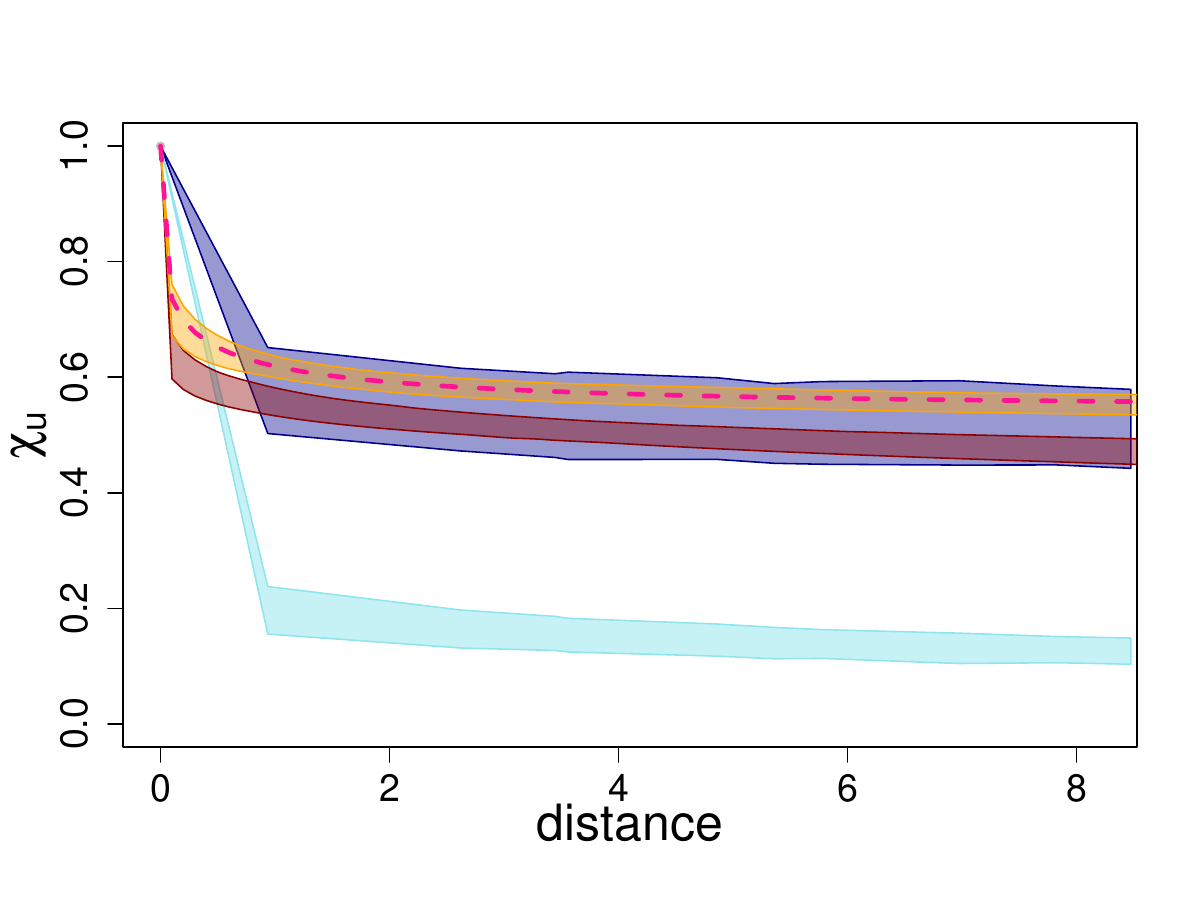}
  \hfill
  \includegraphics[width=0.3\textwidth]{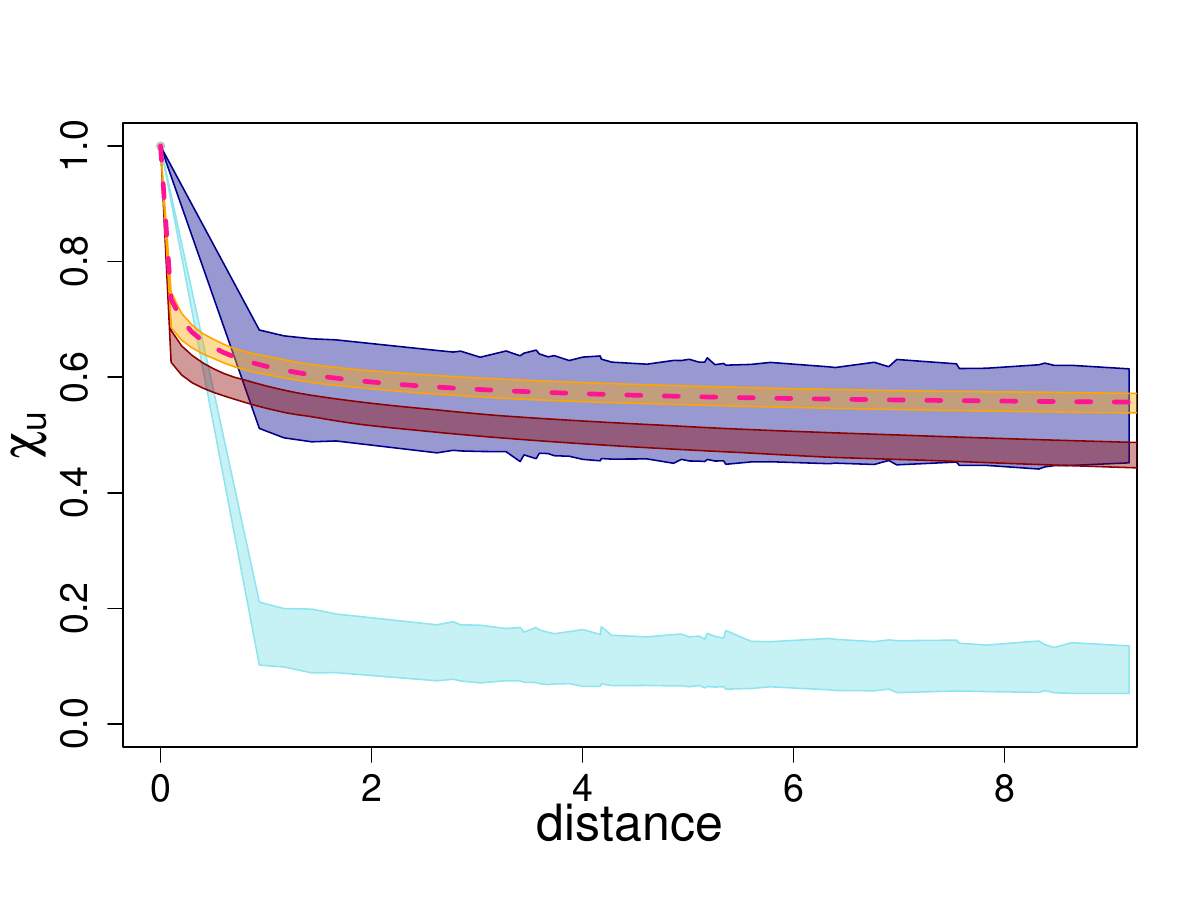}
  \hfill
  \includegraphics[width=0.3\textwidth]{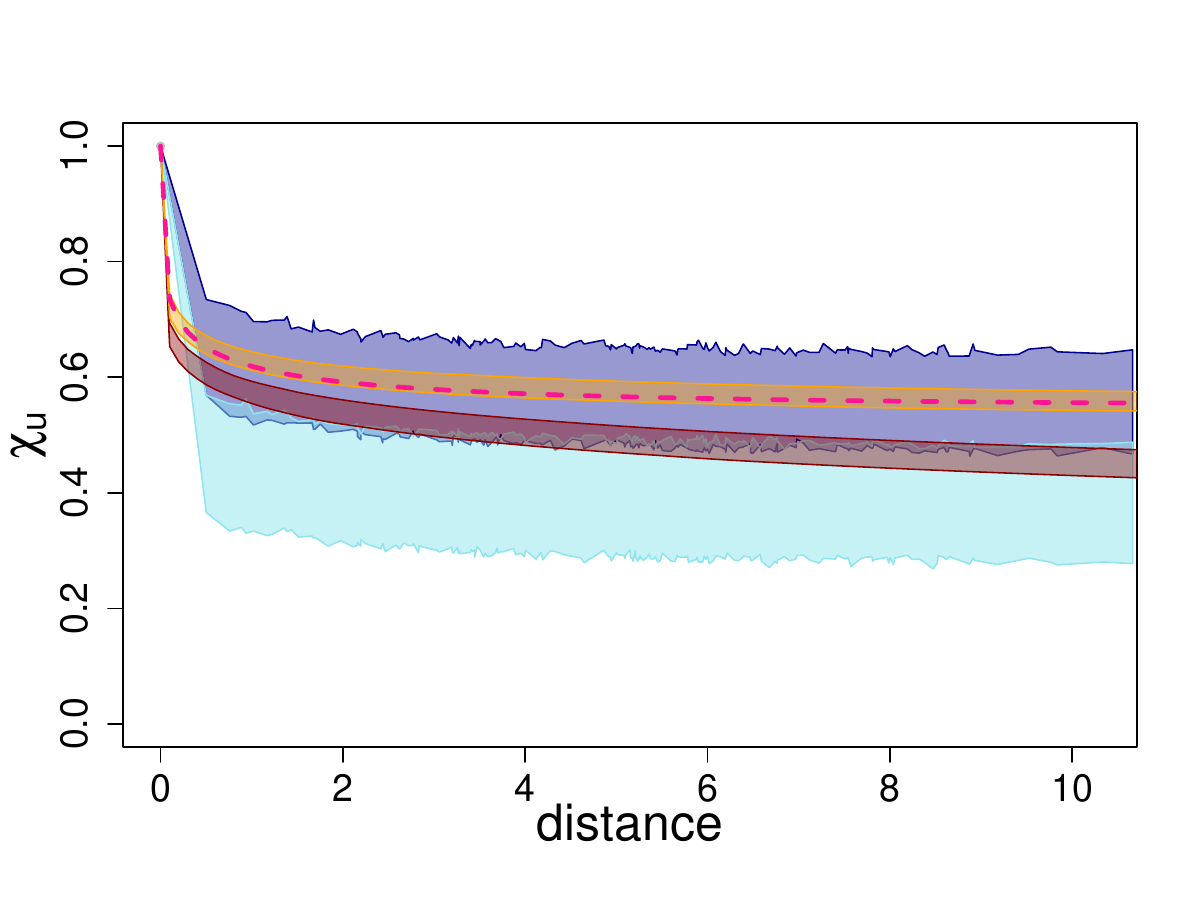}
\caption{$X(\bm{s})$ simulated from a HW process with $\delta=0.6$} \label{fig:cloud295_hw6}
\end{subfigure}

 \caption{Envelope plots of all $\binom{d}{2}$ pairwise $\chi_u$ estimates plotted over distance, calculated for $u=0.95$ using all $200$ simulated datasets. $X(\bm{s})$ is simulated as specified in panels (\subref{fig:cloud295_mvn}) - (\subref{fig:cloud295_hw6}) with parameter $\bm{\theta}_2$. Plots on the left correspond to $d=5$, $d=10$ in the middle and $d=20$ on the right. Envelopes in dark and light blue are obtained via the empirical angular distribution and the angular distribution in \eqref{eq:am1}, respectively. Envelopes in red come from cG fits, while envelopes in orange from HW fits. The dashed pink curve corresponds to the simulated truth.}
 \label{fig:cloud295} 
\end{figure}

\newpage

\begin{figure}[H]
\centering
\begin{subfigure}[t]{1\textwidth}
\centering
  \includegraphics[width=0.3\textwidth]{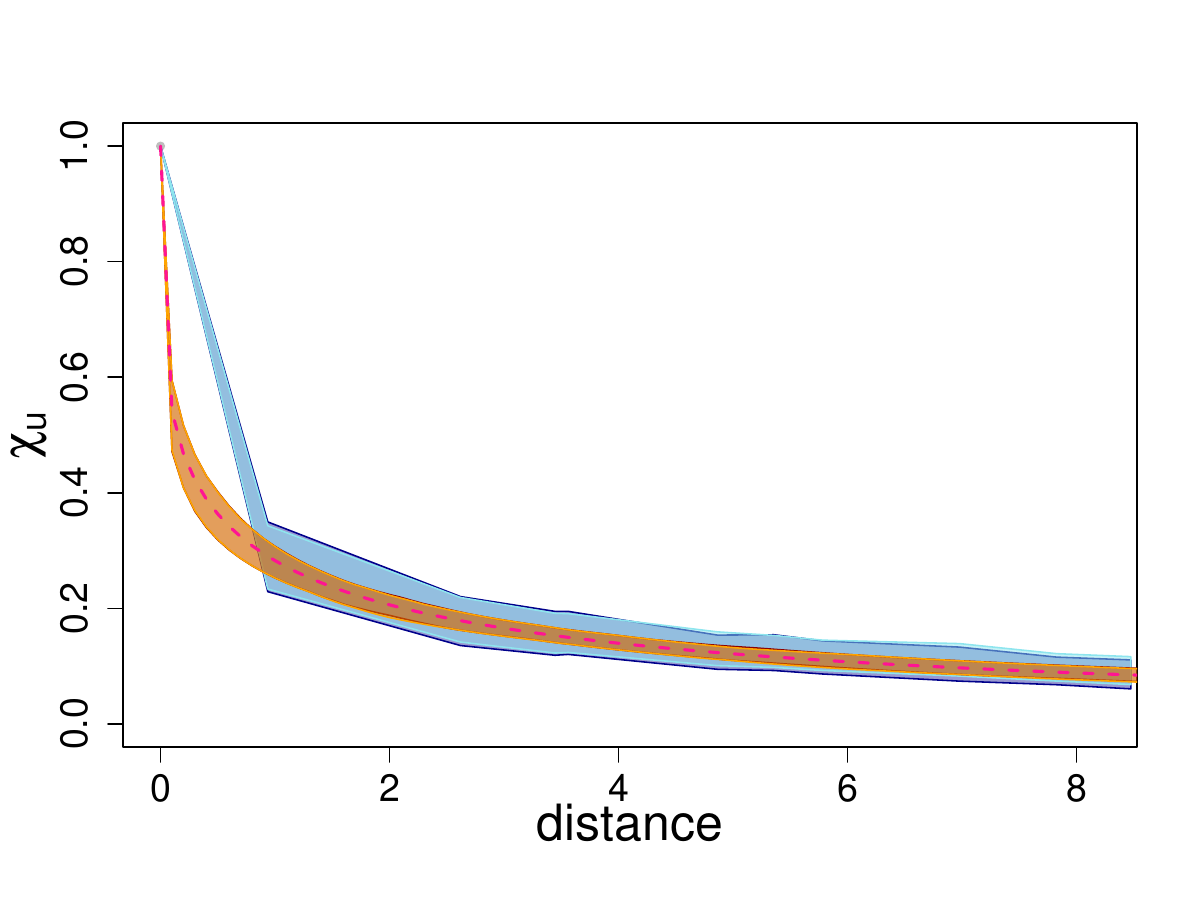}
  \hfill
  \includegraphics[width=0.3\textwidth]{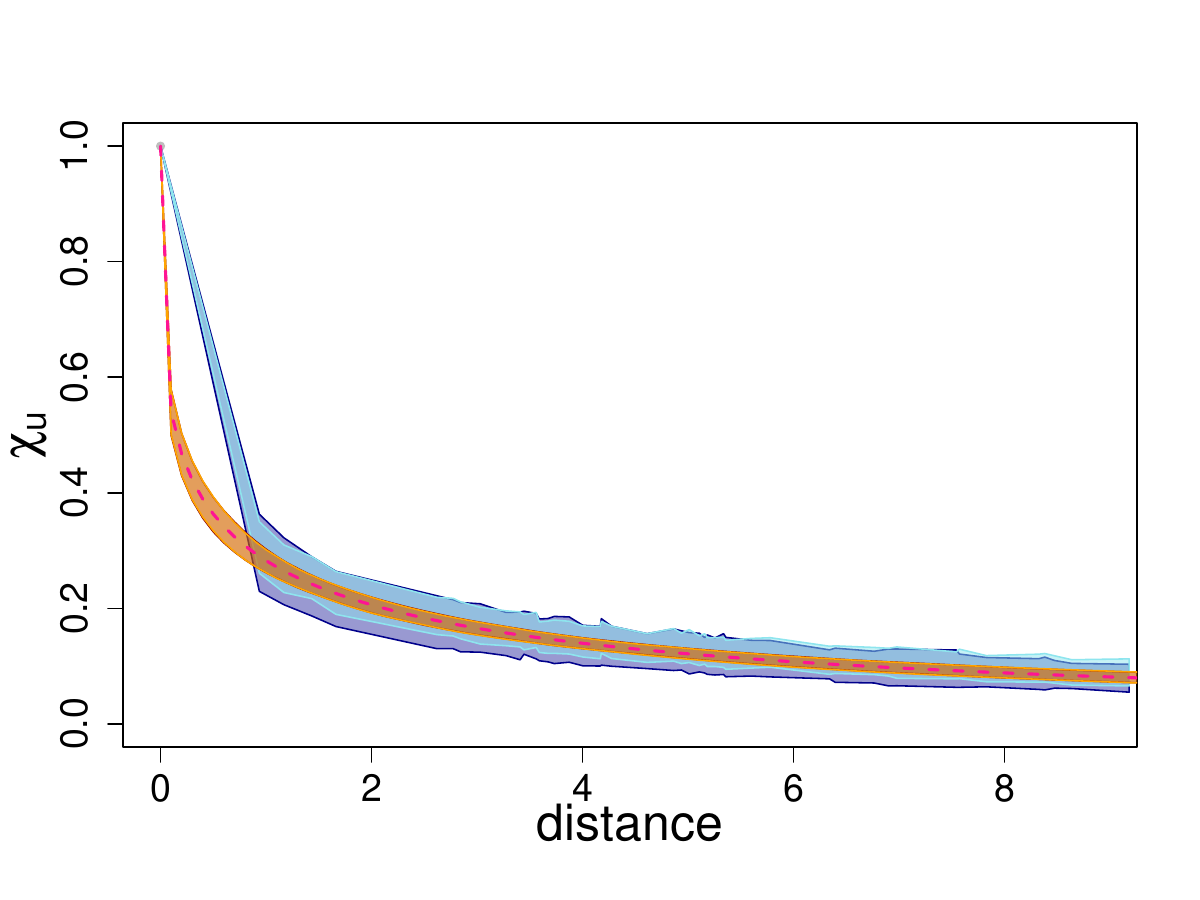}
  \hfill
  \includegraphics[width=0.3\textwidth]{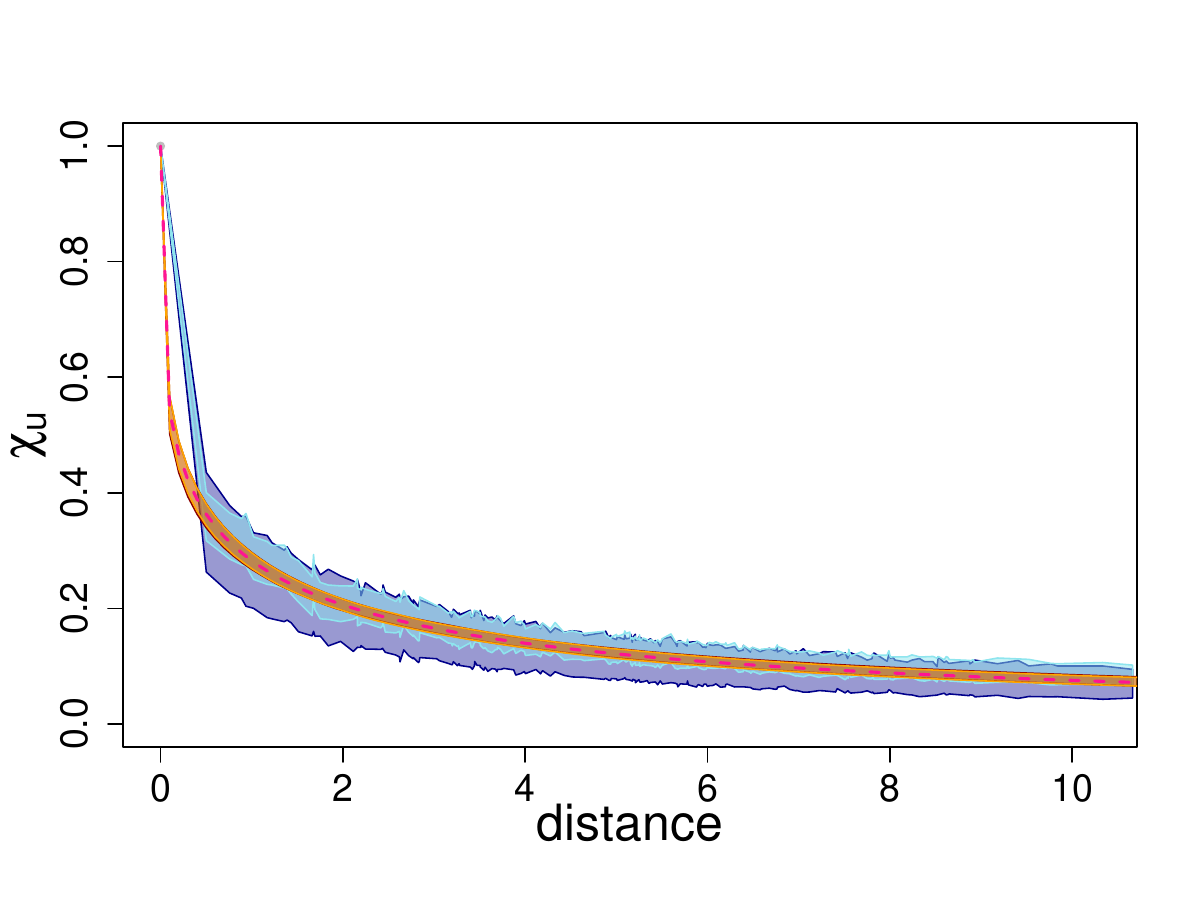}
\caption{$X(\bm{s})$ simulated from a Gaussian process} \label{fig:cloud298_mvn}
\end{subfigure}

\begin{subfigure}[t]{1\textwidth}
\centering
  \includegraphics[width=0.3\textwidth]{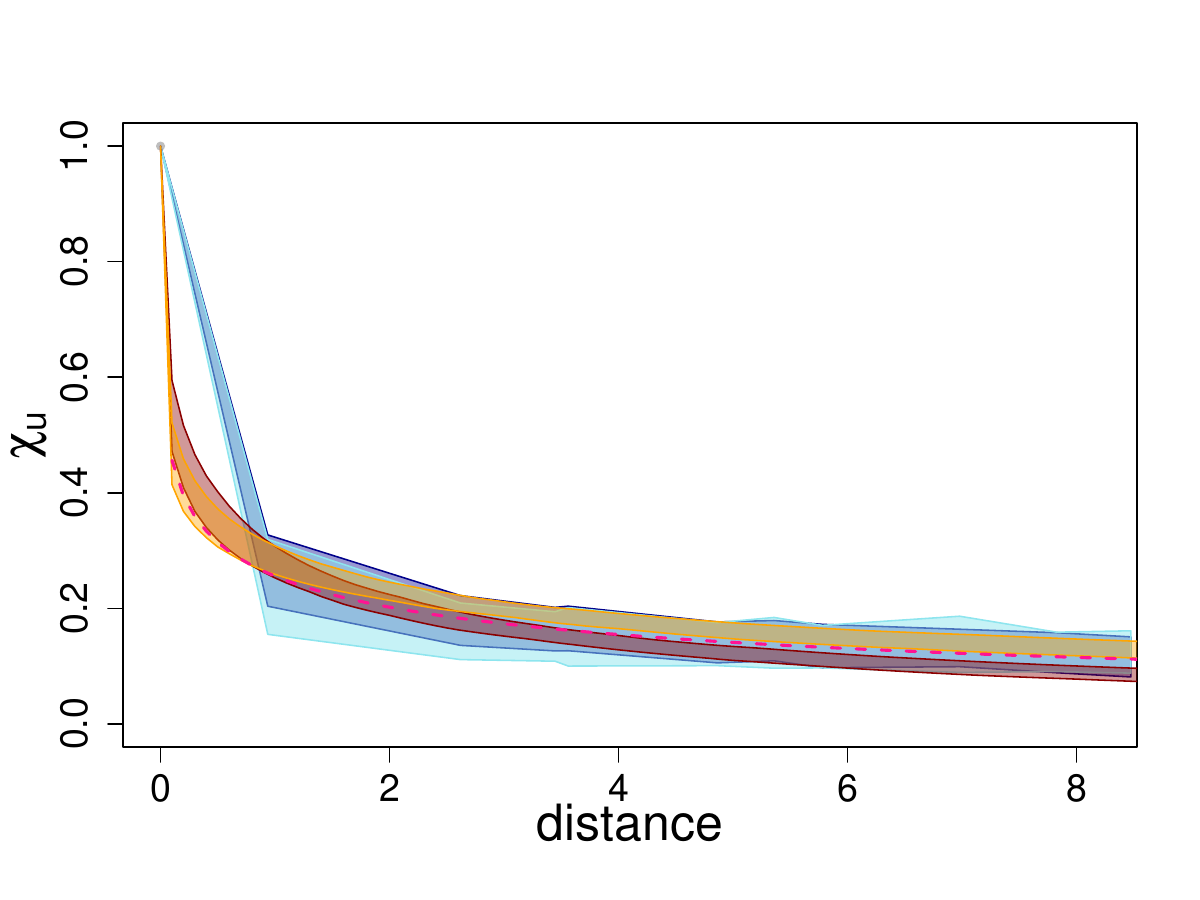}
  \hfill
  \includegraphics[width=0.3\textwidth]{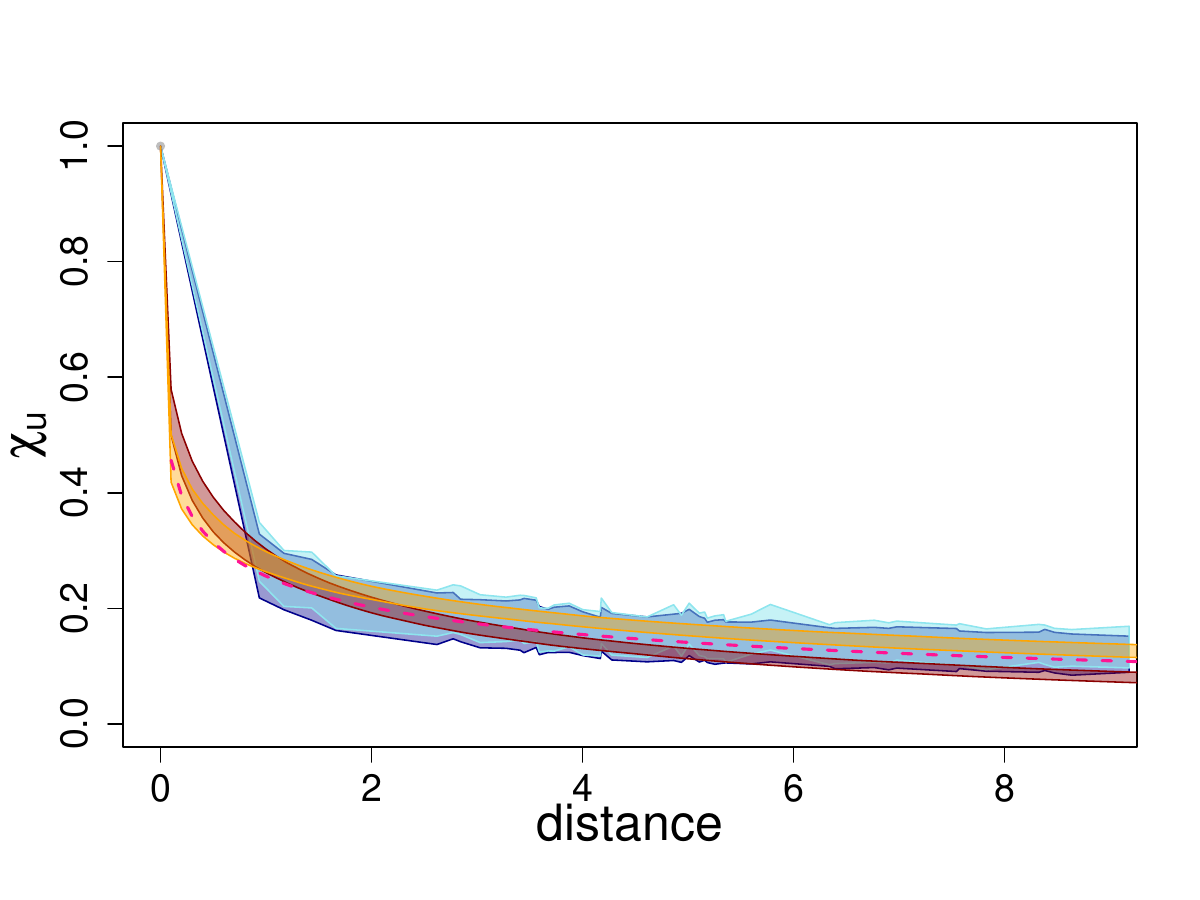}
  \hfill
  \includegraphics[width=0.3\textwidth]{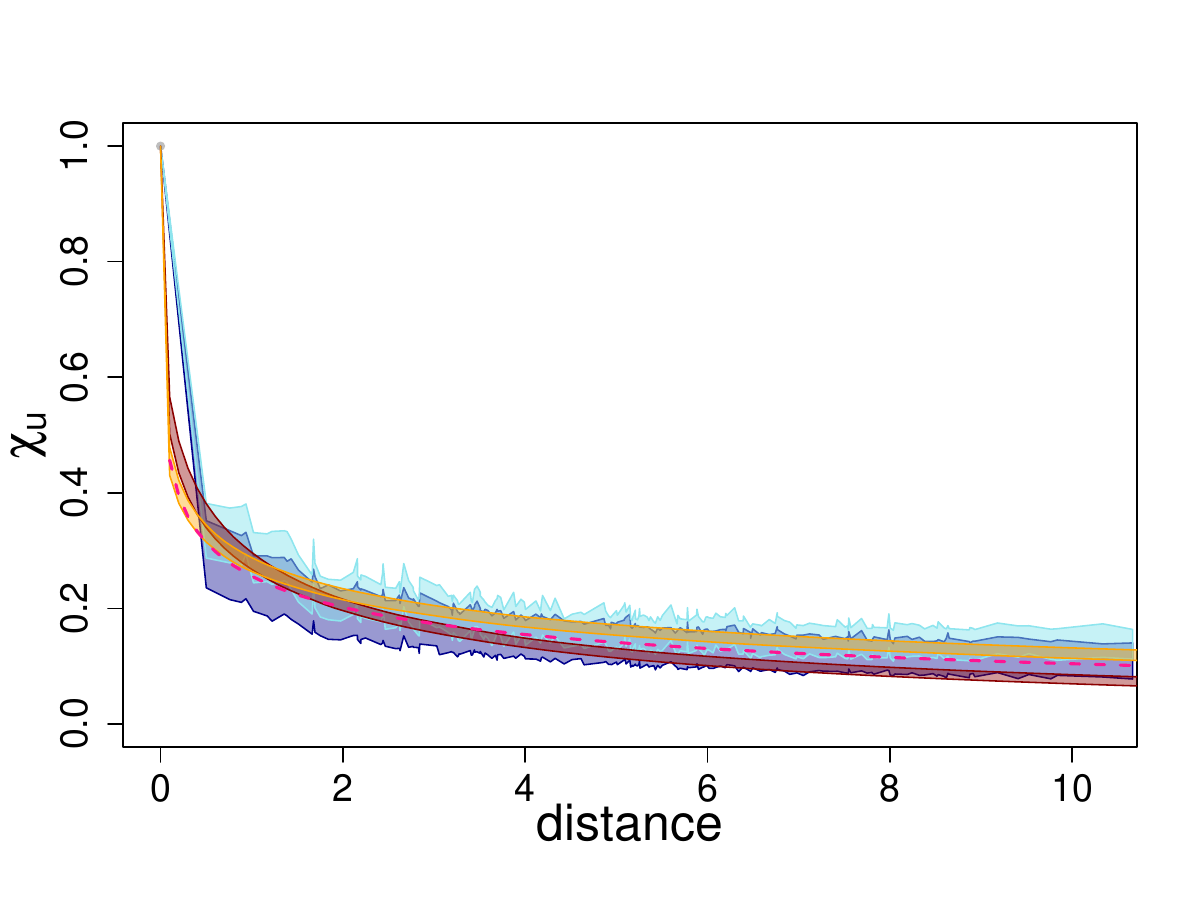} 
\caption{$X(\bm{s})$ simulated from an IBR process} \label{fig:cloud298_ibr}
\end{subfigure}

\begin{subfigure}[t]{1\textwidth}
\centering
  \includegraphics[width=0.3\textwidth]{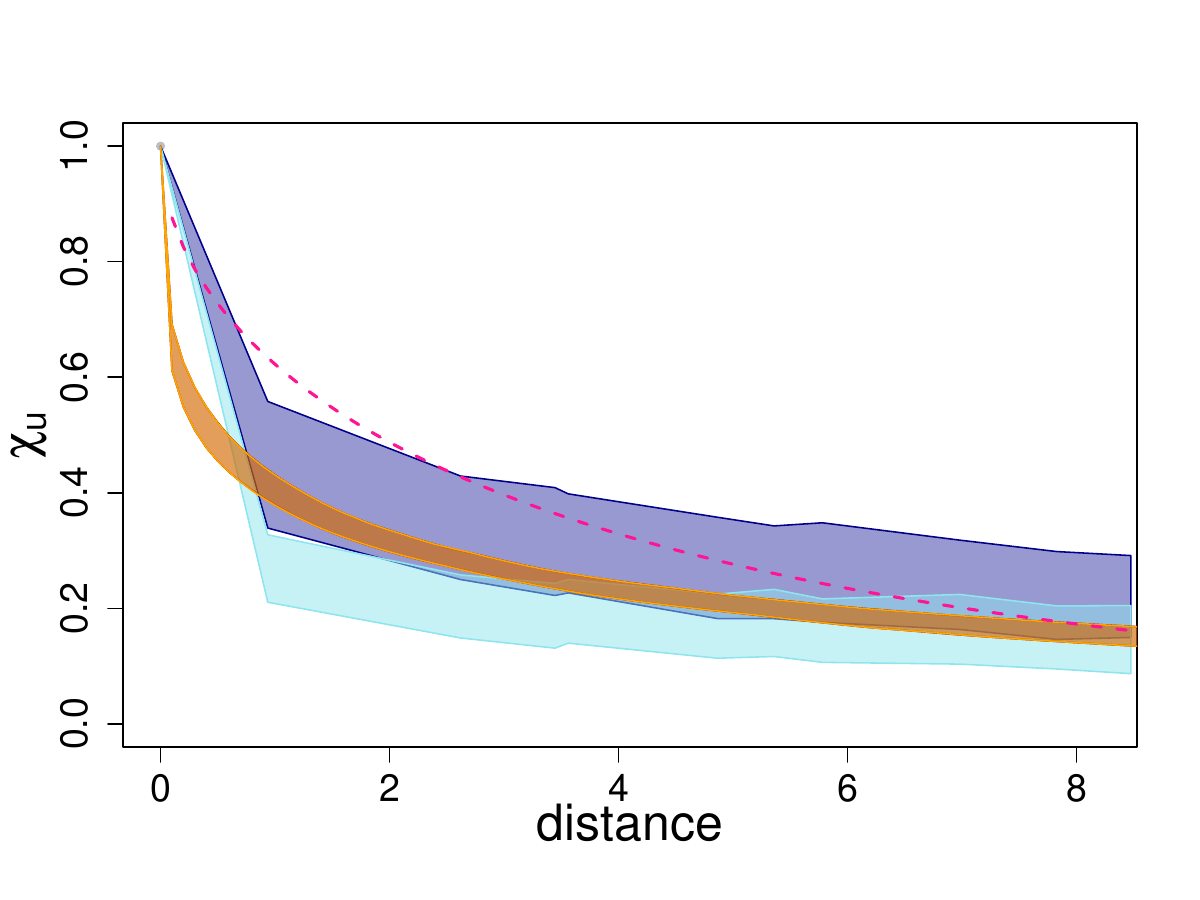}
  \hfill
  \includegraphics[width=0.3\textwidth]{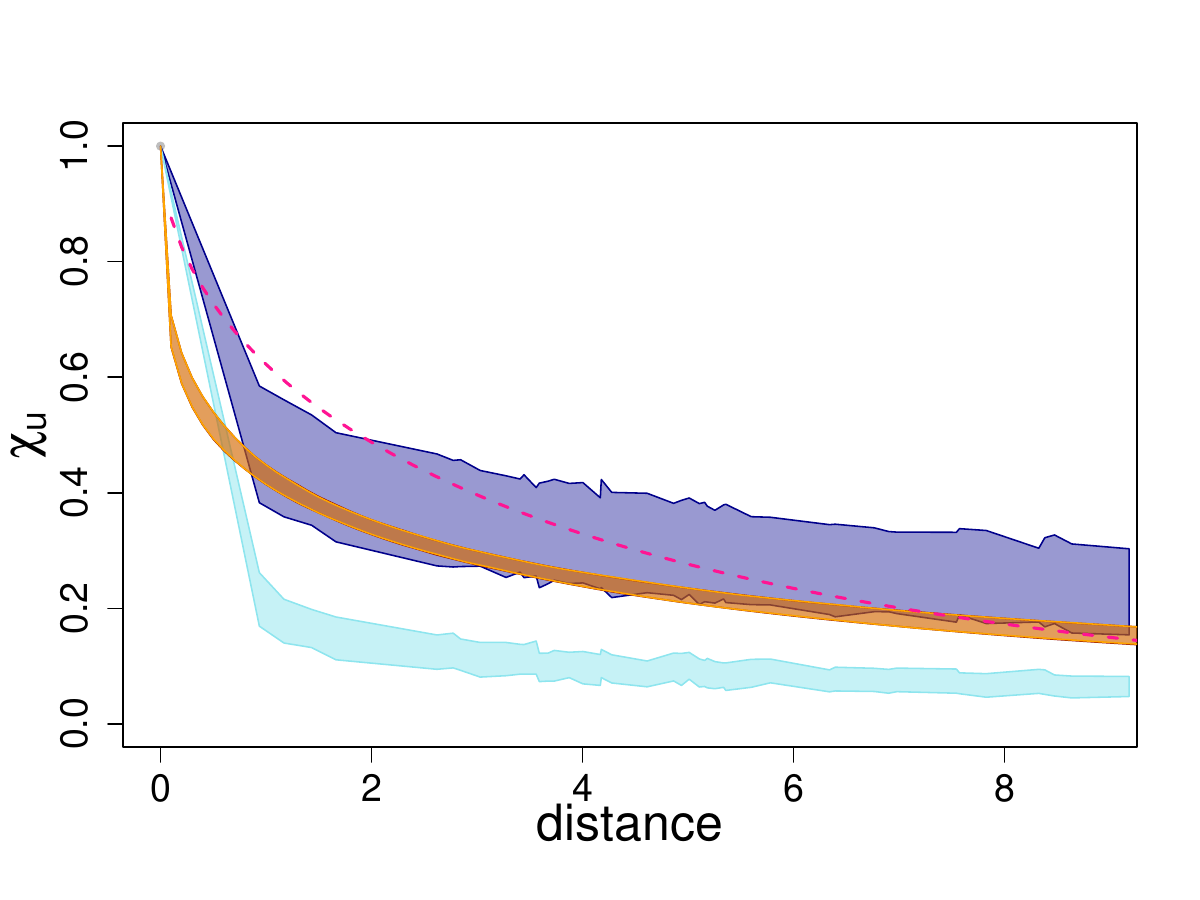}
  \hfill
  \includegraphics[width=0.3\textwidth]{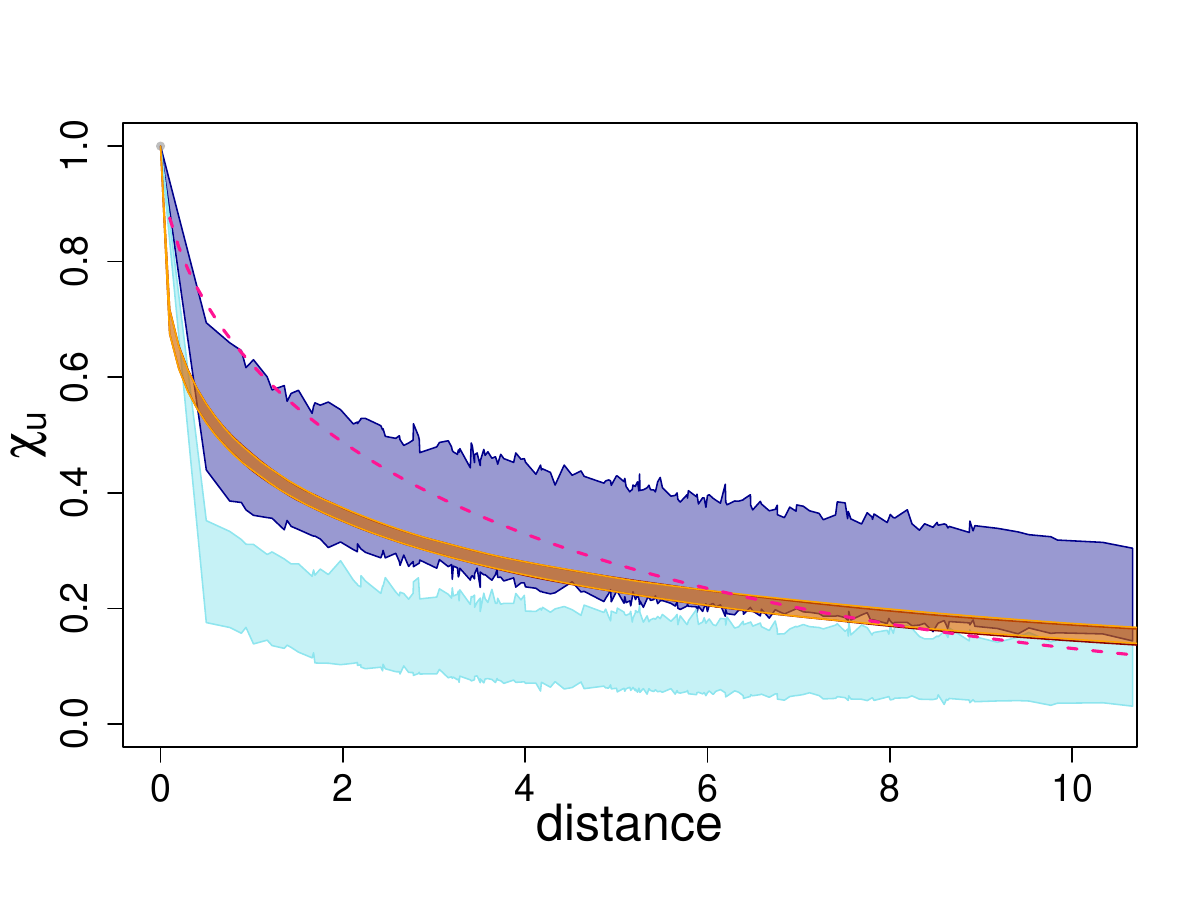}
\caption{$X(\bm{s})$ simulated from a BR process} \label{fig:cloud298_br}
\end{subfigure}

\begin{subfigure}[t]{1\textwidth}
\centering
  \includegraphics[width=0.3\textwidth]{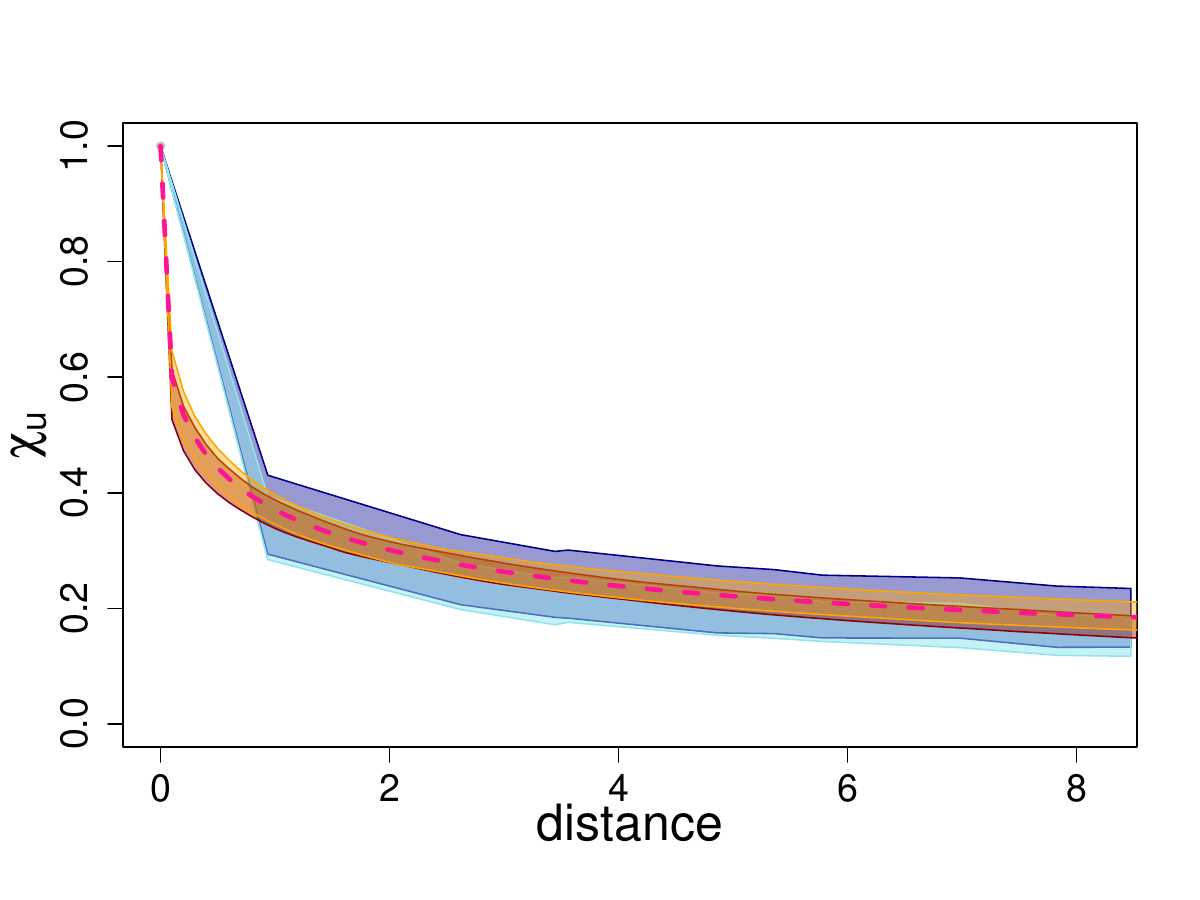}
  \hfill
  \includegraphics[width=0.3\textwidth]{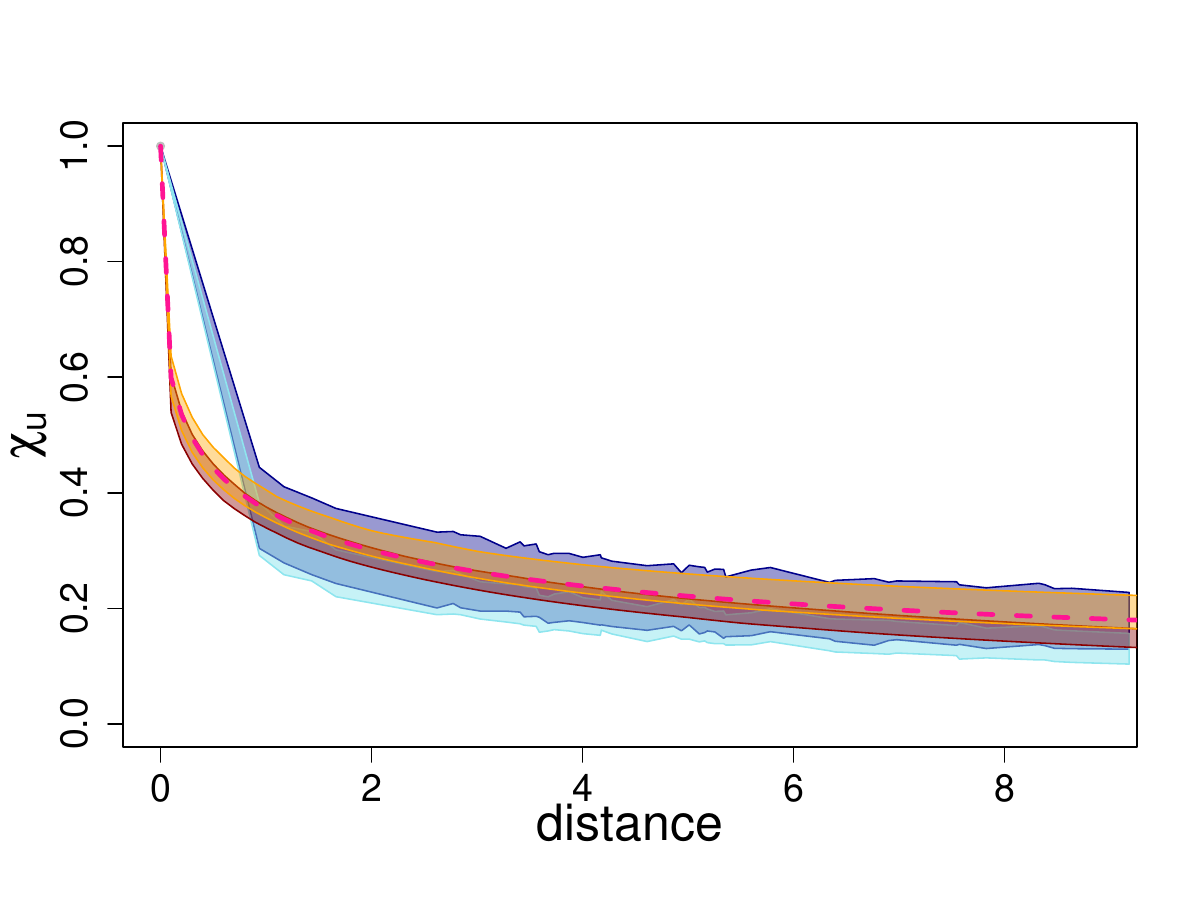}
  \hfill
  \includegraphics[width=0.3\textwidth]{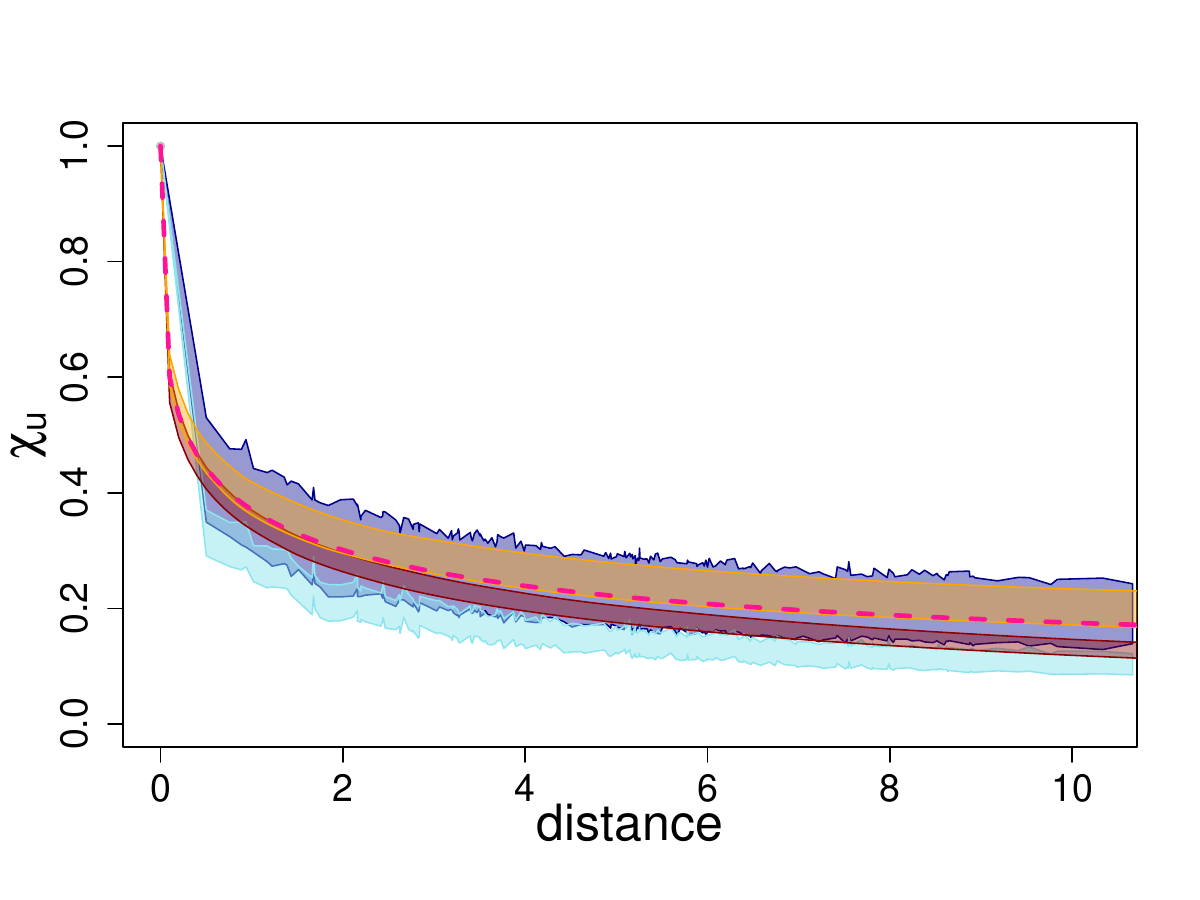}
\caption{$X(\bm{s})$ simulated from a HW process with $\delta=0.4$} \label{fig:cloud298_hw4}
\end{subfigure}

\begin{subfigure}[t]{1\textwidth}
\centering
  \includegraphics[width=0.3\textwidth]{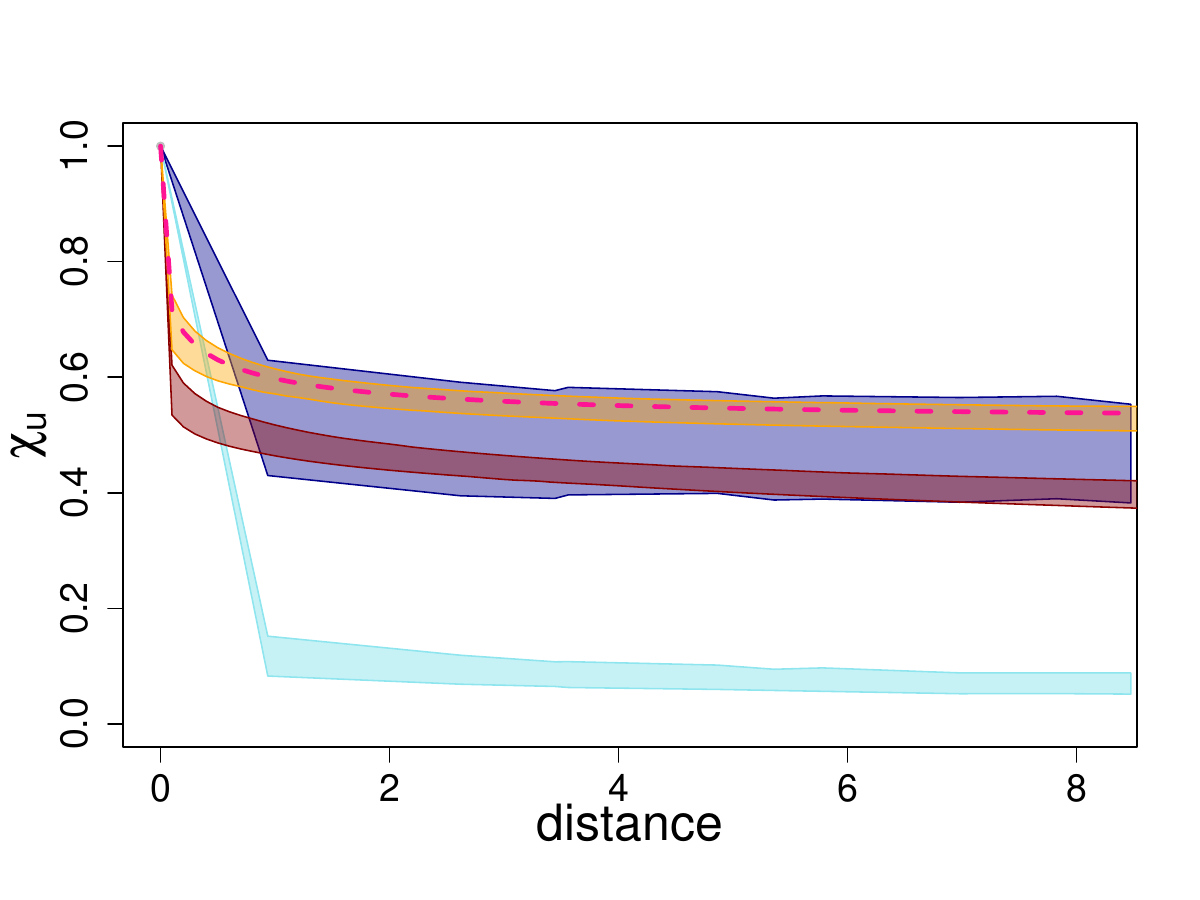}
  \hfill
  \includegraphics[width=0.3\textwidth]{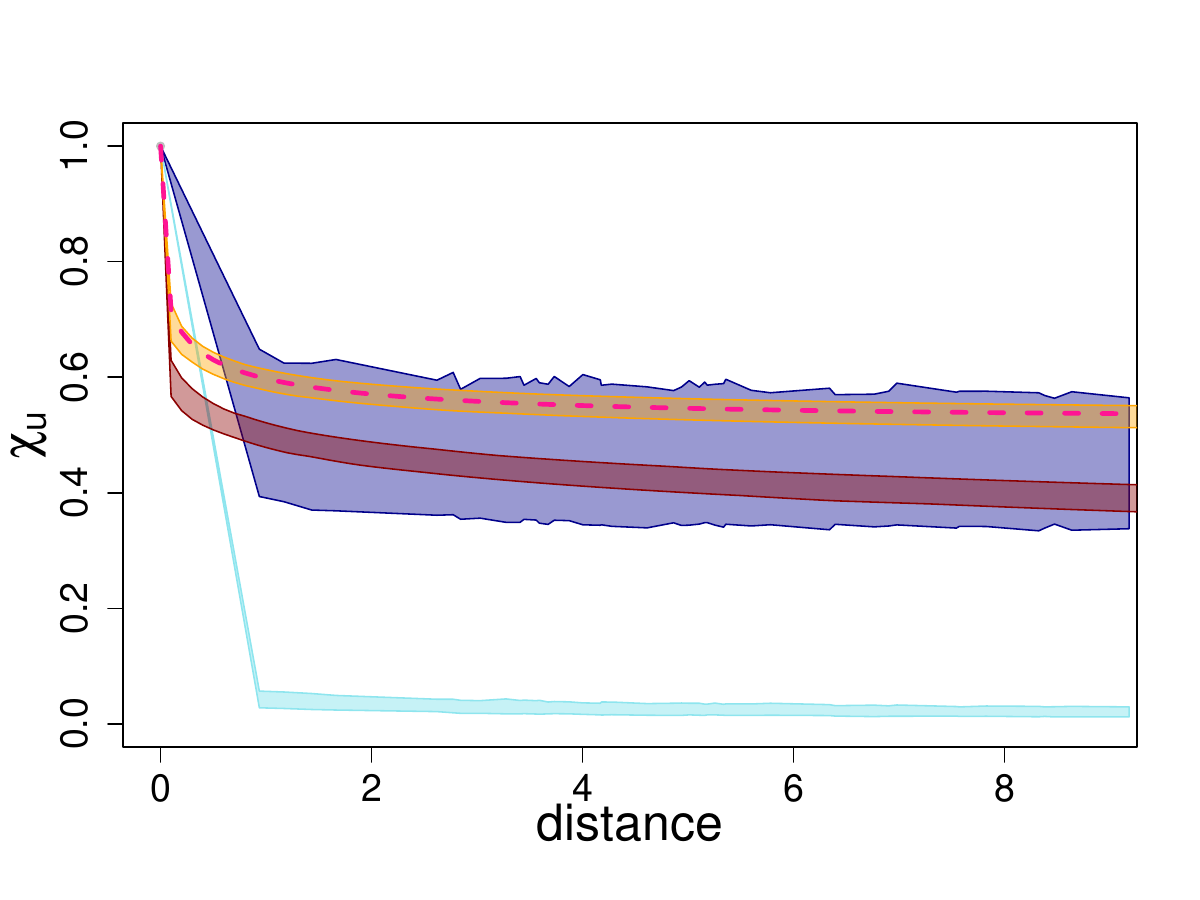}
  \hfill
  \includegraphics[width=0.3\textwidth]{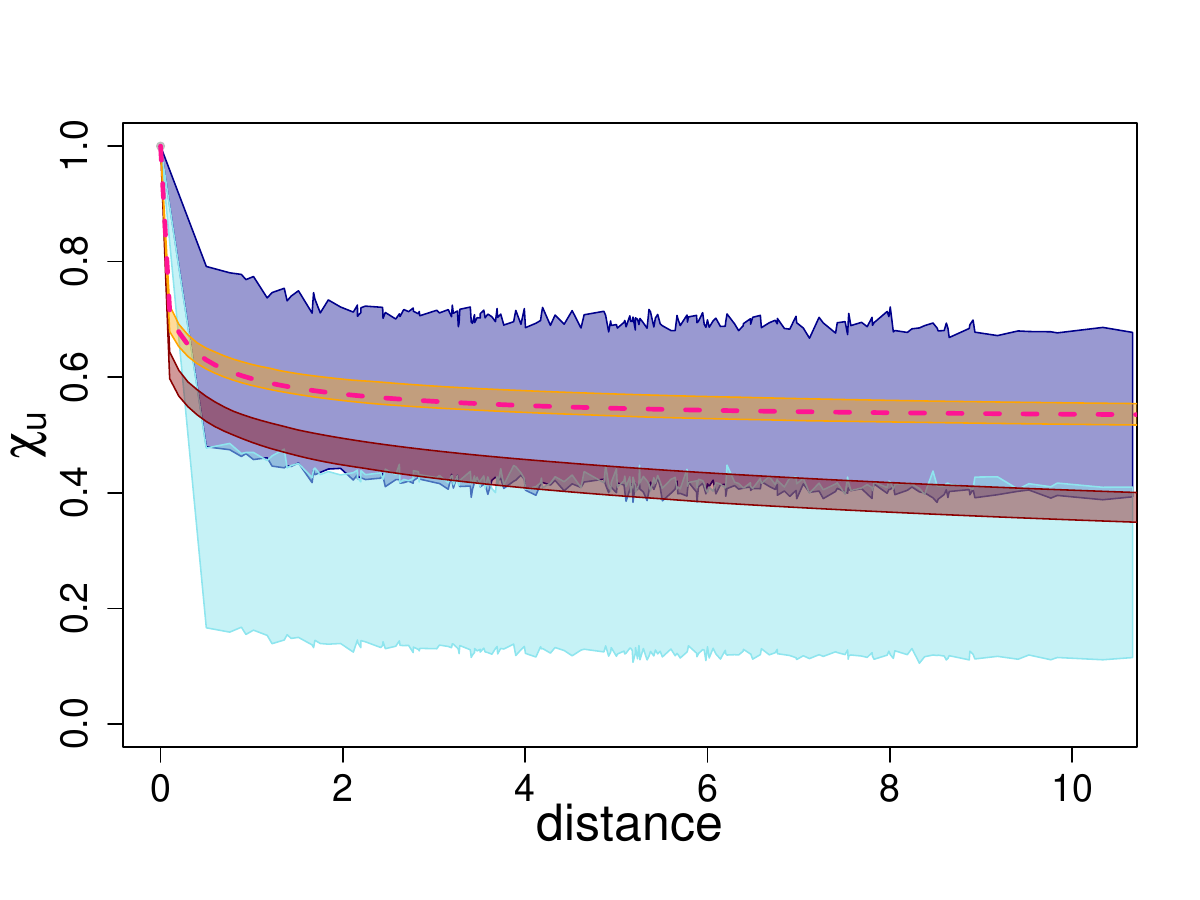}
\caption{$X(\bm{s})$ simulated from a HW process with $\delta=0.6$} \label{fig:cloud298_hw6}
\end{subfigure}

 \caption{Envelope plots of all $\binom{d}{2}$ pairwise $\chi_u$ estimates plotted over distance, calculated for $u=0.98$ using all $200$ simulated datasets. $X(\bm{s})$ is simulated as specified in panels (\subref{fig:cloud298_mvn}) - (\subref{fig:cloud298_hw6}) with parameter $\bm{\theta}_2$. Plots on the left correspond to $d=5$, $d=10$ in the middle and $d=20$ on the right. Envelopes in dark and light blue are obtained via the empirical angular distribution and the angular distribution in \eqref{eq:am1}, respectively. Envelopes in red come from cG fits, while envelopes in orange from HW fits. The dashed pink curve corresponds to the simulated truth.}
 \label{fig:cloud298} 
\end{figure}

\begin{figure}[t!]
\centering
\begin{subfigure}[t]{1\textwidth}
\centering
\includegraphics[width=0.3\textwidth]{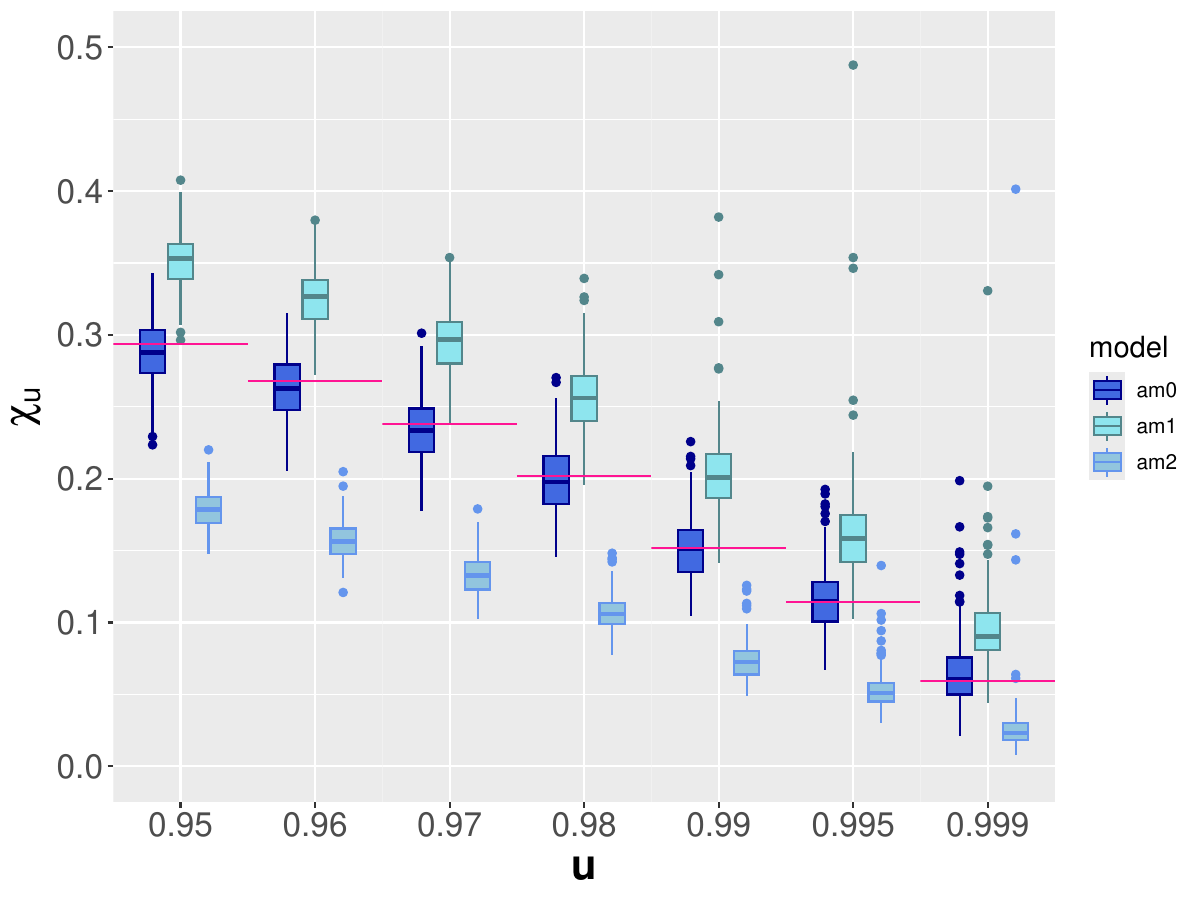}
  \hfill
  \includegraphics[width=0.3\textwidth]{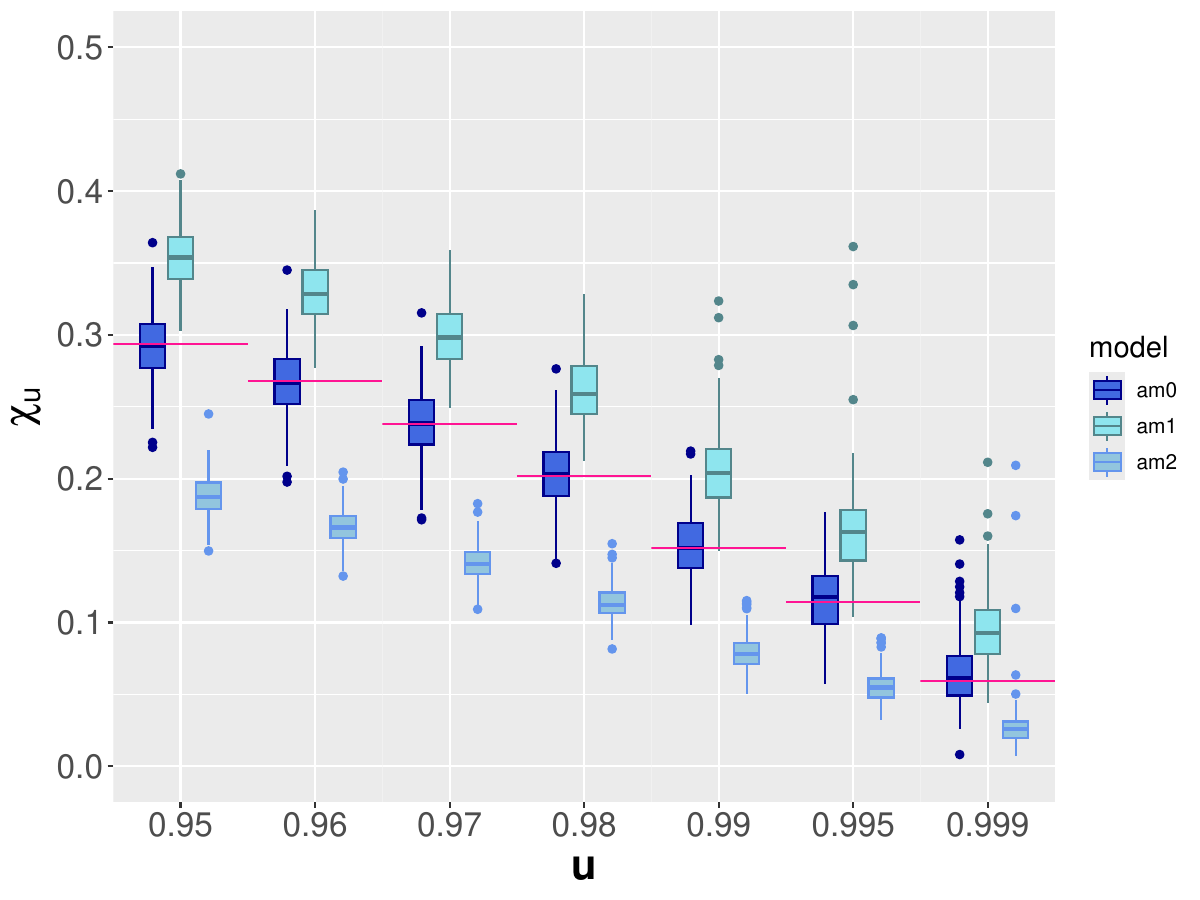}
  \hfill
  \includegraphics[width=0.3\textwidth]{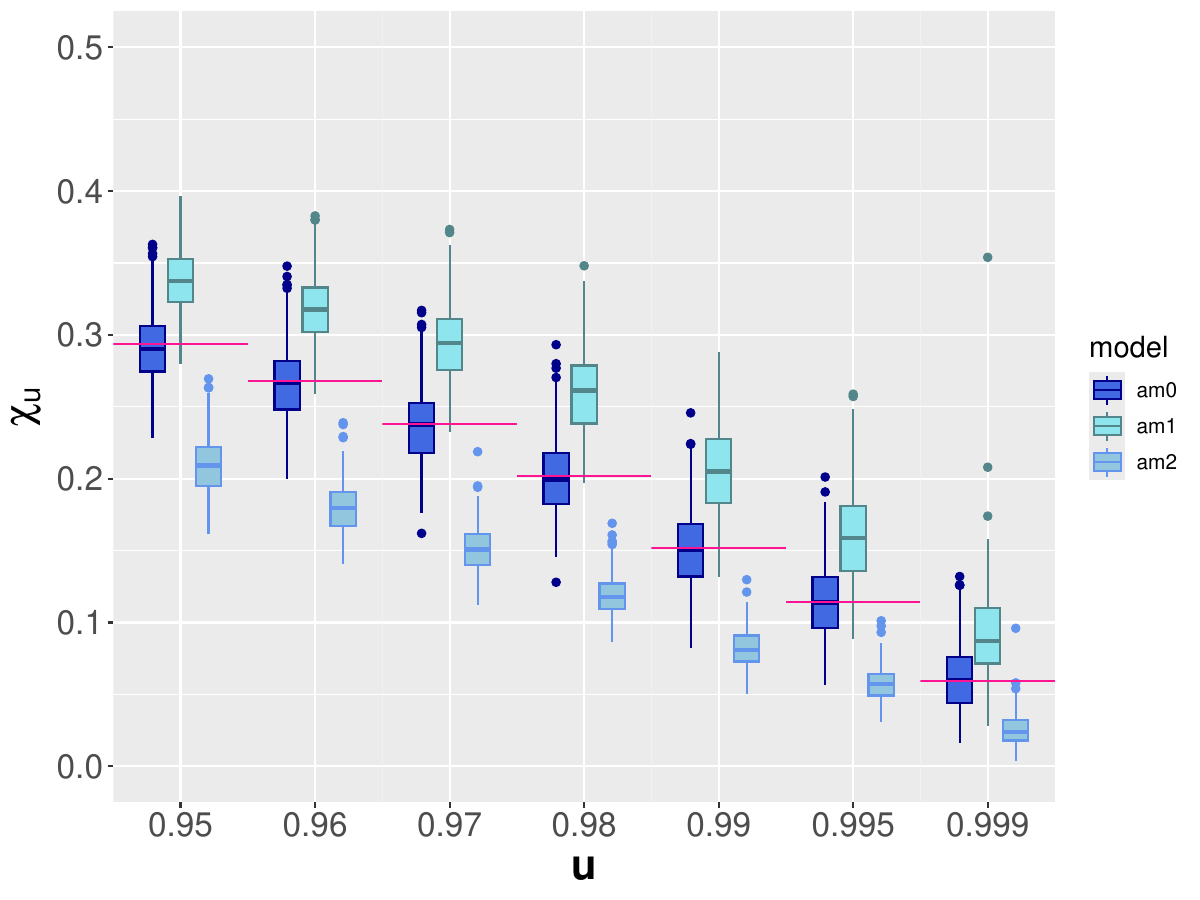} 
\caption{$X(\bm{s})$ simulated from a IBR process with parameters $\bm{\theta}_1$ and $d=20$} 
\end{subfigure}

\begin{subfigure}[h!]{1\textwidth}
\centering
\includegraphics[width=0.3\textwidth]{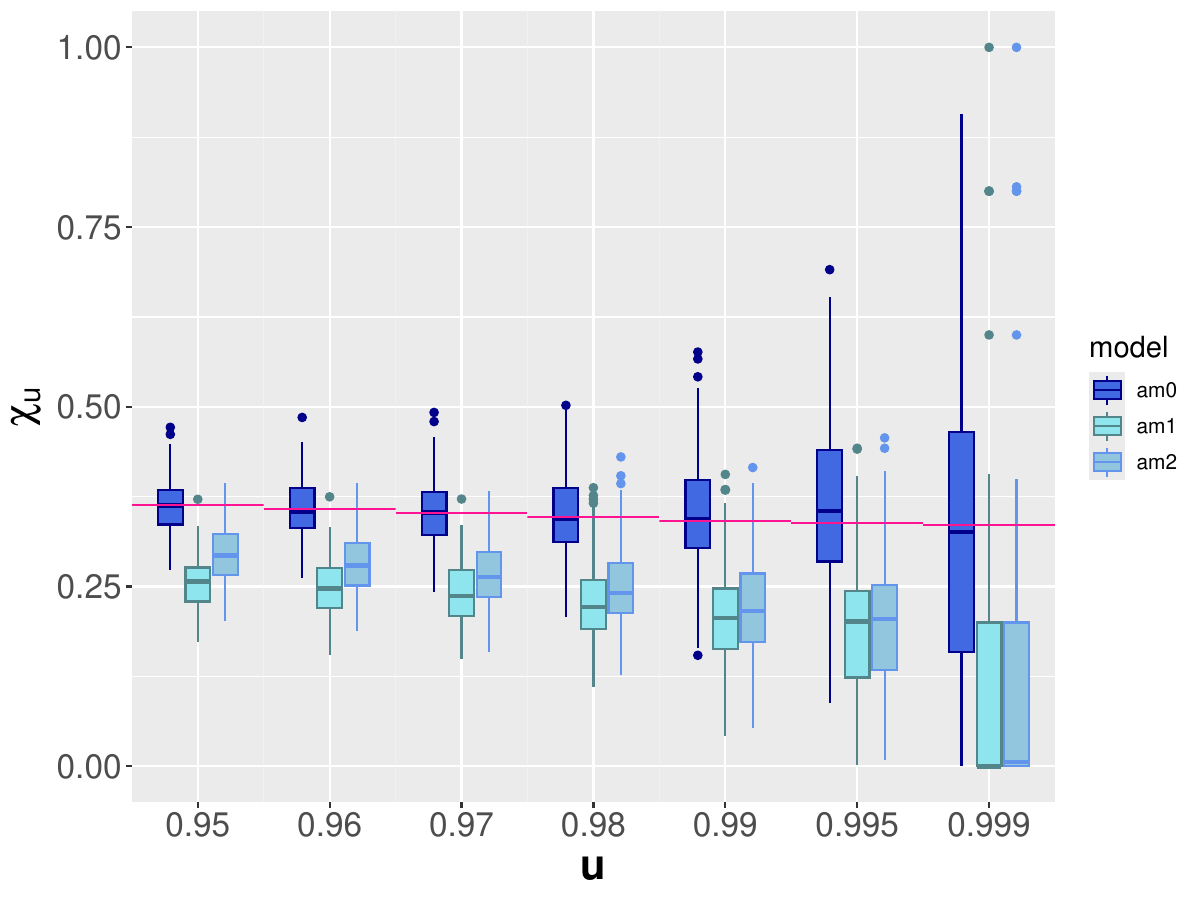}
  \hfill
  \includegraphics[width=0.3\textwidth]{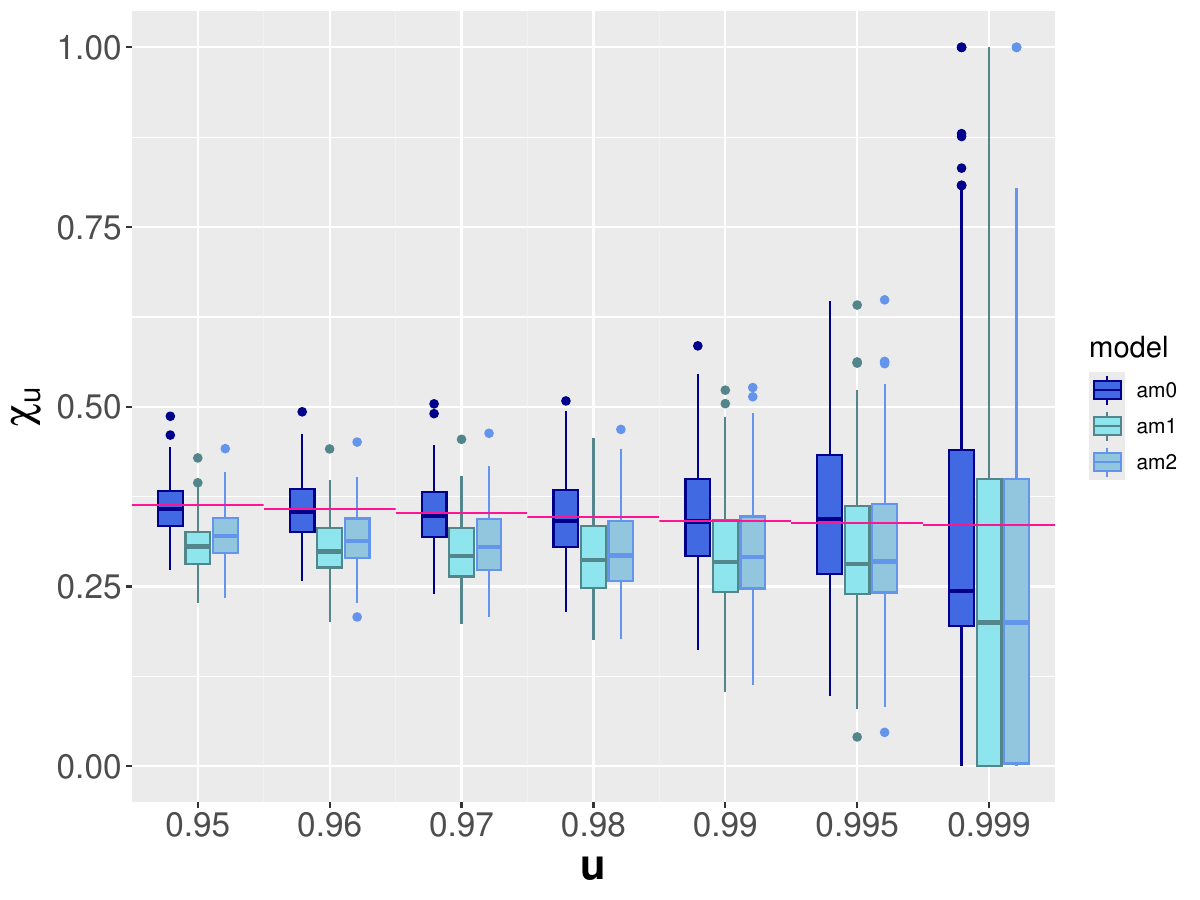}
  \hfill
  \includegraphics[width=0.3\textwidth]{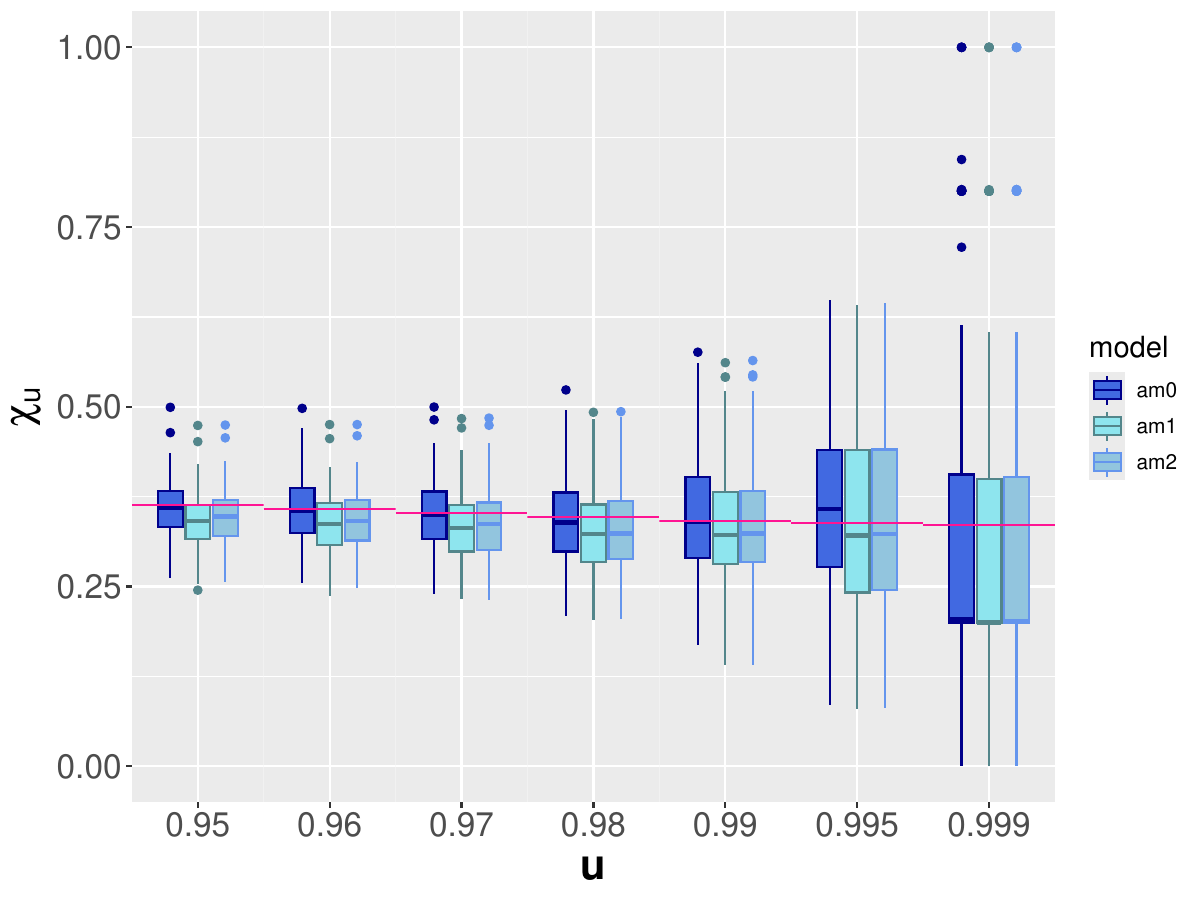}
\caption{$X(\bm{s})$ simulated from an BR process with $\bm{\theta}_1$ and $d=20$} 
\end{subfigure}

\begin{subfigure}[t]{1\textwidth}
\centering
\includegraphics[width=0.3\textwidth]{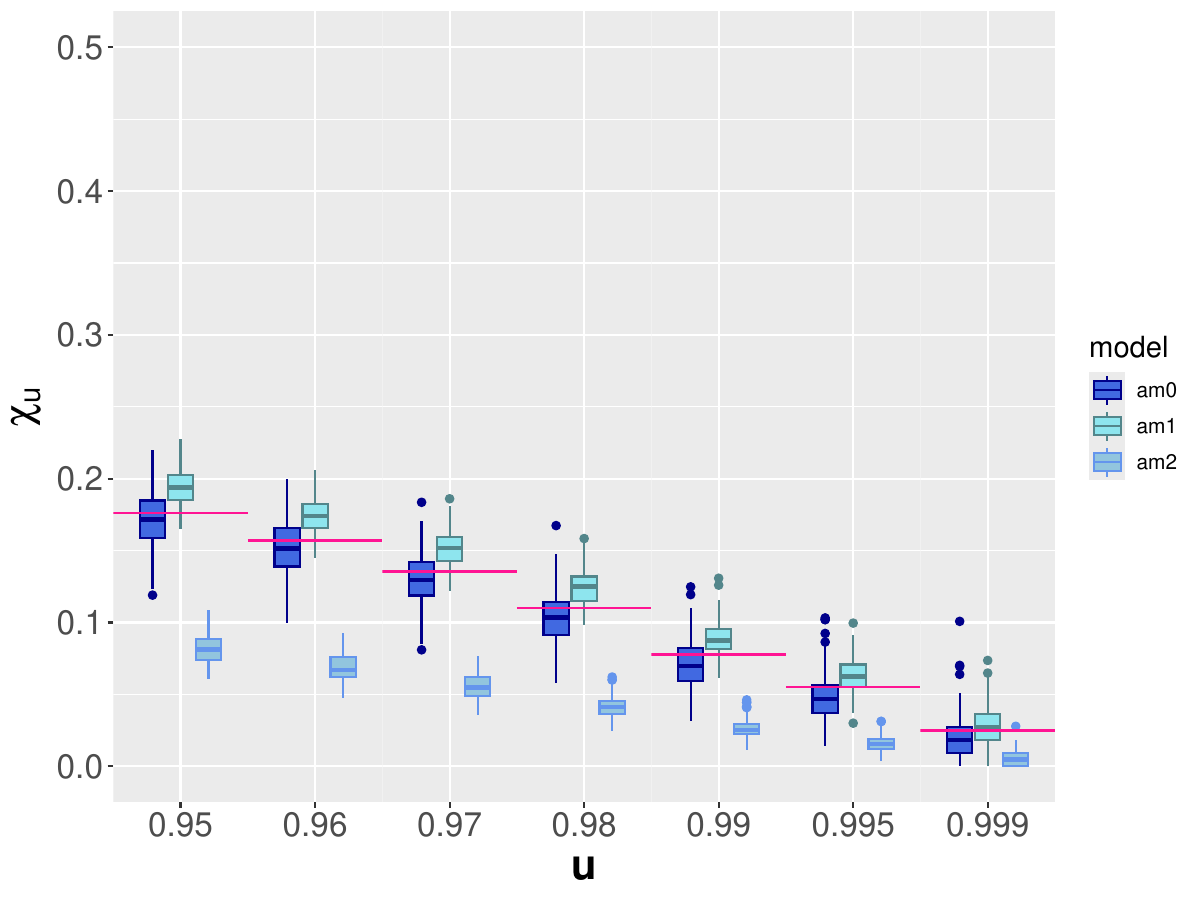}
  \hfill
  \includegraphics[width=0.3\textwidth]{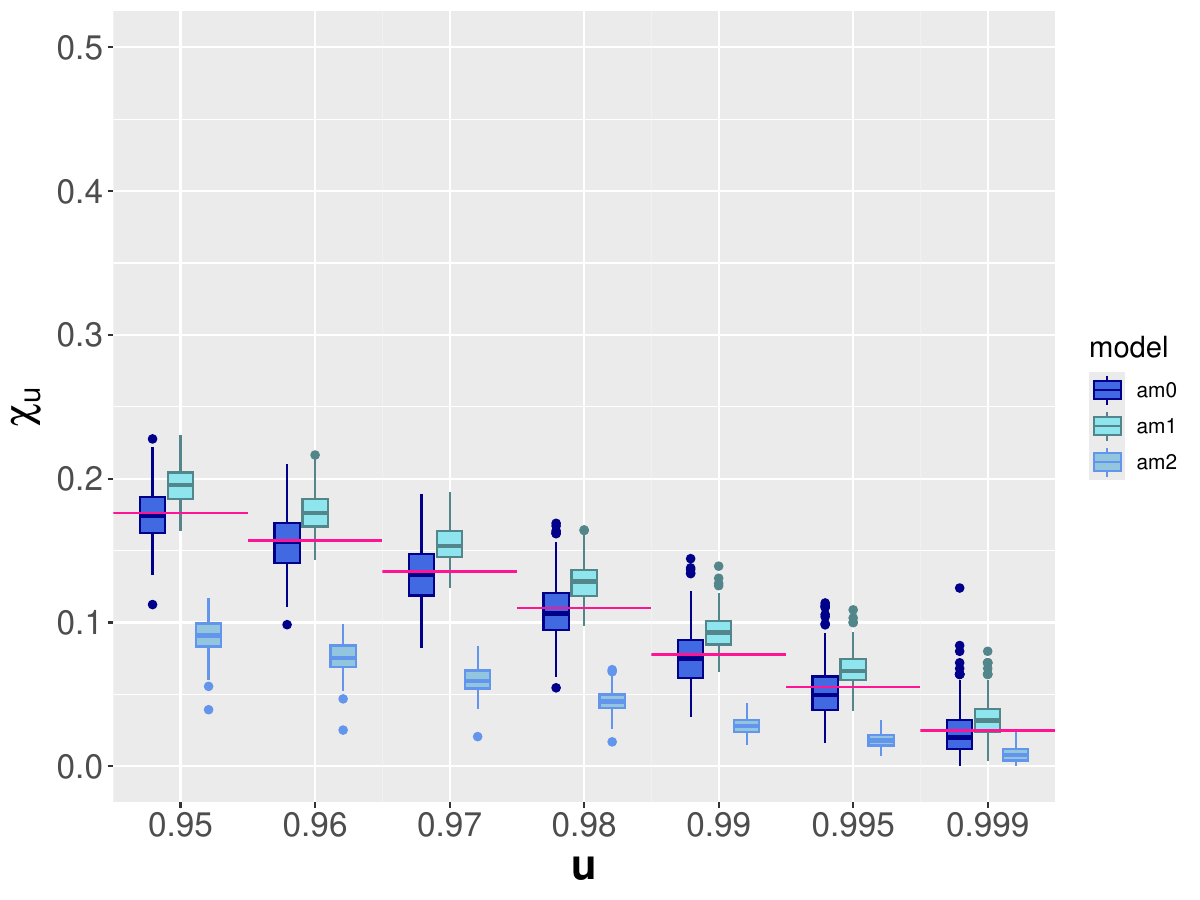}
  \hfill
  \includegraphics[width=0.3\textwidth]{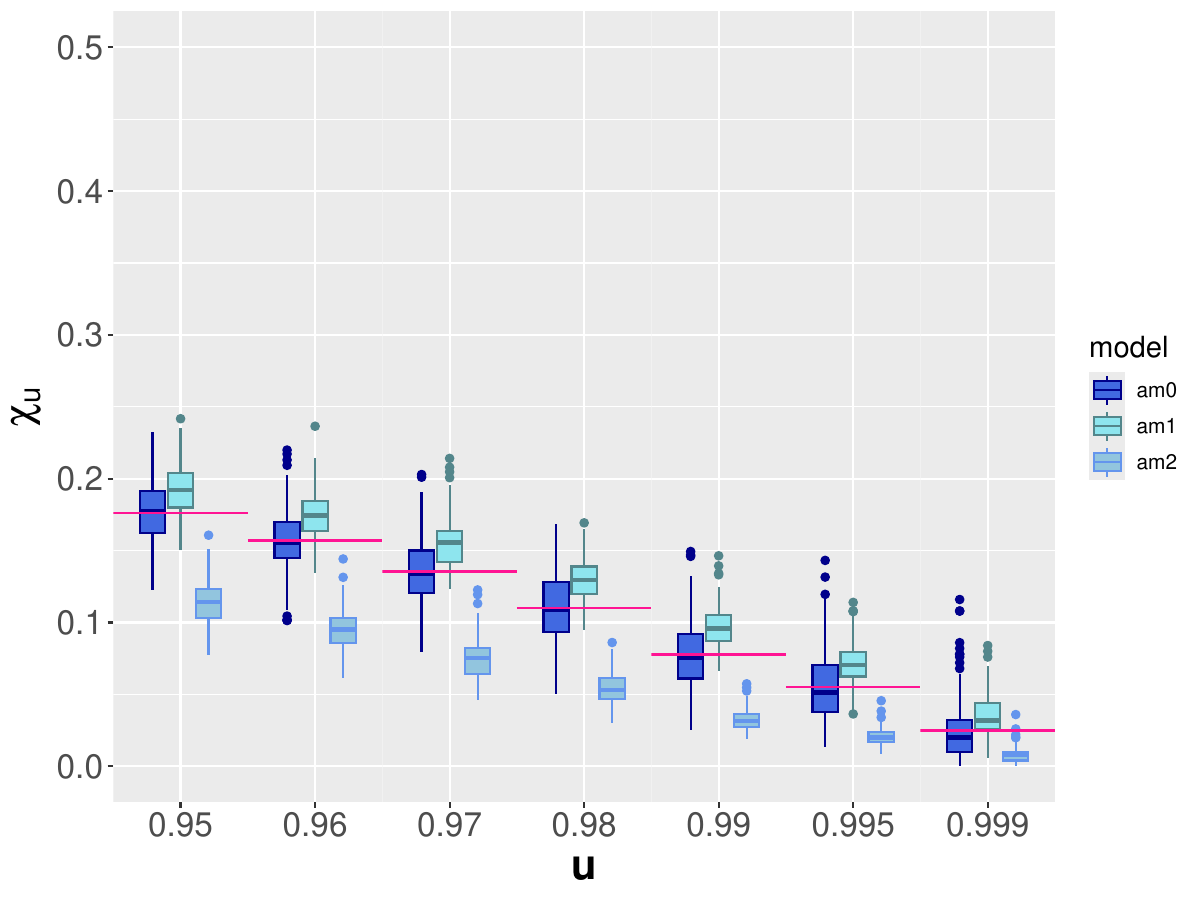}
\caption{$X(\bm{s})$ simulated from a Gaussian process with $\bm{\theta}_2$ and $d=20$} 
\end{subfigure}

 \caption{Boxplots of pairwise $\chi_u$ estimates for the pair $\left(X(\bm{s}_2), X(\bm{s}_3)\right)$ at distance $5.78$ apart, calculated for a range of $u$ values and using all $200$ simulated datasets. Estimates on the left correspond to those in Figures \ref{fig:box1} and \ref{fig:box2} of Section \ref{sec:simstudy} in the main document using $\tau=0.7$. Thresholds $\tau=0.8$ and $\tau=0.9$ are used to obtain the corresponding plots in the middle and right columns, respectively.}
 \label{fig:app_AMthresh}
\end{figure}

\setcounter{equation}{0}
\renewcommand\theequation{S5.1.\arabic{equation}}
\setcounter{figure}{0}       
\renewcommand\thefigure{S5.1.\arabic{figure}} 
\setcounter{table}{0}
\renewcommand\thetable{S5.1.\arabic{table}}

\begin{table}[H]
\centering
\begin{tabular}{cccc}
\toprule
IAGA code & Location & \begin{tabular}[c]{@{}c@{}}Geodetic \\ latitude ($\degree$N)\end{tabular} & \begin{tabular}[c]{@{}c@{}}Geodetic \\ longitude ($\degree$E)\end{tabular} \\ \midrule
BJN & Bj\o rn\o ya, Svalbard, Norway & 74.50 & 19.20 \\
BRW & Utqiagvik, Alaska, USA & 71.30 & -156.75 \\
CHD & Chokurdakh, Russia & 70.62 & 147.89 \\
DMH & Danmarkshavn, Greenland & 76.77 & -18.63 \\
HRN & Hornsund, Svalbard, Norway & 77.00 & 15.60 \\
KUV & Kullorsuaq, Greenland & 74.57 & -57.18 \\
LYR & Longyearbyen, Svalbard, Norway & 78.20 & 15.83 \\
NAL & Ny \AA lesund, Svalbard, Norway & 78.92 & 11.95 \\
RES & Resolute Bay, Canada & 74.69 & -94.89 \\
SCO & Ittoqqortoormiit, Greenland & 70.48 & -21.97 \\
SOR & S\o r\o ya, Norway & 70.54 & 22.22 \\
SVS & Savissivik, Greenland & 76.02 & -65.10 \\
THL & Qaanaaq, Greenland & 77.47 & -69.23 \\
TIK & Tixie, Russia & 71.58 & 129.00 \\
UMQ & Uummannaq, Greenland & 70.68 & -52.13 \\
UPN & Upernavik, Greenland & 72.78 & -56.15 \\ \bottomrule
\end{tabular}
\caption{Location information on the $16$ geomagnetometer stations considered in Section \ref{sec:application}. IAGA is an abbreviation for International Association of Geomagnetism and Aeronomy.}
\label{tab:apl_loc_info}
\end{table}

\begin{figure}[H]
    \centering
    \includegraphics[width=0.75\linewidth]{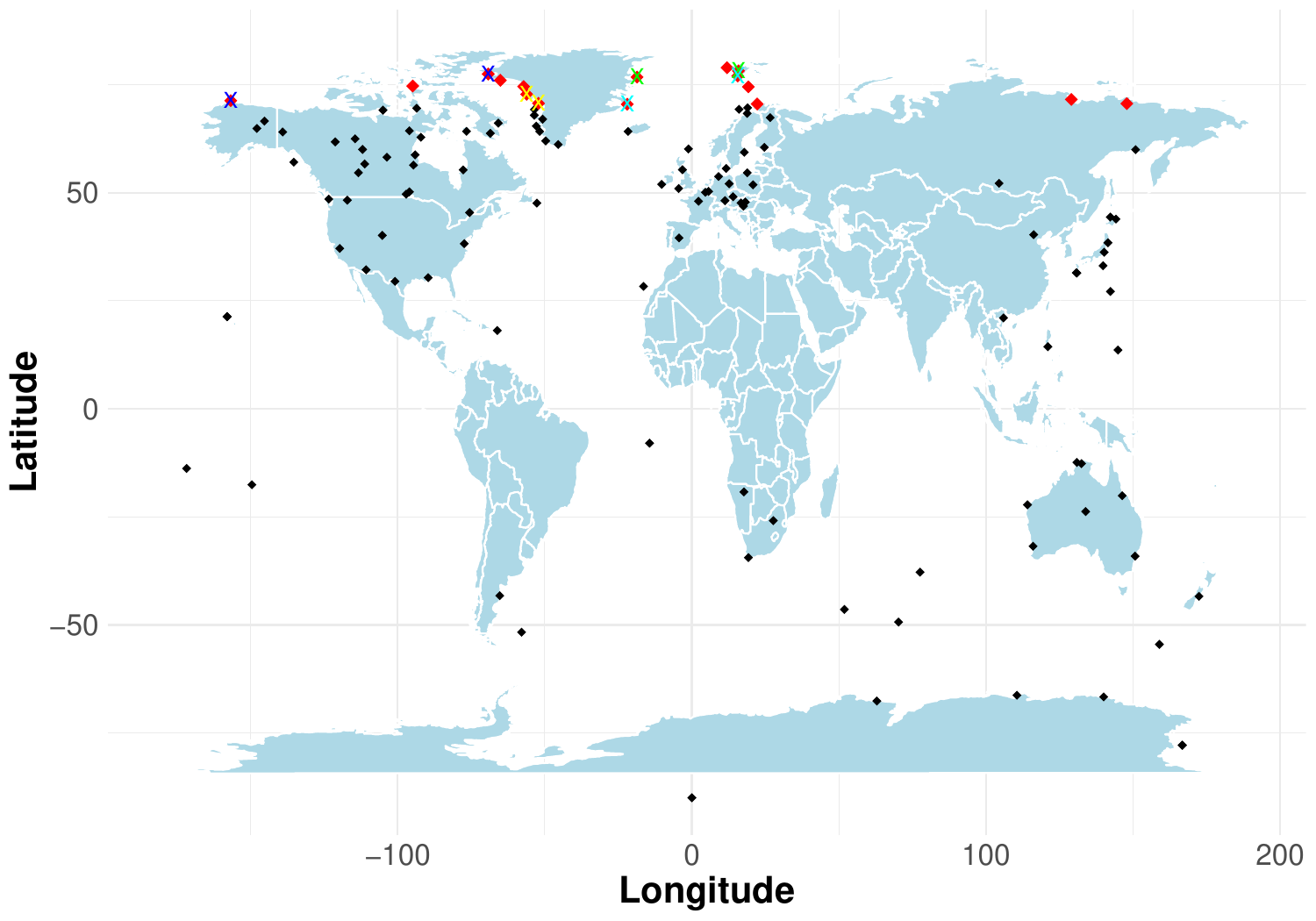}
    \caption{Spatial locations of space weather data. Points in red correspond to latitudes above $70\degree$. Points in black represent the remaining (unused) locations.}
    \label{fig:geomap}
\end{figure}

\setcounter{equation}{0}
\renewcommand\theequation{S5.2.\arabic{equation}}
\setcounter{figure}{0}       
\renewcommand\thefigure{S5.2.\arabic{figure}} 
\setcounter{table}{0}
\renewcommand\thetable{S5.2.\arabic{table}}

\begin{figure}[H]
    \centering
        \includegraphics[width=0.38\linewidth]{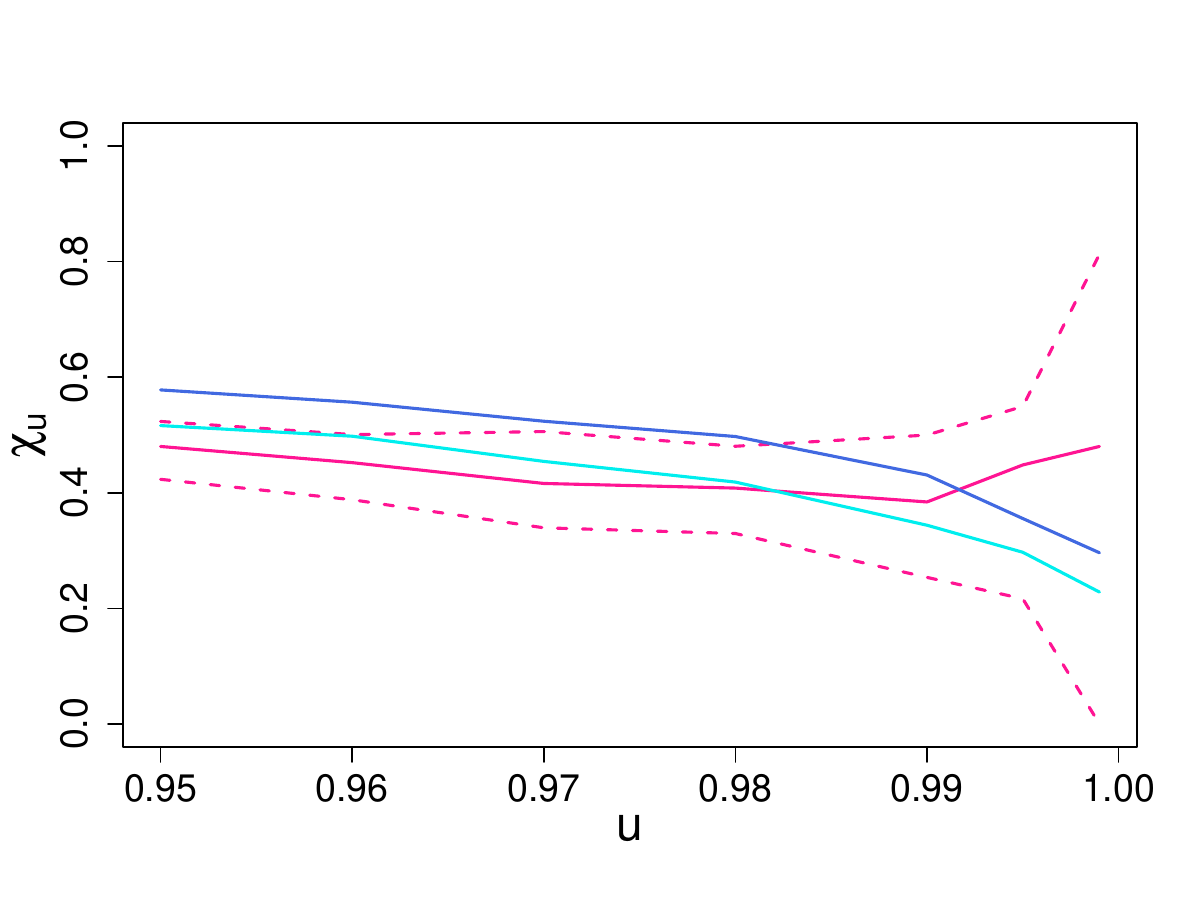}             
        \quad \includegraphics[width=0.38\linewidth]{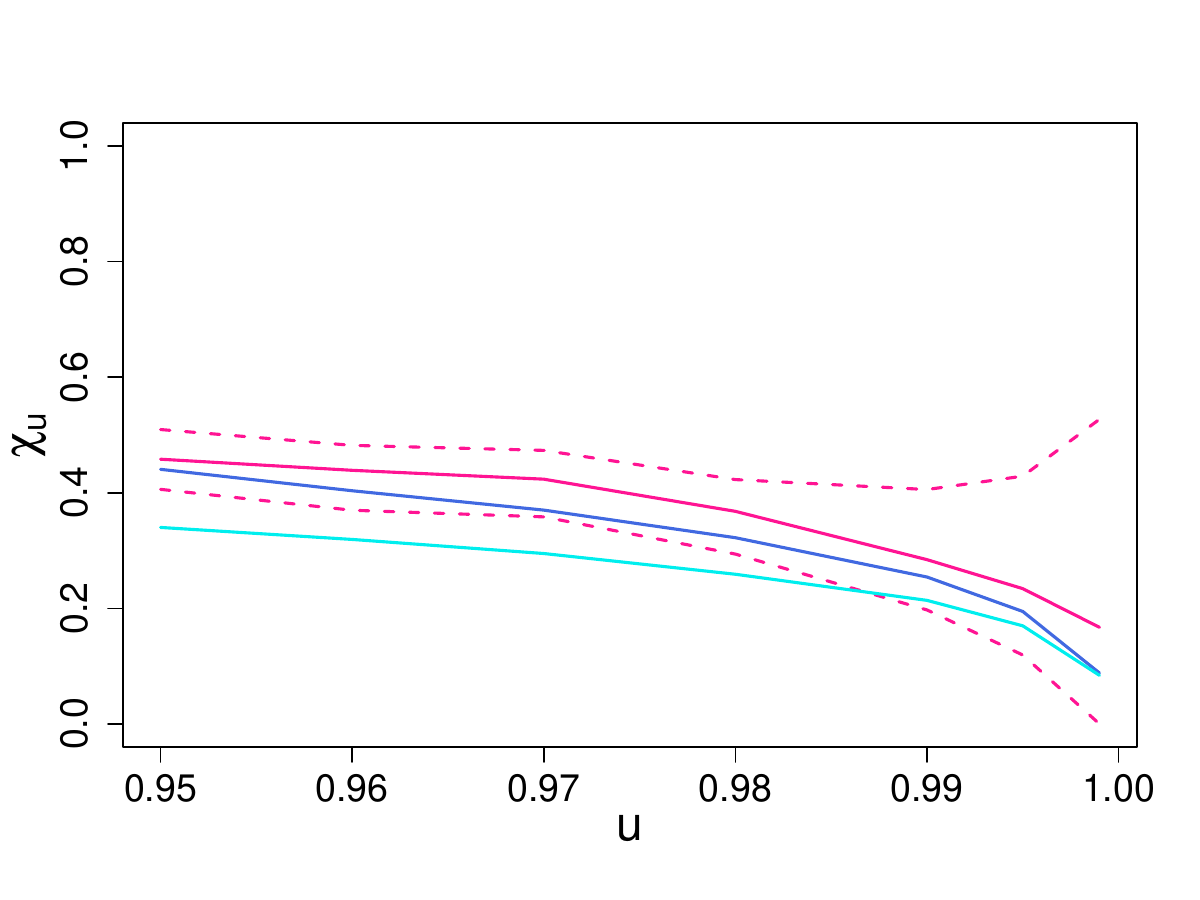}           
        \quad \includegraphics[width=0.38\linewidth]{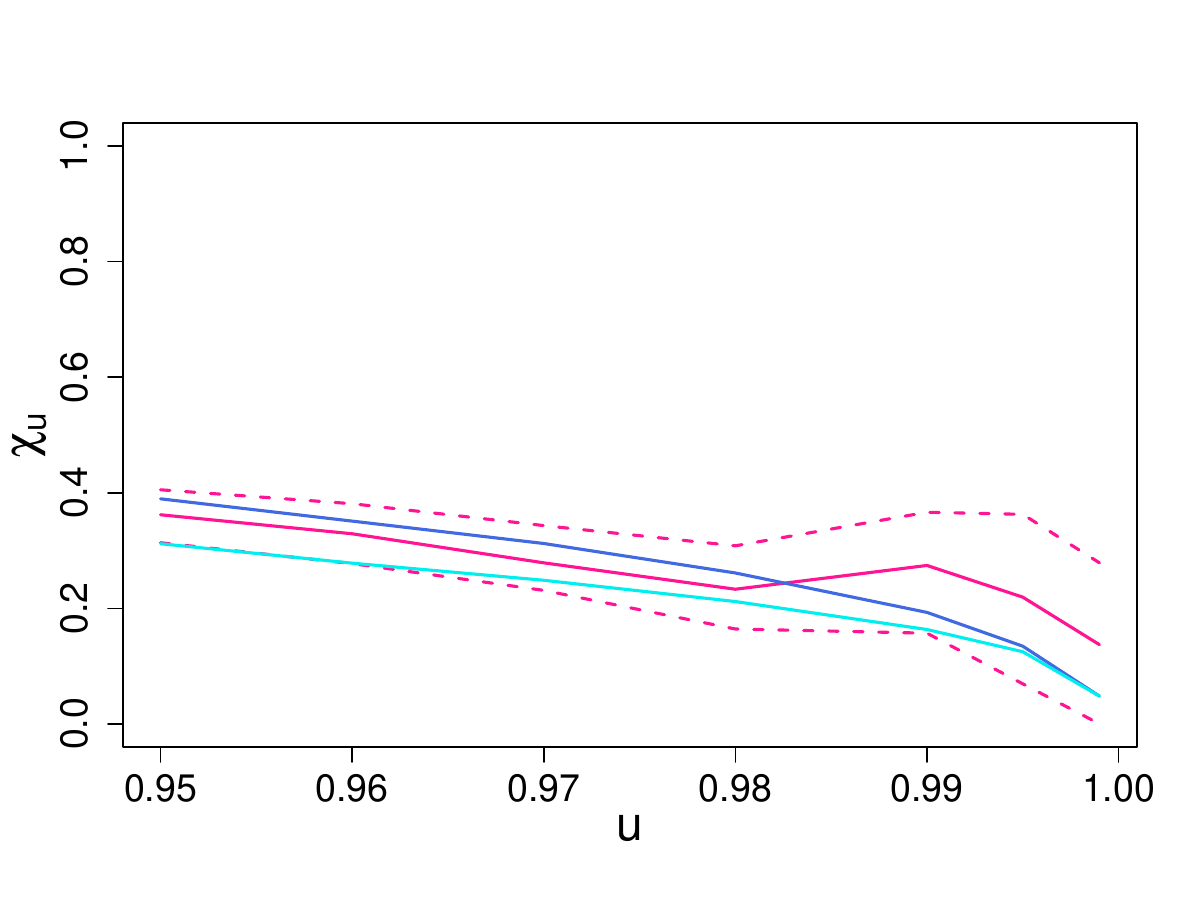}            
        \quad        \includegraphics[width=0.38\linewidth]{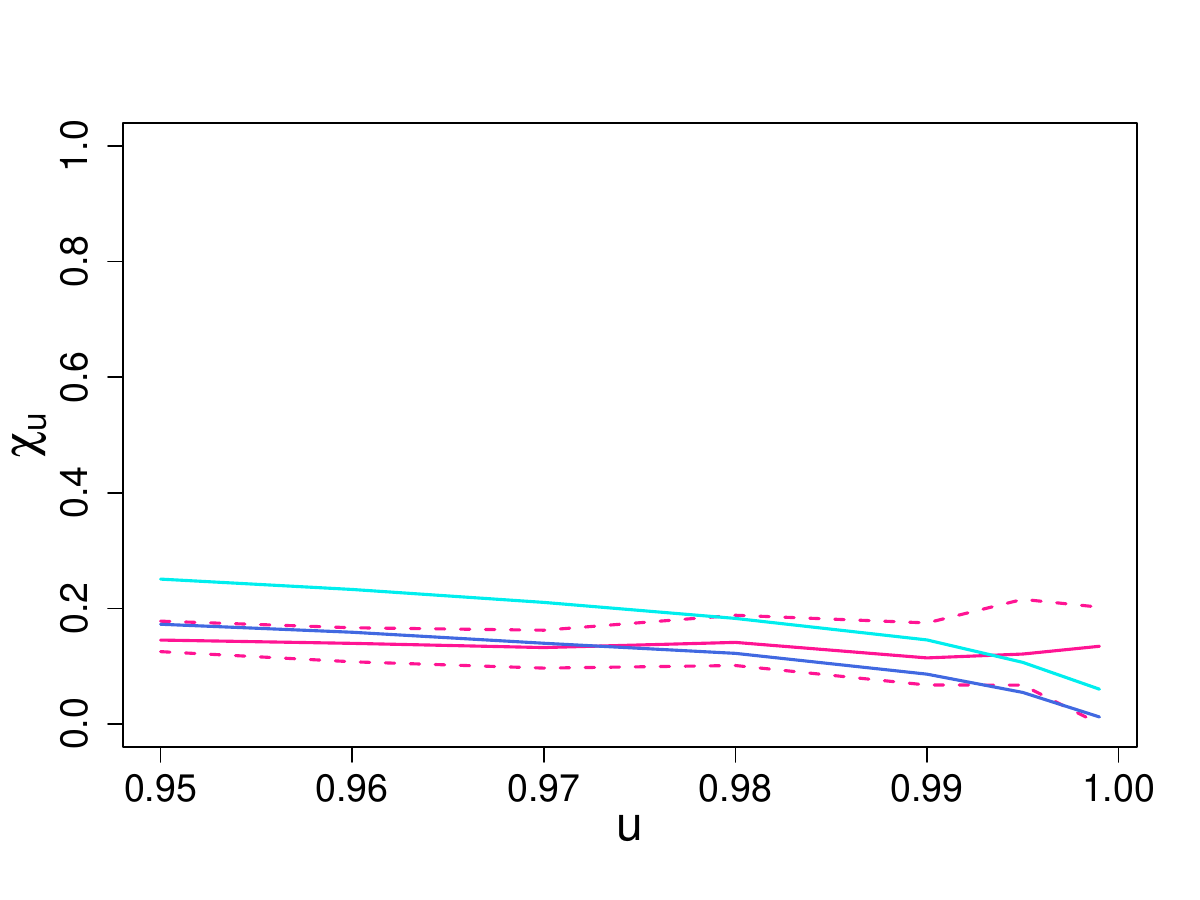}  
\caption{Pairwise $\chi_u(\bm{s}_j,\bm{s}_k)$ estimates over $u$ for four different $(j,k)$ pairs, namely $(15,16)$ (top left), $(4,7)$ (top right), $(5,10)$ (bottom left), $(2,13)$ (bottom right) at increasing distances apart, indicated as coloured (yellow, green, cyan and blue, respectively) crosses on the map of Figure \ref{fig:geomap}. Pink lines correspond to empirical estimates, dark blue to estimates with resampled angles and turquoise to estimates from the angular model in \eqref{eq:am1}. Empirical $95\%$ confidence bounds are also displayed for the empirical estimates, in the dashed pink lines. These were obtained via block bootstrap, using a block length of $30$ days.}
\label{fig:chiuappl}
\end{figure}
%
\setcounter{equation}{0}
\renewcommand\theequation{S5.3.\arabic{equation}}
\setcounter{figure}{0}   
\renewcommand\thefigure{S5.3.\arabic{figure}} 
\setcounter{table}{0}
\renewcommand\thetable{S5.3.\arabic{table}}
\begin{figure}[H]
    \centering
    \includegraphics[width=0.5\linewidth]{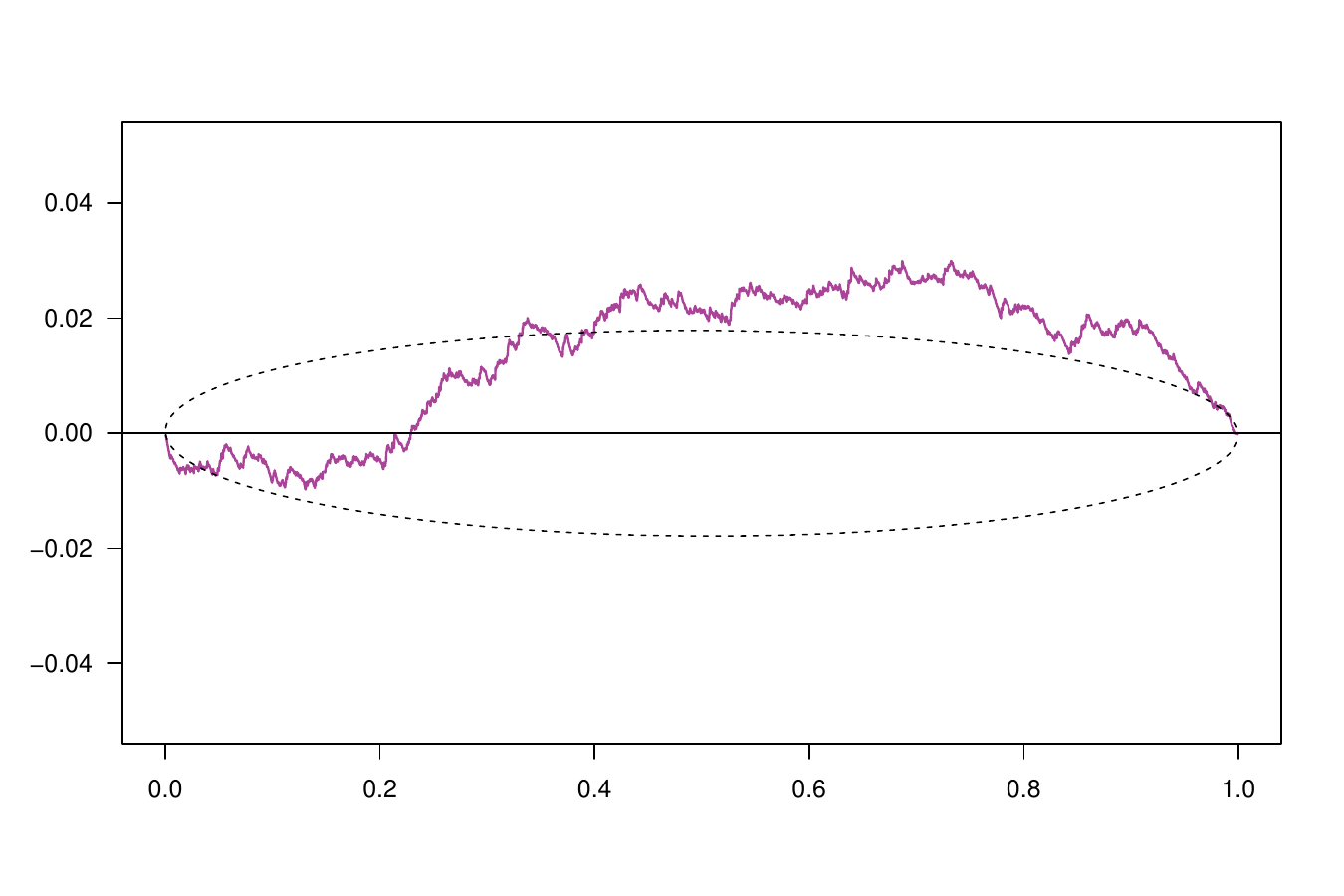} 
    \caption{Rescaled P-P plot for the space weather dataset discussed in Section \ref{sec:application}.}
    \label{fig:app_rescaledPP}
\end{figure} 
%
\setcounter{equation}{0}
\renewcommand\theequation{S6.\arabic{equation}}
\setcounter{figure}{0}    
\renewcommand\thefigure{S6.\arabic{figure}} 
\setcounter{table}{0}
\renewcommand\thetable{S6.\arabic{table}}

\vspace{0.5cm}
\begin{table}[H]
\centering
\begin{tabular}{ccccccccc}
 &  & \multicolumn{5}{c}{Full radial model} &  &  \\ \cline{3-7}
 & CL & G & L & GG & $\text{HW}_{\text{G}}$ & $\text{HW}_{\text{GG}}$ & AM1 & AM2 \\ \cline{2-9} 
\multicolumn{1}{c|}{$d=5$} & 0.138 & 0.109 & 0.134 & 0.202 & 4.05 & 6.36 & 14.4 & 0.0349 \\
\multicolumn{1}{c|}{$d=20$} & 2.15 & 0.528 & 0.538 & 0.712 & 8.99 & 11.9 & 4.15 & 0.0679
\end{tabular}
\caption{Computational time of the composite likelihood model (CL), the full likelihood radial model with different underlying gauges, the process-based angular model (AM1) and the gauge-based angular model (AM2). Times are given in minutes (in $3$ s.f.) and are obtained as averages over ten fits run on a standard laptop. The composite likelihood model consists of $d\choose 2$ fits to approximately $1500$ pairwise threshold exceedances, while all full radial and angular models are fitted to exactly $1500$ threshold exceedances, out of a total $5000$ data points originally simulated.}
\label{tab:comptime}
\end{table}

\end{document}